\documentclass[12pt,oneside,english,reqno]{amsbook}
\usepackage[T1]{fontenc}
\usepackage[latin9]{inputenc}
\usepackage{geometry}
\usepackage{xcolor}
\usepackage{babel}
\usepackage{float}
\usepackage{booktabs}
\usepackage{mathrsfs}
\usepackage{url}
\usepackage{pdfpages}
\usepackage{multirow}
\usepackage{amsbsy}
\usepackage{amstext}
\usepackage{amsthm}
\usepackage{amssymb}
\usepackage{graphicx}
\usepackage{wasysym}
\usepackage{esint}
\usepackage[unicode=true,pdfusetitle,
 bookmarks=true,bookmarksnumbered=false,bookmarksopen=false,
 breaklinks=false,pdfborder={0 0 1},backref=false,colorlinks=false]
 {hyperref}

\makeatletter

\providecommand{\tabularnewline}{\\}

\numberwithin{section}{chapter}
\numberwithin{equation}{section}
\numberwithin{figure}{section}

\usepackage{slashed}
\usepackage{cite}

\newcommand{\TeV}{\textrm{ TeV}}

\makeatother

\begin{document}
\includepdf{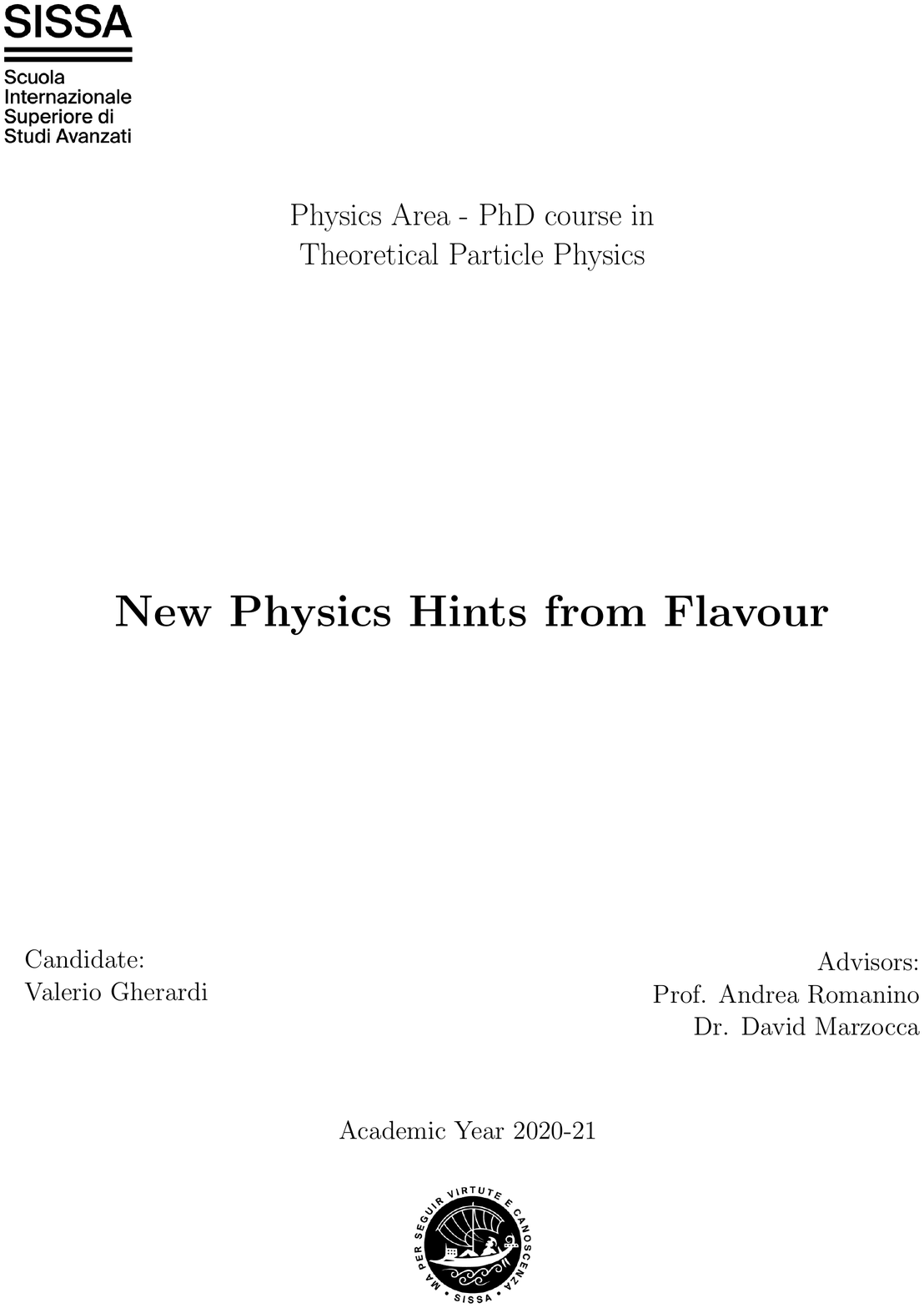}

\newpage{}
\title{New Physics Hints from Flavour}
\begin{abstract}
This Thesis presents my personal contributions to two distinct fields,
namely the recent experimental anomalies in $B$-meson decays, and
the longstanding quest for a theoretical explanation of lepton masses
and mixings, under a unifying umbrella: Flavour Physics. The modern
view of the Standard Model as an effective theory is motivated, and
the dual role of flavour, as a probe of New Physics effects on one
side, and as a piece of the puzzle itself on the other, is emphasized.
By jointly discussing two physics cases of different nature, I attempt
to offer a broader perspective on the current status of Beyond Standard
Model searches.

The discussion begins with a preliminary review of Standard Model
Effective Field Theories, in which some utility results in theory
matching are presented. The focus then moves to $B$-meson decay anomalies:
after introducing Lepton Flavour Universality, we discuss the phenomenology
of a specific Standard Model extension by scalar leptoquarks, and
also present the Effective Field Theory framework of Rank-One Flavor
Violation, providing a less detailed but broader picture for the case
of neutral-current anomalies. Finally, we tackle the Standard Model
flavour puzzle, describing an attempt to address charged lepton mass
hierarchies in the framework of supersymmetric modular invariant models
of lepton flavour.
\end{abstract}

\maketitle
\tableofcontents{}

\chapter*{Introduction}

The Standard Model (SM) of Particle Physics \cite{Glashow:1961tr,Weinberg:1967tq,Salam:1968rm}
is arguably one of the greatest intellectual achievements of modern
science. It synthesizes a whole century of theoretical breakthroughs,
from quantum mechanics and Lorentz invariance to gauge invariance
and spontaneous symmetry breaking through renormalizable field theory,
which were paralleled by a remarkable history of experimental discoveries
that climaxed in 2012, with the detection of the scalar boson \cite{ATLAS:2012yve,CMS:2012qbp}
conjectured about fifty years before by P. Higgs \cite{Higgs:1964pj},
R. Brout and F. Englert \cite{Englert:1964et}, and G. Guralnik, C.
R. Hagen and T. Kibble \cite{Guralnik:1964eu}. The discovery of the
Higgs boson, which was the last angular stone of the theory waiting
for experimental validation, completes a beautiful picture of the
sub-atomic world, which is by far the best available, and to which
the physicist cannot help but look with amazement.

If, on the one hand, the impressive agreement of SM predictions with
experimental data \cite{Zyla:2020zbs} represents a tremendous scientific
success, the same precise feature might also be perceived by some
as a source of uneasiness \cite{Romanino:2016fnz}. In fact, along
with its lengthy record of triumphs, time has brought to surface a
great many pieces of evidence indicating that the SM is, from a strict
point of view, just a phenomenological effective theory - the most
incontrovertible of which is perhaps the existence of unexplained
Dark Matter (see \emph{e.g.} \cite{Gross:2018ivp}). From a Naturalness
\cite{tHooft:1979rat,Giudice:2008bi} standpoint, the fact that the
Higgs discovery would take place in a total absence of signals of
physics Beyond the SM (BSM) was somewhat disappointing, for it opened
the doors to an alternative, rather depressing but logically conceivable
interpretation: that BSM phenomena might be energetically inaccessible
by particle accelerators available at present or foreseeable in a
next future.

While there is no substantial reason suggesting this worst-case scenario
to be the actual case, it is fair to say that direct indications from
data pointing to a concrete BSM threshold energy are still very limited.
In this groping one's way to New Physics (NP), \emph{flavour} plays
the dual role of a glow in the dark on one side, and a long-standing
sphynx on the other. In this Thesis, I discuss both these aspects
through the lens of the topics I have been primarily involved in during
the last three years: the $B$-meson decay anomalies, and the origins
of lepton flavour.

The term ``flavour'' has a long history in Particle Physics, and
its meaning has undergone a certain distortion across the years. Here
and in what follows, the expression refers to all those features of
the SM fermionic matter content (\emph{i.e. }quarks and leptons) which
are not univocally fixed by gauge invariance or other structural requirements:
these include the quark and lepton representations under the SM gauge
group, the number of families, and the mass and mixing patterns of
the two sectors. The dual nature of flavour mentioned above can now
be clarified: the SM flavour structure represents a powerful probe
of NP effects and,\emph{ }at the same time, cries for a theoretical
explanation regarding its origins. The first aspect is what gives
rise to the so-called SM flavour problem, whereas the second one is
what we refer to the SM flavour puzzle \cite{Feruglio:2015jfa}.

Let us briefly introduce the two concrete physics cases addressed
in the present work. The $B$-meson decay anomalies ($B$-anomalies
for short) are a set of discrepancies between SM predictions and experimental
data on semileptonic decays of $B$-mesons, which have by now persisted
altogether for more than seven years. The deviations were initially
reported for $\overline{B}\to D^{(*)}\ell^{-}\nu_{\ell}$ decay \cite{Lees:2012xj,Lees:2013uzd,Aaij:2015yra,Huschle:2015rga,Sato:2016svk,Hirose:2016wfn,Hirose:2017dxl,Aaij:2017uff,Aaij:2017deq,Siddi:2018avt,Belle:2019rba},
which according to the SM are mediated by a tree-level $W^{\pm}$
exchange, and subsequently followed by comparable deviations in $B\to K^{(*)}\ell^{+}\ell^{-}$
decays \cite{Aaij:2013qta,Aaij:2014ora,Aaij:2015oid,Aaij:2015esa,Aaij:2017vad,Aaij:2017vbb,Aaboud:2018mst,Aaij:2019wad,Abdesselam:2019wac,Aaij:2020nrf,LHCb:2021trn},
which is instead a one-loop suppressed flavour-changing neutral-current
process in the SM. The current measurements continue to exhibit a
tension with the SM, with a combined statistical significance of $\approx3$
standard deviations for the data on charged-current decays, wheareas
the global statistical significance of $b\to s\ell^{+}\ell^{-}$ anomalies
has been recently estimated to be 3.9$\sigma$ \cite{Lancierini:2021sdf}. 

Although still in need of experimental confirmation (which, if ever,
might require several years \cite{Bifani:2018zmi}), these experimental
results definitely fuel the hope that NP might be discovered close
to the $\text{TeV}$ scale, in partial rescue of the Naturalness argument.
In fact, the masses of candidate NP mediators for the anomalies can
be roughly estimated from the size of the experimental deviations,
and turn out to be generically in the $1\div100\,\text{TeV}$ range
\cite{DiLuzio:2017chi}, although values could go down to $\mathcal{O}(10)\,\text{GeV}$
in $Z^{\prime}$ models \cite{Nomura:2020vnk}, or $\mathcal{O}(100\,\text{GeV})$
in one-loop models. Moreover, and quite in alignment with the main
topic of the present Thesis, the $B$-anomalies challenge in a notably
coherent way a very special feature of the SM flavour structure: Lepton
Flavour Universality (LFU) \cite{Bifani:2018zmi}, which we will describe
in full detail along this work. Indeed, the most important observables
involved in $B$-anomalies are the so-called LFU ratios, for which
SM predictions are especially clean thanks to LFU.

Precisely concerning lepton flavour structure, from a phenomenological
point of view, this is defined by lepton (charged lepton and neutrino)
masses and mixing parameters. Charged lepton masses are known with
very high precision \cite{Zyla:2020zbs} and, coherently with their
quark counterparts, exhibit magnitude hierarchies between the three
SM replicæ (electron, muon and tau), spanning three orders of magnitude
in the $\text{MeV}\div\text{GeV}$ range. Neutrino oscillation experiments
have measured two independent neutrino squared mass differences and
three mixing angles with an accuracy approaching the percent-level,
and a (Dirac) CP violating phase with $\mathcal{O}(10\%)$ accuracy
\cite{Esteban:2020cvm}. Squared mass differences lie in the $10^{-3}\div10^{-5}\,\text{eV}^{2}$
range, and also exhibit a (mild) relative hierarchy; two out of the
three mixing angles, $\theta_{12}$ and $\theta_{23}$, are of order
unity, while $\theta_{13}$ is comparable in size to the CKM Cabibbo
angle; the Dirac phase $\delta$ is also $\mathcal{O}(1)$. Furthermore,
data from both cosmology \cite{Zyla:2020zbs} and nuclear physics
\cite{KATRIN:2019yun} bounds the absolute scale of neutrino masses,
roughly below the $\text{eV}$ scale. Assuming a framework with three
light active neutrinos, which is coherent with current experimental
data, remaining unknowns are the relative mass ordering (normal or
inverted), the two Majorana CP violating phases of the PMNS matrix
and, of course, the absolute values of neutrino masses.

Lepton flavour is, of course, a BSM issue \emph{per se}, as the mechanism
behind the generation of neutrino masses is not part of the SM (at
least according to its usual definition). From the point of view of
the flavour puzzle, the features which most distinctly call for a
theoretical explanation are the hierarchies in the charged lepton
mass spectrum, the smallness of neutrino masses and the origin of
the PMNS mixing patterns (in the mininal three-neutrino framework).
Addressing all these points in a natural and predictive framework,
and which could hopefully be extended to also describe the quark sector,
represents a challenging open problem. In this Thesis, I describe
some results in this direction, from the recently proposed approach
to the lepton flavour puzzle based on modular invariant supersymmetric
models \cite{Feruglio:2017spp}.

This Thesis is composed of three parts. \emph{Part I}, which has a
somewhat preliminary function with respect to the remaining material,
centers on the so-called SM Effective Field Theory: a non-renormalizable
theory whose renormalizable limit coincides with the familiar SM.
After briefly reviewing the basic formalism and the systematics of
theory matching, I present a personal contribution \cite{Gherardi:2020det},
in which the complete one-loop matching of a phenomenologically motivated
SM extension is performed. In \emph{Part II}, I discuss $B$-anomalies
and my work on the subject \cite{Gherardi:2019zil,Gherardi:2020qhc};
the results of Ref. \cite{Gherardi:2020qhc} leverage on the matching
performed in \cite{Gherardi:2020det}, described in the previous Part.
Finally, \emph{Part III}, whose flavour is more theoretical\footnote{No pun intended.},
tackles the SM flavour puzzle and describes an attempt to address
charged lepton mass hierarchies within the framework of modular invariant
models \cite{Feruglio:2021dte}.

\part{The Standard Model Effective Field Theory}

\chapter{The Standard Model Effective Field Theory}

Effective Field Theory and Renormalization play an essential role
in our current understanding and speculations about Particle Physics.
By taking a bottom-up approach, the Standard Model itself can be fruitfully
characterized within this framework, in which the guiding theoretical
principles of Symmetry and Symmetry Breaking are naturally implemented.
On the other hand, from a top-down perspective, Effective Field Theories
provide a powerful computational tool for studying Standard Model
high-energy extensions.

This Chapter provides an introduction to the Standard Model Effective
Field Theory and sets up the notation to be employed in the subsequent
Chapters of this thesis. Section \ref{sec:The SMEFT lagrangian} introduces
the notations used throughout this work and present the Effective
Field Theory formulation of the Standard Model; Section \ref{sec:Matching}
provides a brief review of the general theory matching procedure;
finally, Section \ref{sec:Advanced matching} discusses some advanced
matching methods.

\section{\label{sec:The SMEFT lagrangian}The Standard Model as an Effective
Field Theory}

The Standard Model (SM) can be succinctly described as the most general
renormalizable theory of quarks, leptons and the Higgs field, invariant
under the electroweak gauge group:

\begin{equation}
G_{\text{SM}}=\text{SU}(3)_{c}\times\text{SU}(2)_{L}\times\text{U}(1)_{Y}.\label{eq:Electroweak group}
\end{equation}
Renormalizability implies scale-independent self-consistency, in the
sense that, from a purely theoretical point of view, the validity
of the SM can be extended up to arbitrarily high energy scales\footnote{This is admittedly an oversimplification, for our discussion ignores
both the problems of vacuum stability \cite{Degrassi:2012ry} and
of potential Landau poles \cite{Gell-Mann:1954yli,Gockeler:1997dn}
in the renormalization group flow of SM couplings. While, from a purely
theoretical point of view, both these issues should be considered
seriously, the involved energy scales are usually several order of
magnitudes higher than effective cut-offs coming from the explicit
introduction of a non-renormalizable interaction in the theory.}. On the other hand, as discussed in the Introduction, there are irrefutable
reasons to believe that the SM is not the ultimate theory of Nature,
which in turn makes renormalizability a dispensable, if not unmotivated,
feature for a realistic theory. Dropping the renormalizability requirement
leads from the ultraviolet complete SM theory to an Effective Field
Theory (EFT) known as Standard Model Effective Field Theory (SMEFT)
\cite{Buchmuller:1985jz,Grzadkowski:2010es}. 

Any quantum field theory with the same light degrees of freedom as
the SM, plus some extra heavy degree of freedom, is correctly described
by SMEFT at sufficiently low energies (that is, below the EFT cut-off,
which is usually of order of the lightest new particle mass). Altough
this case does not cover all conceivable generalizations of the SM\footnote{For instance, theories which extend the SM with light, feebly coupled
degrees of freedom, such as axions or light singlet neutrinos, cannot
be directly described by SMEFT.}, it definitely includes a large majority of phenomenological theories
which attempt to address one or more SM shortcomings.

\renewcommand{\arraystretch}{1.3}
\begin{table}[t]
\begin{centering}
\begin{tabular}{|c|c|c|c|c|}
\hline 
$\text{Field}$ & Lorentz & $\text{SU}(3)_{c}$ & $\text{SU}(2)_{L}$ & $\text{U}(1)_{Y}$\tabularnewline
\hline 
\hline 
$G^{\mu}$ & $\left(\frac{1}{2},\frac{1}{2}\right)$ & $8$ & $1$ & $0$\tabularnewline
\hline 
$W^{\mu}$ & $\left(\frac{1}{2},\frac{1}{2}\right)$ & $1$ & $3$ & $0$\tabularnewline
\hline 
$B^{\mu}$ & $\left(\frac{1}{2},\frac{1}{2}\right)$ & $1$ & $1$ & $0$\tabularnewline
\hline 
$q_{i}\equiv\begin{pmatrix}u_{i} & d_{i}\end{pmatrix}^{T}$ & $\left(0,\frac{1}{2}\right)$ & $3$ & $2$ & $+\frac{1}{6}$\tabularnewline
\hline 
$u_{i}$ & $\left(\frac{1}{2},0\right)$ & $3$ & $1$ & $+\frac{2}{3}$\tabularnewline
\hline 
$d_{i}$ & $\left(\frac{1}{2},0\right)$ & $3$ & $1$ & $-\frac{1}{3}$\tabularnewline
\hline 
$\ell_{\alpha}\equiv\begin{pmatrix}\nu_{\alpha} & e_{\alpha}\end{pmatrix}^{T}$ & $\left(0,\frac{1}{2}\right)$ & $1$ & $2$ & $-\frac{1}{2}$\tabularnewline
\hline 
$e_{\alpha}$ & $\left(\frac{1}{2},0\right)$ & $1$ & $1$ & $+1$\tabularnewline
\hline 
$H$ & $\left(0,0\right)$ & $1$ & $2$ & $+\frac{1}{2}$\tabularnewline
\hline 
\end{tabular}
\par\end{centering}
\caption{\label{tab:SM fields}Standard Model fields. All fermionic fields
are four-component Dirac-Weyl fields and family indexes $i$ or $\alpha$
run from $1$ to $3$}
\end{table}
\renewcommand{\arraystretch}{1.0}

In what follows we review the standard SMEFT construction \cite{Buchmuller:1985jz,Grzadkowski:2010es}
and set our SMEFT notations. The quantum numbers of SM fields under
$G_{\text{SM}}$ are collected in Table \ref{tab:SM fields}, together
with their Lorentz representations. From Table \ref{tab:SM fields},
we can immediately construct the SM lagrangian density (or, simply,
``lagrangian''):

\begin{align}
\mathcal{L}_{\text{SM}} & =\mathcal{L}_{\text{SM}}^{\text{gauge}}+\mathcal{L}_{\text{SM}}^{\text{yuk}}-\mathcal{V}_{\text{SM}}\label{eq:L_SM decomposition}\\
\mathcal{L}_{\text{SM}}^{\text{gauge}} & =-\frac{1}{4}G_{\mu\nu}^{A}G^{A\mu\nu}-\frac{1}{4}W_{\mu\nu}^{I}W^{\mu\nu I}-\frac{1}{4}B_{\mu\nu}B^{\mu\nu}+\label{eq:L_SM gauge}\\
 & +\overline{q}i\slashed Dq+\overline{u}i\slashed Du+\overline{d}i\slashed Dd+\overline{\ell}i\slashed D\ell+\overline{e}i\slashed De+\nonumber \\
 & +(D_{\mu}H)^{\dagger}(D^{\mu}H)+\nonumber \\
 & +\frac{\theta g_{s}^{2}}{64\pi^{2}}\delta_{AB}\varepsilon^{\mu\nu\rho\sigma}G_{\mu\nu}^{A}G_{\rho\sigma}^{B},\nonumber \\
\mathcal{L}_{\text{SM}}^{\text{yuk}} & =-(y_{U})_{ij}\overline{q}_{i}\widetilde{H}u_{j}-(y_{D})_{ij}\overline{q}_{i}Hd_{j}-(y_{E})_{\alpha\beta}\overline{\ell}_{\alpha}He_{\beta}+\text{h.c.},\label{eq:L_SM yuk}\\
\mathcal{V}_{\text{SM}} & =\frac{1}{2}\lambda(H^{\dagger}H)^{2}-m^{2}H^{\dagger}H.\label{eq:V_SM}
\end{align}
In the equations above, $\widetilde{H}$ denotes the conjugate Higgs
field $\widetilde{H}=i\sigma_{2}H^{*}$, and $D_{\mu}$ is the gauge
covariant derivative, defined by:

\begin{equation}
D_{\mu}=\partial_{\mu}+ig_{s}G_{\mu}^{A}T_{A}^{(3)}+igW_{\mu}^{I}T_{I}^{(2)}+ig'B_{\mu}Y,\label{eq:Covariant derivatives}
\end{equation}
where the generators $T^{(3,2)}$ and $Y$ are given in the relevant
representations; we will also occasionally employ the notation $\phi_{1}\overleftrightarrow{D_{\mu}}\phi_{2}=\phi_{1}D_{\mu}\phi_{2}-(D_{\mu}\phi_{1})\phi_{2}$.
All constants appearing in Eqs. (\ref{eq:L_SM gauge})-(\ref{eq:V_SM})
have been experimentally measured \cite{Zyla:2020zbs} (except for
the QCD $\theta$ angle, for which only an upper bound exists \cite{Kim:2008hd}).

A key ingredient in postulating the SM lagrangian (\ref{eq:L_SM decomposition})
is renormalizability, which is simply implemented by the requirement
that all SM operators have mass dimension at most four (the space-time
dimension). Together with gauge invariance, it fully determines the
(finite) list of operators which can appear on the right-hand side
of Eq. (\ref{eq:L_SM decomposition}). On the other hand, if renormalizability
is not required, it does still make sense to consider Eq. (\ref{eq:L_SM decomposition})
as the leading order expansion of a non-renormalizable lagrangian,
the SMEFT lagrangian, whose higher dimensional operator coefficients
are unknown. Concretely, we take:

\begin{equation}
\mathscr{L}_{\text{SMEFT}}=\mathscr{L}_{\text{SM}}+\sum_{d_{\mathcal{O}}>4}\frac{c_{\mathcal{O}}}{\Lambda_{\mathcal{O}}^{d_{\mathcal{O}}-4}}\mathcal{O}.\label{eq:L_SMEFT}
\end{equation}
In the previous equation, the sum extends over all gauge-invariant
operators $\mathcal{O}$ with mass dimension $d_{\mathcal{O}}>4$
which can be built out of the SM fields; $\Lambda_{\mathcal{O}}$
is an energy scale specific to the operator $\mathcal{O}$ and $c_{\mathcal{O}}$
is a dimensionless coefficient of $\mathcal{O}((4\pi)^{-\ell})$,
where $\ell$ is the loop-order at which the operator is generated;
the combination $C_{\mathcal{O}}=c_{\mathcal{O}}\Lambda_{\mathcal{O}}{}^{4-d_{\mathcal{O}}}$
defines the so-called Wilson coefficient of the operator $\mathcal{O}$.
The effects of $\mathcal{O}$ at experimental energies $E$ are suppressed
by a factor of $C_{\mathcal{O}}E^{d_{\mathcal{O}}-4}=c_{\mathcal{O}}(E/\Lambda_{\mathcal{O}})^{d_{\mathcal{O}}-4}$,
which could explain why conclusive evidence of deviations from the
SM is still lacking. The non-renormalizable theory defined by (\ref{eq:L_SMEFT})
has an implicit energy cut-off $\Lambda_{\text{SMEFT}}$, which restricts
its validity to energies $E\leq\Lambda_{\text{SMEFT}}$, and is expected
to be of the same order of the smallest operator scales $\Lambda_{\mathcal{O}}$.

In practice, only operators with mass dimension up to a maximal value
are included in Eq. (\ref{eq:L_SMEFT}). Truncating the sum at dimension
five, the only extra non-renormalizable operator is the well known
Weinberg operator \cite{Weinberg:1979sa}:
\begin{equation}
(\mathcal{O}_{\nu\nu})_{\alpha\beta}=(\widetilde{H}^{\dagger}\ell_{\alpha})^{T}\mathscr{C}(\widetilde{H}^{\dagger}\ell_{\beta}),\label{eq:Weinberg operator}
\end{equation}
together with its hermitian conjugate (here $\mathscr{C}$ denotes
the Dirac charge conjugation matrix). At dimension six, the list of
effective operators is much richer and consists of 59 independent
operators, which we report in Tables \ref{tab:SMEFT-bosonic-operators.}
to \ref{tab:SMEFT-four-fermion-baryon-1} in the so-called Warsaw
basis \cite{Grzadkowski:2010es}.

\renewcommand{\arraystretch}{1.3}
\begin{center}
\begin{table}[H]
\begin{centering}
\begin{tabular}{|c|c|c|c|c|c|}
\hline 
\multicolumn{2}{|c|}{$X^{3}$} & \multicolumn{2}{c|}{$X^{2}H^{2}$} & \multicolumn{2}{c|}{$H^{4}D^{2}$}\tabularnewline
\hline 
$\mathcal{O}_{3G}$ & $f^{ABC}G_{\mu}^{A\nu}G_{\nu}^{B\rho}G_{\rho}^{C\mu}$ & $\mathcal{O}_{HG}$ & $G_{\mu\nu}^{A}G^{A\mu\nu}(H^{\dagger}H)$ & $\mathcal{O}_{H\square}$ & $(H^{\dagger}H)\square(H^{\dagger}H)$\tabularnewline
$\mathcal{O}_{3\widetilde{G}}$ & $f^{ABC}\widetilde{G}_{\mu}^{A\nu}G_{\nu}^{B\rho}G_{\rho}^{C\mu}$ & $\mathcal{O}_{H\widetilde{G}}$ & $\widetilde{G}_{\mu\nu}^{A}G^{A\mu\nu}(H^{\dagger}H)$ & $\mathcal{O}_{HD}$ & $(H^{\dagger}D^{\mu}H)^{\dagger}(H^{\dagger}D_{\mu}H)$\tabularnewline
\cline{5-6} \cline{6-6} 
$\mathcal{O}_{3W}$ & $\epsilon^{IJK}W_{\mu}^{I\nu}W_{\nu}^{J\rho}W_{\rho}^{K\mu}$ & $\mathcal{O}_{HW}$ & $W_{\mu\nu}^{I}W^{I\mu\nu}(H^{\dagger}H)$ & \multicolumn{2}{c|}{$H^{6}$}\tabularnewline
\cline{5-6} \cline{6-6} 
$\mathcal{O}_{3\widetilde{W}}$ & $\epsilon^{IJK}\widetilde{W}_{\mu}^{I\nu}W_{\nu}^{J\rho}W_{\rho}^{K\mu}$ & $\mathcal{O}_{H\widetilde{W}}$ & $\widetilde{W}_{\mu\nu}^{I}W^{I\mu\nu}(H^{\dagger}H)$ & $\mathcal{O}_{H}$ & $(H^{\dagger}H)^{3}$\tabularnewline
 &  & $\mathcal{O}_{HB}$ & $B_{\mu\nu}B^{\mu\nu}(H^{\dagger}H)$ &  & \tabularnewline
 &  & $\mathcal{O}_{H\widetilde{B}}$ & $\widetilde{B}_{\mu\nu}B^{\mu\nu}(H^{\dagger}H)$ &  & \tabularnewline
 &  & $\mathcal{O}_{HWB}$ & $W_{\mu\nu}^{I}B^{\mu\nu}(H^{\dagger}\tau^{I}H)$ &  & \tabularnewline
 &  & $\mathcal{O}_{H\widetilde{W}B}$ & $\widetilde{W}_{\mu\nu}^{I}B^{\mu\nu}(H^{\dagger}\tau^{I}H)$ &  & \tabularnewline
\hline 
\end{tabular}
\par\end{centering}
\caption{\label{tab:SMEFT-bosonic-operators.}SMEFT bosonic operators.}
\end{table}
\par\end{center}

\begin{center}
\begin{table}[H]
\begin{centering}
\begin{tabular}{|c|c|c|c|c|c|}
\hline 
\multicolumn{2}{|c|}{$\psi^{2}XH+\text{h.c.}$} & \multicolumn{2}{c|}{$\psi^{2}H^{3}+\text{h.c.}$} & \multicolumn{2}{c|}{$\psi^{2}DH^{2}$}\tabularnewline
\hline 
$\mathcal{O}_{uG}$ & $(\overline{q}T^{A}\sigma^{\mu\nu}u)\widetilde{H}G_{\mu\nu}^{A}$ & $\mathcal{O}_{uH}$ & $(H^{\dagger}H)\overline{q}\widetilde{H}u$ & $\mathcal{O}_{Hq}^{(1)}$ & $(\overline{q}\gamma^{\mu}q)(H^{\dagger}i\overleftrightarrow{D}_{\mu}H)$\tabularnewline
$\mathcal{O}_{uW}$ & $(\overline{q}\sigma^{\mu\nu}u)\tau^{I}\widetilde{H}W_{\mu\nu}^{I}$ & $\mathcal{O}_{dH}$ & $(H^{\dagger}H)\overline{q}Hd$ & $\mathcal{O}_{Hq}^{(3)}$ & $(\overline{q}\tau^{I}\gamma^{\mu}q)(H^{\dagger}i\overleftrightarrow{D}_{\mu}^{I}H)$\tabularnewline
$\mathcal{O}_{uB}$ & $(\overline{q}\sigma^{\mu\nu}u)\widetilde{H}B_{\mu\nu}$ & $\mathcal{O}_{eH}$ & $(H^{\dagger}H)\overline{\ell}He$ & $\mathcal{O}_{Hu}$ & $(\overline{u}\gamma^{\mu}u)(H^{\dagger}i\overleftrightarrow{D}_{\mu}H)$\tabularnewline
$\mathcal{O}_{dG}$ & $(\overline{q}T^{A}\sigma^{\mu\nu}d)HG_{\mu\nu}^{A}$ &  &  & $\mathcal{O}_{Hd}$ & $(\overline{d}\gamma^{\mu}d)(H^{\dagger}i\overleftrightarrow{D}_{\mu}H)$\tabularnewline
$\mathcal{O}_{dW}$ & $(\overline{q}\sigma^{\mu\nu}d)\tau^{I}HW_{\mu\nu}^{I}$ &  &  & $\mathcal{O}_{Hud}$ & $(\overline{u}\gamma^{\mu}d)(\widetilde{H}^{\dagger}iD_{\mu}H)$\tabularnewline
$\mathcal{O}_{dB}$ & $(\overline{q}\sigma^{\mu\nu}d)HB_{\mu\nu}$ &  &  & $\mathcal{O}_{H\ell}^{(1)}$ & $(\overline{\ell}\gamma^{\mu}\ell)(H^{\dagger}i\overleftrightarrow{D}_{\mu}H)$\tabularnewline
$\mathcal{O}_{eW}$ & $(\overline{\ell}\sigma^{\mu\nu}e)\tau^{I}HW_{\mu\nu}^{I}$ &  &  & $\mathcal{O}_{H\ell}^{(3)}$ & $(\overline{\ell}\tau^{I}\gamma^{\mu}\ell)(H^{\dagger}i\overleftrightarrow{D}_{\mu}^{I}H)$\tabularnewline
$\mathcal{O}_{eB}$ & $(\overline{\ell}\sigma^{\mu\nu}e)HB_{\mu\nu}$ &  &  & $\mathcal{O}_{He}$ & $(\overline{e}\gamma^{\mu}e)(H^{\dagger}i\overleftrightarrow{D}_{\mu}H)$\tabularnewline
\hline 
\end{tabular}
\par\end{centering}
\caption{\label{tab:SMEFT-two-fermion-operators.}SMEFT two-fermion operators.
Fermion family indices are omitted.}
\end{table}
\par\end{center}

\begin{center}
\begin{table}[H]
\begin{centering}
\begin{tabular}{|c|c|c|c|c|c|}
\hline 
\multicolumn{2}{|c|}{Four quark} & \multicolumn{2}{c|}{Four lepton} & \multicolumn{2}{c|}{Semileptonic}\tabularnewline
\hline 
$\mathcal{O}_{qq}^{(1)}$ & $(\overline{q}\gamma^{\mu}q)(\overline{q}\gamma_{\mu}q)$ & $\mathcal{O}_{\ell\ell}$ & $(\overline{\ell}\gamma^{\mu}\ell)(\overline{\ell}\gamma_{\mu}\ell)$ & $\mathcal{O}_{\ell q}^{(1)}$ & $(\overline{\ell}\gamma^{\mu}\ell)(\overline{q}\gamma_{\mu}q)$\tabularnewline
$\mathcal{O}_{qq}^{(3)}$ & $(\overline{q}\gamma^{\mu}\sigma^{I}q)(\overline{q}\gamma_{\mu}\sigma^{I}q)$ & $\mathcal{O}_{ee}$ & $(\overline{e}\gamma^{\mu}e)(\overline{e}\gamma_{\mu}e)$ & $\mathcal{O}_{\ell q}^{(3)}$ & $(\overline{\ell}\gamma^{\mu}\sigma^{I}\ell)(\overline{q}\gamma_{\mu}\sigma^{I}q)$\tabularnewline
$\mathcal{O}_{uu}$ & $(\overline{u}\gamma^{\mu}u)(\overline{u}\gamma_{\mu}u)$ & $\mathcal{O}_{\ell e}$ & $(\overline{\ell}\gamma^{\mu}\ell)(\overline{e}\gamma_{\mu}e)$ & $\mathcal{O}_{eu}$ & $(\overline{e}\gamma^{\mu}e)(\overline{u}\gamma_{\mu}u)$\tabularnewline
$\mathcal{O}_{dd}$ & $(\overline{d}\gamma^{\mu}d)(\overline{d}\gamma_{\mu}d)$ &  &  & $\mathcal{O}_{ed}$ & $(\overline{e}\gamma^{\mu}e)(\overline{d}\gamma_{\mu}d)$\tabularnewline
$\mathcal{O}_{ud}^{(1)}$ & $(\overline{u}\gamma^{\mu}u)(\overline{d}\gamma_{\mu}d)$ &  &  & $\mathcal{O}_{qe}$ & $(\overline{q}\gamma^{\mu}q)(\overline{e}\gamma_{\mu}e)$\tabularnewline
$\mathcal{O}_{ud}^{(8)}$ & $(\overline{u}\gamma^{\mu}T^{A}u)(\overline{d}\gamma_{\mu}T^{A}d)$ &  &  & $\mathcal{O}_{\ell u}$ & $(\overline{\ell}\gamma^{\mu}\ell)(\overline{u}\gamma_{\mu}u)$\tabularnewline
$\mathcal{O}_{qu}^{(1)}$ & $(\overline{q}\gamma^{\mu}q)(\overline{u}\gamma_{\mu}u)$ &  &  & $\mathcal{O}_{\ell d}$ & $(\overline{\ell}\gamma^{\mu}\ell)(\overline{d}\gamma_{\mu}d)$\tabularnewline
$\mathcal{O}_{qu}^{(8)}$ & $(\overline{q}\gamma^{\mu}T^{A}q)(\overline{u}\gamma_{\mu}T^{A}u)$ &  &  & $\mathcal{O}_{\ell edq}$ & $(\overline{\ell}e)(\overline{d}q)$\tabularnewline
$\mathcal{O}_{qd}^{(1)}$ & $(\overline{q}\gamma^{\mu}q)(\overline{d}\gamma_{\mu}d)$ &  &  & $\mathcal{O}_{\ell equ}^{(1)}$ & $(\overline{\ell}^{r}e)\epsilon_{rs}(\overline{q}^{s}u)$\tabularnewline
$\mathcal{O}_{qd}^{(8)}$ & $(\overline{q}\gamma^{\mu}T^{A}q)(\overline{d}\gamma_{\mu}T^{A}d)$ &  &  & $\mathcal{O}_{\ell equ}^{(3)}$ & $(\overline{\ell}^{r}\sigma^{\mu\nu}e)\epsilon_{rs}(\overline{q}^{s}\sigma_{\mu\nu}u)$\tabularnewline
$\mathcal{O}_{quqd}^{(1)}$ & $(\overline{q}^{r}u)\epsilon_{rs}(\overline{q}^{s}d)$ &  &  &  & \tabularnewline
$\mathcal{O}_{quqd}^{(8)}$ & $(\overline{q}^{r}T^{A}u)\epsilon_{rs}(\overline{q}^{s}T^{A}d)$ &  &  &  & \tabularnewline
\hline 
\end{tabular}
\par\end{centering}
\caption{\label{tab:SMEFT-four-fermion-baryon}SMEFT four-fermion baryon number
conserving operators. Fermion family indices are omitted.}
\end{table}
\par\end{center}

\begin{center}
\begin{table}[H]
\begin{centering}
\begin{tabular}{|c|c|}
\hline 
\multicolumn{2}{|c|}{$B$ and $L$ violating}\tabularnewline
\hline 
$\mathcal{O}_{duq}$ & $\varepsilon_{\alpha\beta\gamma}\epsilon_{rs}\left[(d^{\alpha})^{T}\mathscr{C}u^{\beta})\right]\left[(q^{\gamma r})^{T}\mathscr{C}\ell^{s}\right]$\tabularnewline
$\mathcal{O}_{qqu}$ & $\varepsilon_{\alpha\beta\gamma}\epsilon_{rs}\left[(q^{\alpha r})^{T}\mathscr{C}q^{\beta s})\right]\left[(u^{\gamma})^{T}\mathscr{C}e\right]$\tabularnewline
$\mathcal{O}_{qqq}$ & $\varepsilon_{\alpha\beta\gamma}\epsilon_{rs}\epsilon_{pt}\left[(q^{\alpha r})^{T}\mathscr{C}q^{\beta s})\right]\left[(q^{\gamma p})^{T}\mathscr{C}\ell^{t}\right]$\tabularnewline
$\mathcal{O}_{duu}$ & $\varepsilon_{\alpha\beta\gamma}\left[(d^{\alpha})^{T}\mathscr{C}u^{\beta})\right]\left[(u^{\gamma})^{T}\mathscr{C}e\right]$\tabularnewline
\hline 
\end{tabular}
\par\end{centering}
\caption{\label{tab:SMEFT-four-fermion-baryon-1}SMEFT four-fermion baryon
number violating operators. Fermion family indices are omitted. Greek
and latin indices denote color and weak isospin, respectively.}
\end{table}
\par\end{center}

\renewcommand{\arraystretch}{1.0}

The precise sense in which Warsaw basis operators are independent
is the following: none of the operators in Tables \ref{tab:SMEFT-bosonic-operators.}-\ref{tab:SMEFT-four-fermion-baryon-1}
can be obtained from the remaining ones by taking linear combinations,
adding a total divergence, or applying smooth field redefinitions
(as its name suggests, the Warsaw basis is also complete, in the sense
that any dimension six operator can be obtained from Warsaw basis
operators by means of these three operations). The last requirement
follows from a well known theorem \cite{Kamefuchi:1961sb,Bando:1987br}
which states that for any smooth transformation $\phi\mapsto F(\phi)$
of the coordinate fields $\phi$, the two quantum lagrangians $\mathcal{L}$
and $\mathcal{L}^{\prime}=\mathcal{L}\circ F$ give rise to the same
physical $S$ matrix.

To conclude this Section, we observe that the SMEFT formalism we have
just described allows one to obtain low energy parametrizations of,
virtually, any SM extension by heavy new fields. The procedure by
which one can pass from a high-energy ultraviolet complete model to
the low-energy SMEFT description is called \emph{matching}, and is
the object of the next Section.

\section{\label{sec:Matching}Computing with EFTs: theory matching}

The standard matching of EFTs enforces the physical equivalence between
an high-energy theory and an effective theory with a reduced number
of light degrees of freedom, by requiring equality between the Green's
functions\footnote{This is actually stronger than requiring the equality of (physical)
$S$-matrix elements, but does not imply a loss in generality. In
fact, given a theory with lagrangian $\mathcal{L}(\phi,\Phi)$, where
$\phi$ and $\Phi$ denote the set of light and heavy fields respectively,
we can always define a low-energy theory by the lagrangian:
\[
\mathcal{L}_{\text{EFT}}(\phi)=\log\left\{ \intop\mathcal{D}\Phi\,\exp\left[\mathcal{L}(\phi,\Phi)\right]\right\} .
\]

By construction, $\mathcal{L}_{\text{EFT}}$ and $\mathcal{L}$ give
rise to the same $\phi$ Green's functions, and the scale separation
between $\phi$ and $\Phi$ fields ensures that $\mathcal{L}_{\text{EFT}}$
can be expanded in a series of local operators.} of light fields computed in the two theories at a fixed renormalization
scale $\mu$. The latter is typically chosen to be close to the typical
mass scale $M$ of the fields being integrated out\footnote{We assume here that there exists only one such scale. In the presence
of several scales $M_{1}\ll M_{2}\ll\cdots\ll M_{n}$ with sizable
separations, the matching is usually performed in a sequential fashion,
connecting the various scales $M_{i}$ and $M_{i+1}$ using the renormalization
group flow of the intermediate EFTs resulting from the procedure.}, to ensure that the matching conditions can be computed perturbatively,
as $l$-loop diagrams can contribute to the matching equations with
terms proportional to $(4\pi)^{-l}\ln(\mu/M)^{l-i}$ for $i=0,\,1,\,\dots,\,l$
\cite{Georgi:1993mps}.

In practice, it is sufficient to enforce matching for the so-called
1-Light-Particle Irreducible (1LPI) Green's functions of the light
fields, which are defined as sums of connected Feynman graphs which
do not have a single line cut corresponding to an internal light field
exchange. At a fixed maximal power $M^{-k}$ in the EFT expansion,
only a finite number of 1LPI Green's function receive contributions
from integrating out the heavy fields in the high-energy theory, and
each of these gives rise to an independent matching equation for the
Wilson coefficients of EFT operators of dimension $d\leq4+k$. Concretely,
a sample computation would go as follows \cite{Georgi:1993mps,Buras:1998raa}:
\begin{enumerate}
\item Compute the 1LPI Green's function in the high-energy theory, expanding
in powers of $\frac{1}{M}$ up to the desired order.
\item Compute the same function in the EFT, including the contributions
from effective operators with unknown (as yet) Wilson coefficients.
\item Equate the two results.
\end{enumerate}
It goes without mention that steps 1 and 2 above can usually be carried
out only in a perturbative fashion, \emph{i.e.} expanding the corresponding
Green's function to a fixed loop order (which should be the same in
the high-energy and effective theories). The output of this procedure
is the full list of EFT Wilson coefficients at the given loop accuracy,
which are usually reported in a standard basis, such as the Warsaw
basis for SMEFT dimension-six operators.

Adopting a standard operator basis is useful for communicating the
final results of an EFT calculation in a compact and unambiguous manner.
That said, the diagrammatic matching calculations described above
are most naturally performed using a larger effective operator basis,
which I will refer to as the \emph{Green's basis,} following the terminology
of Ref. \cite{Jiang:2018pbd}. In defining Green's basis operators,
we drop the requirement of independence under field redefinitions;
the Green's basis is the correct operator menu to be used in the second
step of the matching procedure described above, since there exists
a bijective linear correspondence between dimension-$d$ Green's operators
and Green's \emph{functions}.

\section{\label{sec:Advanced matching}Advanced matching methods}

Simplifying EFT calculations can provide great help to BSM phenomenologists,
by accelerating the process of model building and evaluation. There
are two main possible sources of simplification, namely: mathematical
methods and automation.

On the mathematical side, the problem of SMEFT tree-level matching
up to dimension-six terms was completely solved in Ref. \cite{deBlas:2017xtg},
which provided the complete dictionary of SMEFT contributions for
all possible tree-level mediators. The key observation behind this
work is that, for any fixed maximal effective operator dimension,
the number of extra fields and couplings which can give rise to SMEFT
operators at low energies is finite, so that the program of Ref. \cite{deBlas:2017xtg}
can, at least in principle, be carried out up to any effective operator
dimension. 

Unfortunately, such a simplification no longer occurs for $\ell\geq1$-loop
contributions, in which case more general methods are required. In
particular, functional methods (focusing on Green's function generating
functionals) have provided significant advances in the context of
one-loop computations, the latest approaches being based on the so-called
Universal One-Loop Effective Action (UOLEA) \cite{Haisch:2020ahr},
which generalises methods based on the Covariant Derivative Expansion
(CDE) \cite{Henning:2014wua}, but whose formalism is still incomplete
(see Ref. \cite{Haisch:2020ahr} and references therein for a recent
discussion, and for an application of the UOLEA formalism).

Concerning automated methods, many general tools aimed at simplifying
EFT calculations are already available (for a comprehensive review,
see \cite{Proceedings:2019rnh} and references therein). For the specific
task of EFT matching, however, the tool-set is still for a large part
under development. We cite here the software packages mentioned in
Ref. \cite{Proceedings:2019rnh}:
\begin{itemize}
\item \texttt{MatchingTools} \cite{Criado:2017khh}, a Python library providing
support for tree-level matching of generic field theories (\emph{i.e.
}not restricted by the SM gauge group and field content). This package
was used, in particular, for an extensive check of the analytic results
provided by Ref. \cite{deBlas:2017xtg} mentioned earlier.
\item \texttt{CoDEx} \cite{DasBakshi:2018vni}, a Mathematica package devoted
to SMEFT matching at one-loop and up to dimension-six operators. \texttt{CoDEx}
is based on the Covariant Derivative Expansions, and the latest release
(v1.0.0) by the time of this writing can correctly handle only one-loop
diagrams not involving light internal particles in the loop.
\item \texttt{MatchMaker} \cite{Proceedings:2019rnh}, a work-in-progress
Python package whose primary purpose is, again, the complete SMEFT
one-loop matching (with the grand goal of extending the tool-set to
arbitrary field theories). Unfortunately, by the time of this writing,
no further update on the progress status of \texttt{MatchMaker} is
available.
\end{itemize}
To conclude, both mathematical and computer-based methods for EFT
calculations still offer large room for improvement, and progress
in this area could definitely provide precious help in our meandering
search for BSM signals.

\chapter{\label{chap:S1S3 leptoquark theory}A case study: the $S_{1}+S_{3}$
Leptoquark model}

This Chapter discusses my contribution \cite{Gherardi:2020det}, in
collaboration with D. Marzocca and E. Venturini, in which we performed
the complete SMEFT one-loop matching for the $S_{1}+S_{3}$ leptoquark
model, up to dimension-six (leading order) in the EFT expansion. 

This model has received attention from the phenomenology community
during recent years, as one of the most promising candidates for the
solution of so-called (neutral- and charged-current) $B$-anomalies,
which will be the central subject of Part II of this Thesis. As was
recently realized, the model can also (simultaneously) provide a good
fit to the long-standing muon $(g-2)_{\mu}$ discrepancy \cite{Gherardi:2020det},
while giving rise to additional further predictions (for \emph{e.g.
}Lepton Flavor Violating observables) which are compatible with current
bounds, but in the ball-park of future prospects. All in all, the
$S_{1}+S_{3}$ model offers a particularly intriguing New Physics
scenario, which is fully amenable to the EFT analysis discussed in
the previous Chapter.

Beyond its phenomenological interest, Ref. \cite{Gherardi:2020det}
provides one of the very few examples of complete SMEFT one-loop matchings,
performed in the standard diagrammatic way, available in the literature.
In \cite{Huo:2015nka,Wells:2017vla} the one-loop matching for bosonic
SMEFT operators from integrating out sfermions in the MSSM is derived,
Refs.~\cite{Jiang:2018pbd,Haisch:2020ahr} perform the complete one-loop
matching for a singlet scalar (see also \cite{Boggia:2016asg}), and
\cite{Chala:2020vqp} considers the SM with an additional light sterile
neutrino and heavy fermions and a scalar singlet. The model considered
in Ref. \cite{Gherardi:2020det}, with two coloured and weakly-charged
states coupled to all SM particles with non-trivial flavour structures,
represents a very rich example of such a matching. In hindsight, results
of this kind will be of value as important cross-checks for more advanced
matching techniques, including functional- and even computer-based
approaches (some ongoing work in this direction was mentioned in Sec.
\ref{sec:Advanced matching}). Indeed, given the relevance of the
subject to model building, automating (or, at least, simplifying)
EFT calculations has a great potential for streamlining the work of
many phenomenologists, and establishing a set of (reasonably complex)
test cases is an unavoidable step in the process of building these
more sophisticated tools.

The complete one-loop matching presented in this Chapter allows for
in-depth phenomenological analyses of the leptoquark model. In Part
II, Chapter \ref{chap:S1S3-phenomenology}, we will describe in detail
the phenomenology of the $S_{1}+S_{3}$ model, as a potential combined
solution of the $B$-decay anomalies and the muon $(g-2)_{\mu}$ discrepancy.

\section{\label{sec:The-S1S3-model}The $S_{1}+S_{3}$ model}

The UV model under consideration is defined by the SM gauge group
and field content, with the addition of two colored scalar \emph{leptoquarks}:

\begin{equation}
S_{1}\sim\left(\overline{\boldsymbol{3}},\,\boldsymbol{1}\right)_{1/3},\qquad S_{3}\sim\left(\overline{\boldsymbol{3}},\,\boldsymbol{3}\right)_{1/3},\label{eq:S1-S3 qnumbers}
\end{equation}
where in parenthesis we indicate the $S_{1,3}$ representations under
$\left(\text{SU}(3)_{c},\,\text{SU}(2)_{L}\right)_{\text{U}(1)_{Y}}$.
At the renormalizable level, the two additional scalars directly couple
to SM fermions through the Yukawa terms:
\begin{equation}
\left((\lambda^{1L})_{i\alpha}\overline{q}_{i}^{c}\epsilon\ell_{\alpha}+(\lambda^{1R})_{i\alpha}\overline{u}_{i}^{c}e_{\alpha}\right)S_{1}+(\lambda^{3L})_{i\alpha}\overline{q}_{i}^{c}\sigma^{I}\epsilon\ell_{\alpha}S_{3}^{I},\label{eq:LQ yukawa couplings}
\end{equation}
(hence the name ``leptoquarks''). The couplings in Eq. (\ref{eq:LQ yukawa couplings})
by themselves do not imply a violation of baryon or lepton number
conservation, as can be seen by the $S_{1/3}$ lepton and baryon number
assignments: 
\begin{equation}
\left(L_{S},B_{S}\right)=\left(-1,+\frac{1}{3}\right).\label{eq:Lepton-Baryon number LQ}
\end{equation}
On the other hand, the following additional ``diquark'' operators
are also allowed by gauge invariance:
\begin{equation}
\overline{u}_{i}d_{j}^{c}S_{1},\qquad\overline{q}_{i}q_{j}^{c}S_{1},\qquad\overline{q}_{i}\epsilon\sigma^{I}q_{j}^{c}S_{3}^{I},\label{eq:S1-S3 diquark ops}
\end{equation}
which violate baryon number conservation according to Eq. (\ref{eq:Lepton-Baryon number LQ}),
and may lead, \emph{e.g.}, to proton decay. As a consequence, since
the phenomenological motivations mentioned above require the introduction
of relatively low-scale NP degrees of freedom (of order $\mathcal{O}(\text{TeV})$,
say), baryon and lepton number conservation is actually postulated
\emph{a priori}\footnote{See Ref. \cite{Davighi:2020qqa} for an extended gauge model involving
$S_{1}$ and $S_{3}$, in which lepton and baryon number are accidental
symmetries, as in the SM.}, along with Eq. (\ref{eq:Lepton-Baryon number LQ}), which allows
the leptoquark couplings (\ref{eq:LQ yukawa couplings}) but forbids
the diquark ones, Eq. (\ref{eq:S1-S3 diquark ops}). Finally, the
addition of $S_{1/3}$ also gives rise to a scalar potential:
\begin{equation}
\mathcal{V}_{S}=M_{1}^{2}\left|S_{1}\right|^{2}+M_{3}^{2}\left|S_{3}\right|^{2}+\Delta\mathcal{V}_{S}(S_{1},S_{3},H),\label{eq:LQ scalar potential}
\end{equation}
where $\Delta\mathcal{V}_{S}$ denotes trilinear and tetralinear terms
in the $S_{1/3}$ and Higgs fields, which we omit here for brevity
(the full expression is reported in Sec. \ref{sec:S1S3-potential}).

The leptoquark couplings in Eq. (\ref{eq:LQ yukawa couplings}), as
well as the omitted potential couplings in Eq. (\ref{eq:LQ scalar potential}))
are assumed to be in the perturbative regime\footnote{See Ref. \cite{DiLuzio:2016sur} for an analysis of perturbativity
bounds from unitarity principles.}, so that theory quantities such as Green's functions admit a sensible
perturbative expansion, which, following standard arguments \cite{Coleman:1973jx},
is equivalent to a loop-wise expansion in Feynman graph calculations.
Assuming the leptoquark masses $M_{1}^{2}$ and $M_{3}^{2}$ to be
of similar order of magnitude (and, of course, large with respect
to the electroweak scale), it is easy\footnote{These can be easily obtained by replacing $S_{1}$ and $S_{3}$ in
the leptoquark model's lagrangian with the corresponding solutions
(in terms of SM fields) of the classical Equations Of Motion (EOM).
In doing this, one should take into account the following simplifications,
which reduce the problem at hand to linear algebra:
\begin{enumerate}
\item Terms in the scalar potential (\ref{eq:LQ scalar potential}) which
are not quadratic in $S_{1/3}$ do not contribute to tree-level matching
at leading dimension, and can be safely omitted from the EOM.
\item Covariant derivative terms (\emph{i.e.} $\left|D_{\mu}S_{1,3}\right|^{2}$)
from the kinetic lagrangian also give rise to higher ($\geq8$) dimensional
operators, and can be disregarded for the purpose of tree-level matching.
\end{enumerate}
} to obtain the tree-level, dimension-six SMEFT matching conditions
for the (lepton-baryon number conserving) $S_{1}+S_{3}$ model:
\begin{align}
\left[C_{\ell q}^{(1)}(\mu)\right]_{\alpha\beta ij}^{(0)} & =\frac{\lambda_{i\alpha}^{1L*}\lambda_{j\beta}^{1L}}{4M_{1}^{2}}+3\frac{\lambda_{i\alpha}^{3L*}\lambda_{j\beta}^{3L}}{4M_{3}^{2}},\label{eq:C_lq1 tree}\\
\left[C_{\ell q}^{(3)}(\mu)\right]_{\alpha\beta ij}^{(0)} & =-\frac{\lambda_{i\alpha}^{1L*}\lambda_{j\beta}^{1L}}{4M_{1}^{2}}+\frac{\lambda_{i\alpha}^{3L*}\lambda_{j\beta}^{3L}}{4M_{3}^{2}},\label{eq:C_lq3 tree}\\
\left[C_{\ell equ}^{(1)}(\mu)\right]_{\alpha\beta ij}^{(0)} & =\frac{\lambda_{i\alpha}^{1L*}\lambda_{j\beta}^{1R}}{2M_{1}^{2}},\label{eq:C_lequ1 tree}\\
\left[C_{\ell equ}^{(3)}(\mu)\right]_{\alpha\beta ij}^{(0)} & =-\frac{\lambda_{i\alpha}^{1L*}\lambda_{j\beta}^{1R}}{8M_{1}^{2}},\label{eq:C_lequ3 tree}\\
\left[C_{eu}^{(1)}(\mu)\right]_{\alpha\beta ij}^{(0)} & =\frac{\lambda_{i\alpha}^{1R*}\lambda_{j\beta}^{1R}}{2M_{1}^{2}}.\label{eq:C_eu tree}
\end{align}
The Wilson coefficient notation refers to the Warsaw basis (Tables
\ref{tab:SMEFT-bosonic-operators.}-\ref{tab:SMEFT-four-fermion-baryon-1}).
The suffixes $^{(0)}$ indicate the accuracy (in terms of loop orders,
here zero) of Eqs. (\ref{eq:C_lq1 tree})-(\ref{eq:C_eu tree}), and
the RG sliding scale $\mu$ is assumed to be close in magnitude to
$M_{1,3}^{2}$ (the running UV couplings in the right-hand sides of
Eqs. (\ref{eq:C_lq1 tree})-(\ref{eq:C_eu tree}) are understood to
be evaluated at the same scale).

\section{One-loop matching}

In general, taking perturbativity as given, Wilson coefficients admit
a loop-wise expansion of the form:
\begin{equation}
C_{\mathcal{O}}(\mu)=C_{\mathcal{O}}(\mu)^{(0)}+\frac{1}{(4\pi)^{2}}C_{\mathcal{O}}(\mu)^{(1)}+\frac{1}{(4\pi)^{4}}C_{\mathcal{O}}(\mu)^{(2)}+\cdots.\label{eq:Loopwise expansion of WC}
\end{equation}
Going beyond the tree-level matching, apart from the (obviously) increased
computational complexity, a few other technicalities need to be taken
into account:
\begin{enumerate}
\item Loop computations require the introduction of a renormalization scheme,
which must be entirely specified along with the final (renormalized)
results. We employ the $\overline{\text{MS}}$ scheme with the Naive
Dimensional Regularization (NDR) prescription for analytically extending
Dirac matrices to arbitrary real dimensions $d=4-2\epsilon$. Furthermore,
all computations are performed in a generic $R_{\xi}$ gauge, in order
to check for the expected elision of $\xi$-dependencies in the Wilson
coefficients of gauge invariant operators.
\item A further scheme-dependence can (and does, in our case) arise from
the prescriptions adopted for evanescent operators \cite{Herrlich:1994kh},
which are Dirac operators that vanish in four dimension, but may be
non-vanishing in $d=4-2\epsilon$\footnote{An instructive example is given by the two Dirac tensors:
\begin{align*}
\mathcal{E}_{1} & =P_{L}\gamma^{\mu}\gamma^{\nu}P_{L}\otimes P_{L}\gamma_{\mu}\gamma_{\nu}P_{L}-4P_{L}\otimes P_{L}+P_{L}\sigma^{\mu\nu}P_{L}\otimes P_{L}\sigma_{\mu\nu}P_{L},\\
\mathcal{E}_{2} & =P_{L}\gamma^{\mu}\gamma^{\nu}P_{L}\otimes P_{R}\gamma_{\mu}\gamma_{\nu}P_{R}-4P_{L}\otimes P_{R}.
\end{align*}
Following the NDR prescription for the Dirac algebra in dimension
$d=4-2\epsilon$:
\[
\left\{ \gamma^{\mu},\gamma^{\nu}\right\} =2\eta^{\mu\nu},\qquad\left\{ \gamma^{\mu},\gamma^{5}\right\} =0,\qquad\text{Tr}(\eta)=4-2\epsilon,
\]
one can easily work out the result $\mathcal{E}_{1}=-2\epsilon P_{L}\otimes P_{L}$.
However, the operator $\mathcal{E}_{2}$ is not univocally fixed by
the NDR rules, and $\mathcal{E}_{2}$ is usually expressed in terms
of canonical Dirac tensors as follows \cite{Dekens:2019ept}:
\[
\mathcal{E}_{2}=4a_{\text{ev}}\epsilon P_{L}\otimes P_{R}+E_{LR}^{(2)}(a_{\text{ev}}),
\]
where the actual form of $E_{LR}^{(2)}(a_{\text{ev}})$ in the right-hand
side depends on the value of $a_{\text{ev}}$, which is completely
arbitrary and should be regarded as part of the regularization scheme
(for $a_{\text{ev}}=-\frac{1}{2}$ one gets $E_{LR}^{(2)}(a_{\text{ev}})=P_{L}\sigma^{\mu\nu}P_{L}\otimes P_{R}\sigma_{\mu\nu}P_{R}$,
cf. Ref. \cite{Dekens:2019ept}).}. We follow the conventions of Ref. \cite{Dekens:2019ept}, keeping
the scheme-defining coefficients $\left\{ a_{\text{ev}},\,b_{\text{ev}},\,...\right\} $
arbitrary in our computations (in practice, the only coefficient entering
one-loop computations is $a_{\text{ev}}$).
\item As discussed in Sec. \ref{sec:Matching}, diagrammatic matching computations
are naturally performed in the Green's basis. An important result
of Ref. \cite{Gherardi:2020det} was thus (i) to classify all SMEFT
dimension-six Green's basis operators, and (ii) to provide a set of
(linear) reduction equations which allow to pass from the Green's
basis to the standard Warsaw basis, after field redefinition redundancy
is properly taken into account.
\item Integrating out the leptoquarks at one loop also generates contributions
to SM renormalizable operators and, in particular, fermion kinetic
terms. Such modifications can be undone by suitable field and SM coupling
redefinitions, which however also introduce additional contributions
to tree-level generated WCs\footnote{Since field redefinitions arise at one loop in our model, only tree-level
WCs are affected. In general, any tree-level shift in SM couplings
and wave-function renormalizations that could influence loop-generated
coefficients should be taken into account, see e.g. \cite{Jiang:2018pbd}.}. In our case only fermion kinetic terms (i.e. wave-functions renormalizations)
are relevant, as the tree-level WCs in Eqs. (\ref{eq:C_lq1 tree})-(\ref{eq:C_eu tree})
do not depend on any SM coupling. The one-loop formulas below include
the contributions due to fermion field renormalization.
\end{enumerate}
In what follows we illustrate the diagrammatic matching steps with
a fully detailed example, and subsequently review the procedures used
in Ref. \cite{Gherardi:2020det} to obtain a dimension-six SMEFT Green's
basis and the corresponding equations for reduction onto the Warsaw
basis. For additional details, including the final complete results
for the SMEFT Wilson coefficients of the $S_{1}+S_{3}$ model, we
address the reader to the original reference \cite{Gherardi:2020det}.

\section{One-loop matching example}

In this Section we discuss in some details the matching of a specific
Green's function, in order to illustrate some of the most relevant
aspects of our computation.
\begin{center}
\begin{figure}[H]
\begin{centering}
\includegraphics{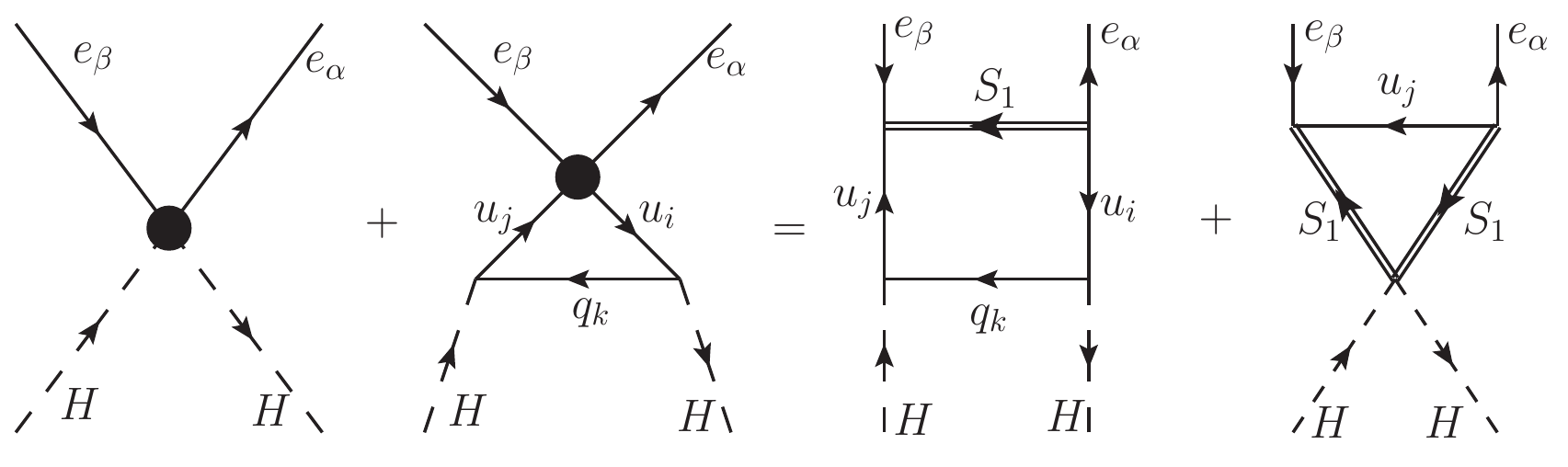}
\par\end{centering}
\caption{\label{fig:Matching Ex. - Diagrams}Diagrams of the matching of the
$\langle e\overline{e}HH^{\dagger}\rangle$ Green's function.}
\end{figure}
\par\end{center}

Let us consider the off-shell Green's function $\mathcal{G}\equiv\langle e_{\beta}(p_{1})\bar{e}_{\alpha}(p_{2})H_{b}(q_{1})H_{a}^{\dagger}(q_{2})\rangle$,
where all momenta are incoming and $a,\,b$ are $\text{SU}(2)_{L}$
indices. The matching conditions for this correlator are depicted
diagrammatically in Fig. \ref{fig:Matching Ex. - Diagrams}, where
the left and right hand-side show the EFT and UV contributions, respectively.
We briefly comment on the various steps of this computation.

We begin by listing the various contributions to $\mathcal{G}$, both
in the SMEFT and the leptoquark model. The SMEFT operators which contribute
at tree level to $\mathcal{G}$ are (cf. Table \ref{tab:Two-fermion-Green's-basis}
for the notation):
\begin{align}
[\mathcal{O}_{He}]_{\alpha\beta} & =(\bar{e}_{\alpha}\gamma^{\mu}e_{\beta})(H^{\dagger}i\overleftrightarrow{D}_{\mu}H),\label{eq:Ops matching example, 1}\\{}
[\mathcal{O}_{He}^{\prime}]_{\alpha\beta} & =(\bar{e}_{\alpha}i\gamma_{\mu}\overleftrightarrow{D^{\mu}}e_{\beta})(H^{\dagger}H),\label{eq:Ops matching example, 2}\\{}
[\mathcal{O}_{He}^{\prime\prime}]_{\alpha\beta} & =(\bar{e}_{\alpha}\gamma^{\mu}e_{\beta})\partial_{\mu}(H^{\dagger}H)\label{eq:Ops matching example, 3}
\end{align}

Moreover, we must take into account a one-loop contribution from $\mathcal{O}_{eu}$,
which is generated at the tree-level in our model according to Eq.
(\ref{eq:C_eu tree}). Since this tree-level WC is fixed, the matching
of $\mathcal{G}$ allows us to fix the coefficients of the operators
in Eqs. (\ref{eq:Ops matching example, 1})-(\ref{eq:Ops matching example, 3}),
see the left-hand side of Fig. \ref{fig:Matching Ex. - Diagrams}.
In the leptoquark model there are two diagrams contributing to $\mathcal{G}$,
both mediated by $S_{1}$, shown on the right-hand side of Fig. \ref{fig:Matching Ex. - Diagrams}:
a box diagram proportional to (schematically) $y_{U}y_{U}^{\dagger}\lambda^{1R}\lambda^{1R}{}^{\dagger}$,
and a triangle diagram proportional to $\lambda_{H1}\lambda^{1R}{}^{\dagger}\lambda^{1R}$
(the coupling $\lambda_{H1}$ is defined in Eq. (\ref{eq:S1S3-potential})).

By total momentum conservation, only three out of the four momenta
$p_{1},\,p_{2},\,q_{1},\,q_{2}$ are independent. Writing $(p_{1},p_{2},q_{1},q_{2})=(p-r,\,-p-r,\,q+r,\,-q+r)$,
the tree-level contributions from the operators in Eqs. (\ref{eq:Ops matching example, 1})-(\ref{eq:Ops matching example, 3})
read: 
\begin{equation}
[\mathcal{G}_{\text{EFT}}^{\text{tree}}(\mu_{M})]_{\alpha\beta}=2\slashed q[G_{He}(\mu_{M})]_{\alpha\beta}+2\slashed p[G_{He}^{\prime}(\mu_{M})]_{\alpha\beta}-2i\slashed r[G_{He}^{\prime\prime}(\mu_{M})]_{\alpha\beta}\label{eq:Matching Ex. - Tree EFT}
\end{equation}
where we drop here and below a global $\delta_{ab}$ factor, and we
denote Green's basis WCs by $G_{i}$. The UV and EFT one-loop contributions
are more easily computed when only one of the independent momenta
$p,\,q,\,r$ is non-vanishing, and yield respectively: 
\begin{align}
\left[\mathcal{G}_{\text{UV}}^{\text{1-loop}}(\mu_{M})\right]_{\alpha\beta}^{q=r=0} & =-\slashed{p}\frac{N_{c}(\lambda^{1R\dagger}y_{U}^{T}y_{U}^{*}\lambda^{1R})_{\alpha\beta}}{(4\pi)^{2}2M_{1}^{2}}+\slashed{p}\frac{N_{c}\lambda_{H1}(\lambda^{1R\dagger}\lambda^{1R})_{\alpha\beta}}{(4\pi)^{2}2M_{1}^{2}},\label{eq:Matching Ex. - 1Loop UV, 1}\\
\left[\mathcal{G}_{\text{UV}}^{\text{1-loop}}(\mu_{M})\right]_{\alpha\beta}^{p=r=0} & =-\slashed{q}\frac{N_{c}(\lambda^{1R\dagger}y_{U}^{T}y_{U}^{*}\lambda^{1R})_{\alpha\beta}}{(4\pi)^{2}M_{1}^{2}}\log\frac{-q^{2}}{M_{1}^{2}},\label{eq:Matching Ex. - 1Loop UV, 2}\\
\left[\mathcal{G}_{\text{UV}}^{\text{1-loop}}(\mu_{M})\right]_{\alpha\beta}^{p=q=0} & =0\label{eq:Matching Ex. - 1Loop UV, 3}
\end{align}
and 
\begin{align}
\left[\mathcal{G}_{\text{EFT}}^{\text{1-loop}}(\mu_{M})\right]_{\alpha\beta}^{q=r=0} & =0\label{eq:Matching Ex. - 1Loop EFT, 1}\\
\left[\mathcal{G}_{\text{EFT}}^{\text{1-loop}}(\mu_{M})\right]_{\alpha\beta}^{p=r=0} & =\slashed{q}\frac{N_{c}(\lambda^{1R\dagger}y_{U}^{T}y_{U}^{*}\lambda^{1R})_{\alpha\beta}}{(4\pi)^{2}M_{1}^{2}}\left(1+\log\frac{\mu_{M}^{2}}{-q^{2}}\right)\label{eq:Matching Ex. - 1Loop EFT, 2}\\
\left[\mathcal{G}_{\text{EFT}}^{\text{1-loop}}(\mu_{M})\right]_{\alpha\beta}^{p=q=0} & =0,\label{eq:Matching Ex. - 1Loop EFT, 3}
\end{align}
 where we employed the tree-level value of $[C_{eu}]_{\alpha\beta}^{(0)}$
given in Eq. (\ref{eq:C_eu tree}). Notice that the EFT computation
presents an ultraviolet divergence, which we regulate in the $\overline{\text{MS}}$
scheme at renormalization scale $\mu_{M}$. On the other hand, on
the basis of renormalizability, the UV contribution must be (and is)
finite. Finally, both EFT and UV diagrams present an infrared divergence,
corresponding to the $\log(-q^{2})$ terms in Eqs. (\ref{eq:Matching Ex. - 1Loop UV, 2})
and (\ref{eq:Matching Ex. - 1Loop EFT, 2}). The agreement of these
two terms, which is guaranteed by the EFT construction, provides a
further check of validity of the computation.

Requiring $\mathcal{G}_{\text{EFT}}(\mu_{M})=\mathcal{G}_{\text{UV}}(\mu_{M})$,
we finally obtain the matching conditions: 
\[
[G_{He}(\mu_{M})]_{\alpha\beta}=-\frac{N_{c}(\lambda^{1R\dagger}y_{U}^{T}y_{U}^{*}\lambda^{1R})_{\alpha\beta}}{32\pi^{2}M_{1}^{2}}\left(1+\log\frac{\mu_{M}^{2}}{M_{1}^{2}}\right)
\]
\[
[G_{He}^{\prime}(\mu_{M})]_{\alpha\beta}=-\frac{N_{c}(\lambda^{1R\dagger}y_{U}^{T}y_{U}^{*}\lambda^{1R})_{\alpha\beta}}{64\pi^{2}M_{1}^{2}}+\frac{N_{c}\lambda_{H1}(\lambda^{1R\dagger}\lambda^{1R})_{\alpha\beta}}{64\pi^{2}M_{1}^{2}}
\]
\[
[G_{He}^{\prime\prime}(\mu_{M})]_{\alpha\beta}=0
\]
As a cross-check, we observe that the $\mu_{M}$ dependence of $[G_{He}(\mu_{M})]_{\alpha\beta}$
corresponds to the SMEFT RG running of $C_{He}$ due to $C_{eu}$
\cite{Jenkins:2013wua} 
\[
(4\pi)^{2}\mu\dfrac{d[C_{He}]_{\alpha\beta}}{d\mu}=-2N_{c}[C_{eu}]_{\alpha\beta ij}(y_{U}^{T}y_{U}^{*})_{ij},
\]
once Eq. (\ref{eq:C_eu tree}) is taken into account. 

\section{A SMEFT dimension-six Green's basis}

As discussed in Sec. \ref{sec:Matching}, a Green's basis of fixed
dimension $d$ consists of a maximal set of $d$-dimensional operators
independent by linear combinations and addition of total divergences.
The strategy used in Ref. \cite{Gherardi:2020det} for obtaining a
SMEFT dimension-six Green's basis is mainly a re-adaptation of the
line of reasoning of Ref. \cite{Grzadkowski:2010es} (to which we
refer the reader for further clarification), with the important exception
that field redefinitions are not allowed for removing operator redundancies.
We simply examine all possible Lorentz-invariant combinations of gauge
field strengths, covariant derivatives, Standard Model fermions and
the Higgs field, denoted $X$, $D$, $\psi$ and $H$ respectively.
Tables \ref{tab:Bosonic-Green's-basis}-\ref{tab:BLviol-Green's-basis}
list the operators of a dimension-six SMEFT Green's basis obtained
in this way, defined in such a way that the Warsaw basis forms a proper
subset of it.
\begin{center}
\renewcommand{\arraystretch}{1.3}
\begin{table}[H]
\begin{centering}
\begin{tabular}{|c|c||c|c||c|c|}
\hline 
\multicolumn{2}{|c||}{$X^{3}$} & \multicolumn{2}{c||}{$X^{2}H^{2}$} & \multicolumn{2}{c|}{$H^{2}D^{4}$}\tabularnewline
\hline 
\textcolor{blue}{$\mathcal{O}_{3G}$} & \textcolor{blue}{$f^{ABC}G_{\mu}^{A\nu}G_{\nu}^{B\rho}G_{\rho}^{C\mu}$} & \textcolor{blue}{$\mathcal{O}_{HG}$} & \textcolor{blue}{$G_{\mu\nu}^{A}G^{A\mu\nu}(H^{\dagger}H)$} & $\mathcal{O}_{DH}$ & $(D_{\mu}D^{\mu}H)^{\dagger}(D_{\nu}D^{\nu}H)$\tabularnewline
\cline{5-6} \cline{6-6} 
\textcolor{blue}{$\mathcal{O}_{3\widetilde{G}}$} & \textcolor{blue}{$f^{ABC}\widetilde{G}_{\mu}^{A\nu}G_{\nu}^{B\rho}G_{\rho}^{C\mu}$} & \textcolor{blue}{$\mathcal{O}_{H\widetilde{G}}$} & \textcolor{blue}{$\widetilde{G}_{\mu\nu}^{A}G^{A\mu\nu}(H^{\dagger}H)$} & \multicolumn{2}{c|}{$H^{4}D^{2}$}\tabularnewline
\cline{5-6} \cline{6-6} 
\textcolor{blue}{$\mathcal{O}_{3W}$} & \textcolor{blue}{$\epsilon^{IJK}W_{\mu}^{I\nu}W_{\nu}^{J\rho}W_{\rho}^{K\mu}$} & \textcolor{blue}{$\mathcal{O}_{HW}$} & \textcolor{blue}{$W_{\mu\nu}^{I}W^{I\mu\nu}(H^{\dagger}H)$} & \textcolor{blue}{$\mathcal{O}_{H\square}$} & \textcolor{blue}{$(H^{\dagger}H)\square(H^{\dagger}H)$}\tabularnewline
\textcolor{blue}{$\mathcal{O}_{3\widetilde{W}}$} & \textcolor{blue}{$\epsilon^{IJK}\widetilde{W}_{\mu}^{I\nu}W_{\nu}^{J\rho}W_{\rho}^{K\mu}$} & \textcolor{blue}{$\mathcal{O}_{H\widetilde{W}}$} & \textcolor{blue}{$\widetilde{W}_{\mu\nu}^{I}W^{I\mu\nu}(H^{\dagger}H)$} & \textcolor{blue}{$\mathcal{O}_{HD}$} & \textcolor{blue}{$(H^{\dagger}D^{\mu}H)^{\dagger}(H^{\dagger}D_{\mu}H)$}\tabularnewline
\cline{1-2} \cline{2-2} 
\multicolumn{2}{|c||}{$X^{2}D^{2}$} & \textcolor{blue}{$\mathcal{O}_{HB}$} & \textcolor{blue}{$B_{\mu\nu}B^{\mu\nu}(H^{\dagger}H)$} & $\mathcal{O}_{HD}^{\prime}$ & $(H^{\dagger}H)(D_{\mu}H)^{\dagger}(D^{\mu}H)$\tabularnewline
\cline{1-2} \cline{2-2} 
$\mathcal{O}_{2G}$ & $-\frac{1}{2}(D_{\mu}G^{A\mu\nu})(D^{\rho}G_{\rho\nu}^{A})$ & \textcolor{blue}{$\mathcal{O}_{H\widetilde{B}}$} & \textcolor{blue}{$\widetilde{B}_{\mu\nu}B^{\mu\nu}(H^{\dagger}H)$} & $\mathcal{O}_{HD}^{\prime\prime}$ & $(H^{\dagger}H)D_{\mu}(H^{\dagger}i\overleftrightarrow{D}^{\mu}H)$\tabularnewline
\cline{5-6} \cline{6-6} 
$\mathcal{O}_{2W}$ & $-\frac{1}{2}(D_{\mu}W^{I\mu\nu})(D^{\rho}W_{\rho\nu}^{I})$ & \textcolor{blue}{$\mathcal{O}_{HWB}$} & \textcolor{blue}{$W_{\mu\nu}^{I}B^{\mu\nu}(H^{\dagger}\tau^{I}H)$} & \multicolumn{2}{c|}{$H^{6}$}\tabularnewline
\cline{5-6} \cline{6-6} 
$\mathcal{O}_{2B}$ & $-\frac{1}{2}(\partial_{\mu}B^{\mu\nu})(\partial^{\rho}B_{\rho\nu})$ & \textcolor{blue}{$\mathcal{O}_{H\widetilde{W}B}$} & \textcolor{blue}{$\widetilde{W}_{\mu\nu}^{I}B^{\mu\nu}(H^{\dagger}\tau^{I}H)$} & \textcolor{blue}{$\mathcal{O}_{H}$} & \textcolor{blue}{$(H^{\dagger}H)^{3}$}\tabularnewline
\cline{3-4} \cline{4-4} 
 &  & \multicolumn{2}{c||}{$H^{2}XD^{2}$} &  & \tabularnewline
\cline{3-4} \cline{4-4} 
 &  & $\mathcal{O}_{WDH}$ & $D_{\nu}W^{I\mu\nu}(H^{\dagger}i\overleftrightarrow{D}_{\mu}^{I}H)$ &  & \tabularnewline
 &  & $\mathcal{O}_{BDH}$ & $\partial_{\nu}B^{\mu\nu}(H^{\dagger}i\overleftrightarrow{D}_{\mu}H)$ &  & \tabularnewline
\hline 
\end{tabular}
\par\end{centering}
\caption{\label{tab:Bosonic-Green's-basis}Bosonic Green's basis operators.
Operators colored in blue are also included in the Warsaw basis.}
\end{table}
\renewcommand{\arraystretch}{1.0}
\par\end{center}

\pagebreak{}
\begin{center}
\renewcommand{\arraystretch}{1.3}
\begin{table}[H]
\begin{centering}
\begin{tabular}{|c|c||c|c||c|c|}
\hline 
\multicolumn{2}{|c|}{$\psi^{2}D^{3}$} & \multicolumn{2}{c|}{$\psi^{2}XD$} & \multicolumn{2}{c|}{$\psi^{2}DH^{2}$}\tabularnewline
\hline 
$\mathcal{O}_{qD}$ & $\frac{i}{2}\overline{q}\left\{ D_{\mu}D^{\mu},\slashed D\right\} q$ & $\mathcal{O}_{Gq}$ & $(\overline{q}T^{A}\gamma^{\mu}q)D^{\nu}G_{\mu\nu}^{A}$ & \textcolor{blue}{$\mathcal{O}_{Hq}^{(1)}$} & \textcolor{blue}{$(\overline{q}\gamma^{\mu}q)(H^{\dagger}i\overleftrightarrow{D}_{\mu}H)$}\tabularnewline
$\mathcal{O}_{uD}$ & $\frac{i}{2}\overline{u}\left\{ D_{\mu}D^{\mu},\slashed D\right\} u$ & $\mathcal{O}_{Gq}^{\prime}$ & $\frac{1}{2}(\overline{q}T^{A}\gamma^{\mu}i\overleftrightarrow{D}^{\nu}q)G_{\mu\nu}^{A}$ & $\mathcal{O}_{Hq}^{\prime(1)}$ & $(\overline{q}i\overleftrightarrow{\slashed D}q)(H^{\dagger}H)$\tabularnewline
$\mathcal{O}_{dD}$ & $\frac{i}{2}\overline{d}\left\{ D_{\mu}D^{\mu},\slashed D\right\} d$ & $\mathcal{O}_{\widetilde{G}q}^{\prime}$ & $\frac{1}{2}(\overline{q}T^{A}\gamma^{\mu}i\overleftrightarrow{D}^{\nu}q)\widetilde{G}_{\mu\nu}^{A}$ & $\mathcal{O}_{Hq}^{\prime\prime(1)}$ & $(\overline{q}\gamma^{\mu}q)\partial_{\mu}(H^{\dagger}H)$\tabularnewline
$\mathcal{O}_{\ell D}$ & $\frac{i}{2}\overline{\ell}\left\{ D_{\mu}D^{\mu},\slashed D\right\} \ell$ & $\mathcal{O}_{Wq}$ & $(\overline{q}\tau^{I}\gamma^{\mu}q)D^{\nu}W_{\mu\nu}^{I}$ & \textcolor{blue}{$\mathcal{O}_{Hq}^{(3)}$} & \textcolor{blue}{$(\overline{q}\tau^{I}\gamma^{\mu}q)(H^{\dagger}i\overleftrightarrow{D}_{\mu}^{I}H)$}\tabularnewline
$\mathcal{O}_{eD}$ & $\frac{i}{2}\overline{e}\left\{ D_{\mu}D^{\mu},\slashed D\right\} e$ & $\mathcal{O}_{Wq}^{\prime}$ & $\frac{1}{2}(\overline{q}\tau^{I}\gamma^{\mu}i\overleftrightarrow{D}^{\nu}q)W_{\mu\nu}^{I}$ & $\mathcal{O}_{Hq}^{\prime(3)}$ & $(\overline{q}i\overleftrightarrow{\slashed D}^{I}q)(H^{\dagger}\tau^{I}H)$\tabularnewline
\cline{1-2} \cline{2-2} 
\multicolumn{2}{|c||}{$\psi^{2}HD^{2}+\text{h.c.}$} & $\mathcal{O}_{\widetilde{W}q}^{\prime}$ & $\frac{1}{2}(\overline{q}\tau^{I}\gamma^{\mu}i\overleftrightarrow{D}^{\nu}q)\widetilde{W}_{\mu\nu}^{I}$ & $\mathcal{O}_{Hq}^{\prime\prime(3)}$ & $(\overline{q}\tau^{I}\gamma^{\mu}q)D_{\mu}(H^{\dagger}\tau^{I}H)$\tabularnewline
\cline{1-2} \cline{2-2} 
$\mathcal{O}_{uHD1}$ & $(\overline{q}u)D_{\mu}D^{\mu}\widetilde{H}$ & $\mathcal{O}_{Bq}$ & $(\overline{q}\gamma^{\mu}q)\partial^{\nu}B_{\mu\nu}$ & \textcolor{blue}{$\mathcal{O}_{Hu}$} & \textcolor{blue}{$(\overline{u}\gamma^{\mu}u)(H^{\dagger}i\overleftrightarrow{D}_{\mu}H)$}\tabularnewline
$\mathcal{O}_{uHD2}$ & $(\overline{q}i\sigma_{\mu\nu}D^{\mu}u)D^{\nu}\widetilde{H}$ & $\mathcal{O}_{Bq}^{\prime}$ & $\frac{1}{2}(\overline{q}\gamma^{\mu}i\overleftrightarrow{D}^{\nu}q)B_{\mu\nu}$ & $\mathcal{O}_{Hu}^{\prime}$ & $(\overline{u}i\overleftrightarrow{\slashed D}u)(H^{\dagger}H)$\tabularnewline
$\mathcal{O}_{uHD3}$ & $(\overline{q}D_{\mu}D^{\mu}u)\widetilde{H}$ & $\mathcal{O}_{\widetilde{B}q}^{\prime}$ & $\frac{1}{2}(\overline{q}\gamma^{\mu}i\overleftrightarrow{D}^{\nu}q)\widetilde{B}_{\mu\nu}$ & $\mathcal{O}_{Hu}^{\prime\prime}$ & $(\overline{u}\gamma^{\mu}u)\partial_{\mu}(H^{\dagger}H)$\tabularnewline
$\mathcal{O}_{uHD4}$ & $(\overline{q}D_{\mu}u)D^{\mu}\widetilde{H}$ & $\mathcal{O}_{Gu}$ & $(\overline{u}T^{A}\gamma^{\mu}u)D^{\nu}G_{\mu\nu}^{A}$ & \textcolor{blue}{$\mathcal{O}_{Hd}$} & \textcolor{blue}{$(\overline{d}\gamma^{\mu}d)(H^{\dagger}i\overleftrightarrow{D}_{\mu}H)$}\tabularnewline
$\mathcal{O}_{dHD1}$ & $(\overline{q}d)D_{\mu}D^{\mu}H$ & $\mathcal{O}_{Gu}^{\prime}$ & $\frac{1}{2}(\overline{u}T^{A}\gamma^{\mu}i\overleftrightarrow{D}^{\nu}u)G_{\mu\nu}^{A}$ & $\mathcal{O}_{Hd}^{\prime}$ & $(\overline{d}i\overleftrightarrow{\slashed D}d)(H^{\dagger}H)$\tabularnewline
$\mathcal{O}_{dHD2}$ & $(\overline{q}i\sigma_{\mu\nu}D^{\mu}d)D^{\nu}H$ & $\mathcal{O}_{\widetilde{G}u}^{\prime}$ & $\frac{1}{2}(\overline{u}T^{A}\gamma^{\mu}i\overleftrightarrow{D}^{\nu}u)\widetilde{G}_{\mu\nu}^{A}$ & $\mathcal{O}_{Hd}^{\prime\prime}$ & $(\overline{d}\gamma^{\mu}d)\partial_{\mu}(H^{\dagger}H)$\tabularnewline
$\mathcal{O}_{dHD3}$ & $(\overline{q}D_{\mu}D^{\mu}d)H$ & $\mathcal{O}_{Bu}$ & $(\overline{u}\gamma^{\mu}u)\partial^{\nu}B_{\mu\nu}$ & \textcolor{blue}{$\mathcal{O}_{Hud}$} & \textcolor{blue}{$(\overline{u}\gamma^{\mu}d)(\widetilde{H}^{\dagger}iD_{\mu}H)$}\tabularnewline
$\mathcal{O}_{dHD4}$ & $(\overline{q}D_{\mu}d)D^{\mu}H$ & $\mathcal{O}_{Bu}^{\prime}$ & $\frac{1}{2}(\overline{u}\gamma^{\mu}i\overleftrightarrow{D}^{\nu}u)B_{\mu\nu}$ & \textcolor{blue}{$\mathcal{O}_{H\ell}^{(1)}$} & \textcolor{blue}{$(\overline{\ell}\gamma^{\mu}\ell)(H^{\dagger}i\overleftrightarrow{D}_{\mu}H)$}\tabularnewline
$\mathcal{O}_{eHD1}$ & $(\overline{\ell}e)D_{\mu}D^{\mu}H$ & $\mathcal{O}_{\widetilde{B}u}^{\prime}$ & $\frac{1}{2}(\overline{u}\gamma^{\mu}i\overleftrightarrow{D}^{\nu}u)\widetilde{B}_{\mu\nu}$ & $\mathcal{O}_{H\ell}^{\prime(1)}$ & $(\overline{\ell}i\overleftrightarrow{\slashed D}\ell)(H^{\dagger}H)$\tabularnewline
$\mathcal{O}_{eHD2}$ & $(\overline{\ell}i\sigma_{\mu\nu}D^{\mu}e)D^{\nu}H$ & $\mathcal{O}_{Gd}$ & $(\overline{d}T^{A}\gamma^{\mu}d)D^{\nu}G_{\mu\nu}^{A}$ & $\mathcal{O}_{H\ell}^{\prime\prime(1)}$ & $(\overline{\ell}\gamma^{\mu}\ell)\partial_{\mu}(H^{\dagger}H)$\tabularnewline
$\mathcal{O}_{eHD3}$ & $(\overline{\ell}D_{\mu}D^{\mu}e)H$ & $\mathcal{O}_{Gd}^{\prime}$ & $\frac{1}{2}(\overline{d}T^{A}\gamma^{\mu}i\overleftrightarrow{D}^{\nu}d)G_{\mu\nu}^{A}$ & \textcolor{blue}{$\mathcal{O}_{H\ell}^{(3)}$} & \textcolor{blue}{$(\overline{\ell}\tau^{I}\gamma^{\mu}\ell)(H^{\dagger}i\overleftrightarrow{D}_{\mu}^{I}H)$}\tabularnewline
$\mathcal{O}_{eHD4}$ & $(\overline{\ell}D_{\mu}e)D^{\mu}H$ & $\mathcal{O}_{\widetilde{G}d}^{\prime}$ & $\frac{1}{2}(\overline{d}T^{A}\gamma^{\mu}i\overleftrightarrow{D}^{\nu}d)\widetilde{G}_{\mu\nu}^{A}$ & $\mathcal{O}_{H\ell}^{\prime(3)}$ & $(\overline{\ell}i\overleftrightarrow{\slashed D}^{I}\ell)(H^{\dagger}\tau^{I}H)$\tabularnewline
\cline{1-2} \cline{2-2} 
\multicolumn{2}{|c||}{$\psi^{2}XH+\text{h.c.}$} & $\mathcal{O}_{Bd}$ & $(\overline{d}\gamma^{\mu}d)\partial^{\nu}B_{\mu\nu}$ & $\mathcal{O}_{H\ell}^{\prime\prime(3)}$ & $(\overline{\ell}\tau^{I}\gamma^{\mu}\ell)\partial_{\mu}(H^{\dagger}H)$\tabularnewline
\cline{1-2} \cline{2-2} 
\textcolor{blue}{$\mathcal{O}_{uG}$} & \textcolor{blue}{$(\overline{q}T^{A}\sigma^{\mu\nu}u)\widetilde{H}G_{\mu\nu}^{A}$} & $\mathcal{O}_{Bd}^{\prime}$ & $\frac{1}{2}(\overline{d}\gamma^{\mu}i\overleftrightarrow{D}^{\nu}d)B_{\mu\nu}$ & \textcolor{blue}{$\mathcal{O}_{He}$} & \textcolor{blue}{$(\overline{e}\gamma^{\mu}e)(H^{\dagger}i\overleftrightarrow{D}_{\mu}H)$}\tabularnewline
\textcolor{blue}{$\mathcal{O}_{uW}$} & \textcolor{blue}{$(\overline{q}\sigma^{\mu\nu}u)\tau^{I}\widetilde{H}W_{\mu\nu}^{I}$} & $\mathcal{O}_{\widetilde{B}d}^{\prime}$ & $\frac{1}{2}(\overline{d}\gamma^{\mu}i\overleftrightarrow{D}^{\nu}d)\widetilde{B}_{\mu\nu}$ & $\mathcal{O}_{He}^{\prime}$ & $(\overline{e}i\overleftrightarrow{\slashed D}e)(H^{\dagger}H)$\tabularnewline
\textcolor{blue}{$\mathcal{O}_{uB}$} & \textcolor{blue}{$(\overline{q}\sigma^{\mu\nu}u)\widetilde{H}B_{\mu\nu}$} & $\mathcal{O}_{W\ell}$ & $(\overline{\ell}\tau^{I}\gamma^{\mu}\ell)D^{\nu}W_{\mu\nu}^{I}$ & $\mathcal{O}_{He}^{\prime\prime}$ & $(\overline{e}\gamma^{\mu}e)\partial_{\mu}(H^{\dagger}H)$\tabularnewline
\cline{5-6} \cline{6-6} 
\textcolor{blue}{$\mathcal{O}_{dG}$} & \textcolor{blue}{$(\overline{q}T^{A}\sigma^{\mu\nu}d)HG_{\mu\nu}^{A}$} & $\mathcal{O}_{W\ell}^{\prime}$ & $\frac{1}{2}(\overline{\ell}\tau^{I}\gamma^{\mu}i\overleftrightarrow{D}^{\nu}\ell)W_{\mu\nu}^{I}$ & \multicolumn{2}{c|}{$\psi^{2}H^{3}+\text{h.c.}$}\tabularnewline
\cline{5-6} \cline{6-6} 
\textcolor{blue}{$\mathcal{O}_{dW}$} & \textcolor{blue}{$(\overline{q}\sigma^{\mu\nu}d)\tau^{I}HW_{\mu\nu}^{I}$} & $\mathcal{O}_{\widetilde{W}\ell}^{\prime}$ & $\frac{1}{2}(\overline{\ell}\tau^{I}\gamma^{\mu}i\overleftrightarrow{D}^{\nu}\ell)\widetilde{W}_{\mu\nu}^{I}$ & \textcolor{blue}{$\mathcal{O}_{uH}$} & \textcolor{blue}{$(H^{\dagger}H)\overline{q}\widetilde{H}u$}\tabularnewline
\textcolor{blue}{$\mathcal{O}_{dB}$} & \textcolor{blue}{$(\overline{q}\sigma^{\mu\nu}d)HB_{\mu\nu}$} & $\mathcal{O}_{B\ell}$ & $(\overline{\ell}\gamma^{\mu}\ell)\partial^{\nu}B_{\mu\nu}$ & \textcolor{blue}{$\mathcal{O}_{dH}$} & \textcolor{blue}{$(H^{\dagger}H)\overline{q}Hd$}\tabularnewline
\textcolor{blue}{$\mathcal{O}_{eW}$} & \textcolor{blue}{$(\overline{\ell}\sigma^{\mu\nu}e)\tau^{I}HW_{\mu\nu}^{I}$} & $\mathcal{O}_{B\ell}^{\prime}$ & $\frac{1}{2}(\overline{\ell}\gamma^{\mu}i\overleftrightarrow{D}^{\nu}\ell)B_{\mu\nu}$ & \textcolor{blue}{$\mathcal{O}_{eH}$} & \textcolor{blue}{$(H^{\dagger}H)\overline{\ell}He$}\tabularnewline
\textcolor{blue}{$\mathcal{O}_{eB}$} & \textcolor{blue}{$(\overline{\ell}\sigma^{\mu\nu}e)HB_{\mu\nu}$} & $\mathcal{O}_{\widetilde{B}\ell}^{\prime}$ & $\frac{1}{2}(\overline{\ell}\gamma^{\mu}i\overleftrightarrow{D}^{\nu}\ell)\widetilde{B}_{\mu\nu}$ &  & \tabularnewline
 &  & $\mathcal{O}_{Be}$ & $(\overline{e}\gamma^{\mu}e)\partial^{\nu}B_{\mu\nu}$ &  & \tabularnewline
 &  & $\mathcal{O}_{Be}^{\prime}$ & $\frac{1}{2}(\overline{e}\gamma^{\mu}i\overleftrightarrow{D}^{\nu}e)B_{\mu\nu}$ &  & \tabularnewline
 &  & $\mathcal{O}_{\widetilde{B}e}^{\prime}$ & $\frac{1}{2}(\overline{e}\gamma^{\mu}i\overleftrightarrow{D}^{\nu}e)\widetilde{B}_{\mu\nu}$ &  & \tabularnewline
\hline 
\end{tabular}
\par\end{centering}
\caption{\label{tab:Two-fermion-Green's-basis}Two-fermion Green's basis operators.
Operators colored in blue are also included in the Warsaw basis. Fermion
family indices are omitted.}
\end{table}
\renewcommand{\arraystretch}{1.0}
\par\end{center}

\pagebreak{}
\begin{center}
\renewcommand{\arraystretch}{1.3}
\begin{table}[H]
\begin{centering}
\begin{tabular}{|c|c||c|c||c|c|}
\hline 
\multicolumn{2}{|c|}{Four quark} & \multicolumn{2}{c|}{Four lepton} & \multicolumn{2}{c|}{Semileptonic}\tabularnewline
\hline 
\textcolor{blue}{$\mathcal{O}_{qq}^{(1)}$} & \textcolor{blue}{$(\overline{q}\gamma^{\mu}q)(\overline{q}\gamma_{\mu}q)$} & \textcolor{blue}{$\mathcal{O}_{\ell\ell}$} & \textcolor{blue}{$(\overline{\ell}\gamma^{\mu}\ell)(\overline{\ell}\gamma_{\mu}\ell)$} & \textcolor{blue}{$\mathcal{O}_{\ell q}^{(1)}$} & \textcolor{blue}{$(\overline{\ell}\gamma^{\mu}\ell)(\overline{q}\gamma_{\mu}q)$}\tabularnewline
\textcolor{blue}{$\mathcal{O}_{qq}^{(3)}$} & \textcolor{blue}{$(\overline{q}\gamma^{\mu}\sigma^{I}q)(\overline{q}\gamma_{\mu}\sigma^{I}q)$} & \textcolor{blue}{$\mathcal{O}_{ee}$} & \textcolor{blue}{$(\overline{e}\gamma^{\mu}e)(\overline{e}\gamma_{\mu}e)$} & \textcolor{blue}{$\mathcal{O}_{\ell q}^{(3)}$} & \textcolor{blue}{$(\overline{\ell}\gamma^{\mu}\sigma^{I}\ell)(\overline{q}\gamma_{\mu}\sigma^{I}q)$}\tabularnewline
\textcolor{blue}{$\mathcal{O}_{uu}$} & \textcolor{blue}{$(\overline{u}\gamma^{\mu}u)(\overline{u}\gamma_{\mu}u)$} & \textcolor{blue}{$\mathcal{O}_{\ell e}$} & \textcolor{blue}{$(\overline{\ell}\gamma^{\mu}\ell)(\overline{e}\gamma_{\mu}e)$} & \textcolor{blue}{$\mathcal{O}_{eu}$} & \textcolor{blue}{$(\overline{e}\gamma^{\mu}e)(\overline{u}\gamma_{\mu}u)$}\tabularnewline
\textcolor{blue}{$\mathcal{O}_{dd}$} & \textcolor{blue}{$(\overline{d}\gamma^{\mu}d)(\overline{d}\gamma_{\mu}d)$} &  &  & \textcolor{blue}{$\mathcal{O}_{ed}$} & \textcolor{blue}{$(\overline{e}\gamma^{\mu}e)(\overline{d}\gamma_{\mu}d)$}\tabularnewline
\textcolor{blue}{$\mathcal{O}_{ud}^{(1)}$} & \textcolor{blue}{$(\overline{u}\gamma^{\mu}u)(\overline{d}\gamma_{\mu}d)$} &  &  & \textcolor{blue}{$\mathcal{O}_{qe}$} & \textcolor{blue}{$(\overline{q}\gamma^{\mu}q)(\overline{e}\gamma_{\mu}e)$}\tabularnewline
\textcolor{blue}{$\mathcal{O}_{ud}^{(8)}$} & \textcolor{blue}{$(\overline{u}\gamma^{\mu}T^{A}u)(\overline{d}\gamma_{\mu}T^{A}d)$} &  &  & \textcolor{blue}{$\mathcal{O}_{\ell u}$} & \textcolor{blue}{$(\overline{\ell}\gamma^{\mu}\ell)(\overline{u}\gamma_{\mu}u)$}\tabularnewline
\textcolor{blue}{$\mathcal{O}_{qu}^{(1)}$} & \textcolor{blue}{$(\overline{q}\gamma^{\mu}q)(\overline{u}\gamma_{\mu}u)$} &  &  & \textcolor{blue}{$\mathcal{O}_{\ell d}$} & \textcolor{blue}{$(\overline{\ell}\gamma^{\mu}\ell)(\overline{d}\gamma_{\mu}d)$}\tabularnewline
\textcolor{blue}{$\mathcal{O}_{qu}^{(8)}$} & \textcolor{blue}{$(\overline{q}\gamma^{\mu}T^{A}q)(\overline{u}\gamma_{\mu}T^{A}u)$} &  &  & \textcolor{blue}{$\mathcal{O}_{\ell edq}$} & \textcolor{blue}{$(\overline{\ell}e)(\overline{d}q)$}\tabularnewline
\textcolor{blue}{$\mathcal{O}_{qd}^{(1)}$} & \textcolor{blue}{$(\overline{q}\gamma^{\mu}q)(\overline{d}\gamma_{\mu}d)$} &  &  & \textcolor{blue}{$\mathcal{O}_{\ell equ}^{(1)}$} & \textcolor{blue}{$(\overline{\ell}^{r}e)\epsilon_{rs}(\overline{q}^{s}u)$}\tabularnewline
\textcolor{blue}{$\mathcal{O}_{qd}^{(8)}$} & \textcolor{blue}{$(\overline{q}\gamma^{\mu}T^{A}q)(\overline{d}\gamma_{\mu}T^{A}d)$} &  &  & \textcolor{blue}{$\mathcal{O}_{\ell equ}^{(3)}$} & \textcolor{blue}{$(\overline{\ell}^{r}\sigma^{\mu\nu}e)\epsilon_{rs}(\overline{q}^{s}\sigma_{\mu\nu}u)$}\tabularnewline
\textcolor{blue}{$\mathcal{O}_{quqd}^{(1)}$} & \textcolor{blue}{$(\overline{q}^{r}u)\epsilon_{rs}(\overline{q}^{s}d)$} &  &  &  & \tabularnewline
\textcolor{blue}{$\mathcal{O}_{quqd}^{(8)}$} & \textcolor{blue}{$(\overline{q}^{r}T^{A}u)\epsilon_{rs}(\overline{q}^{s}T^{A}d)$} &  &  &  & \tabularnewline
\hline 
\end{tabular}
\par\end{centering}
\caption{\label{tab:BLcons-Green's-basis}Baryon and lepton number conserving
four-fermion Green's basis operators. All operators are included in
the Warsaw basis. Fermion family indices are omitted. Indices $r,\,s,\,p,\,t,\,...$
denote the $\text{SU}(2)_{L}$ fundamental representations.}
\end{table}
\renewcommand{\arraystretch}{1.0}
\par\end{center}

\medskip{}

\begin{center}
\renewcommand{\arraystretch}{1.3}
\begin{table}[H]
\begin{centering}
\begin{tabular}{|c|c|}
\hline 
\multicolumn{2}{|c|}{$B$ and $L$ violating}\tabularnewline
\hline 
\textcolor{blue}{$\mathcal{O}_{duq}$} & \textcolor{blue}{$\varepsilon_{\alpha\beta\gamma}\epsilon_{rs}\left[(d^{\alpha})^{T}\mathscr{C}u^{\beta})\right]\left[(q^{\gamma r})^{T}\mathscr{C}\ell^{s}\right]$}\tabularnewline
\textcolor{blue}{$\mathcal{O}_{qqu}$} & \textcolor{blue}{$\varepsilon_{\alpha\beta\gamma}\epsilon_{rs}\left[(q^{\alpha r})^{T}\mathscr{C}q^{\beta s})\right]\left[(u^{\gamma})^{T}\mathscr{C}e\right]$}\tabularnewline
\textcolor{blue}{$\mathcal{O}_{qqq}$} & \textcolor{blue}{$\varepsilon_{\alpha\beta\gamma}\epsilon_{rs}\epsilon_{pt}\left[(q^{\alpha r})^{T}\mathscr{C}q^{\beta s})\right]\left[(q^{\gamma p})^{T}\mathscr{C}\ell^{t}\right]$}\tabularnewline
\textcolor{blue}{$\mathcal{O}_{duu}$} & \textcolor{blue}{$\varepsilon_{\alpha\beta\gamma}\left[(d^{\alpha})^{T}\mathscr{C}u^{\beta})\right]\left[(u^{\gamma})^{T}\mathscr{C}e\right]$}\tabularnewline
\hline 
\end{tabular}
\par\end{centering}
\caption{\label{tab:BLviol-Green's-basis}Baryon and lepton number violating
four-fermion Green's basis operators. All operators are included in
the Warsaw basis. Fermion family indices are omitted. Indices $r,\,s,\,p,\,t,\,...$
and $a,\,b,\,c,\,...$ denote the $\text{SU}(2)_{L}$ and $\text{SU}(3)_{c}$
fundamental representations, respectively. $C$ is the Dirac charge
conjugation matrix.}
\end{table}
\renewcommand{\arraystretch}{1.0}
\par\end{center}

\pagebreak{}

In order to obtain the reduction equations from the Green's to the
Warsaw basis, one must apply the SM equations of motion \cite{Grzadkowski:2010es}
to the additional (not marked in blue) Green's operators in Tables
\ref{tab:Bosonic-Green's-basis}-\ref{tab:BLviol-Green's-basis},
which results in a set of linear equations in the form $C_{i}=\sum_{j}a_{ij}G_{j}$,
where $C_{i}$ and $G_{j}$ are the Warsaw and Green's basis Wilson
coefficients, respectively, and the $a_{ij}$ are functions of SM
couplings. The full set of $a_{ij}$ coefficients is available from
our original reference \cite{Gherardi:2020det}.

\section{\label{sec:S1S3-potential}Appendix: $S_{1}+S_{3}$ scalar potential}

We report here, for completeness, the expression of the $S_{1}+S_{3}$
scalar potential used in Ref. \cite{Gherardi:2020det}:
\begin{align}
V & =M_{1}^{2}\left|S_{1}\right|^{2}+M_{3}^{2}\left|S_{3}\right|^{2}+\label{eq:S1S3-potential}\\
 & +\lambda_{H1}|H|^{2}|S_{1}|^{2}+\lambda_{H3}|H|^{2}|S_{3}^{I}|^{2}+\left(\lambda_{H13}(H^{\dagger}\sigma^{I}H)S_{3}^{I\dagger}S_{1}+\text{h.c.}\right)+\nonumber \\
 & +\lambda_{\epsilon H3}i\epsilon^{IJK}(H^{\dagger}\sigma^{I}H)S_{3}^{J\dagger}S_{3}^{K}+\nonumber \\
 & +\frac{c_{1}}{2}(S_{1}^{\dagger}S_{1})^{2}+c_{13}^{(1)}(S_{1}^{\dagger}S_{1})(S_{3}^{\dagger}S_{3})+c_{13}^{(8)}(S_{1}^{\dagger}T^{A}S_{1})(S_{3}^{\dagger}T^{A}S_{3})+\nonumber \\
 & +\frac{c_{3}^{(1)}}{2}(S_{3}^{\dagger}S_{3})(S_{3}^{\dagger}S_{3})+\frac{c_{3}^{(3)}}{2}(S_{3}^{I\dagger}\epsilon^{IJK}S_{3}^{J})(S_{3}^{L\dagger}\epsilon^{LMK}S_{3}^{M})+\nonumber \\
 & +\frac{c_{3}^{(5)}}{2}\left[\frac{(S_{3}^{I\dagger}S_{3}^{J})(S_{3}^{I\dagger}S_{3}^{J})+(S_{3}^{I\dagger}S_{3}^{J})(S_{3}^{J\dagger}S_{3}^{I})}{2}-\frac{1}{3}(S_{3}^{\dagger}S_{3})(S_{3}^{\dagger}S_{3})\right]\nonumber 
\end{align}

\part{New Physics signals from flavour: the $B$-anomalies}

\chapter{\label{chap:Flavour-structures in HE-th}Introduction}

The rich variety of phenomena predicted by the SM is due, to a large
extent, to its non-trivial flavour structure. The particular flavour
patterns in physical observables predicted by the SM can be turned
into a powerful probe of NP interactions, which are a priori restricted
by no means to follow the same SM patterns. In this Chapter, we discuss
a set of experimental results, collectively known as $B$-anomalies,
which put into question SM predictions precisely from this point of
view.

\section{\label{sec:The-SM-flavor structure}Low-energy implications of SM
flavour}

The phenomenological peculiarities of SM quarks and leptons originate
from the Yukawa part of the SM lagrangian (cf. Eq. (\ref{eq:L_SM decomposition})):
\begin{equation}
\mathcal{L}_{\text{SM}}^{\text{yuk}}=-(y_{U})_{ij}\overline{q_{i}}\widetilde{H}u{}_{j}-(y_{D})_{ij}\overline{q_{i}}Hd_{j}-(y_{E})_{\alpha\beta}\overline{\ell_{\alpha}}He_{\beta}+\text{h.c.}.\label{eq:SM Yukawa lagrangian}
\end{equation}
In the absence of these terms the whole SM would be symmetric under
a global flavour symmetry group:
\begin{equation}
G_{F}=\text{U}(3)^{5}\equiv\text{U}(3)_{q}\times\text{U}(3)_{u}\times\text{U}(3)_{d}\times\text{U}(3)_{\ell}\times\text{U}(3)_{e},\label{eq:U(3)^5}
\end{equation}
which acts by independent unitary transformations of the SM electroweak
multiplets:
\begin{equation}
\psi_{n}\stackrel{G_{F}}{\mapsto}(U^{\psi}){}_{n}^{\ m}\psi_{m}\qquad(\psi=q,\,u,\,d,\,\ell,\,e).\label{eq:Action U(3)^5}
\end{equation}
It is the breaking of $G_{F}$ induced by $\mathcal{L}_{\text{SM}}^{\text{yuk}}$
which gives rise to the actual properties of quarks and leptons which
we observe in low-energy experiments. For convenience, we will discuss
separately the breakings of the quark $\text{U}(3)_{\mathcal{Q}}^{3}\equiv\text{U}(3)_{q}\times\text{U}(3)_{u}\times\text{U}(3)_{d}$
and lepton $\text{U}(3)_{\mathcal{L}}^{2}\equiv\text{U}(3)_{\ell}\times\text{U}(3)_{e}$
part of $G_{F}$.

Concerning the quark sector, we can assume without loss of generality
that the Yukawa matrices $y_{U,D}$ take the following forms\footnote{The special forms in Eq. (\ref{eq:y_UD in down basis}) can always
be achieved through an appropriate redefinition of the basic quark
fields $q$, $u$, $d$ (this is sometimes referred to in the literature
as the ``down-quark basis''). The field redefinition can be made
to employ $G_{F}$ transformations only, which, by definition of $G_{F}$
itself, leave the rest of the SM lagrangian invariant. }:
\begin{equation}
y_{D}=\text{diag}(y_{d},\,y_{s},\,y_{b}),\qquad y_{U}=V^{\dagger}\text{diag}(y_{u},\,y_{c},\,y_{t}),\label{eq:y_UD in down basis}
\end{equation}
where $V$ can be identified with the Cabibbo-Kobayashi-Maskawa (CKM)
matrix, and the singular values $y_{\bullet}$ are proportional to
quark masses. The minimal residual symmetry group resulting from Eqs.
(\ref{eq:y_UD in down basis}) is a $\text{U}(1)_{\mathcal{B}}$ group,
whose corresponding conserved charge is the total baryon number $\mathcal{B}$,
and this is what is actually realized in Nature: since all quark masses
are different, and $V$ is experimentally known to be non-trivial
(\emph{i.e. }not equivalent to the identity matrix), the only subgroup
of $G_{F}$ which leaves $y_{D}$ and $y_{U}$ simultaneously invariant
consists of $\text{U}(1)$ transformations of the form:
\begin{equation}
q\to e^{i\alpha}q,\qquad u\to e^{i\alpha}u,\qquad d\to e^{i\alpha}d.\label{eq:Baryon number transformations}
\end{equation}
Quark masses $m_{\bullet}\propto y_{\bullet}$ and mixing parameters
contained in the CKM matrix $V$ give rise to the different phenomenological
properties of the six (up and down) quarks.

A corresponding discussion for the lepton sector is unavoidably made
somewhat fuzzier by our limited knowledge regarding the mechanism
generating neutrino masses. The charged lepton Yukawa matrix in Eq.
(\ref{eq:SM Yukawa lagrangian}) can, again without loss of generality,
be taken to be diagonal:
\begin{equation}
y_{E}=\text{diag}(y_{e},\,y_{\mu},\,y_{\tau}),\label{eq:y_E in charged lepton basis}
\end{equation}
and the resulting minimal residual symmetry group is $\text{U}(1)_{e}\times\text{U}(1)_{\mu}\times\text{U}(1)_{\tau}$,
whose corresponding conserved charges are the total numbers of electrons,
muons and tauons. Here, the only source of breaking of the original
$\text{U}(3)_{\mathcal{L}}^{2}$ are the differences in lepton masses
$m_{e,\mu,\tau}\propto y_{e,\mu,\tau}$. This description is, of course,
incomplete, for it does not take into account neutrino masses and
mixings, whose well-established measurements \cite{NuFITv50} provide
conclusive evidence for the violation of individual (flavour specific)
lepton numbers, and arguably also hint to total lepton number violation
\cite{Weinberg:1979sa}. However, the extra (external to the SM) sources
of this further breaking are expected to be feebly coupled to the
SM, either because of heavy mediators or small couplings, as suggested
by the extreme smallness of neutrino masses. As a consequence, in
many experimental settings, the SM description with massless neutrinos
is sufficient for all practical purposes.

An important SM prediction concerning leptonic flavour observables
is Lepton Flavor Universality\emph{ }(LFU), which is the mere observation
that the full leptonic $\text{U}(3)_{\mathcal{L}}^{2}$ symmetry is
restored in the limit of vanishing lepton masses\footnote{The same observation is, of course, valid for quark flavour observables.
For example, the SM decay widths $\Gamma(Z\to q\overline{q})$ are
essentially equal for all light down quarks $q=d,\,s,\,b$, an instance
of ``Down Quark Flavor Universality''.}. Such a symmetry manifests itself in actual experiments as a degeneracy
in physical processes involving different charged leptons and/or neutrinos
in their final state. As a relevant example, consider the differential
branching fractions:
\begin{equation}
\frac{\text{d}}{\text{d}q^{2}}\text{Br}(B\to K\ell^{+}\ell^{-}),\qquad(\ell=e,\,\mu),\label{eq:dB/dq=0000B2 R_K}
\end{equation}
where $q^{2}$ denotes the invariant mass of the charged lepton pair.
This process is at the core of neutral current $B$-anomalies, which
we discuss below. For sufficiently large $q^{2}$, say $q^{2}\geq(1\,\text{GeV})^{2}$,
the final state lepton masses are practically negligible, and it turns
out that:
\begin{equation}
\frac{\text{d}}{\text{d}q^{2}}\text{Br}(B\to Ke^{+}e^{-})\approx\frac{\text{d}}{\text{d}q^{2}}\text{Br}(B\to K\mu^{+}\mu^{-})\label{eq:LFU in B -> K l l}
\end{equation}
(we will formalize the approximate equalities in Eq. (\ref{eq:LFU in B -> K l l})
in the following Section).

\subsection{\label{subsec:Flavor-from-symmetries}Flavor from symmetries}

We make a small digression to discuss some prominent ideas connected
with the SM flavour structure, and with the quark sector in particular.
Explaining the SM flavour structure is the main topic of Chapter \ref{chap:Explaining-the-SM},
but the ideas presented here also play an important role in model
building for SM deviations.

The SM breaking pattern $\text{U}(3)_{\mathcal{Q}}^{3}\to\text{U}(1)_{\mathcal{B}}$,
parametrized by quark masses and the CKM matrix, works remarkably
well from the phenomenological point of view, perhaps\emph{ }beyond
reasonable expectations, if one takes the view that there exists NP
lying not too far from the TeV scale, \emph{and} whose quark flavour
structure is completely unrelated to the SM one. A patent example
of this fact is provided by the bounds on NP Wilson coefficients coming
from $\Delta F=2$ neutral meson mixing processes \cite{Bona:2007vi,Bona:2017cxr}:
if one assumes arbitrary NP flavour structures, these bounds constrain
the $\Delta F=2$ effective operator scales to orders of magnitude
such as $10^{3}\div10^{5}\,\text{TeV}$, compared to the $1\div10^{2}\,\text{TeV}$
scales which are affordable for an SM-like flavour structure. The
conclusion is that flavour precision observables (such as $\Delta F=2$
amplitudes) naturally require low-scale SM extensions to align, to
some extent, to the SM flavour structure.

A theoretically robust way of obtaining such an alignment makes use
of \emph{flavour symmetries}: one postulates that some subgroup of
$\text{U}(3)_{\mathcal{Q}}^{3}$ is an actual high-energy, spontaneously
broken symmetry group, and tries to correlate the SM and NP flavour
structures by assuming that both these arise from a shared, limited
set of fields with symmetry breaking expectation values, called flavons.
An example is provided by the Minimal Flavor Violation (MFV) framework
\cite{DAmbrosio:2002vsn}, in which the symmetry group is the whole
$\text{U}(3)_{\mathcal{Q}}^{3}$, and the flavons are the two SM Yukawa
matrices themselves, which carry the following representations: 
\begin{equation}
y_{U}\sim\boldsymbol{3}_{q}\otimes\overline{\boldsymbol{3}}_{u},\qquad y_{D}\sim\boldsymbol{3}_{q}\otimes\overline{\boldsymbol{3}}_{d},\qquad(\text{\text{U}(3\ensuremath{)_{\mathcal{Q}}^{3}}})\label{eq:MFV Yukawa q-numbers}
\end{equation}
In MFV, $y_{U}$ and $y_{D}$ are assumed to be the only sources of
$\text{U}(3)_{\mathcal{Q}}^{3}$ breaking for the SM \emph{and }for
NP, which provides the desired alignment in, \emph{e.g.}, $\Delta F=2$
amplitudes. A less stringent example is provided by $\text{U}(2)_{\mathcal{Q}}^{3}$
symmetry \cite{Barbieri:2011ci}, which only acts on light generation
fermions. One can again decompose the Yukawa matrices in terms of
$\text{U}(2)_{\mathcal{Q}}^{3}$ representations:
\begin{equation}
y_{U}\sim\boldsymbol{1}\oplus\boldsymbol{2}_{q}\oplus\boldsymbol{2}_{u}\oplus\boldsymbol{2}_{q}\otimes\boldsymbol{2}_{u},\quad y_{D}\sim\boldsymbol{1}\oplus\boldsymbol{2}_{q}\oplus\boldsymbol{2}_{d}\oplus\boldsymbol{2}_{q}\otimes\boldsymbol{2}_{d}\qquad(\text{U}(2)_{\mathcal{Q}}^{3}).\label{eq:U(2)quark Yukawa q-numbers}
\end{equation}
Actually, in the minimal $\text{U}(2)_{\mathcal{Q}}^{3}$ setup \cite{Buttazzo:2017ixm},
we can restrict ourselves to the following set of flavons:
\begin{equation}
\Delta_{u(d)}\sim\boldsymbol{2}_{q}\otimes\boldsymbol{2}_{u(d)},\qquad\boldsymbol{V}_{q}\sim\boldsymbol{2}_{q},\label{eq:Flavons U(2)^3}
\end{equation}
in terms of which the Yukawa matrices are given by:
\begin{equation}
y_{U}=y_{t}\begin{pmatrix}\Delta_{u} & c_{U}\boldsymbol{V}_{q}\\
0 & 1
\end{pmatrix},\qquad y_{D}=y_{b}\begin{pmatrix}\Delta_{d} & c_{D}\boldsymbol{V}_{q}\\
0 & 1
\end{pmatrix}.\label{eq:yU,D minimal U(2)}
\end{equation}
NP couplings involving quarks must also be expressed in terms of $\Delta_{u,d}$
and $\boldsymbol{V}_{q}$, which again yields correlations between
the SM and NP flavour structure. In fact,\emph{ }for instance, $\boldsymbol{V}_{q}$
in Eq. (\ref{eq:yU,D minimal U(2)}) can be shown to be approximately
proportional to $\left(V_{td}^{*},V_{ts}^{*}\right)^{T}$ in the down
quark mass basis \cite{Buttazzo:2017ixm}.

The above discussion can be extended as a whole (including the MFV
and $\text{U}(2)$ symmetry examples) to the leptonic sector, in which
case, however, correlations tend to be looser than in the quark case,
the main source of uncertainty being again the unknown neutrino mass
generation mechanism.

\section{\label{sec:B-anomalies}The $B$-meson decay anomalies}

The SM provides an excellent description of physical phenomena in
a wide range of energies and scales. Despite no direct evidence for
new physics emerged in direct searches at the LHC, for several years
now some low energy measurements continue to show significant deviations
from the respective SM predictions, which fuel the hope that some
New Physics (NP) might be lurking somewhere at the TeV scale. In this
Chapter, we discuss a set of measurements in $B$-meson decays, which
exhibit a very explicit tension with SM predictions and, in particular,
with LFU. 

Specifically, important deviations from the SM have been observed
within the following three set of observables:
\begin{itemize}
\item The neutral current LFU ratios \cite{Aaij:2013qta,Aaij:2014ora,Aaij:2017vbb,Aaij:2019wad,Abdesselam:2019wac,LHCb:2021trn}:
\begin{equation}
R(K^{(*)})=\left.\frac{\text{Br}(B\to K^{(*)}\mu^{+}\mu^{-})}{\text{Br}(B\to K^{(*)}e^{+}e^{-})}\right|_{q^{2}\in\left[1.1,\,6\right]\,\text{GeV}^{2}}.\label{eq:R(K)}
\end{equation}
Here $q^{2}$ denotes the invariant mass of the dilepton pairs in
the final states (more on this below).
\item The differential angular distribution in $B\to K^{*}\mu^{+}\mu^{-}$,
as well as several branching fractions of $b\to s\mu^{+}\mu^{-}$
processes \cite{Aaij:2015oid,Aaij:2015esa,Aaij:2017vad,Aaboud:2018mst,Aaij:2020nrf}.
\item The charged current LFU ratios \cite{Lees:2012xj,Lees:2013uzd,Aaij:2015yra,Huschle:2015rga,Sato:2016svk,Hirose:2016wfn,Hirose:2017dxl,Aaij:2017uff,Aaij:2017deq,Siddi:2018avt,Belle:2019rba}:
\begin{equation}
R(D^{(*)})=\frac{\text{Br}(B\to D^{(*)}\tau\nu)}{\text{Br}(B\to D^{(*)}l\nu)}\qquad(l=\mu,\,e).\label{eq:R(D)}
\end{equation}
\end{itemize}
While the muon specific observables of the second point provide an
important piece of information for disentangling NP effects in $b\to s\mu^{+}\mu^{-}$
and $b\to se^{+}e^{-}$ in the $R(K^{(*)})$ observables (which, as
we will shortly see, both exhibit a deficiency with respect to SM
predictions), for the present discussion I will mainly focus on the
LFU ratios (\ref{eq:R(K)}) and (\ref{eq:R(D)}). For these observables,
the SM predictions are particularly clean, since the theoretical uncertainties
coming from the hadronic $B\to D^{(*)}$ and $B\to K^{(*)}$ integrated
form factors cancel out in large part in the $R(K^{(*)})$ and $R(D^{(*)})$
ratios, respectively.

Let us first consider the theoretical predictions for Eqs. (\ref{eq:R(K)})
and (\ref{eq:R(D)}). The SM predictions for $R(K^{(*)})$ are particularly
simple:
\begin{equation}
R(K^{(*)})_{\text{SM}}=1\label{eq:R(K) in the SM}
\end{equation}
where the theoretical relative uncertainties, due to neglected electromagnetic
corrections, are of order $\mathcal{O}(1\%)$ \cite{Bordone:2016gaq}
and can be safely ignored for phenomenological purposes. This simple
result can be understood from the viewpoint of LFU: given the considered
energy range (cf. Eq. (\ref{eq:R(K)})), both the final state electrons
and muons can be effectively taken to be massless; therefore, due
to LFU, electrons or muons are both kinematically and dynamically
equivalent in the decay, and the rates for the two different leptonic
channels are equal. As a side note, we observe that the $q^{2}$ bin
in Eq. (\ref{eq:R(K)}) is chosen in such a way that both the $\rho$
($m_{\rho}^{2}\sim0.6\,\text{GeV}^{2})$ and $J/\psi$ ($m_{J/\psi}^{2}\sim9\,\text{GeV}^{2}$)
resonances lie far away from the $q^{2}$ range; this ensures that
the $B\to K^{(*)}\ell^{+}\ell^{-}$ SM decays are dominated by short
distance (\emph{i.e.} electroweak) interactions, whose contributions
can be computed with relatively good accuracy.

The story is slightly more complicated for $R(D^{(*)})$. The SM predictions
read \cite{Amhis:2016xyh}:
\begin{align}
R(D) & =0.299,\label{eq:R(D) SM}\\
R(D^{*}) & =0.258,\label{eq:R(D*) SM}
\end{align}
where we again neglected theoretical uncertainties, which are of order
a few percents. In this case lepton masses are definitely relevant,
since the final states with $\tau$ or $\mu$/$e$ leptons have very
different configuration spaces, and the cancellation between the hadronic
$B\to D^{(*)}$ form factors between the numerators and denominators
of $R(D^{(*)})$ is only partial. In spite of this, the hadronic form
factors can be computed with good accuracy in the SM \cite{Amhis:2016xyh},
and moreover the calculation is in this case free of complications
from long-distance effects.

We now compare the theoretical predictions (\ref{eq:R(K) in the SM}),
(\ref{eq:R(D) SM}) and (\ref{eq:R(D*) SM}) with the corresponding
experimental results, which are collected in Table \ref{tab:Exp LFU ratios}.
We observe in the first place that all experimental relative uncertainties
range between $5\%$ and $10\%$, justifying our neglect of theoretical
uncertainties. The experimental results for $R(K^{(*)})$ both show
a deficit with respect to the SM prediction: if combined with the
$b\to s\mu^{+}\mu^{-}$ data previously mentioned, the global significance
of the deviation is of about $3.9$ standard deviations \cite{Lancierini:2021sdf}\footnote{Such an estimate differs from (and is more robust than) the SM pulls
found by previous fits (see \emph{e.g.} Refs. \cite{Alguero:2019ptt,Alguero:2019aa,Geng:2021nhg,Altmannshofer:2021qrr},
and the additional references in \cite{Lancierini:2021sdf}), which
would seem to report higher significances, in that it does not assume
any special direction in NP Wilson Coefficient space - \emph{i.e.
}it does not enforce a priori any NP alternative hypothesis using
the bias from the $B$-anomalies data.}. The measurements of $R(D^{(*)})$ show, instead, an enhancement
with respect to the SM, with a combined significance of $\approx3$
standard deviations.

One thing to notice is that the quark level transitions underlying
$R(D^{(*)})$ (\emph{i.e. }$b\to c\ell\nu)$ and $R(K^{(*)})$ (\emph{i.e.}
$b\to s\ell\ell$) are a tree-level and a one-loop process in the
SM, respectively. Since, on the other hand, all experimental deviations
are roughly of the same relative order, this implies that in order
to provide a \emph{combined} NP explanation of both anomalies, one
of the two following conditions must hold:
\begin{itemize}
\item The NP mediators giving rise to the deviation in $R(D^{(*)})$ must
be lighter (roughly by a factor of ten) than those contributing to
$R(K^{(*)})$, or
\item The NP effective couplings affecting the $b\to c\ell\nu$ transition
must be enhanced with respect to those affecting $b\to s\ell\ell$.
\end{itemize}
The second scenario is clearly more appealing, if one assumes that
$R(D^{(*)})$ and $R(K^{(*)})$ do not have independent origins, and
can in principle be naturally realized, for instance under the framework
of flavour symmetries \cite{Buttazzo:2017ixm}; a different possibility,
which we also include under the second scenario, is that $R(D^{(*)})$
is modified at the tree-level, whereas $R(K^{(*)})$ is contributed
only at one-loop, see e.g. Refs. \cite{Gripaios:2015gra}\cite{Coy:2019rfr}
for some relevant work in this direction.

\renewcommand{\arraystretch}{1.3}
\begin{table}
\begin{centering}
\begin{tabular}{|c|c|c|}
\hline 
Observable & Experimental value & Ref.\tabularnewline
\hline 
\hline 
$R(K)$ & $0.846_{-0.041}^{+0.044}$ & \cite{LHCb:2021trn}\tabularnewline
\hline 
$R(K^{*})$ & $0.69_{-0.09}^{+0.12}$ & \cite{Aaij:2017vbb,Abdesselam:2019wac}\tabularnewline
\hline 
$R(D)$ & $0.340\pm0.030$ & \cite{Amhis:2016xyh}\tabularnewline
\hline 
$R(D^{*})$ & $0.295\pm0.014$ & \cite{Amhis:2016xyh}\tabularnewline
\hline 
\end{tabular}
\par\end{centering}
\caption{\label{tab:Exp LFU ratios}Experimental results for LFU ratios.}
\end{table}
\renewcommand{\arraystretch}{1.0}

The experimental discrepancies discussed in this Chapter have been
for many years object of great interest from both the phenomenological
and experimental communities. The upcoming years will be critical
in revealing the true nature of the anomalies, which may either be
recognized as genuine New Physics or turn out to be mere statistical
fluctuations and/or to result from uncontrolled systematics. Even
though the precise evolution of statistical significances critically
depends on central values, it is highly likely that the next five/ten
years, with upcoming data from LHCb, Belle II, ATLAS and CMS, should
provide definitive answers in these respects.

\chapter{\label{chap:S1S3-phenomenology}$S_{1}+S_{3}$ model's phenomenology}

We now come back to the leptoquark theory discussed in Sec. \ref{chap:S1S3 leptoquark theory}.
As we already commented there, this model provides one of the most
promising candidate solutions to $B$-anomalies, as well as to the
longstanding $(g-2)_{\mu}$ anomaly \cite{Muong-2:2006rrc,Muong-2:2021ojo,Aoyama:2020ynm}.
This Chapter presents results from my work \cite{Gherardi:2020qhc},
in collaboration with D. Marzocca and E. Venturini, in which we performed
a detailed phenomenological study of the $S_{1}+S_{3}$ model, including
all one-loop contributions to a selected list of observables. The
study leverages on our previous work \cite{Gherardi:2020det}, described
in Chap. \ref{chap:S1S3 leptoquark theory}, in which the full one-loop
matching of the model onto the SMEFT was performed (see Chap. \ref{chap:S1S3 leptoquark theory}
for more details).

The logic behind the $S_{1}+S_{3}$ model is quite simple: it was
early observed that the leptoquarks $S_{1}$ and $S_{3}$ separately
provided a solution to the $R(D^{(*)})$ and $R(K^{(*)})$ anomalies,
respectively, and one might naively think that a model featuring both
of them would provide a combined solution for both anomalies. The
primary purpose of our work was to validate such intuition, through
a global fit which included all observables which could provide relevant
phenomenological constraints.

The goal of Ref. \cite{Gherardi:2020qhc} was to find interesting
scenarios, within the $S_{1}+S_{3}$ setup, capable of addressing
one or more of the anomalies listed above, find the preferred region
in parameter space, and discuss the most important experimental constraints
in each case. Specifically, we first aimed to quantify how well single
leptoquark models are able to address the various anomalies, then
we discussed combined explanations with both leptoquarks. Thanks to
the complete one-loop matching, we also discussed limits on leptoquark
couplings to the SM Higgs boson, arising from electroweak precision
data and Higgs measurements.

We found that models involving only the $S_{1}$ leptoquark can consistently
address $R(D^{(*)})$ and $(g-2)_{\mu}$ anomalies, while a fully-satisfactory
solution for $b\to s\ell\ell$ anomalies is prevented by the combination
of constraints from $B_{s}$-mixing and LFU in $\tau$ decays. Conversely,
the $S_{3}$ leptoquark when taken alone can only address neutral-current
$B$-meson anomalies. A model with both $S_{1}$ and $S_{3}$, and
only left-handed couplings for $S_{1}$, can address both $B$-anomalies
but not the muon magnetic moment. Finally, allowing for right handed
$S_{1}$ couplings makes it possible to fit also $(g-2)_{\mu}$. Ref.
\cite{Gherardi:2020qhc} also examined the prospects for both the
LF conserving branching fraction $\text{Br}(B\to K\tau\tau)$ and
the LFV one $\text{Br}(B\to K\tau\mu)$, which are found to be in
the ballpark of the future expected sensitivity of Belle-II and LHCb. 

The $S_{1}+S_{3}$ model discussed in this Chapter was defined in
Sec. \ref{sec:The-S1S3-model}. In the following Sections, we describe
the fit methodology of Ref. \cite{Gherardi:2020qhc}, the observables
considered in the fit, and the detailed results (including future
prospects) for the various scenarios considered in our work.

\section{Methodology}

Our goal is to study the phenomenology of the $S_{1}+S_{3}$ model
described in the Sec. \ref{sec:The-S1S3-model}, expressing the low-energy
observables as functions of the UV parameters at one-loop level. Given
the separation of scales between the LQ masses, assumed to be at the
TeV scale, and the typical energy scales of the observables considered,
the EFT approach is particularly suited for this goal. In fact, it
allows to separate the complete procedure in a sequence of steps,
which can be generalised to be applicable also to other UV scenarios.
Going from the ultraviolet to the infrared, the matching procedure
allows to pass physical thresholds, i.e. to integrate out heavy fields
while defining a new EFT for that energy range, while the renormalization
group evolution (RGE) allows to change the scale within an EFT approach. 

In our specific case, we have the following steps: 
\begin{itemize}
\item The one-loop matching for the $S_{1}+S_{3}$ model onto the SMEFT,
up to dimension-six operators, resulting by integrating out the two
scalar leptoquarks at a scale of the order of their masses $\mu_{M}\sim M_{1},M_{3}$.
The complete set of matching conditions, obtained with $\overline{\text{MS}}$
renormalization scheme, has been discussed in Chap. \ref{chap:S1S3 leptoquark theory}. 
\item The RGE of the SMEFT Wilson coefficients from the UV matching scale
$\mu_{M}$ down to the electroweak scale \cite{Jenkins:2013zja,Jenkins:2013wua,Alonso:2013hga}.
\item The one-loop matching between the SMEFT and the EFT valid below the
electroweak scale, known as Low Energy EFT (LEFT). This results from
integrating out the Higgs, the massive electroweak gauge bosons and
the top quark and has been done in \cite{Dekens:2019ept}. 
\item The RGE of the LEFT Wilson coefficients \cite{Jenkins:2017dyc} from
the electroweak scale to the relevant scales of the processes; 
\item The expression of the low-energy observables and pseudo-observables
in terms of the LEFT Wilson coefficients, taking into account contributions
that arise at one-loop level within the LEFT, from the operators generated
already at the tree-level.\footnote{In case of observables at the electroweak scale, such as the measurements
of $Z$ couplings, the observables can be expressed directly in terms
of SMEFT Wilson coefficients at the elecroweak scales, so that the
last three steps do not need to be considered.}
\end{itemize}
By combining everything, we obtain expressions for the observables
as a function of the parameters of the scalar leptoquark model at
the TeV scale; in such a way, experimental bounds on low-energy data
can be used to set constraints on the $S_{1,3}$ couplings. On the
other hand, the intermediate steps provide model-independent expressions
for observables in terms of EFT Wilson coefficients, which might be
exploited in other NP scenarios.

For a generic EFT coefficient we can separate a contribution arising
at the tree-level from one arising at one-loop $C_{i}=C_{i}^{(0)}+(4\pi)^{-2}C_{i}^{(1)}$.
Working at one-loop accuracy, the RGE, one-loop matching between SMEFT
and LEFT, and the one-loop matrix elements to the observables, should
only be considered for tree-level generated coefficients, $C_{i}^{(0)}$
(in our case, those in Eqs. (\ref{eq:C_lq1 tree})-(\ref{eq:C_eu tree})).
For the loop-generated coefficients, $C_{i}^{(1)}$, only the tree-level
matching conditions from SMEFT to LEFT, and tree-level matrix elements
should be included, the other contributions giving terms which are
formally of two-loop order and that could be of the same order as
neglected two-loop matching conditions. 

The exception to this is in the RGE due to QCD from the TeV to the
GeV scale, for example in four-quark operators contributing to $\Delta F=2$
observables. In this case the RGE contribution is well known to be
important, also due to the large separation of scales, which gives
to this effect a parametric enhancement with respect to the neglected
two-loop corrections even if four-quark operators are generated at
one-loop. 

\section{Observables}

\begin{table}[t]
\centering{}%
\begin{tabular}{|c|c|c|}
\hline 
Observable & SM prediction & Experimental bounds\tabularnewline
\hline 
\hline 
\multicolumn{3}{|c|}{$b\to s\ell\ell$ observables}\tabularnewline
\hline 
$\Delta\mathcal{C}_{9}^{sb\mu\mu}$ & 0 & $-0.43\pm0.09$ \cite{Aebischer:2019mlg}\tabularnewline
\hline 
$\mathcal{C}_{9}^{\text{univ}}$ & 0 & $-0.48\pm0.24$ \cite{Aebischer:2019mlg}\tabularnewline
\hline 
\hline 
\multicolumn{3}{|c|}{$b\to c\tau(\ell)\nu$ observables}\tabularnewline
\hline 
$R_{D}$ & $0.299\pm0.003$ \cite{Amhis:2016xyh} & $0.34\pm0.027\pm0.013$ \cite{Amhis:2016xyh}\tabularnewline
\hline 
$R_{D}^{*}$ & $0.258\pm0.005$ \cite{Amhis:2016xyh} & $0.295\pm0.011\pm0.008$ \cite{Amhis:2016xyh}\tabularnewline
\hline 
$P_{\tau}^{D^{*}}$ & $-0.488\pm0.018$ \cite{Bordone:2019vic} & $-0.38\pm0.51\pm0.2\pm0.018$ \cite{Hirose:2017dxl}\tabularnewline
\hline 
$F_{L}$ & $0.470\pm0.012$ \cite{Bordone:2019vic} & $0.60\pm0.08\pm0.038\pm0.012$ \cite{Abdesselam:2019wbt}\tabularnewline
\hline 
$\text{Br}(B_{c}^{+}\to\tau^{+}\nu)$ & $2.3\%$ & $<10\%$ (95\% CL) \cite{Akeroyd:2017mhr}\tabularnewline
\hline 
$R_{D}^{\mu/e}$ & 1 & $0.978\pm0.035$ \cite{Aubert:2008yv,Glattauer:2015teq}\tabularnewline
\hline 
\hline 
\multicolumn{3}{|c|}{$D$ leptonic decay}\tabularnewline
\hline 
$\text{Br}(D_{s}\to\tau\nu)$ & $(5.169\pm0.004)\times10^{-2}$~\cite{Aoki:2016frl} & $(5.48\pm0.23)\times10^{-2}$~\cite{Tanabashi:2018oca}\tabularnewline
\hline 
\hline 
\multicolumn{3}{|c|}{$b\to s\nu\nu$ and $s\to d\nu\nu$}\tabularnewline
\hline 
$R_{K}^{\nu}$ & 1 \cite{Buras:2014fpa} & $<4.65$ ~\cite{Grygier:2017tzo}\tabularnewline
\hline 
$R_{K^{*}}^{\nu}$ & 1 \cite{Buras:2014fpa} & $<3.22$ ~\cite{Grygier:2017tzo}\tabularnewline
\hline 
$\text{Br}(K^{+}\to\pi^{+}\nu\nu)$ & $8.64\times10^{-11}$~\cite{Buras:1998raa} & $(11.0\pm4.0)\times10^{-11}$ \cite{CortinaGil:2020vlo}\tabularnewline
\hline 
$\text{Br}(K_{L}\to\pi^{0}\nu\nu)$ & $3.4\times10^{-11}$ ~\cite{Buras:1998raa} & $<3.57\times10^{-9}$ \cite{Ahn:2018mvc}\tabularnewline
\hline 
\hline 
\multicolumn{3}{|c|}{$B$ LFV decays}\tabularnewline
\hline 
$\text{Br}(B_{d}\to\tau^{\pm}\mu^{\mp})$ & 0 & $<1.4\times10^{-5}$ ~\cite{Aaij:2019okb}\tabularnewline
\hline 
$\text{Br}(B_{s}\to\tau^{\pm}\mu^{\mp})$ & 0 & $<4.2\times10^{-5}$ ~\cite{Aaij:2019okb}\tabularnewline
\hline 
$\text{Br}(B^{+}\to K^{+}\tau^{-}\mu^{+})$ & 0 & $<5.4\times10^{-5}$ ~\cite{Lees:2012zz}\tabularnewline
\hline 
\multirow{2}{*}{$\text{Br}(B^{+}\to K^{+}\tau^{+}\mu^{-})$} & \multirow{2}{*}{0} & $<3.3\times10^{-5}$ ~\cite{Lees:2012zz}\tabularnewline
 &  & $<4.5\times10^{-5}$ ~\cite{Aaij:2020mqb}\tabularnewline
\cline{3-3} 
\end{tabular}\caption{\label{tab:S1S3-obs-semileptonic}Low-energy \emph{semileptonic} observables
with their SM predictions and experimental bounds. Upper limits correspond
to 95\%CL.}
\end{table}

\begin{table}[b]
\centering{}%
\begin{tabular}{|c|c|c|}
\hline 
Observable & SM prediction & Experimental bounds\tabularnewline
\hline 
\hline 
\multicolumn{3}{|c|}{$\Delta F=2$ processes}\tabularnewline
\hline 
$B^{0}-\overline{B}^{0}$: $|C_{B_{d}}^{1}|$ & 0 & $<9.11\times10^{-7}$ TeV$^{-2}$ \cite{Bona:2007vi,UTFIT:2016}\tabularnewline
\hline 
$B_{s}^{0}-\overline{B}_{s}^{0}$: $|C_{B_{s}}^{1}|$ & 0 & $<2.01\times10^{-5}$ TeV$^{-2}$ \cite{Bona:2007vi,UTFIT:2016}\tabularnewline
\hline 
$K^{0}-\overline{K}^{0}$: Re{[}$C_{K}^{1}${]} & 0 & $<8.04\times10^{-7}$ TeV$^{-2}$ \cite{Bona:2007vi,UTFIT:2016}\tabularnewline
\hline 
$K^{0}-\overline{K}^{0}$: Im{[}$C_{K}^{1}${]} & 0 & $<2.95\times10^{-9}$ TeV$^{-2}$ \cite{Bona:2007vi,UTFIT:2016}\tabularnewline
\hline 
$D^{0}-\overline{D}^{0}$: Re{[}$C_{D}^{1}${]} & 0 & $<3.57\times10^{-7}$ TeV$^{-2}$ \cite{Bona:2007vi,UTFIT:2016}\tabularnewline
\hline 
$D^{0}-\overline{D}^{0}$: Im{[}$C_{D}^{1}${]} & 0 & $<2.23\times10^{-8}$ TeV$^{-2}$ \cite{Bona:2007vi,UTFIT:2016}\tabularnewline
\hline 
$D^{0}-\overline{D}^{0}$: Re{[}$C_{D}^{4}${]} & 0 & $<3.22\times10^{-8}$ TeV$^{-2}$ \cite{Bona:2007vi,UTFIT:2016}\tabularnewline
\hline 
$D^{0}-\overline{D}^{0}$: Im{[}$C_{D}^{4}${]} & 0 & $<1.17\times10^{-9}$ TeV$^{-2}$ \cite{Bona:2007vi,UTFIT:2016}\tabularnewline
\hline 
$D^{0}-\overline{D}^{0}$: Re{[}$C_{D}^{5}${]} & 0 & $<2.65\times10^{-7}$ TeV$^{-2}$ \cite{Bona:2007vi,UTFIT:2016}\tabularnewline
\hline 
$D^{0}-\overline{D}^{0}$: Im{[}$C_{D}^{5}${]} & 0 & $<1.11\times10^{-8}$ TeV$^{-2}$ \cite{Bona:2007vi,UTFIT:2016}\tabularnewline
\hline 
\hline 
\multicolumn{3}{|c|}{LFU in $\tau$ decays}\tabularnewline
\hline 
$|g_{\mu}/g_{e}|^{2}$ & 1 & $1.0036\pm0.0028$~\cite{Pich:2013lsa}\tabularnewline
\hline 
$|g_{\tau}/g_{\mu}|^{2}$ & 1 & $1.0022\pm0.0030$~\cite{Pich:2013lsa}\tabularnewline
\hline 
$|g_{\tau}/g_{e}|^{2}$ & 1 & $1.0058\pm0.0030$~\cite{Pich:2013lsa}\tabularnewline
\hline 
\hline 
\multicolumn{3}{|c|}{LFV observables}\tabularnewline
\hline 
$\text{Br}(\tau\to\mu\phi)$ & 0 & $<1.00\times10^{-7}$ \cite{Miyazaki:2011xe}\tabularnewline
\hline 
$\text{Br}(\tau\to3\mu)$ & 0 & $<2.5\times10^{-8}$ \cite{Hayasaka:2010np}\tabularnewline
\hline 
$\text{Br}(\mu\to e\gamma)$ & 0 & $<5.00\times10^{-13}$ \cite{TheMEG:2016wtm}\tabularnewline
\hline 
$\text{Br}(\tau\to\mu\gamma)$ & 0 & $<5.24\times10^{-8}$ \cite{Aubert:2009ag}\tabularnewline
\hline 
$\text{Br}(\tau\to e\gamma)$ & 0 & $<3.93\times10^{-8}$ \cite{Aubert:2009ag}\tabularnewline
\hline 
\hline 
\multicolumn{3}{|c|}{EDMs}\tabularnewline
\hline 
$d_{e}$ & $<10^{-44}\,e\,cm$~\cite{Pospelov:2013sca,Smith:2017dtz} & $<1.1\times10^{-29}\,e\,cm$~\cite{Andreev:2018ayy}\tabularnewline
\hline 
$d_{\mu}$ & $<10^{-42}\,e\,cm$~\cite{Smith:2017dtz} & $<1.9\times10^{-19}\,e\,cm$~\cite{Bennett:2008dy}\tabularnewline
\hline 
$d_{\tau}$ & $<10^{-41}\,e\,cm$~\cite{Smith:2017dtz} & $(1.15\pm1.70)\times10^{-17}\,e\,cm$~\cite{Inami:2002ah}\tabularnewline
\hline 
\hline 
\multicolumn{3}{|c|}{Anomalous Magnetic Moments}\tabularnewline
\hline 
$a_{e}-a_{e}^{SM}$ & $\pm2.3\times10^{-13}$~\cite{Keshavarzi:2019abf,Parker:2018vye} & $(-8.9\pm3.6)\times10^{-13}$~\cite{Hanneke:2008tm}\tabularnewline
\hline 
$a_{\mu}-a_{\mu}^{SM}$ & $\pm43\times10^{-11}$ \cite{Aoyama:2020ynm} & $(279\pm76)\times10^{-11}$~\cite{Bennett:2006fi,Aoyama:2020ynm}\tabularnewline
\hline 
$a_{\tau}-a_{\tau}^{SM}$ & $\pm3.9\times10^{-8}$~\cite{Keshavarzi:2019abf} & $(-2.1\pm1.7)\times10^{-7}$~\cite{Abdallah:2003xd}\tabularnewline
\hline 
\end{tabular}\caption{\label{tab:S1S3-obs-DF2}Meson-mixing and leptonic observables, with
their SM predictions and experimental bounds. Upper limits correspond
to 95\%CL.}
\end{table}

\begin{table}[t]
\centering{}%
\begin{tabular}{|c|c|}
\hline 
Observable & Experimental bounds\tabularnewline
\hline 
\hline 
\multicolumn{2}{|c|}{$Z$ boson couplings}\tabularnewline
\hline 
$\delta g_{\mu_{L}}^{Z}$ & $(0.3\pm1.1)10^{-3}$ \cite{ALEPH:2005ab}\tabularnewline
\hline 
$\delta g_{\mu_{R}}^{Z}$ & $(0.2\pm1.3)10^{-3}$ \cite{ALEPH:2005ab}\tabularnewline
\hline 
$\delta g_{\tau_{L}}^{Z}$ & $(-0.11\pm0.61)10^{-3}$ \cite{ALEPH:2005ab}\tabularnewline
\hline 
$\delta g_{\tau_{R}}^{Z}$ & $(0.66\pm0.65)10^{-3}$ \cite{ALEPH:2005ab}\tabularnewline
\hline 
$\delta g_{b_{L}}^{Z}$ & $(2.9\pm1.6)10^{-3}$ \cite{ALEPH:2005ab}\tabularnewline
\hline 
$\delta g_{c_{R}}^{Z}$ & $(-3.3\pm5.1)10^{-3}$ \cite{ALEPH:2005ab}\tabularnewline
\hline 
$N_{\nu}$ & $2.9963\pm0.0074$ \cite{Janot:2019oyi}\tabularnewline
\hline 
\end{tabular}\caption{\label{tab:S1S3-obs-Zbos}Limits on the deviations in $Z$ boson couplings
to fermions from LEP I.}
\end{table}

One of our main goals is to provide, with the $S_{1}+S_{3}$ model,
a combined explanation for the hints of LFU violation in the neutral
and charged current semileptonic $B$-meson decays, namely to account
for the experimental measurements of $R(K^{(*)})$ and $R(D^{(*)})$,
and of the deviation in the muon anomalous magnetic moment $(g-2)_{\mu}$.
The leptoquark couplings involved in these observable enter also in
the other low-energy observables (or pseudo observables), both at
tree-level or one-loop level. Therefore, to quantify how the $S_{1}+S_{3}$
model can consistently explain the observed anomalies, one should
take into account a set of low-energy data as complete as possible.
In Tables \ref{tab:S1S3-obs-semileptonic}, \ref{tab:S1S3-obs-DF2}
and \ref{tab:S1S3-obs-Zbos} we show the list of low-energy observables
that we analyze, together with their SM predictions and experimental
bounds. 

The detailed computations for each observables are contained in the
Appendix of the original reference \cite{Gherardi:2020qhc}; here
we limit ourselves to quote the relevant results for the global fit.
In our numerical analysis, the full set of one-loop corrections to
each observable was considered. Some observables vanish or are flavour-suppressed
at tree-level, for example meson-mixing $\Delta F=2$ processes, $\tau\to3\mu$
and $\tau\to\mu\gamma$ LFV interactions or $\tau\to\mu\phi(\eta,\eta^{\prime})$
decay; in such cases the inclusion of one-loop contributions is relevant
and might bring non negligible changes in a global fit of the low-energy
data. 

From the observables listed above, and their expression in terms of
the parameters of the model, LQ couplings and masses, we build a global
likelihood as: 
\begin{equation}
-2\log\mathcal{L}\equiv\chi^{2}(\lambda_{x},M_{x})=\sum_{i}\frac{\left(\mathcal{O}_{i}(\lambda_{x},M_{x})-\mu_{i}\right)^{2}}{\sigma_{i}^{2}},\label{eq:S1S3 Likelihood}
\end{equation}
where $\mathcal{O}_{i}(\lambda_{x},M_{x})$ is the expression of the
observable as function of the model parameters, $\mu_{i}$ its experimental
central value, and $\sigma_{i}$ the uncertainty. From the $\chi^{2}$
built in this way, in each scenario considered we obtain the maximum
likelihood point by minimizing the $\chi^{2}$, which we use to compute
the $\Delta\chi^{2}\equiv\chi^{2}-\chi_{{\rm min}}^{2}$. This allows
us to obtain the $68$, $95$, and $99\,\%$ CL regions. In the Standard
Model limit we get a $\chi_{\text{SM}}^{2}=101.0$, for 50 observables.

For each scenario we get the CL regions in the plane of two real couplings,
by profiling the likelihood over all the other couplings. We are often
also interested in the values of some observables corresponding to
these CL regions. To obtain this, we perform a numerical scan over
all the parameter space\footnote{For each numerical scan we collected $O(10^{4})$ benchmark points.
For our more complex models (\emph{i.e.} with up to ten parameters),
this is quite demanding from the computational point of view; in order
to efficiently scan the high-dimensional parameter spaces, we employ
a Markov Chain Monte Carlo algorithm (Hastings-Metropolis) for the
generation of trial points.} and select only the points with a $\Delta\chi^{2}$ less than the
one corresponding to $68$ and $95\,\%$ CL. The points obtained in
this way also reproduce the CL regions in parameter space obtained
by profiling. With this set of parameter-space points we can then
plot any observable evaluated on them. 

\subsection{Collider constraints}

Leptoquarks are also actively searched for at high-energy colliders.
Their most important signatures can be classified in three categories:
\emph{i)} pair production, \emph{ii)} resonant single-production,
and \emph{iii)} off-shell $t$-channel exchange in Drell-Yan processes,
$pp\to\ell^{+}\ell^{-}$ or $\ell\nu$. See e.g. Refs. \cite{Dorsner:2018ynv,Schmaltz:2018nls}
for reviews.

The pair production cross section is mostly independent on the LQ
couplings to fermions, unless some are very large, and thus provides
limits which depend only on the LQ mass and the branching ratios in
the relevant search channels. We refer to Refs. \cite{Marzocca:2018wcf,Angelescu:2018tyl,Schmaltz:2018nls,Saad:2020ihm}
for reviews of such searches. Once the branching ratios are taken
into account, the most recent ATLAS and CMS searches using an integrated
luminosity of $\sim36\text{fb}^{-1}$ put a lower bound on the $S_{1}$
and $S_{3}$ masses at $\approx1\,\text{TeV}$ or less. At present,
limits from single production are not competitive with those from
pair production and Drell-Yan \cite{Schmaltz:2018nls}.

Leptoquarks can also be exchanged off-shell in the $t$-channel in
Drell-Yan processes. The final states most relevant to our setup are
$\tau\bar{\tau}$, $\tau\bar{\nu}$, and $\mu\bar{\mu}$. The limits
on LQ couplings as a function of their mass from neutral-current processes
can be taken directly from \cite{Faroughy:2016osc,Angelescu:2018tyl,Schmaltz:2018nls}
(see also \cite{Greljo:2017vvb,Afik:2018nlr,Afik:2019htr,Angelescu:2020uug}
for other studies of dilepton tails in relation with $B$-anomalies)
while the mono-tau channel in relation to the $B$-anomalies has been
studied in \cite{Altmannshofer:2017poe,Greljo:2018tzh,Abdullah:2018ets,Brooijmans:2020yij,Fuentes-Martin:2020lea,Marzocca:2020ueu}
and at present it doesn't exclude the region of interest. Using the
results from \cite{Angelescu:2018tyl} we get the following $95\,\%$
CL upper limits on the couplings relevant to our model for $M_{1,3}=1\,\text{TeV}$,
taken one at a time: 

\begin{align}
\lambda_{c\tau}^{1R} & <1.62,\\
\lambda_{c\mu}^{1R} & <0.90,\\
\lambda_{s\tau}^{1L} & <1.66,\\
\lambda_{s\mu}^{1L} & <0.91,\\
\lambda_{b\tau}^{3L} & <1.40,\\
\lambda_{s\tau}^{3L} & <0.97,\\
\lambda_{b\mu}^{3L} & <0.77,\\
\lambda_{s\mu}^{3L} & <0.56
\end{align}

\section{Scenarios and results\label{sec:S1S3-Scenarios-and-results}}

In this Section we discuss several minimal models within the $S_{1}+S_{3}$
setup, and how well (or bad) each of them is able to address the charged
and/or neutral current anomalies, while remaining compatible with
all the other experimental constraints. We denote the leptoquark couplings
to fermions by: 
\begin{equation}
\lambda^{1R}=\begin{pmatrix}\lambda_{ue}^{1R} & \lambda_{u\mu}^{1R} & \lambda_{u\tau}^{1R}\\
\lambda_{ce}^{1R} & \lambda_{c\mu}^{1R} & \lambda_{c\tau}^{1R}\\
\lambda_{te}^{1R} & \lambda_{t\mu}^{1R} & \lambda_{t\tau}^{1R}
\end{pmatrix},\quad\lambda^{1L}=\begin{pmatrix}\lambda_{de}^{1L} & \lambda_{d\mu}^{1L} & \lambda_{d\tau}^{1L}\\
\lambda_{se}^{1L} & \lambda_{s\mu}^{1L} & \lambda_{s\tau}^{1L}\\
\lambda_{be}^{1L} & \lambda_{b\mu}^{1L} & \lambda_{b\tau}^{1L}
\end{pmatrix},\quad\lambda^{3L}=\begin{pmatrix}\lambda_{de}^{3L} & \lambda_{d\mu}^{3L} & \lambda_{d\tau}^{3L}\\
\lambda_{se}^{3L} & \lambda_{s\mu}^{3L} & \lambda_{s\tau}^{3L}\\
\lambda_{be}^{3L} & \lambda_{b\mu}^{3L} & \lambda_{b\tau}^{3L}
\end{pmatrix}.\label{eq:Leptoquark couplings}
\end{equation}

The main experimental anomalies driving the fit can be split in three
categories: 
\begin{itemize}
\item \emph{CC}: deviations in $b\to c\tau\nu$ transitions; 
\item \emph{NC}: deviations in $b\to s\mu\mu$ transitions; 
\item \emph{$(g-2)_{\mu}$}: deviation in the muon magnetic moment. 
\end{itemize}
While our setup allows to keep all the above couplings in a completely
general analysis, given the large number of parameters this would
preclude a clear understanding of the physics underlying the fit.
Furthermore, it can be interesting to consider only one leptoquark
or to focus on one specific experimental anomaly. For these reasons
we take a step-by-step approach by starting with single-leptoquark
scenarios and switching on the couplings needed to fit a given set
of anomalies. In all cases, we keep the complete likelihood described
in the previous Section, with all the observables. For instance, if
the couplings to muons are set to zero, neutral-current $B$-anomalies
and the muon anomalous magnetic moment are automatically \emph{frozen}
to the corresponding Standard Model values and do not impact the final
fit.

The models are thus defined by the leptoquark content and the set
of active couplings, which, for simplicity, we assume to be real.
We have considered the models detailed in Table \ref{tab:Summary of models},
for each of which we allow the couplings listed in the third column
of the table to be non-vanishing in our global fit. We first analyze
single mediator models and study their potential to address as many
anomalies as possible. In each case we point out the main tensions
which prevent a combined explanation of all anomalies. Then, we move
on to study models involving both leptoquarks. In the first we only
allow left-handed couplings, $\lambda^{1L}$ and $\lambda^{3L}$,
as this possibility has better chances to find motivation in a scenario
in which the flavour structure is determined by a flavour symmetry,
see e.g. \cite{Buttazzo:2017ixm,Marzocca:2018wcf}. In the second
we switch on also some of the $S_{1}$ couplings to right-handed fermions,
and aim to provide a combined explanation for all three anomalies.
Finally, we study the limits on the leptoquark potential couplings
to the Higgs, which is an analysis largely independent on the couplings
to fermions and requires to consider different observables than those
studied in the main fit, see also \cite{Crivellin:2020ukd}.

\begin{table}[t]
\begin{tabular}{|c|l|c|c|c|}
\hline 
Model & Couplings & CC & NC & $(g-2)_{\mu}$\tabularnewline
\hline 
\hline 
\multirow{2}{*}{$S_{1}^{~(CC)}$} & \multirow{2}{*}{$\lambda_{c\tau}^{1R},\lambda_{b\tau}^{1L}$} & \multirow{2}{*}{\textcolor{green}{$\checked$}} & \multirow{2}{*}{\textcolor{red}{$\times$}} & \multirow{2}{*}{\textcolor{red}{$\times$}}\tabularnewline
 &  &  &  & \tabularnewline
\multirow{2}{*}{$S_{1}^{~(a_{\mu})}$} & \multirow{2}{*}{$\lambda_{t\mu}^{1R},\lambda_{b\mu}^{1L}$} & \multirow{2}{*}{\textcolor{red}{$\times$}} & \multirow{2}{*}{\textcolor{red}{$\times$}} & \multirow{2}{*}{\textcolor{green}{$\checked$}}\tabularnewline
 &  &  &  & \tabularnewline
 &  &  &  & \tabularnewline
\hline 
\multirow{2}{*}{$S_{3}^{~(CC+NC)}$} & \multirow{2}{*}{$\lambda_{b\tau}^{3L},\lambda_{s\tau}^{3L},\lambda_{b\mu}^{3L},\lambda_{s\mu}^{3L}$} & \multirow{2}{*}{\textcolor{red}{$\times$}} & \multirow{2}{*}{\textcolor{green}{$\checked$}} & \multirow{2}{*}{\textcolor{red}{$\times$}}\tabularnewline
 &  &  &  & \tabularnewline
\hline 
\multirow{2}{*}{$S_{1}+S_{3}^{{\rm ~(LH)}}$} & \multirow{2}{*}{$\lambda_{b\tau}^{1L},\lambda_{s\tau}^{1L},\lambda_{b\tau}^{3L},\lambda_{s\tau}^{3L},\lambda_{b\mu}^{3L},\lambda_{s\mu}^{3L}$} & \multirow{2}{*}{\textcolor{green}{$\checked$}} & \multirow{2}{*}{\textcolor{green}{$\checked$}} & \multirow{2}{*}{\textcolor{red}{$\times$}}\tabularnewline
 &  &  &  & \tabularnewline
\hline 
\multirow{2}{*}{$S_{1}+S_{3}^{{\rm ~(all)}}$} & \multirow{2}{*}{$\lambda_{b\tau}^{1L},\lambda_{s\tau}^{1L},\lambda_{b\mu}^{1L},\lambda_{t\tau}^{1R},\lambda_{c\tau}^{1R},\lambda_{t\mu}^{1R},\lambda_{b\tau}^{3L},\lambda_{s\tau}^{3L},\lambda_{b\mu}^{3L},\lambda_{s\mu}^{3L}$} & \multirow{2}{*}{\textcolor{green}{$\checked$}} & \multirow{2}{*}{\textcolor{green}{$\checked$}} & \multirow{2}{*}{\textcolor{green}{$\checked$}}\tabularnewline
 &  &  &  & \tabularnewline
\hline 
\multirow{2}{*}{$S_{1}+S_{3}^{{\rm ~(pot)}}$} & \multirow{2}{*}{$\lambda_{H1},\lambda_{H3},\lambda_{H13},\lambda_{\epsilon H3}$} & \multirow{2}{*}{--} & \multirow{2}{*}{--} & \multirow{2}{*}{--}\tabularnewline
 &  &  &  & \tabularnewline
\end{tabular}

\caption{\label{tab:Summary of models}Summary of leptoquark models considered
in this Chapter. The third columns lists the couplings we allow to
be different from zero in our global fit. The last three columns indicate
whether the models provide a satisfying fit of each set of anomalies,
respectively.}
\end{table}

In any given model there is, of course, no particular reason to expect
the exact flavour structures implied by Table \ref{tab:Summary of models}.
For instance, the couplings we set to zero will be radiatively generated.
In our bottom-up approach we assume them to be small enough at the
matching scale that the observables in the fit are not impacted in
a sizeable way. In a more top-down approach one might have expectations
on the size of these terms based on the UV picture, such as due to
the presence of approximate flavour symmetries or other flavour-protection
mechanisms \cite{Marzocca:2021miv} (see Subsec. \ref{sec:S1S3-Light-generations}).

In the numerical analysis we fix for concreteness values of leptoquark
masses equal to $M_{1}=M_{3}=1\,\text{TeV}$. While this is borderline
with the exclusion limits from pair production, discussed previously,
the results do not change qualitatively by increasing slightly the
masses. Since most of the observables driving the fits scale as $\lambda^{2}/M^{2}$,
with a good approximation this scaling can be used to adapt our fits
to slightly larger masses.\footnote{The exception to this scaling are $\Delta F=2$ observables, which
scale as $\lambda^{4}/M^{4}$, but are relevant only for the fits
of Sec.~\ref{sec:S1S3-models-S3CCNC} and \ref{sec:S1S3-models-S1S3LH}.} We note that the future limits on LQ masses from HL-LHC are expected
to not go much above $1.5\,\text{TeV}$ \cite{Marzocca:2018wcf}.

Concerning our specific benchmarks, the choice of active couplings
in each case is guided by some simple phenomenological observations
(more details on each concrete model can be found in the relevant
Subsections below): 
\begin{enumerate}
\item Since the observed deviations in $B$-decays involve LQ couplings
to second and third generation, and given the strong constraints on
$s\leftrightarrow d$ quark flavour transitions, couplings to first
generation of down quarks can only play a minor role in the fit of
$B$-anomalies and are thus set to zero (see however Sec. \ref{sec:S1S3-Light-generations};
note that even in our case, due to the CKM matrix, effects in up-quark
observables are present, for instance $D$-meson mixing). 
\item Hints to LFU violation in rare $B$-decays, combined with the deviations
observed in $B\to K^{*}\mu^{+}\mu^{-}$, suggest that the LQ couplings
to muons should be larger than those to electrons. We consider, for
simplicity the case in which $b\to s\ell\ell$ anomalies are entirely
explained by muon couplings and set to zero the couplings to electrons. 
\item The $S_{1}$ couplings to $\mu_{R}$ and $c_{R}$ or $t_{R}$ do not
contribute to $b\to s\ell\ell$, nor to $b\to c(\tau/\ell)\nu$, however
are relevant for fitting the observed anomaly in the muon anomalous
magnetic moment, which gets the main contribution from the couplings
to $b_{L}\mu_{L}$ and $t_{R}\mu_{R}$, as shown in the original reference
\cite{Gherardi:2020qhc}.
\end{enumerate}
Details for all models are given in the following Subsections.

\subsection{\label{sec:S1S3-models-S1}Single-leptoquark $S_{1}$}

\subsubsection{Addressing CC anomalies}

\begin{figure}[b]
\centering \includegraphics[scale=0.6]{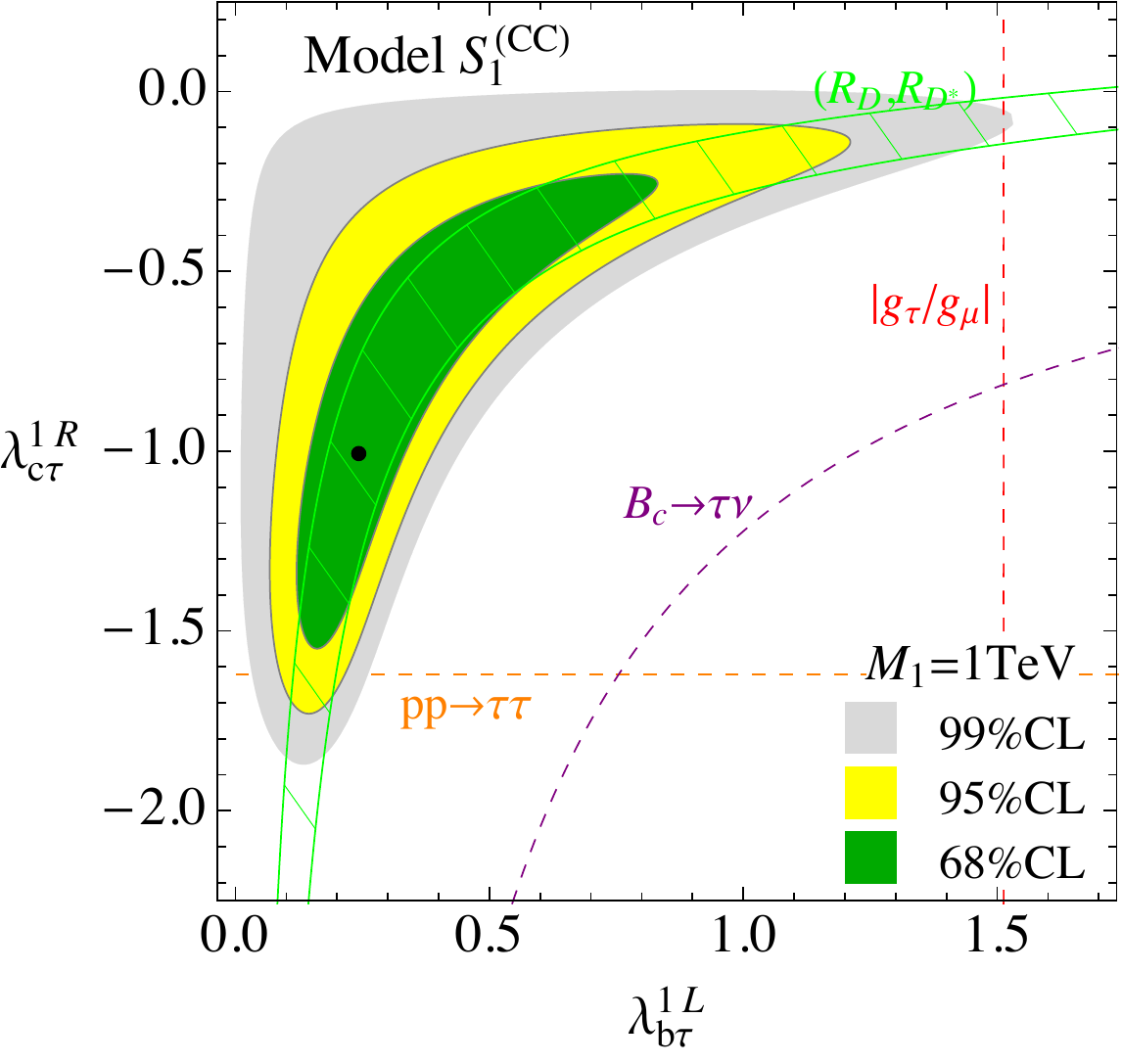}
~ \includegraphics[scale=0.67]{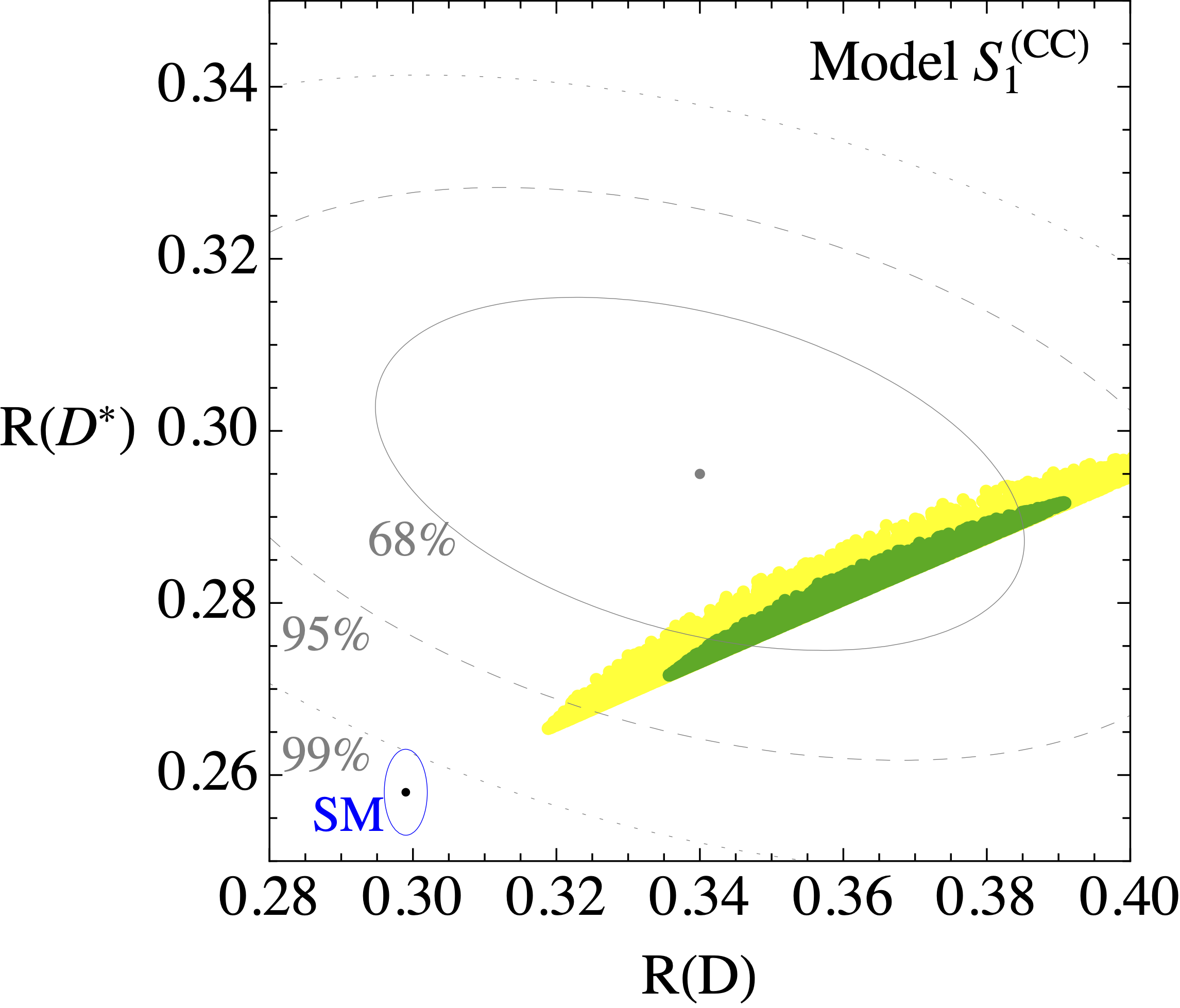}
\caption{\label{fig:S1S3-models-S1CC}{\small{}Result from a fit in the two
$S_{1}$ couplings $\lambda_{b\tau}^{1L}$ and $\lambda_{c\tau}^{1R}$,
for a leptoquark of 1 TeV. In the left panel we show the preferred
regions in the plane of the two couplings, and the individual $2\sigma$
limits from the most relevant observables (except for $(R(D),R(D^{*}))$,
for which we show the $68\%$CL region). The black dot corresponds
to the best-fit point. In the right panel we show where this preferred
region is mapped in the plane of $R(D)-R(D^{*})$, together with the
experimental combination from HFLAV \cite{Amhis:2016xyh}.}}
\end{figure}

This LQ can address the deviations in $R(D)$ and $R(D^{*})$ with
only two couplings: $\lambda_{b\tau}^{1L}$ and $\lambda_{c\tau}^{1R}$.
They generate at tree-level a contribution to the semileptonic scalar
and tensor operators $C_{lequ}^{(1,3)}$ at the UV matching scale,
Eqs.~(\ref{eq:C_lequ1 tree}) and (\ref{eq:C_lequ3 tree}), which
then run down to the GeV scale. The best fit region is entirely determined
by the following few observables: $R(D)$, $R(D^{*})$, $\text{Br}(B_{c}^{+}\to\tau^{+}\nu)$,
$|g_{\tau}/g_{\mu}|$, and the constraints from $pp\to\tau^{+}\tau^{-}$.

The results from the fit, assuming real couplings and $M_{1}=1\,\text{TeV}$,
can be seen in Fig.~\ref{fig:S1S3-models-S1CC}. Since all the relevant
low-energy observables scale with $\lambda/M$, the fit can be easily
adapted to other masses. The left panel shows confidence level regions
for the two couplings. The dashed lines are 95\% CL constraints from
single observables, and help illustrate the role of each observable
within the global fit.

In the right panel we show how the 68\% and 95\% CL region from the
global fit of the left panel maps in the $R(D^{(*)})$ plane. This
is almost degenerate, due to the approximate linear relationship between
these two observables in the present model. We overlay as gray lines
the CL ellipses (for 2 degrees of freedom) from the HFLAV global fit
of the two observables \cite{Amhis:2016xyh} (specifically, the Spring
2019 update).

The model is successful in fitting the deviation in $R(D^{(*)})$
within the 68\%CL level, with smaller values of $R(D^{*})$ (or larger
$R(D)$) preferred by the fit. Improved measurements of $R(D^{(*)})$
can test this setup due to the precise linear relationship among the
two modes predicted by the model, as well as improved Drell-Yan constraints.

The best-fit point, for $M_{1}=1\,\text{TeV}$, is found for $\lambda_{b\tau}^{1L}\approx0.24$,
$\lambda_{c\tau}^{1R}\approx-1.00$.

\subsubsection{Addressing NC anomalies}

\begin{figure}[t]
\centering \includegraphics[scale=0.6]{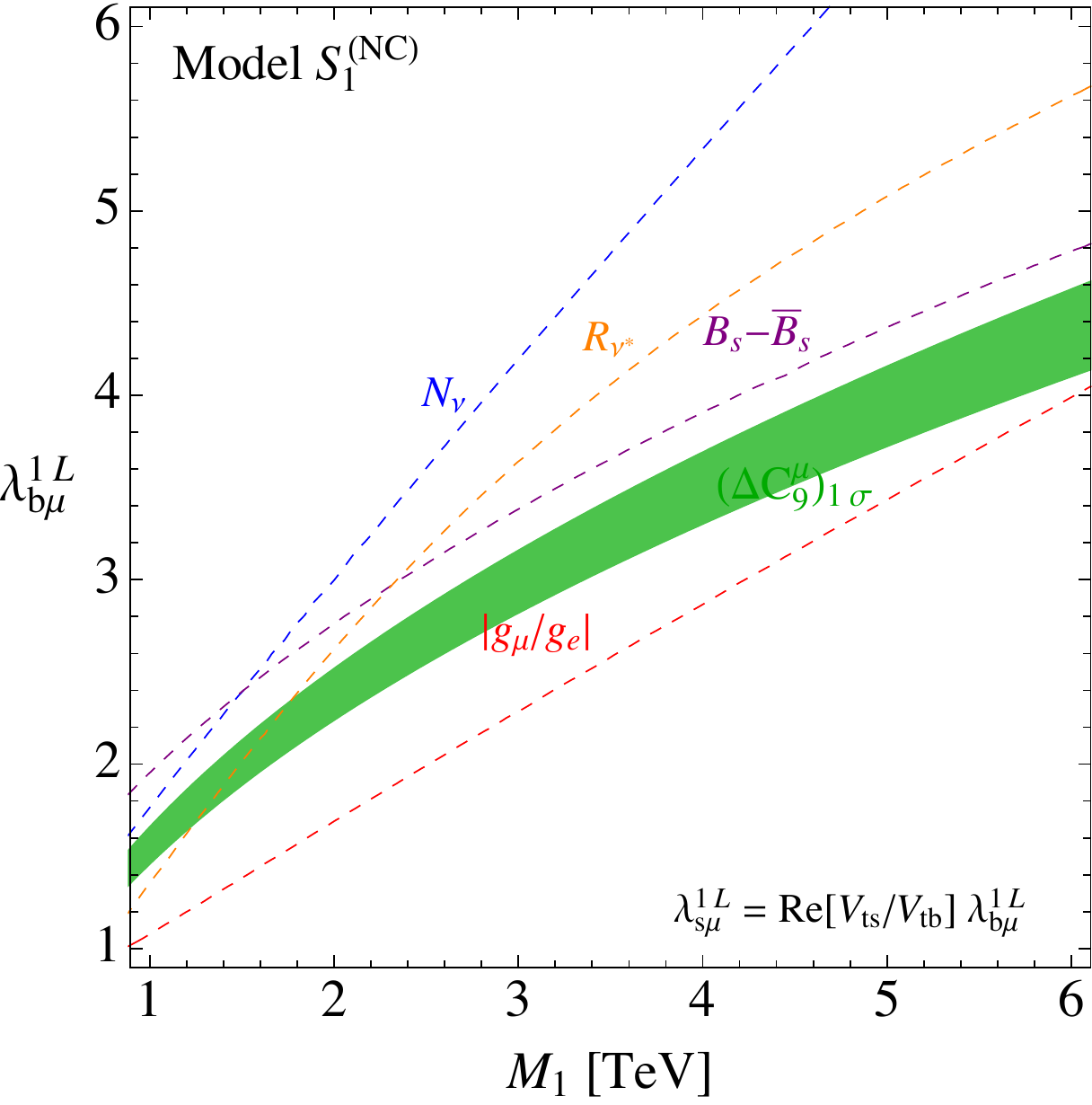}~
\includegraphics[scale=0.63]{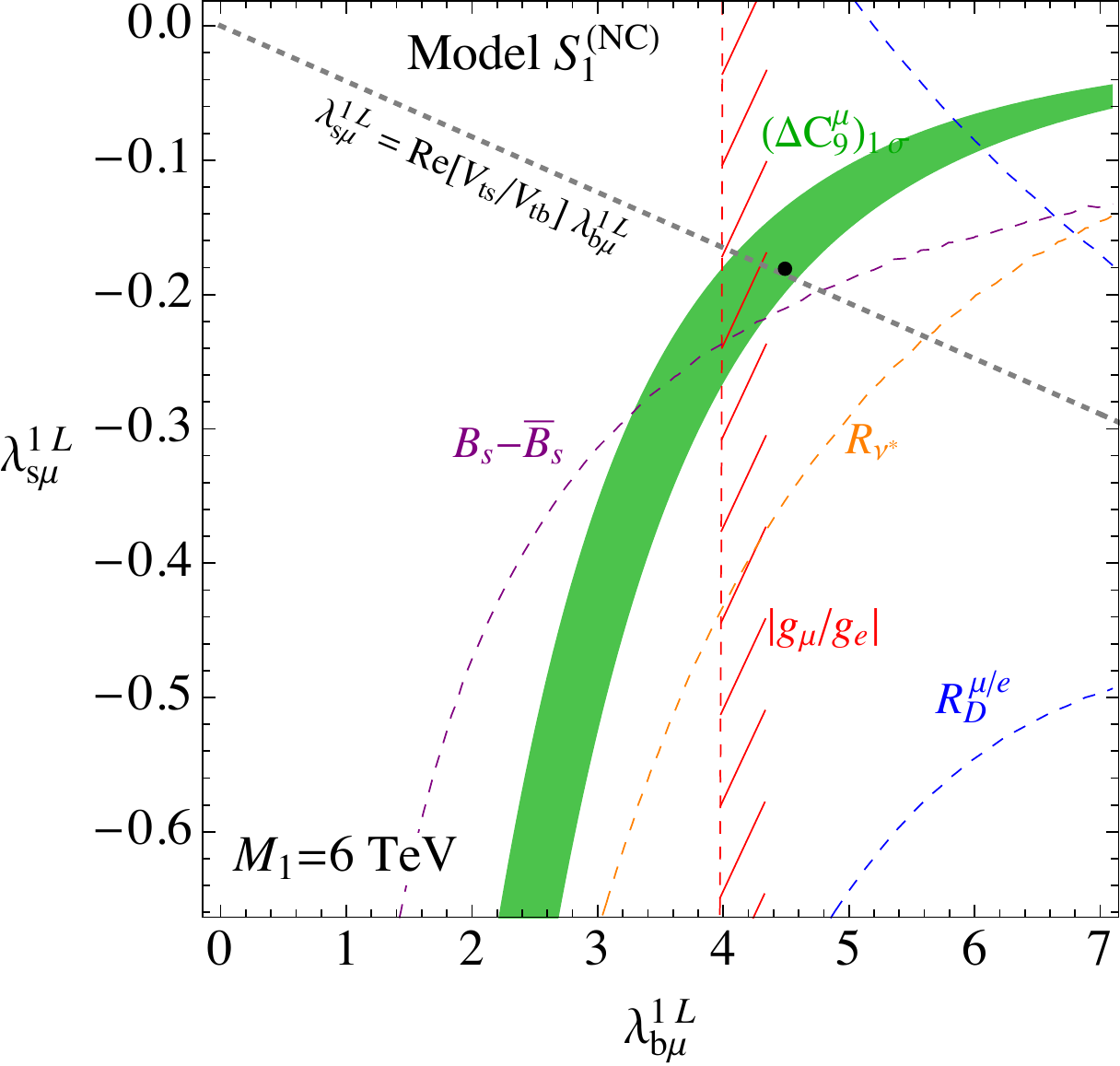} \caption{\label{fig:S1S3-models-S1NCbounds}{\small{}Left: 95\% CL limits in
the plane of $\lambda_{b\mu}^{1L}-M_{1}$, for $\lambda_{s\mu}^{1L}=-\Re[V_{cb}/V_{cs}]\lambda_{b\mu}^{1L}$.
Right: 95\% CL limits from individual observables in the plane of
$\lambda_{b\mu}^{1L}-\lambda_{s\mu}^{1L}$, fixing $M_{1}=6\,\text{TeV}$.
In both plots the green region is the $1\sigma$ favourite one from
$\Delta C_{9}^{\mu}$.}}
\end{figure}

One may attempt to fit neutral-current anomalies in $b\to s\ell\ell$
from the one-loop contributions from $S_{1}$. This scenario has been
considered for the first time in \cite{Bauer:2015knc}. Significant
contributions to $\Delta C_{9}^{\mu}$ may only come from the two
muon couplings $\lambda_{b\mu}^{1L}$ and $\lambda_{s\mu}^{1L}$,
whereas the universal contribution is always negligible ($C_{9}^{u}\approx0$).

$B_{s}$-mixing and $B\to K^{(*)}\nu\nu$ put strong constraints on
the product of the two couplings \cite{Gherardi:2020qhc}. Thanks
to the different scaling of these observables and $\Delta C_{9}^{\mu}$
on the leptoquark couplings, the limits can be avoided by a suitably
large leptoquark mass, as can be seen in Fig.~\ref{fig:S1S3-models-S1NCbounds}
(left). For $M_{1}\gtrsim3\,\text{TeV}$ it is possible to avoid these
limits while having couplings still in the perturbative range (see
also \cite{Cai:2017wry,Azatov:2018kzb}). Nevertheless, even while
marginally evading the $B_{s}$-mixing constraint, the $\Delta C_{9}^{\mu}$
deviation remains in $\approx2\sigma$ tension with the bound on $\lambda_{b\mu}^{1L}$
arising from the LFU limit from $\tau$ decays, $|g_{\mu}/g_{e}|$.
The situation is illustrated in the right panel of Fig.~\ref{fig:S1S3-models-S1NCbounds}.
This is slightly exacerbated by a $\sim1\sigma$ deviation in the
opposite direction measured in $|g_{\mu}/g_{e}|$.

We thus conclude that the $S_{1}$ leptoquark is not able to fit neutral-current
anomalies while remaining completely consistent with all other constraints.
The situation regarding NC anomalies is not modified significantly
by letting also the other couplings vary in the fit. This issue could
be avoided by allowing a mild cancellation by tuning a further contribution
to this observable, possibly arising from some other state. Fixing
$M_{1}=6\,\text{TeV}$ we find the best-fit point for: $\lambda_{b\mu}^{1L}\approx4.5$,
$\lambda_{s\mu}^{1L}\approx-0.18$.

\subsubsection{Addressing $(g-2)_{\mu,e}$}

$S_{1}$ is also a good candidate to address the observed deviation
in the muon anomalous magnetic moment. Ref. \cite{Gherardi:2020qhc}
finds that the leading contribution is given numerically by: 
\begin{equation}
\Delta a_{\mu}\approx8.23\times10^{-7}\frac{\lambda_{b\mu}^{1L}\lambda_{t\mu}^{1R}}{M_{1}^{2}/(1\,\text{TeV})^{2}}(1+0.53\log\frac{M_{1}^{2}}{(1\,\text{TeV})^{2}})
\end{equation}
The observed deviation can thus be addressed for small couplings,
and no other observable is influenced significantly. Analogously,
it is possible to address the (smaller and less significant) deviation
in the electron magnetic moment, see Table~\ref{tab:S1S3-obs-DF2}.

A combined explanations of both deviations with a single mediator
was thought not to be viable, due to the very strong constraint from
$\mu\to e\gamma$, see e.g. Refs.~\cite{Crivellin:2018qmi}. More
recently, in the updated versions of \cite{Bigaran:2020jil,Dorsner:2020aaz}
it was realized that a possible way out is to align the $S_{1}$ leptoquark-muon
couplings to the top quark, while the leptoquark-electron ones to
the charm\footnote{We thank the authors of \cite{Bigaran:2020jil} and \cite{Dorsner:2020aaz}
for pointing this out to us.}. In our formalism this can be achieved aligning the $S_{1}$ couplings
as $\lambda_{ie}^{1L}=V_{ci}\lambda_{e}^{1L}$ and $\lambda_{i\mu}^{1L}=V_{ti}\lambda_{\mu}^{1L}$.
In this way one finds \cite{Gherardi:2020qhc}: 
\begin{align}
\Delta a_{e} & \approx1.9\times10^{-10}\frac{\lambda_{e}^{1L}\lambda_{ce}^{1R}}{M_{1}^{2}/(1\,\text{TeV})^{2}}(1+0.09\log\frac{M_{1}^{2}}{(1\,\text{TeV})^{2}}),\\
\Delta a_{\mu} & \approx8.23\times10^{-7}\frac{\lambda_{\mu}^{1L}\lambda_{t\mu}^{1R}}{M_{1}^{2}/(1\,\text{TeV})^{2}}(1+0.53\log\frac{M_{1}^{2}}{(1\,\text{TeV})^{2}}),\\
\text{Br}(\mu\to e\gamma) & =0.
\end{align}
On the one hand, the strong limit from $\mu\to e\gamma$ implies that
the alignment described above must be held with high accuracy. On
the other hand, radiative corrections to leptoquark couplings from
SM Yukawas are expected to induce deviations from it. For this reason
we will not investigate this direction further.

\subsubsection{\label{sec:S1S3-models-S1CCamu}Addressing CC and $(g-2)_{\mu}$}

\begin{figure}[H]
\centering \includegraphics[height=6.5cm]{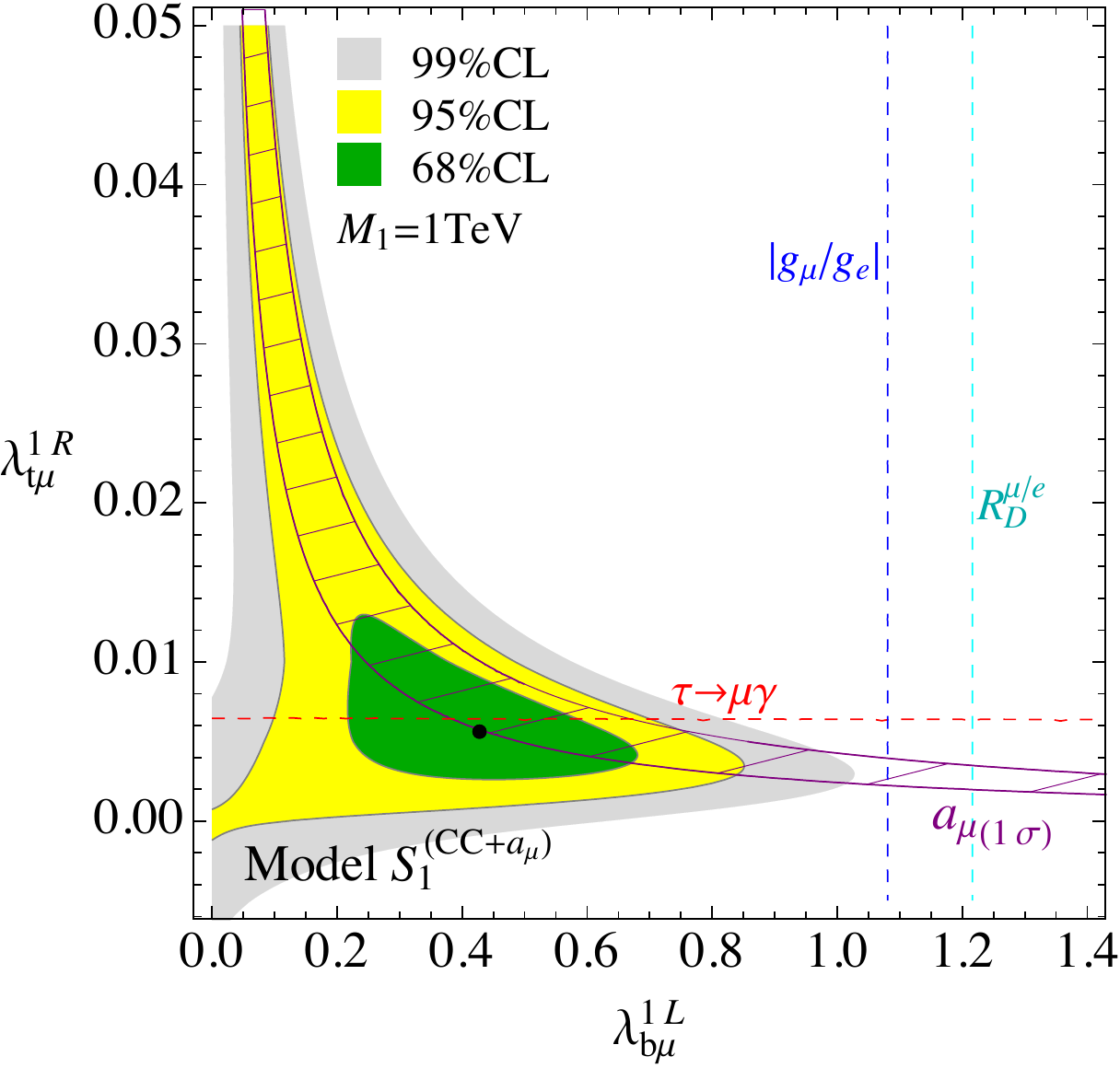}~
\includegraphics[height=6.5cm]{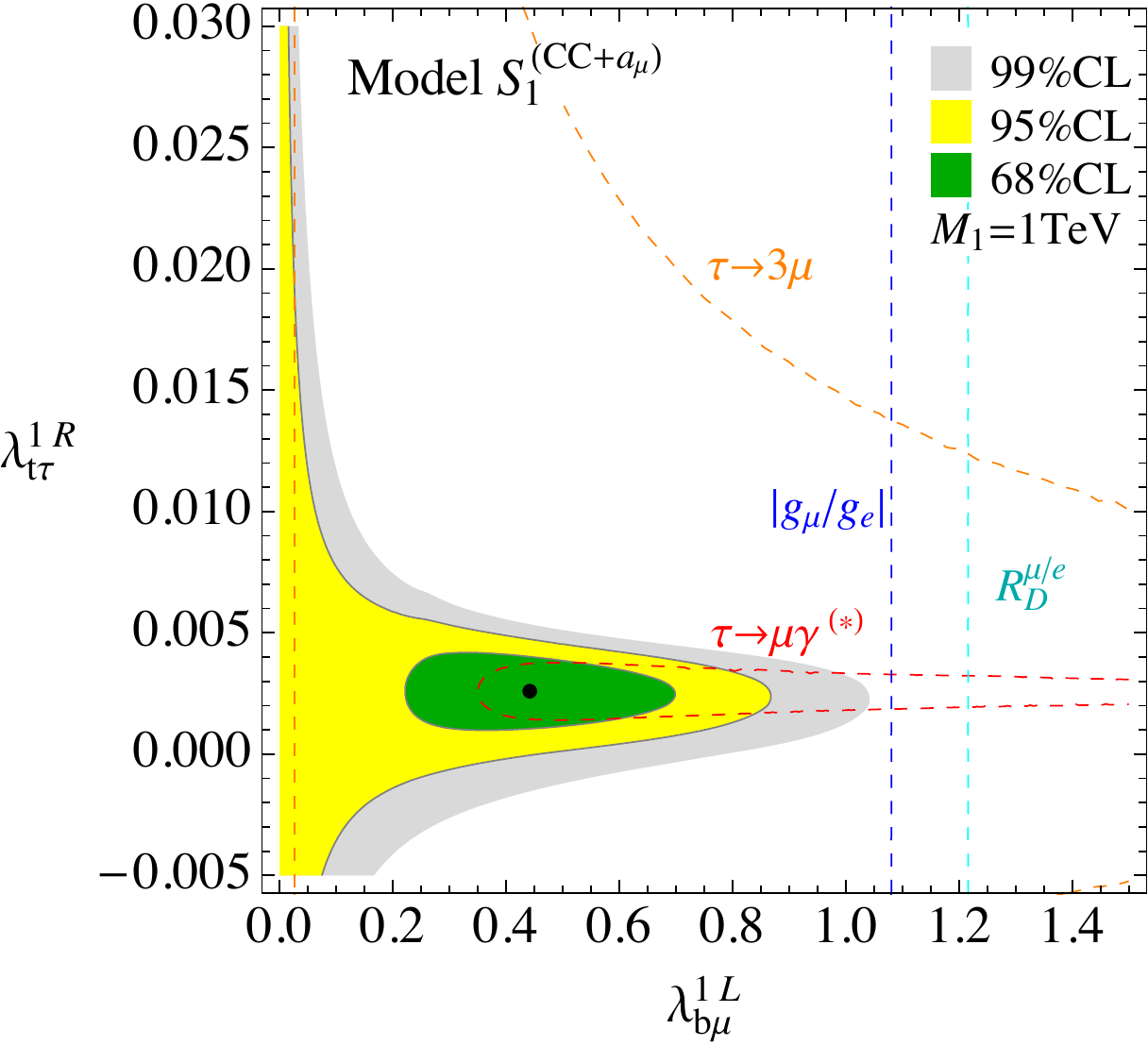}
\\
 \includegraphics[height=6.5cm]{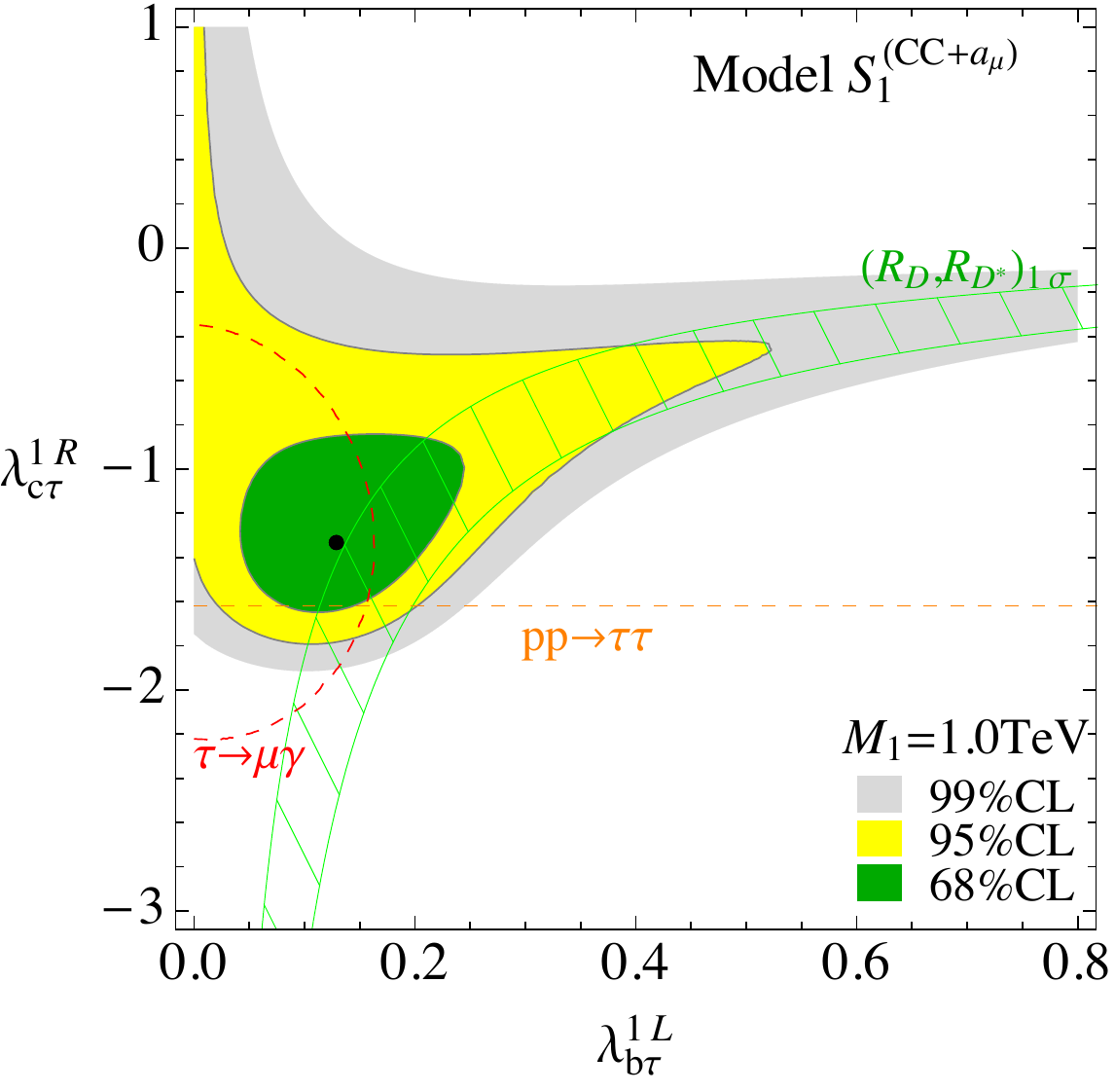}
~ \includegraphics[height=6.5cm]{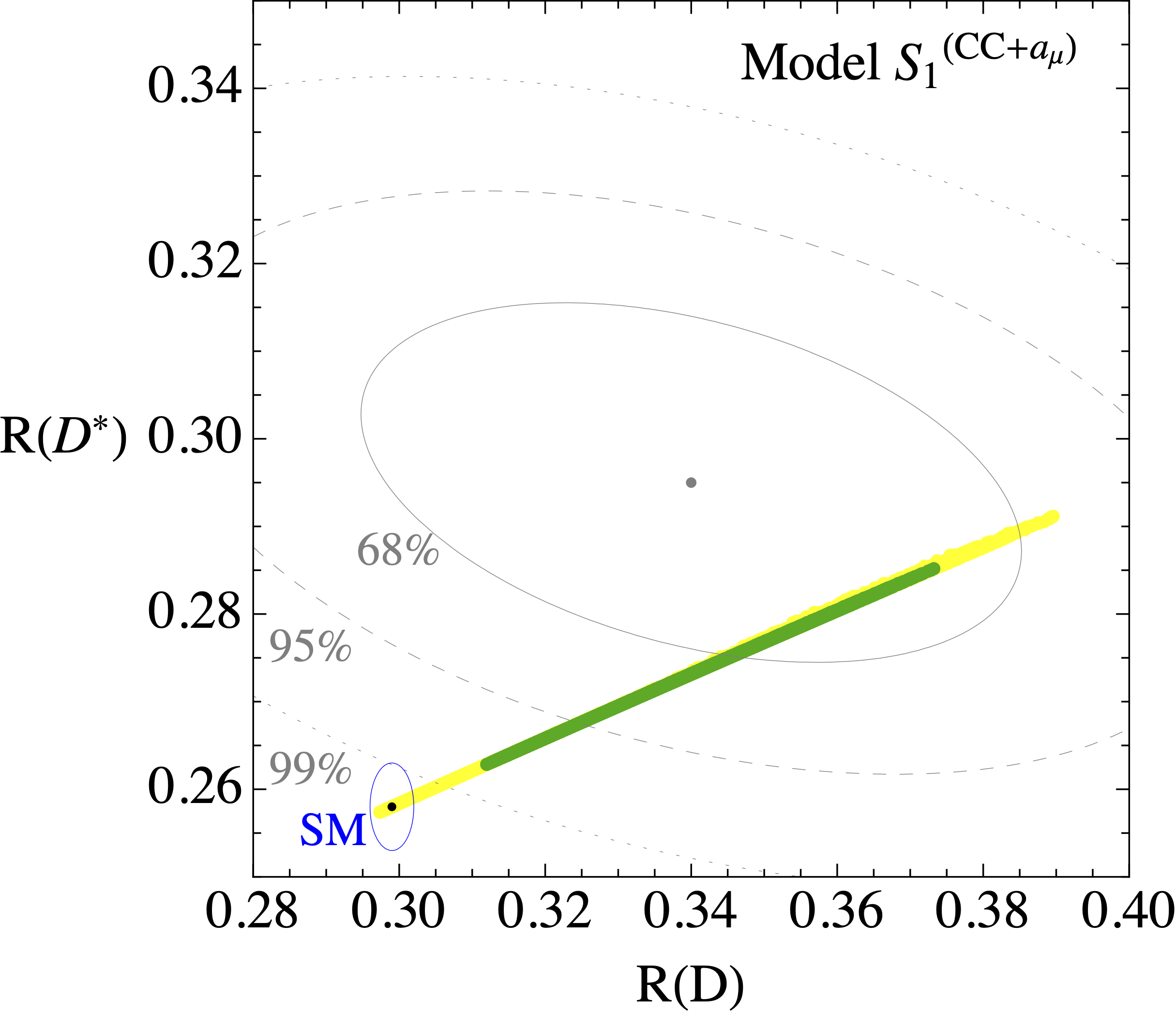}
\caption{{\small{}\label{fig:S1S3-models-S1CCamu}Result from the fit in the
$S_{1}$ model aimed to addressing the CC and $(g-2)_{\mu}$ anomalies.
We show the preferred regions in the planes of two couplings, where
those not shown are profiled over. The dashed lines show, for illustrative
purposes, $2\sigma$ limits from individual observables where the
other couplings are fixed at the best-fit point (black dot). In the
lower-right panel we show where the preferred region is mapped in
the plane of $R(D)-R(D^{*})$.}}
\end{figure}

Following the previous Subsections, the next natural step is to attempt
to fit both $R(D^{(*)})$ and $a_{\mu}$ anomalies with $S_{1}$.
The relevant couplings are $\lambda_{b\tau}^{1L}$, $\lambda_{b\mu}^{1L}$,
$\lambda_{t\tau}^{1R}$, $\lambda_{c\tau}^{1R}$, $\lambda_{t\mu}^{1R}$.

This setup is not simply the combination of those discussed previously.
Indeed, due to the $\lambda_{b\tau}^{1L}$ and $\lambda_{c\tau}^{1R}$
couplings on the one hand (required to fit the charged-current $B$-anomaly),
and $\lambda_{t\mu}^{1R}$, $\lambda_{b\mu}^{1L}$ on the other hand
(necessary to fit $a_{\mu}$), sizeable contributions to $\tau\to\mu\gamma$
are generated at one-loop. The values of $\lambda_{b\tau,b\mu}^{1L}$
and $\lambda_{c\tau,t\mu}^{1R}$ required to fit $R(D^{(*)})$ and
$(g-2)_{\mu}$ would induce a too large contribution to this LFV decay.
However, we find that the large contribution to $\tau\to\mu\gamma$
arising from the product of $\lambda_{b\mu}^{1L}\lambda_{c\tau}^{1R}$
can be mostly cancelled by the $\lambda_{b\mu}^{1L}\lambda_{t\tau}^{1R}$
term with $\lambda_{t\tau}^{1R}\sim O(10^{-3})$ (requiring a mild
fine-tuning of roughly one part in five). Such a small value does
not affect any other observable in our fit. It should be noted that
the only effect of this coupling is to tune this observable.

We show the preferred region in parameters space, obtained from our
fit, in Fig.~\ref{fig:S1S3-models-S1CCamu}. In the upper two panels
we show the fit in the $(\lambda_{b\mu}^{1L},\lambda_{t\mu}^{1R})$
and $(\lambda_{b\mu}^{1L},\lambda_{t\tau}^{1R})$ planes, including
the preferred region from the global fit as well as the individual
95\% CL limits from single observables, to help illustrating the physics
behind the analysis. It can be noted that the observed value of the
muon magnetic moment can be reproduced, and that $\tau\to\mu\gamma$
requires a non-zero value of $\lambda_{t\tau}^{1R}$, as discussed
above. Regarding the top-right panel, the single-observable constraint
from $\tau\to\mu\gamma$ (red dashed line) is shown by imposing that
$a_{\mu}$ is fixed to the value we get at the best-fit point. In
the lower-left panel we show the fit in the couplings contributing
to $R(D^{(*)})$, $(\lambda_{b\tau}^{1L},\lambda_{c\tau}^{1R})$,
and how this preferred region is mapped in the plane of $R(D)-R(D^{*})$
(lower-right panel). Comparing to the allowed region in the same plane
in the model studied in Fig.~\ref{fig:S1S3-models-S1CC}, we see
that $\tau\to\mu\gamma$ strongly reduced the allowed region, while
still not preventing a good fit of the charged-current $B$-anomalies.
Due to this reduction of the allowed $(\lambda_{b\tau}^{1L},\lambda_{c\tau}^{1R})$
parameter space, specifically with smaller values of $\lambda_{b\tau}^{1L}$
, the points in the $R(D^{*})-R(D)$ plane line up more closely in
a line than what is observed in Fig.~\ref{fig:S1S3-models-S1CC}.

We conclude that the $R(D^{(*)})$ and $(g-2)_{\mu}$ anomalies can
be addressed by the $S_{1}$ leptoquark, for perturbative couplings
and TeV-scale mass. We find the best-fit point, with $M_{1}=1\,\text{TeV}$,
for $\lambda_{b\tau}^{1L}\approx0.13$, $\lambda_{b\mu}^{1L}\approx0.44$,
$\lambda_{t\tau}^{1R}\approx0.0026$, $\lambda_{c\tau}^{1R}\approx-1.33$,
$\lambda_{t\mu}^{1R}\approx0.0051$. 

\subsection{\label{sec:S1S3-models-S3CCNC}Single-leptoquark $S_{3}$}

\begin{figure}[b]
\centering \includegraphics[height=6.5cm]{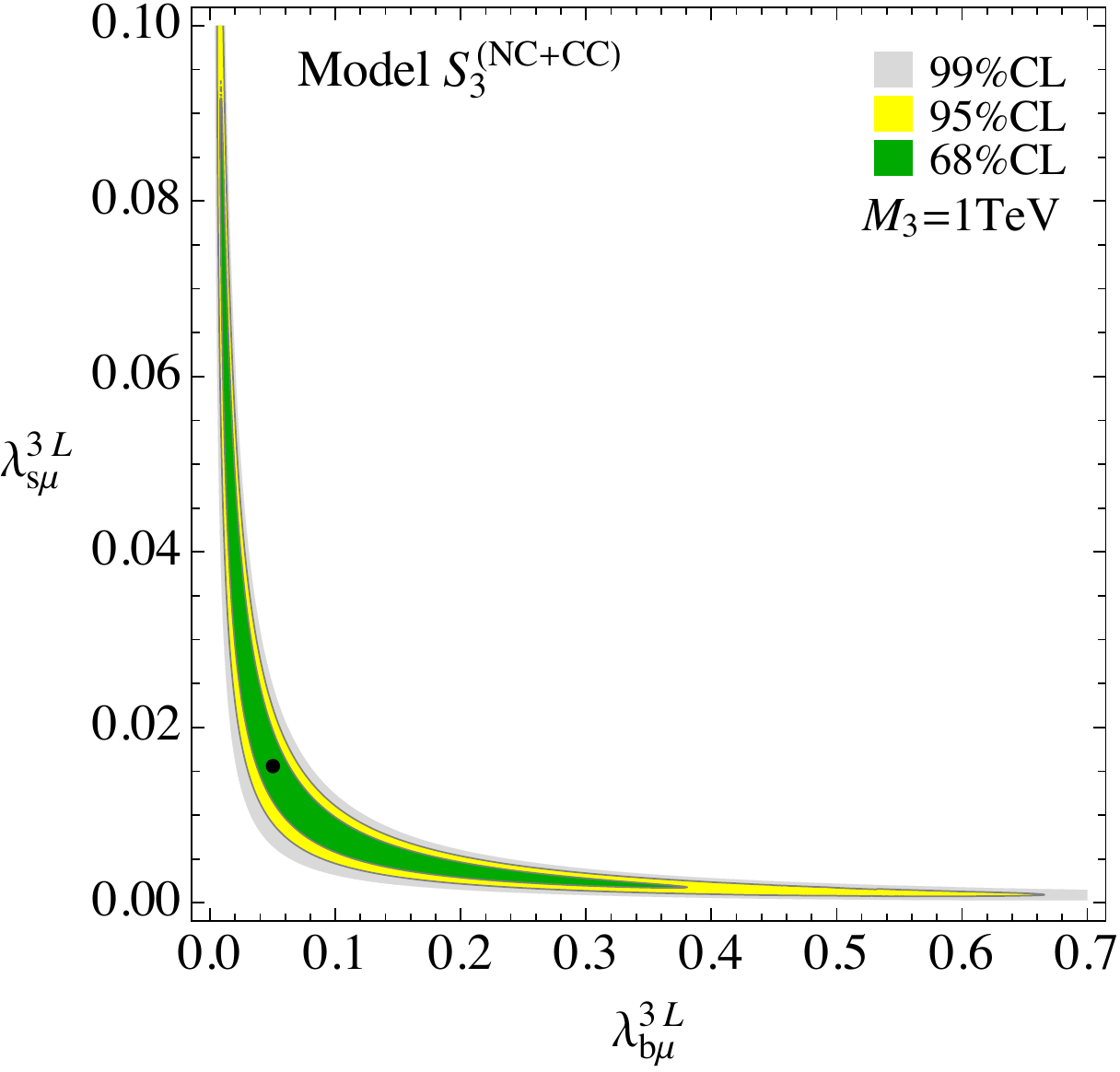}~
\includegraphics[height=6.5cm]{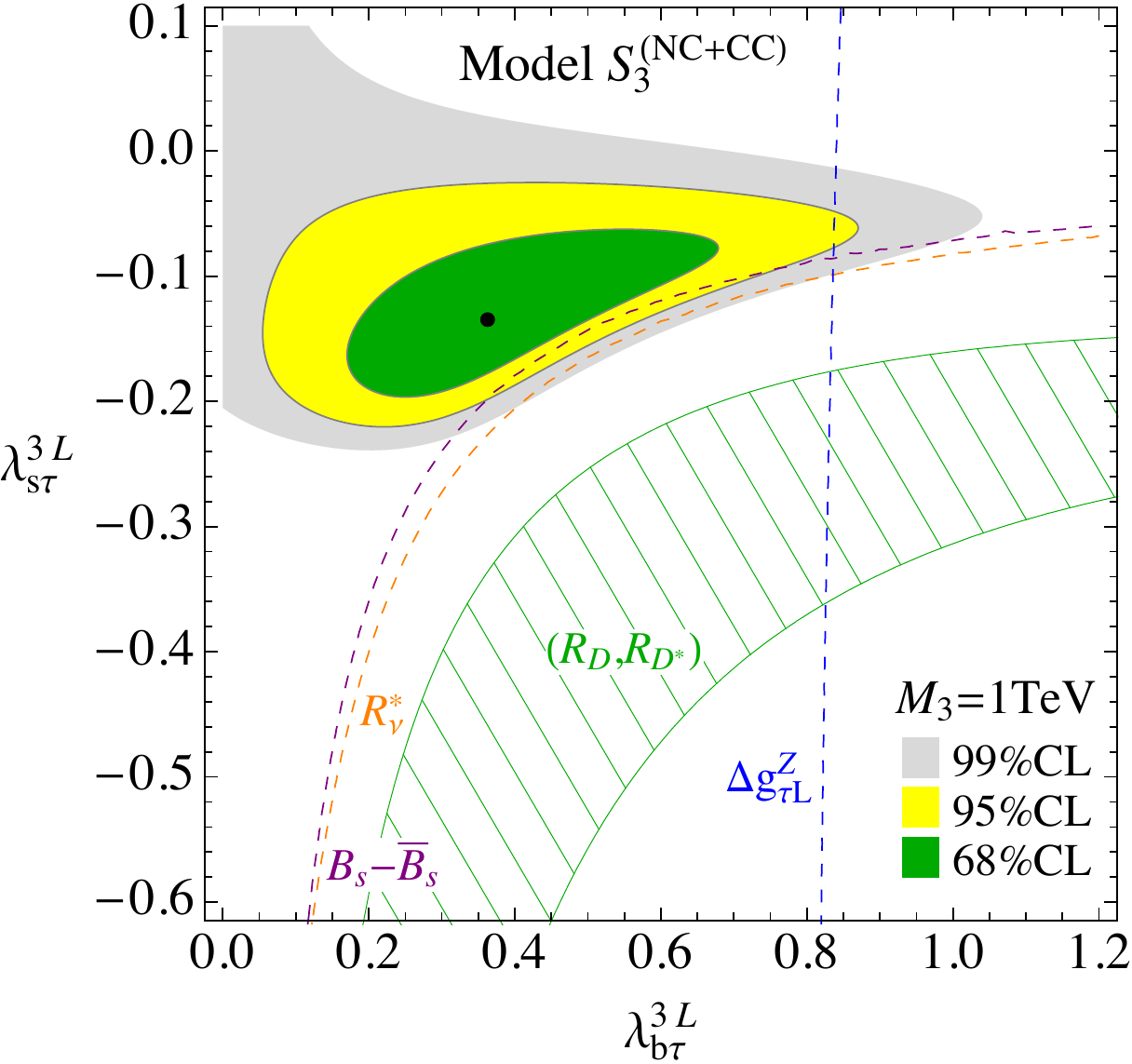}
\\
 \includegraphics[height=6.5cm]{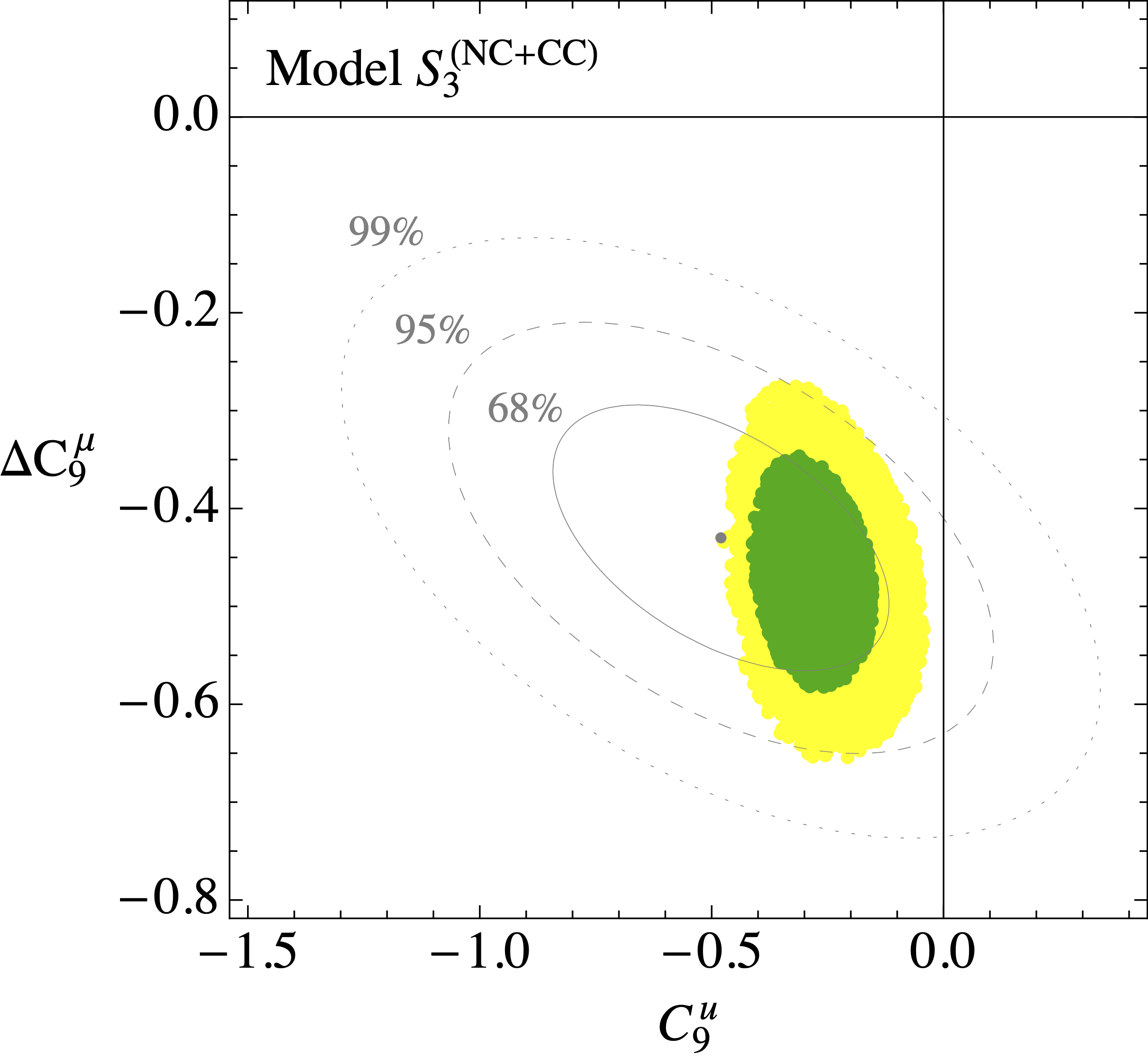}
~ \includegraphics[height=6.5cm]{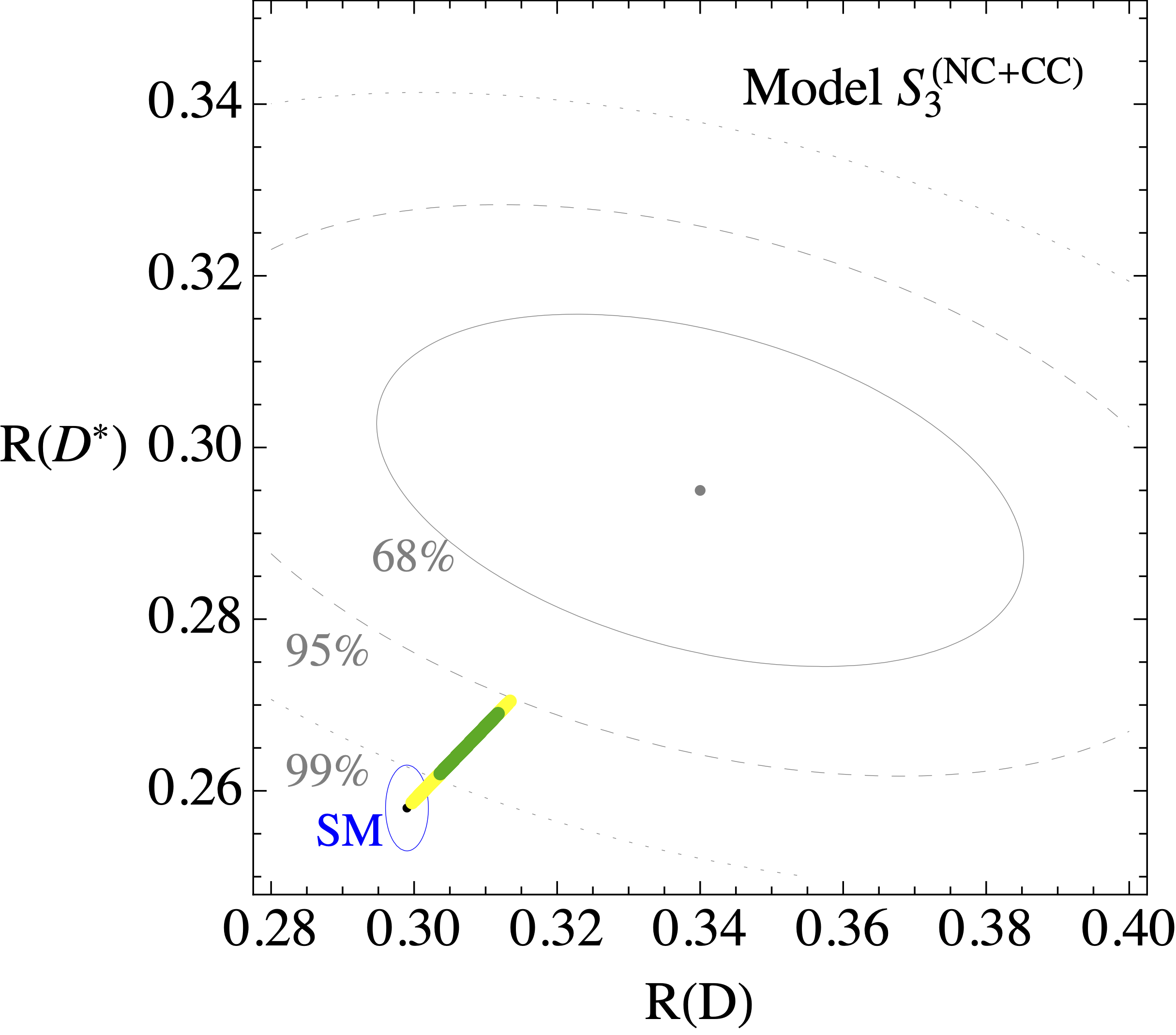}
\\
 \caption{\label{fig:S1S3-models-S3CCNC}{\small{}Result from the fit in the
$S_{3}$ model. In the upper panels we show the preferred regions
in the planes of two couplings, where the two not shown are profiled
over. The dashed lines show, for illustrative purposes, $2\sigma$
limits from individual observables where the other couplings are fixed
at the best-fit point (black dot). In the lower panels we show where
the preferred region is mapped in the planes of the neutral and charged-current
anomalies.}}
\end{figure}

We move on to examine $S_{3}$, and we attempt directly a combined
explanation of charged and neutral current anomalies. It is well known
that $S_{3}$ provides a simple and good explanation for the deviations
observed in $b\to s\ell\ell$, thanks to its tree-level contribution
to the partonic process. The couplings required are $\lambda_{b\mu}^{3L}$-$\lambda_{s\mu}^{3L}$,
with small enough values that other observables do not pose relevant
constraints. The leading contribution to both $R(D^{(*)})$ and $C_{9}^{u}$,
instead, arises via the $\lambda_{b\tau}^{3L}$-$\lambda_{s\tau}^{3L}$
couplings. For concreteness we fix $M_{3}=1\,\text{TeV}$, but the
fit would be very similar for a slightly larger mass.

Our results can be seen in Fig.~\ref{fig:S1S3-models-S3CCNC}. As
expected, the model is successful in fitting $\Delta C_{9}^{\mu}$.
The couplings to the tau allow to also fit $C_{9}^{u}$, while charged-current
anomalies cannot be reproduced. The main limiting observables are
$B_{s}$-mixing and $B\to K^{(*)}\nu\nu$, as can be seen from the
top-right panel.

The best-fit point, for $M_{3}=1\TeV$, is found for $\lambda_{b\tau}^{3L}\approx0.36,$
$\lambda_{s\tau}^{3L}\approx-0.13,$ $\lambda_{b\mu}^{3L}\approx0.050$,
$\lambda_{s\mu}^{3L}\approx0.015$.

\subsection{\label{sec:S1S3-models-S1S3LH}$S_{1}+S_{3}$ with LH couplings only}

\begin{figure}[t]
\centering \includegraphics[height=6.5cm]{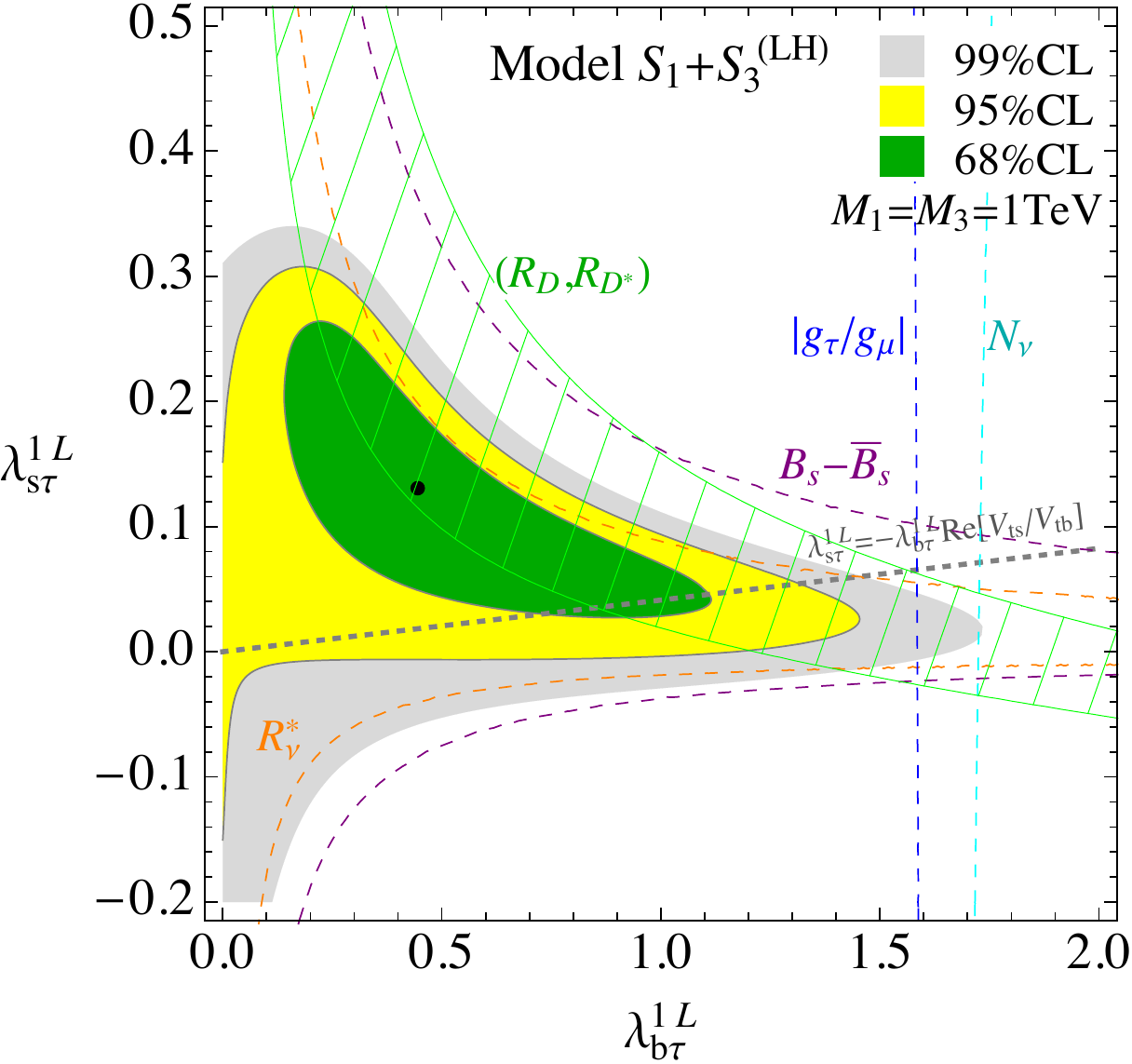}~
\includegraphics[height=6.5cm]{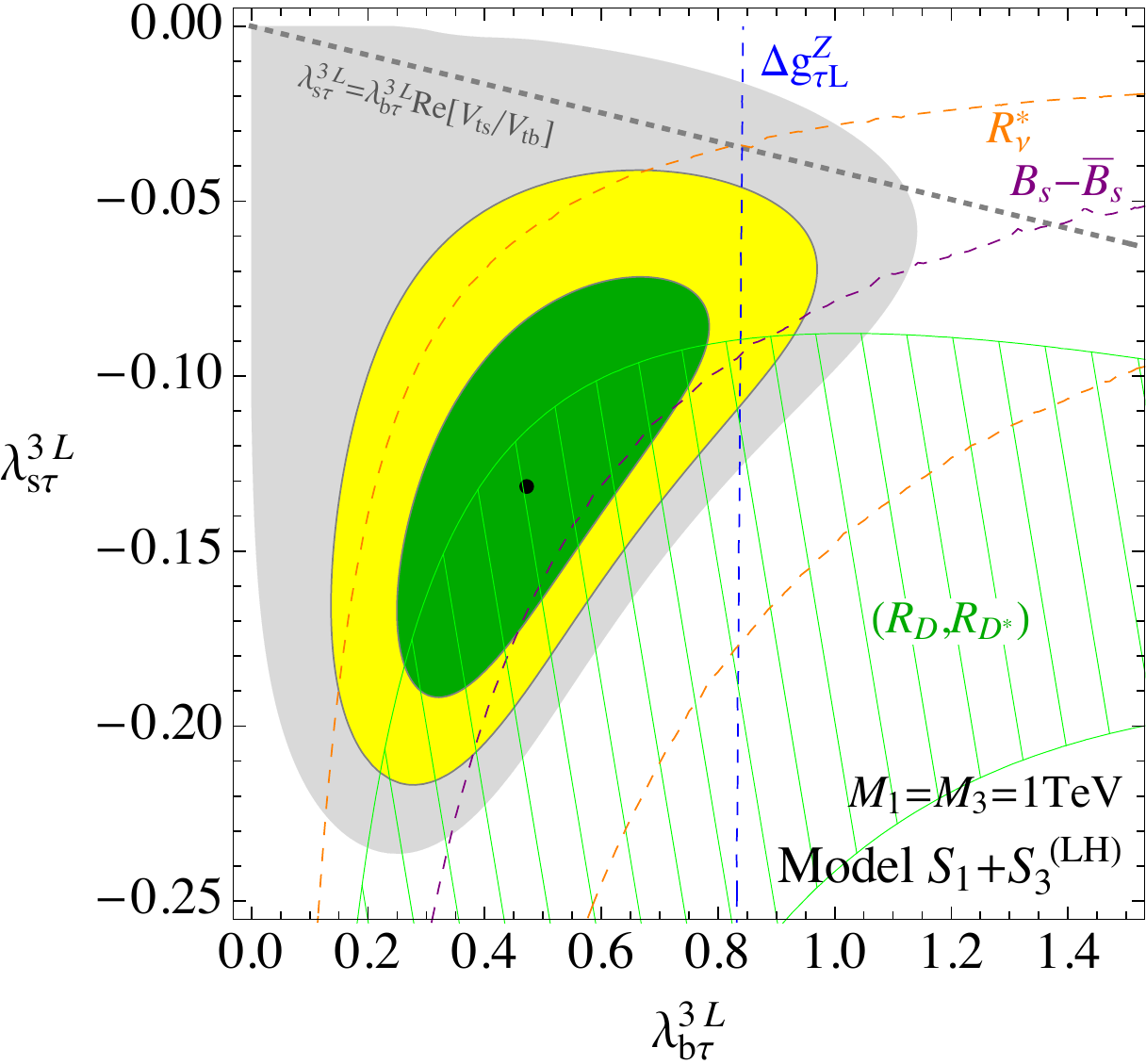}
\\[5pt] \includegraphics[height=6.5cm]{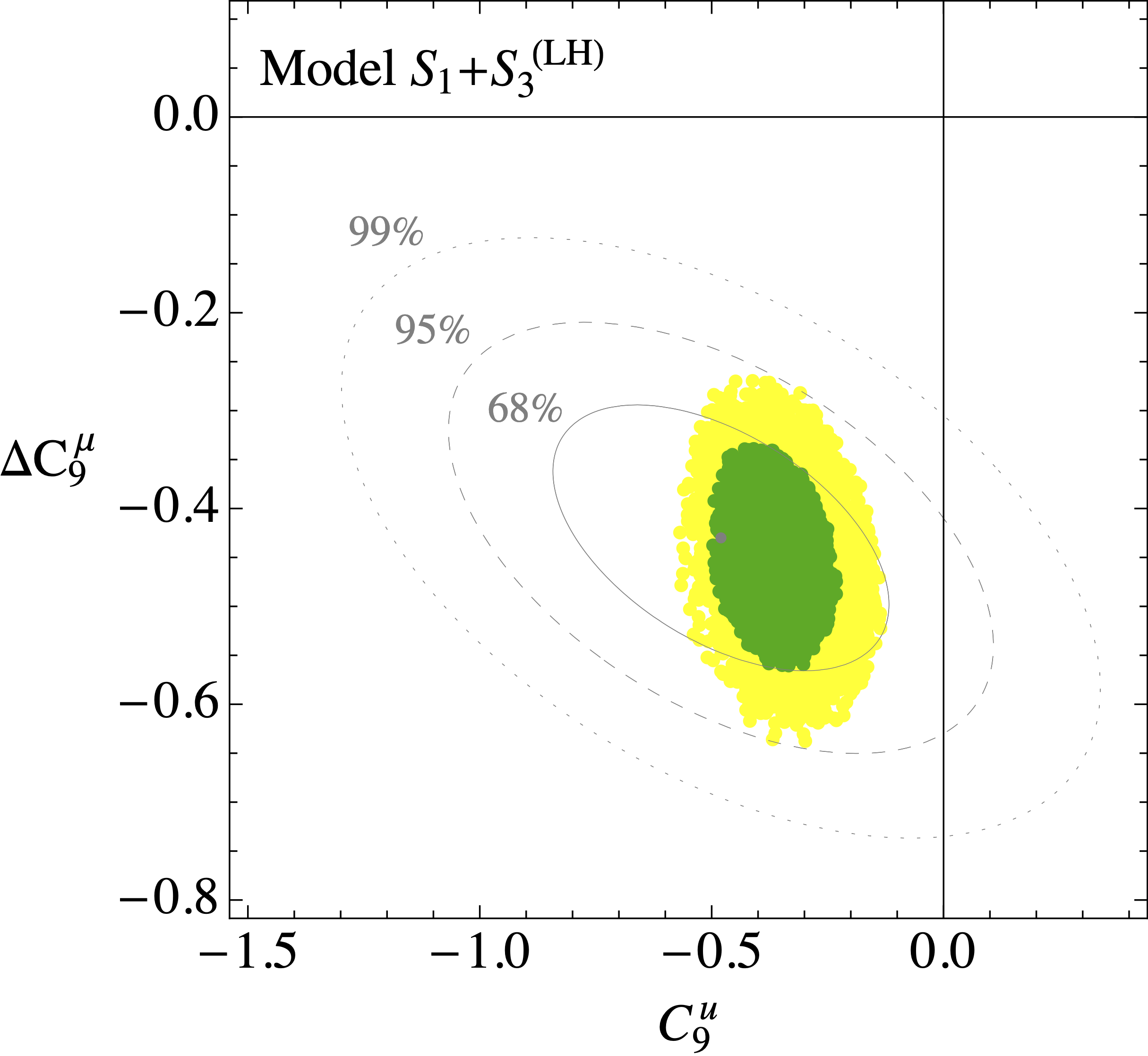}
~ \includegraphics[height=6.5cm]{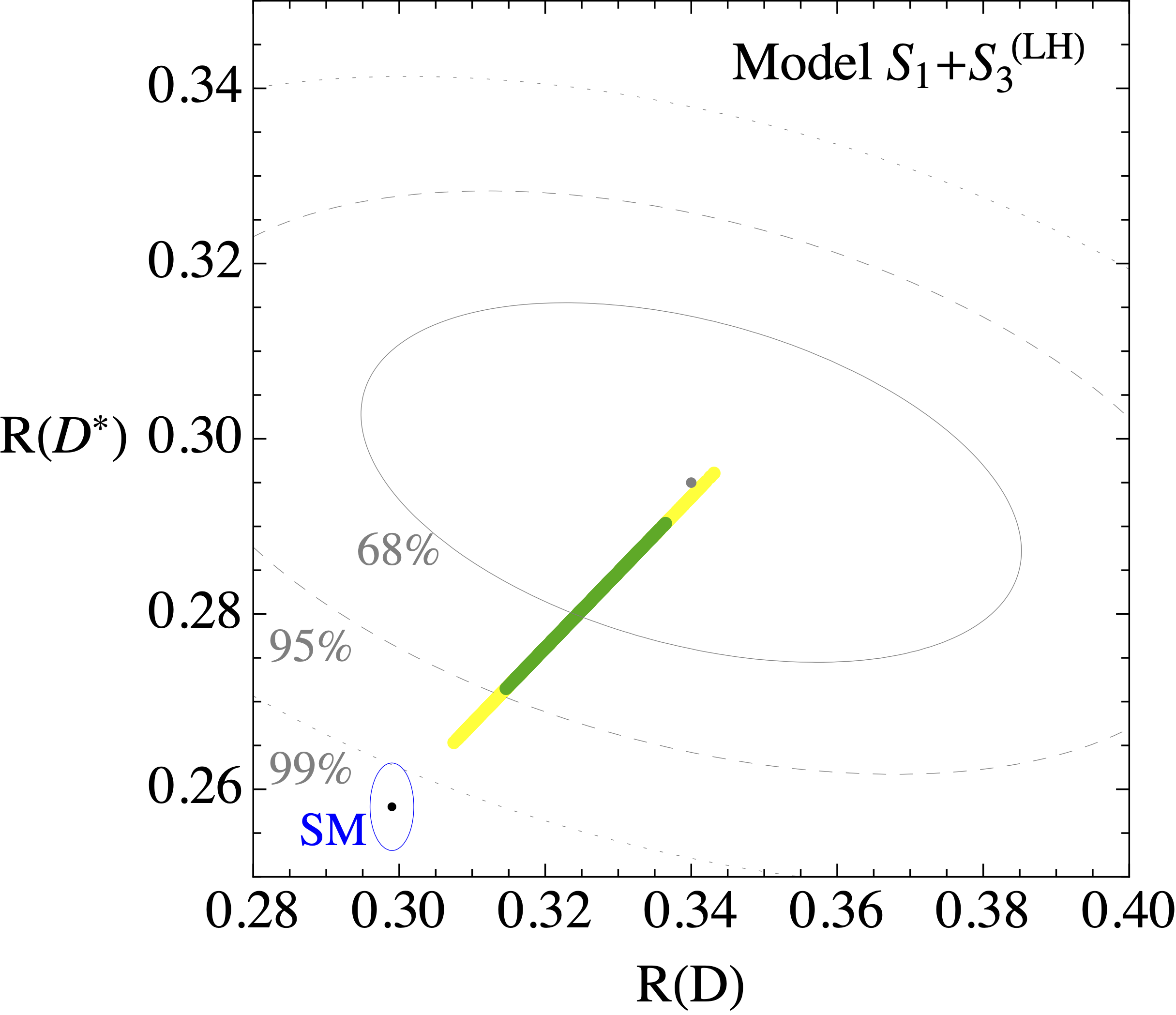}
\\
 \caption{\label{fig:S1S3-models-S1S3LH}{\small{}Result from the fit in the
$S_{1}+S_{3}\,^{(\text{LH})}$ model, with only left-handed couplings.
In the upper panels we show the preferred regions in the planes of
two couplings, where the two not shown are profiled over. The dashed
lines show, for illustrative purposes, $2\sigma$ limits from individual
observables where the other couplings are fixed at the best-fit point
(black dot). In the lower panels we show where the preferred region
is mapped in the planes of the neutral and charged-current anomalies.}}
\end{figure}

Models involving $S_{1}$ and $S_{3}$ with left-handed couplings
have been first considered in \cite{Crivellin:2017zlb,Buttazzo:2017ixm,Marzocca:2018wcf}.
In particular, in \cite{Buttazzo:2017ixm,Marzocca:2018wcf} it was
shown how this setup could fit both charged- and neutral-current anomalies
with couplings compatible with a minimally broken $\text{U}(2)^{5}$
flavour symmetry, albeit with a tension between $R(D^{(*)})$ and
the $B_{s}$-mixing constraint. Since then, new experimental updates
on $R(D^{(*)})$ pushed the preferred region closer to the SM, thus
also alleviating the tension with meson mixing. Here, we update the
fit for this scenario, without assuming a priori a specific flavour
structure for the relevant couplings.

The relevant couplings are $\lambda_{[bs]\tau}^{1L}$, $\lambda_{[bs]\mu}^{3L}$,
and $\lambda_{[bs]\tau}^{3L}$. A first qualitative understanding
of the model can be obtained by noticing the main roles of the various
couplings with regard to the anomalies: 
\[
\lambda_{[b,s]\mu}^{3L}\to\Delta C_{9}^{\mu},\quad\lambda_{[b,s]\tau}^{3L}\to C_{9}^{u},\quad(\lambda_{[b,s]\tau}^{1L},\,\lambda_{[b,s]\tau}^{3L})\to R(D^{(*)})
\]
In this model, the relative deviation in $R(D)$ and $R(D^{*})$ from
the respective SM values is predicted to be exactly the same, since
it is only due to the same left-handed vector-vector operator generated
in the SM.

The most salient features of the fit are summarized in Fig.~\ref{fig:S1S3-models-S1S3LH}.
In the top two panels we show the preferred regions in the $\lambda_{b\tau}^{1L}-\lambda_{s\tau}^{1L}$
and $\lambda_{b\tau}^{3L}-\lambda_{s\tau}^{3L}$ planes, together
with the single-observable $2\sigma$ limits obtained fixing the other
couplings to the global best-fit value. The favoured region in the
$\lambda_{b\mu}^{3L}-\lambda_{s\mu}^{3L}$ plane is very similar to
the one of model $S_{3}^{\,(CC+NC)}$ (Fig.~\ref{fig:S1S3-models-S3CCNC}
top-left), thus we do not show it again. The constraint from $B\to K^{(*)}\nu\nu$
is avoided thanks to a slight cancellation between the tree-level
contributions of the two leptoquarks \cite{Buttazzo:2017ixm}. There
is a (small) leftover tension in the $R(D^{(*)})$ fit, due to constraints
from $B_{s}$-mixing. It should be noted that this tension grows with
larger LQ masses (thus larger required couplings) since the deviation
in $R(D^{(*)})$ scales as $\lambda^{2}/M^{2}$ while the contribution
to meson mixing goes as $\lambda^{4}/M^{2}$.

We also point out that the parameter-region preferred by the fit is
compatible with the relations between couplings predicted by a minimally-broken
$\text{U}(2)^{5}$ flavour symmetry, $\lambda_{s\alpha}=c_{\text{U}(2)}V_{ts}\lambda_{b\alpha}$,
with $c_{\text{U}(2)}$ an $\mathcal{O}(1)$ complex parameter, see
e.g. \cite{Buttazzo:2017ixm} and references therein. The case with
$|c_{\text{U}(2)}|=1$ is shown with grey dashed lines in the upper
panels.

In the lower two panels we show how the preferred regions in parameter
space maps into the anomalous $B$-decay observables. As can be seen,
it is possible to reproduce them within $1\sigma$. The best-fit point,
for $M_{1}=M_{3}=1\TeV$, is found for $\lambda_{b\tau}^{3L}\approx0.47$,
$\lambda_{s\tau}^{3L}\approx-0.13$, $\lambda_{b\mu}^{3L}\approx0.056$,
$\lambda_{s\mu}^{3L}\approx0.014$, $\lambda_{b\tau}^{1L}\approx0.45$,
$\lambda_{s\tau}^{1L}\approx0.13$.

\subsection{\label{sec:S1S3-models-S1S3amu}$S_{1}+S_{3}$ addressing CC, NC,
and $(g-2)_{\mu}$}

\begin{figure}[t]
\centering \includegraphics[height=6.5cm]{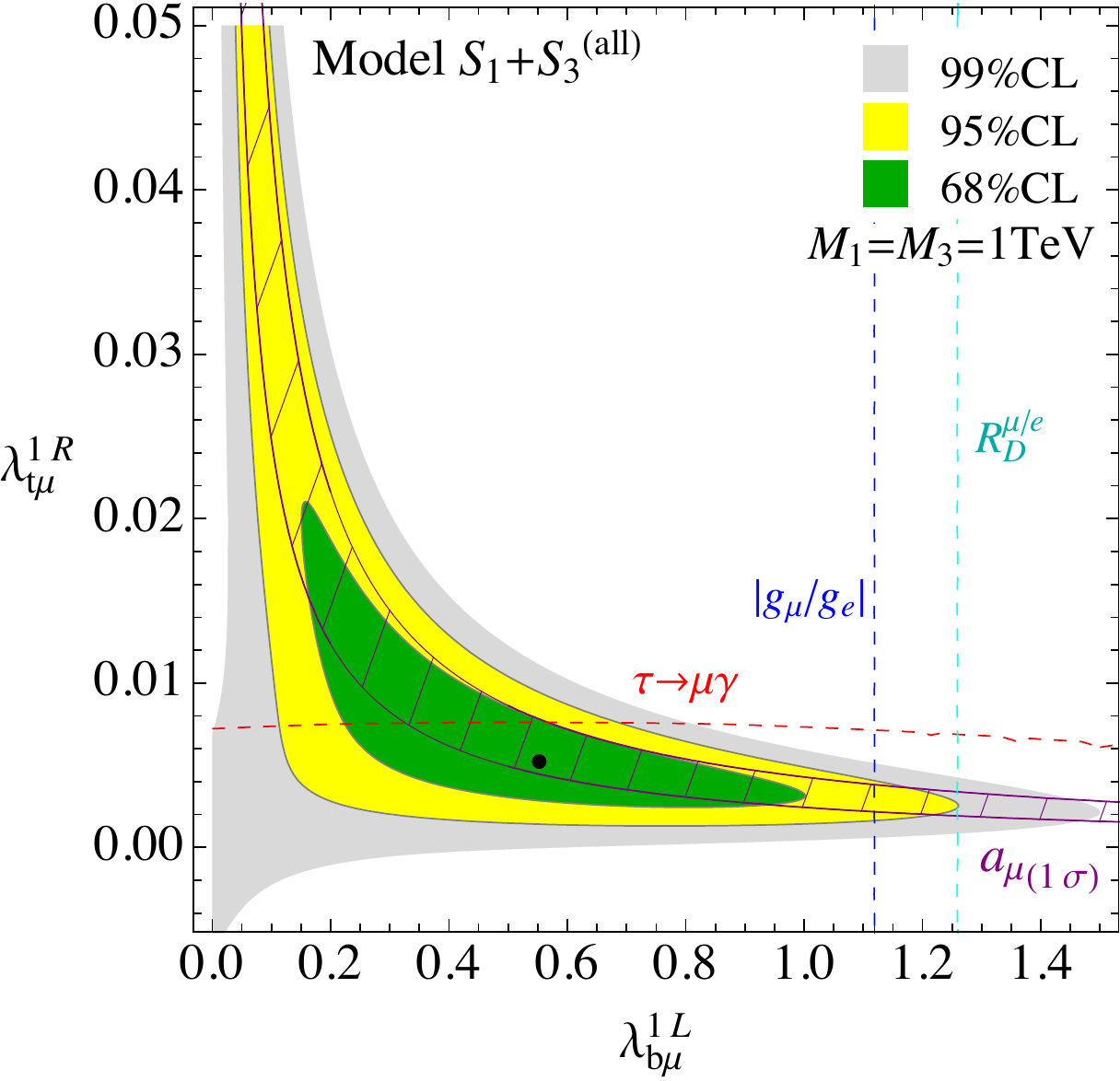}~
\includegraphics[height=6.5cm]{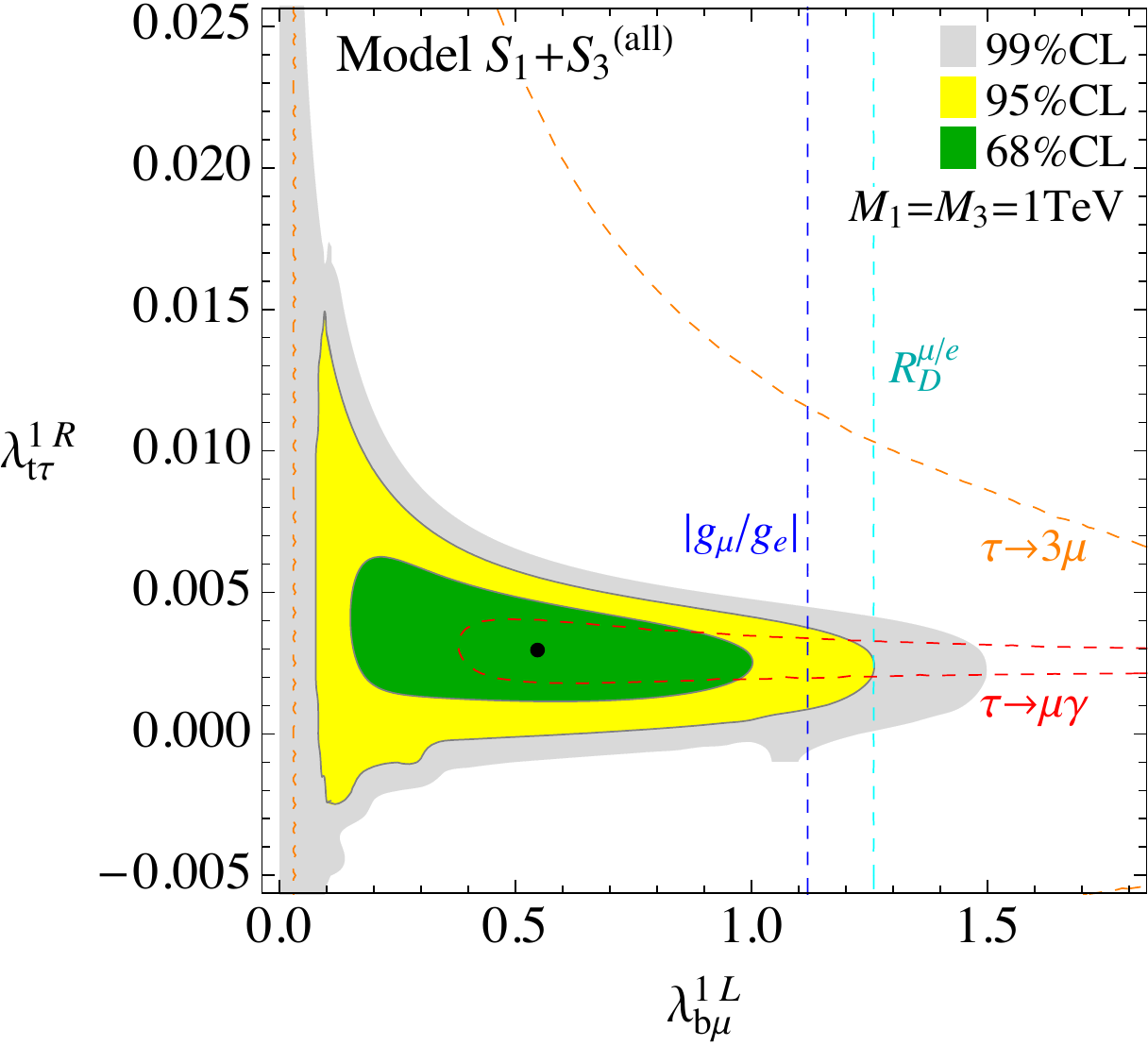}
\\[5pt] \includegraphics[height=6cm]{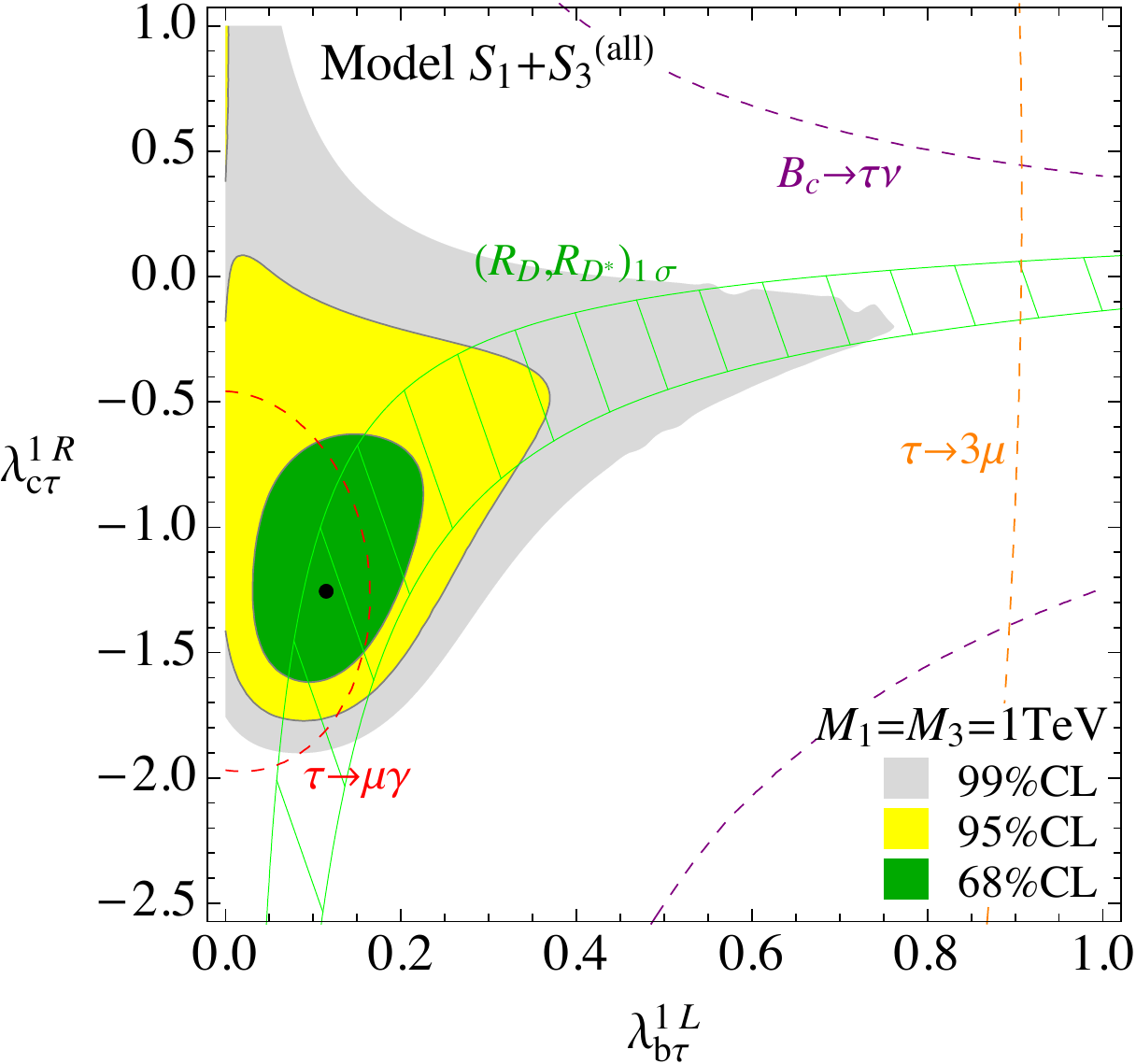}~
\includegraphics[height=6cm]{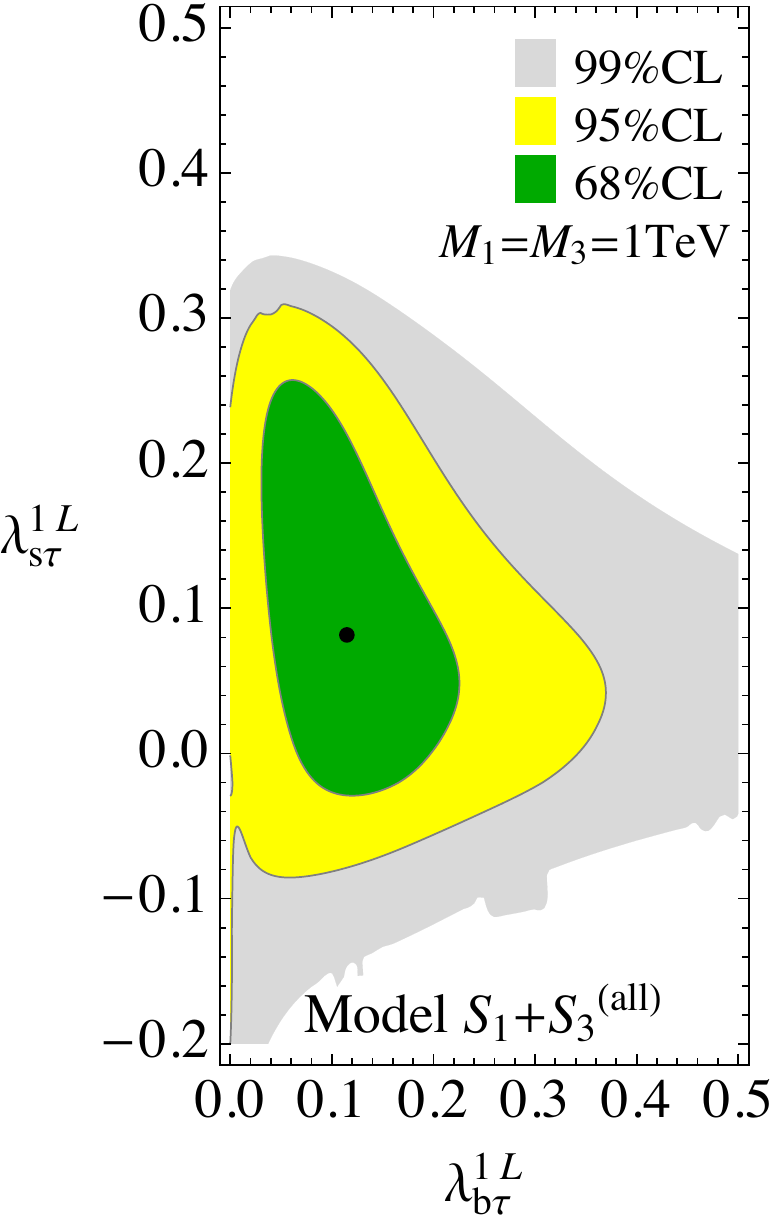}
~ \includegraphics[height=6cm]{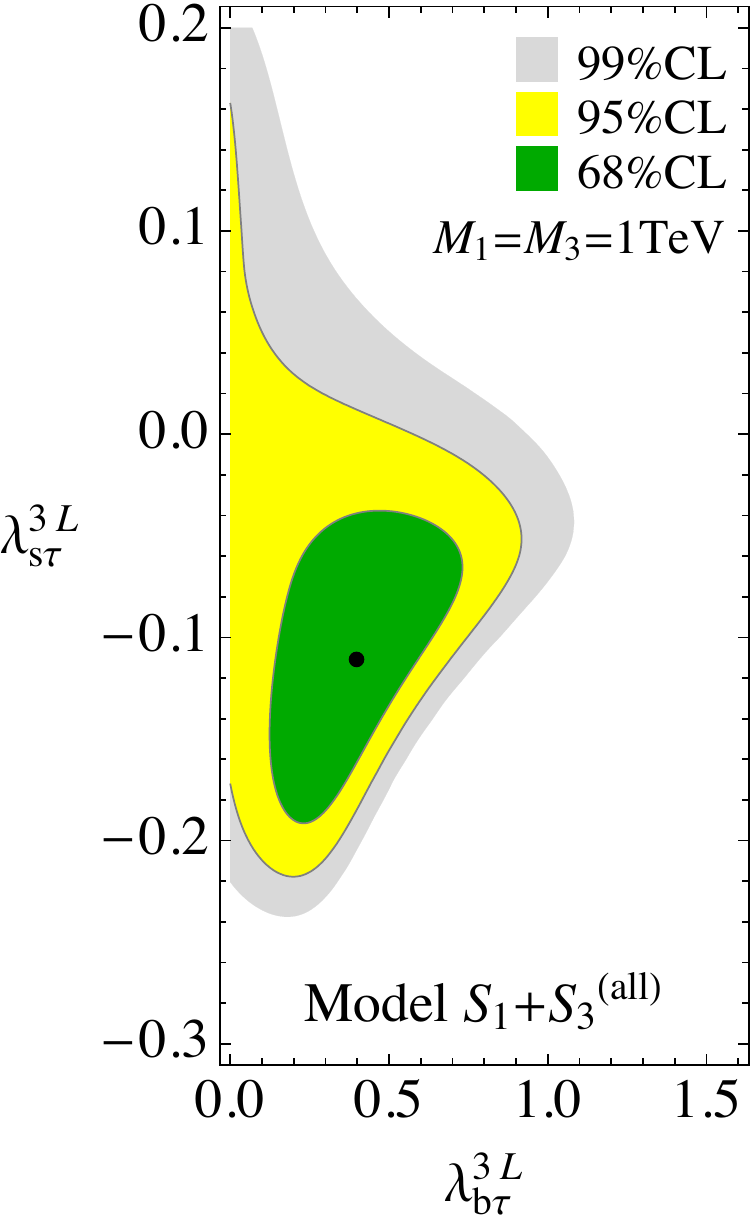}
\\[5pt] \includegraphics[height=6.5cm]{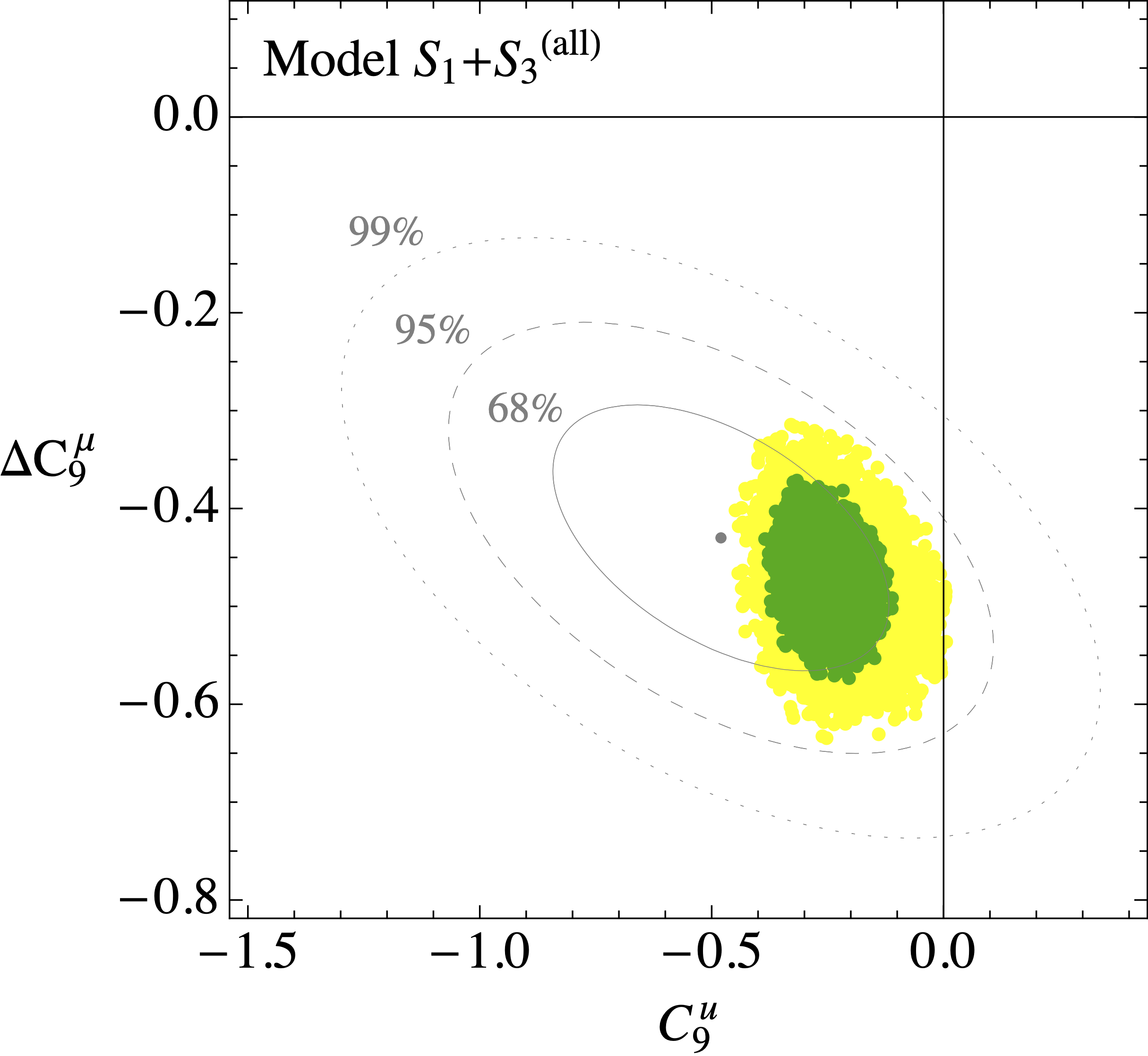}
~ \includegraphics[height=6.5cm]{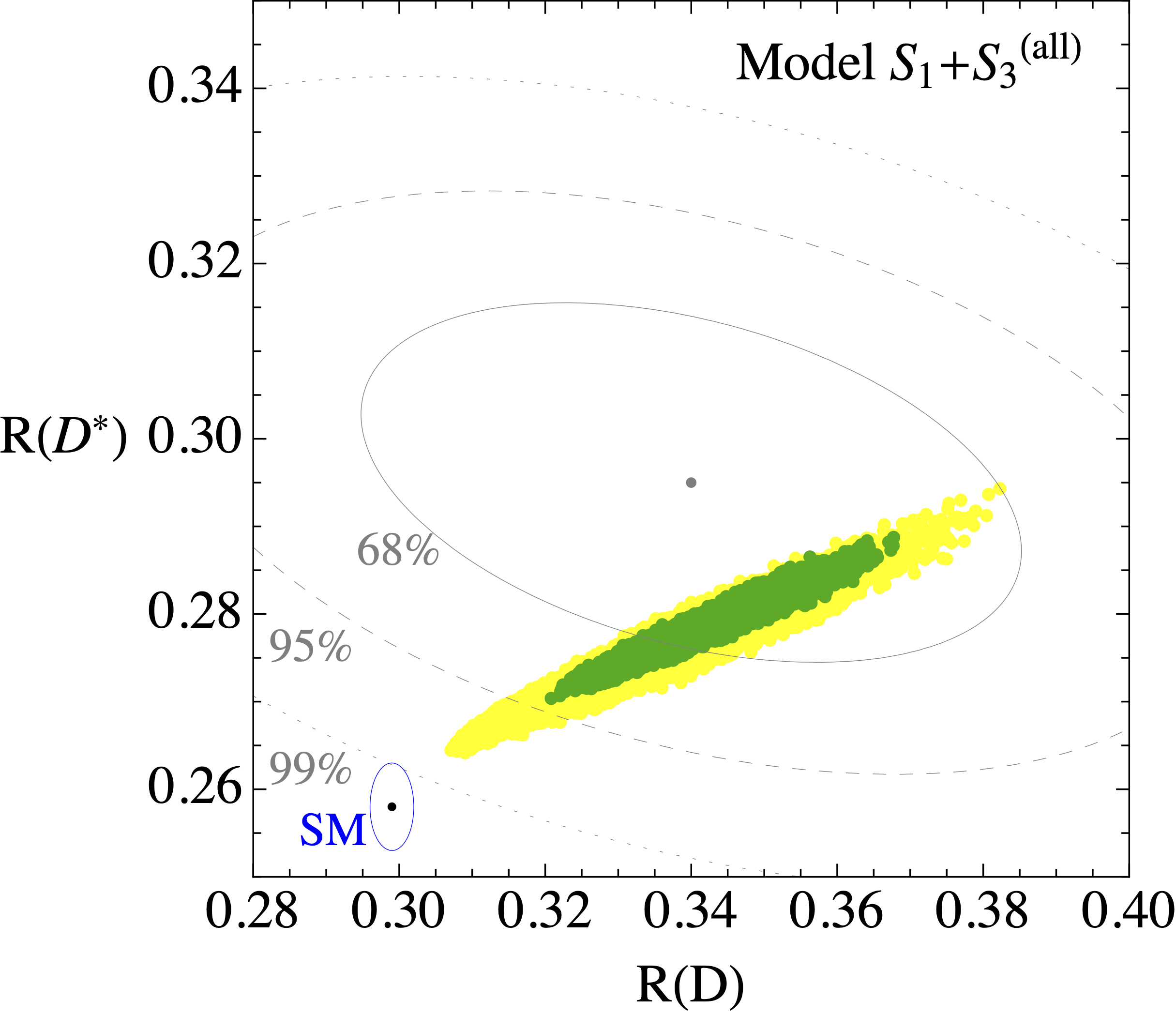}
\\
 \caption{\label{fig:S1S3-models-S1S3all}{\small{}Result from the fit in the
$S_{1}+S_{3}\,^{(\text{all})}$ model, aimed at addressing all anomalies
(see description in the text).}}
\end{figure}

From the previous Sections it is clear that in order to address all
anomalies, both $S_{1}$ and $S_{3}$ leptoquarks are required. NC
anomalies are addressed only by $S_{3}$, the muon anomalous magnetic
moment only by $S_{1}$, while $R(D^{(*)})$ receives sizeable contributions
from both. For our most general analysis we keep ten active couplings:
$\lambda_{[b,s]\tau}^{3L}$, $\lambda_{[b,s]\mu}^{3L}$, $\lambda_{[b,s]\tau}^{1L}$,
$\lambda_{b\mu}^{1L}$, $\lambda_{[t,c]\tau}^{1R}$, $\lambda_{t\mu}^{1R}$.
The results of our fit are shown in Fig.~\ref{fig:S1S3-models-S1S3all}.

In the first row of Fig.~\ref{fig:S1S3-models-S1S3all} we show the
preferred regions for the couplings relevant for the $a_{\mu}$ fit.
The situation is very similar to what already discussed for model
$S_{1}^{\,(CC+a_{\mu})}$, Sec.~\ref{sec:S1S3-models-S1CCamu}.

The couplings relevant for the $R(D^{(*)})$ fit are shown in the
second row. They show a behavior very similar to the one already seen
in the models $S_{1}^{\,(CC+a_{\mu})}$ and $S_{1}+S_{3}^{\,(LH)}$.
The main contribution is due to the scalar+tensor operators generated
via the $\lambda_{c\tau}^{1R}\lambda_{b\tau}^{1L}$ couplings, but
a sizeable contribution, which helps to improve the fit with respect
to model $S_{1}^{\,(CC)}$, is induced via the left-handed couplings
$\lambda_{[b,s]\tau}^{1L}$ and $\lambda_{[b,s]\tau}^{3L}$, analogously
to what we saw in model $S_{1}+S_{3}^{\,(LH)}$. Contrary to that
case, however, here the preferred region avoids any tension with both
$B_{s}$-mixing and $B\to K^{(*)}\nu\nu$.

We do not show in Fig.~\ref{fig:S1S3-models-S1S3all} the preferred
values for $\lambda_{[s,b]\mu}^{3L}$, which are necessary to fit
$\Delta C_{9}^{\mu}$, since they are analogous to what we saw for
model $S_{3}^{\,(CC+NC)}$ (see Fig.~\ref{fig:S1S3-models-S3CCNC}
top-left).

We conclude that all the anomalies in $R(D^{(*)})$, $b\to s\mu\mu$,
and $(g-2)_{\mu}$, can be completely addressed in this model, for
perturbative couplings and TeV-scale leptoquark masses. The best-fit
point, for $M_{1}=M_{3}=1\TeV$, is found for $\lambda_{b\tau}^{3L}\approx0.40$,
$\lambda_{s\tau}^{3L}\approx-0.11$, $\lambda_{b\mu}^{3L}\approx0.31$,
$\lambda_{s\mu}^{3L}\approx0.0024$, $\lambda_{b\tau}^{1L}\approx0.11$,
$\lambda_{s\tau}^{1L}\approx0.082$, $\lambda_{b\mu}^{1L}\approx0.55$,
$\lambda_{t\tau}^{1R}\approx0.0029$, $\lambda_{c\tau}^{1R}\approx-1.26$,
$\lambda_{t\mu}^{1R}\approx0.0052$.

\subsection{Leptoquark potential couplings}

\label{sec:S1S3Higgs}

In this Section we study available constraints for the potential couplings
of leptoquark with the Higgs boson from the leptoquark potential,
Eq. (\ref{eq:S1S3-potential}). There are four such couplings: $\lambda_{H1}$,
$\lambda_{H3}$, $\lambda_{\epsilon H3}$, and $\lambda_{H13}$. All
contribute only at one-loop level in the matching to SMEFT operators,
therefore possible phenomenological effects are suppressed both by
a loop factor and by the LQ mass scale. We focus on effects of these
couplings which are independent on the LQ couplings to fermions. We
thus need precisely measured quantities in the bosonic sector of the
SM.

\begin{table}[t]
\centering{}%
\begin{tabular}{|c|cl|c|}
\hline 
Observable & \multicolumn{2}{c|}{Measurement} & Reference\tabularnewline
\hline 
\hline 
$S$ & $0.04\pm0.08$ &  & \cite{Haller:2018nnx}\tabularnewline
\hline 
$T$ & $0.08\pm0.07$ & ($\rho_{S,T}=0.92$) & \cite{Haller:2018nnx}\tabularnewline
\hline 
$\kappa_{g}$ & $1.00\pm0.06$ &  & \cite{Aad:2019mbh}\tabularnewline
\hline 
$\kappa_{\gamma}$ & $1.03\pm0.07$ & ($\rho_{\gamma,g}=-0.44$) & \cite{Aad:2019mbh}\tabularnewline
\hline 
$\sigma/\sigma_{\text{SM}}(Z\gamma)$ & $2.0_{-0.9}^{+1.0}$ & (ATLAS) & \cite{Aad:2020plj}\tabularnewline
\hline 
$\sigma/\sigma_{\text{SM}}(Z\gamma)$ & $<3.9$ @ 95\% CL & (CMS) & \cite{Sirunyan:2018tbk}\tabularnewline
\hline 
\end{tabular}\caption{Bosonic observables for the LQ potential couplings. \label{tab:obsBosonic}}
\end{table}

Obvious candidates are the gauge-boson oblique corrections measured
at LEP \cite{Barbieri:2004qk}: $\hat{S}$, $\hat{T}$, $Y$, $W$,
as well as the analogous effect for QCD, $Z$. All these parameters
are measured at the per-mille level, and are able to constrain multi-TeV
scale physics. Given the expressions in the Warsaw basis of \cite{Wells:2015uba}
and our one-loop matching of the SMEFT to the LQ model, we find:
\begin{align}
\hat{S} & =\frac{\alpha}{4s_{W}^{2}}S=-\frac{g^{2}N_{c}v^{2}Y_{S_{3}}}{48\pi^{2}}\frac{\lambda_{\epsilon H3}}{M_{3}^{2}}\approx-5.4\times10^{-5}\frac{\lambda_{\epsilon H3}}{m^{2}}\\
\hat{T} & =\alpha T=\frac{N_{c}v^{2}\lambda_{\epsilon H3}^{2}}{48\pi^{2}M_{3}^{2}}+\frac{N_{c}v^{2}}{16\pi^{2}}|\lambda_{H13}|^{2}\frac{M_{1}^{4}-M_{3}^{4}-2M_{1}^{2}M_{3}^{2}\log M_{1}^{2}/M_{3}^{2}}{(M_{1}^{2}-M_{3}^{2})^{3}}\\
 & \approx3.8\times10^{-4}\lambda_{\epsilon H3}^{2}/m^{2}+3.8\times10^{-4}|\lambda_{H13}|^{2}/m^{2}~\nonumber 
\end{align}
where in the numerical expressions for simplicity we fixed $M_{1}=M_{3}=m\TeV$.
The contributions to $Y$, $W$, and $Z$ are instead at, or below,
the $10^{-6}$ level and thus completely negligible given the present
experimental precision. The constraints on $S$ and $T$ from \cite{Haller:2018nnx}
are reported in Table \ref{tab:obsBosonic}. The contribution to the
$T$ parameter from the $\lambda_{H13}$ coupling has been also studied
in \cite{Dorsner:2019itg}, albeit not in the EFT approach. We checked
that we agree once the EFT limit is taken into account.

\begin{figure}[t]
\includegraphics[scale=0.6]{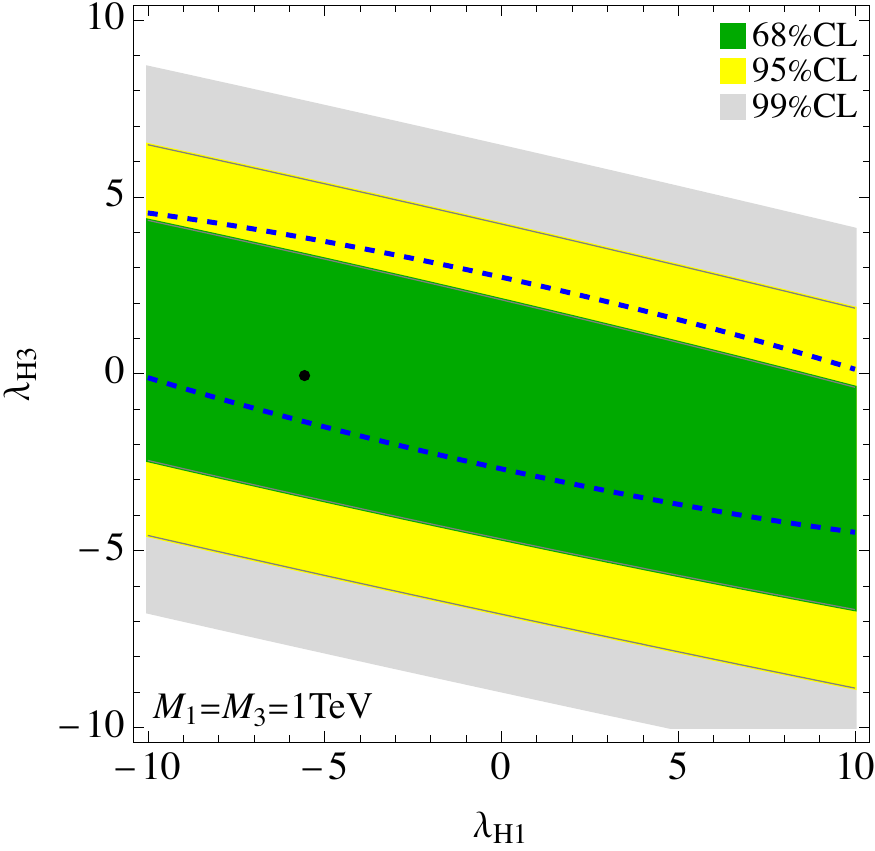}\includegraphics[scale=0.6]{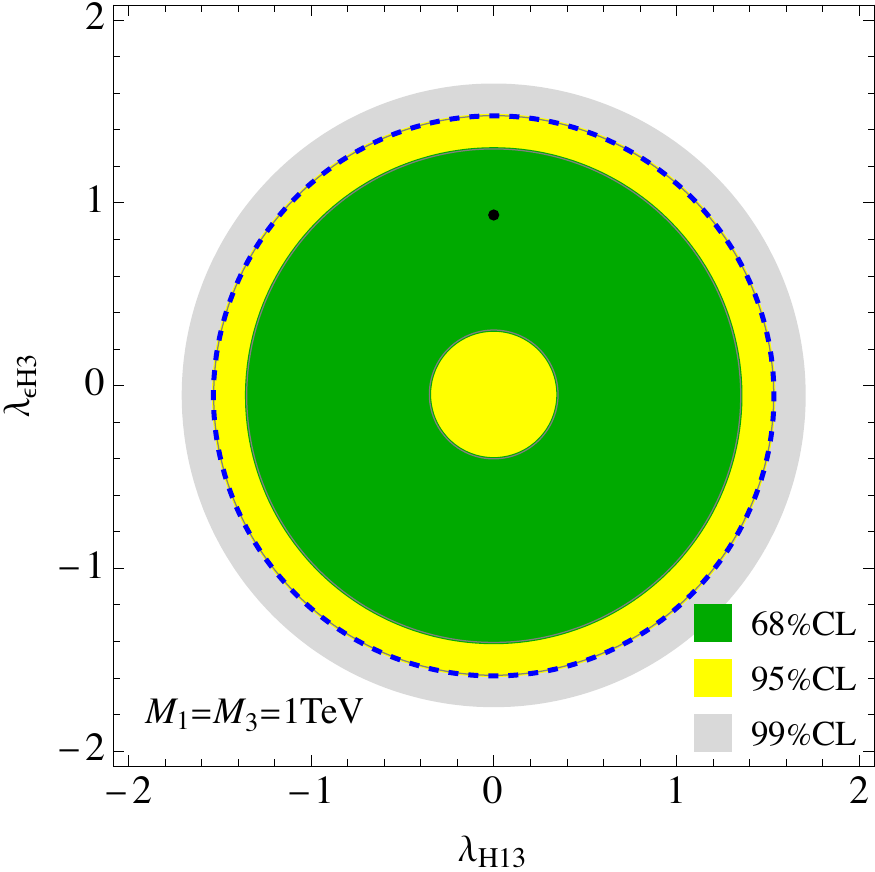}
\caption{\label{fig:S1S3-models-potential}{\small{}Limits on LQ potential
couplings from oblique corrections and Higgs measurements. In each
panel, the other two couplings have been marginalised. The black point
represents the best-fit point while the dashed blue contours are the
prospects for 95\%CL limits after HL-LHC.}}
\end{figure}

The LQ couplings to the Higgs also generate at one-loop contributions
to $hgg$, $h\gamma\gamma$, and $hZ\gamma$ couplings. Since these
are also loop-generated in the SM, the percent-level precision presently
available for the Higgs couplings to photons and gluons couplings
allows to probe heavy new physics. Loop contributions to other couplings,
which arise at tree-level in the SM, are instead too small to have
a sizeable impact. We thus consider the combined fit of Higgs couplings
in the $\kappa$-framework where only $\kappa_{\gamma}$ and $\kappa_{g}$
are left free, and a constraint on $\sigma/\sigma_{\text{SM}}(Z\gamma)=\kappa_{g}^{2}\kappa_{Z\gamma}^{2}$,
which is however still not precisely measured, see Table~\ref{tab:obsBosonic}.
The approximate contributions to these parameters in our model are
given by: 
\begin{align}
\kappa_{g}-1 & =-(3.51\lambda_{H3}+1.17\lambda_{H1})\times10^{-2}/m^{2},\label{eq:S1S3-kappa_g}\\
\kappa_{\gamma}-1 & =-(2.32\lambda_{H3}+0.66\lambda_{\epsilon H3}-0.11\lambda_{H1})\times10^{-2}/m^{2},\label{eq:S1S3-kappa_gamma}
\end{align}
\begin{equation}
\kappa_{Z\gamma}-1=-(1.89\lambda_{H3}+0.23\lambda_{\epsilon H3}-0.033\lambda_{H1})\times10^{-2}/m^{2}.\label{eq:S1S3-kappa_Zgamma}
\end{equation}
Analogously to what presented above for flavour observables, we combine
Higgs couplings and oblique constraints in a global likelihood. From
this we find the maximum likelihood point and construct the 68, 95,
and 99\% CL regions in planes of two couplings, where the other two
are marginalised. The results in the ($\lambda_{H1}$,$\lambda_{H3}$)
and ($\lambda_{H13}$,$\lambda_{\epsilon H3}$) planes are shown in
Fig.~\ref{fig:S1S3-models-potential} for $M_{1}=M_{3}=1\TeV$. We
observe that a limit of about $1.5$ can be put on both $\lambda_{H13}$
and $\lambda_{\epsilon H3}$ (right panel). This comes mainly from
the contribution to the $\hat{T}$ parameter, which is quadratic in
the two couplings and thus allows to constrain both at the same time.
The $\lambda_{H1}$ and $\lambda_{H3}$ couplings, instead, are constrained
mainly from their contribution to the $h\gamma\gamma$ and $hgg$
couplings, Eqs.~(\ref{eq:S1S3-kappa_g}) and (\ref{eq:S1S3-kappa_gamma}).
We see that with present experimental accuracy the limits are still
rather weak, and there is an approximate flat direction which doesn't
allow to put any relevant bound on $\lambda_{H1}$.

This situation will marginally improve with the more precise Higgs
measurements from HL-LHC \cite{Cepeda:2019klc}. The future expected
95\%CL contours are shown as dashed blue lines. This however has no
appreciable effect on the limits shown in the right panel, since those
are dominated by the constraint on the $T$ parameter, which will
instead improve substantially from measurements on the $Z$ pole at
FCC-ee. A more detailed analysis of FCC prospects are however beyond
the scope of this paper.

\section{Prospects\label{sec:S1S3-Prospects}}

In this Section, we discuss the implications of future Belle II measurements
of 
\begin{itemize}
\item LFV $B$ decays induced at parton level by $b\to s\tau\mu$; 
\item $B$ decays induced at parton level by $b\to s\tau\tau$. 
\end{itemize}
These processes, in fact, are particularly interesting for leptoquark
scenarios aiming at addressing both neutral and charged-current $B$-anomalies.
Both are induced at tree-level by $S_{3}$ and, by $\text{SU}(2)_{L}$
relations, the $b\to c\tau\nu_{\tau}$ transition, tree-level in the
SM, is related to the FCNC transition $b\to s\tau\tau$. Also, LFV
is a natural consequence of leptoquark couplings once also the coupling
to muons is considered, as required by neutral-current anomalies.
While the LFV $B$-meson decays are already included in the global
fits described in the previous Sections, the current bounds on $b\to s\tau\tau$
observables are of the order of $\sim10^{-3}$ and thus too weak to
set constraints on the model parameters. However, Belle II, with $50$ab$^{-1}$
of luminosity, will strongly improve the sensitivity, in particular
for the branching fraction of the semileptonic decays. On the other
hand, the Upgrade II of LHCb will set competitive bounds on the leptonic
decay $B_{s}\to\tau\tau$. The relevant future expected limits at
$95\%$ C.L. for Belle II \cite{Kou:2018nap} and LHCb \cite{Bediaga:2018lhg}
are summarised in Table~\ref{tab:BelleIIprosp}.

\begin{table}[t]
\centering{}%
\begin{tabular}{|c|c|c|c|}
\hline 
Observable & Present limit & Belle II $(5)50$ab$^{-1}$ & LHCb Up.-II\tabularnewline
\hline 
\hline 
\multicolumn{4}{|c|}{$b\to s\tau\mu$ observables}\tabularnewline
\hline 
Br$(B^{+}\to K^{+}\tau^{\pm}\mu^{\mp})$ & $<3.3(5.4)\times10^{-5}$ \cite{Lees:2012zz,Aaij:2020mqb} & $3.9\times10^{-6}$ & $\mathcal{O}(10^{-6})$\tabularnewline
\hline 
Br$(B_{s}\to\tau^{\pm}\mu^{\mp})$ & $<4.2\times10^{-5}$ \cite{Aaij:2019okb} & $\sim4\times10^{-6}$ & $\sim1\times10^{-5}$\tabularnewline
\hline 
\hline 
\multicolumn{4}{|c|}{$b\to s\tau\tau$ observables}\tabularnewline
\hline 
Br$(B^{+}\to K^{+}\tau^{+}\tau^{-})$ & $<2.8\times10^{-3}$ \cite{TheBaBar:2016xwe} & $(7.7)\,2.4\times10^{-5}$ & -\tabularnewline
\hline 
Br$(B_{s}\to\tau^{+}\tau^{-})$ & $<6.8\times10^{-3}$ \cite{Aaij:2017xqt} & $(9.7)3\times10^{-4}$ & $5\times10^{-4}$\tabularnewline
\hline 
\end{tabular}\caption{Future Belle II and LHCb sensitivities, at $95\%$ C.L., for $b\to s\tau\mu$
and $b\to s\tau\tau$ observables. \label{tab:BelleIIprosp}}
\end{table}

\begin{figure}[t]
\centering \includegraphics[scale=0.6]{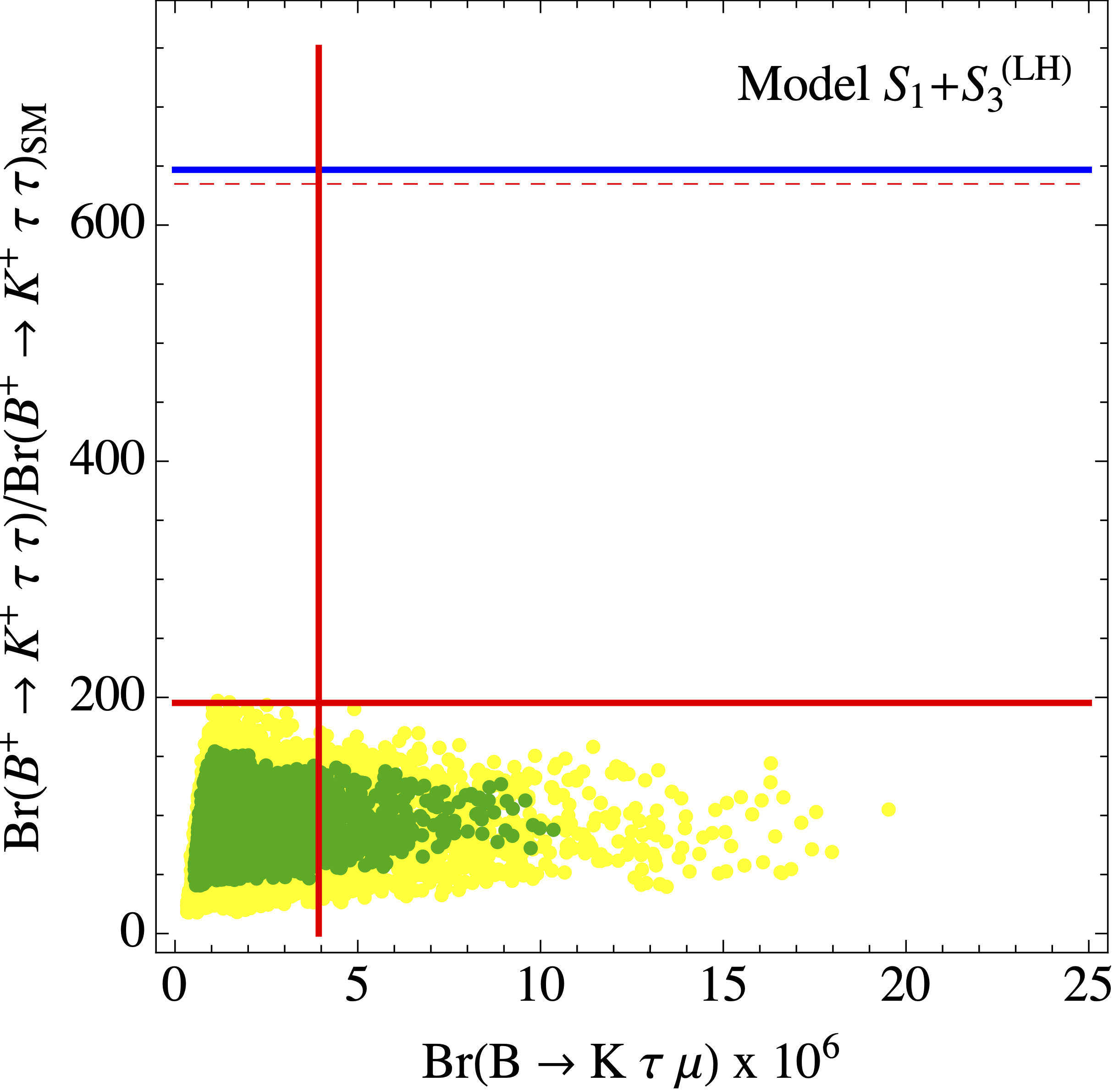}\quad{}\includegraphics[scale=0.6]{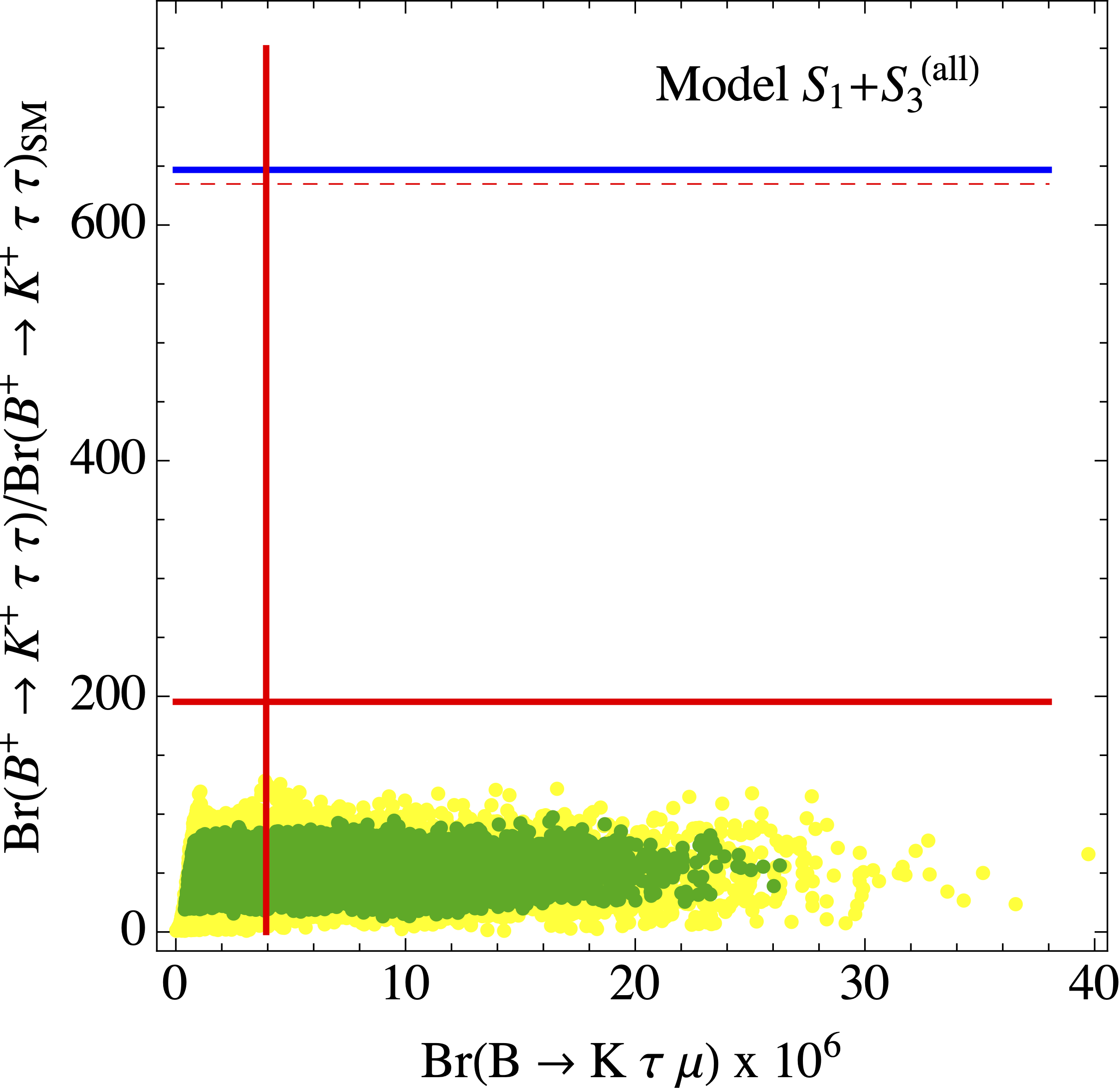}\\[10pt]
\caption{{\small{}Results from the fit in the $S_{1}+S_{3}^{\,{\rm (LH)}}$
model (left panel) and in the $S_{1}+S_{3}^{\,{\rm (all)}}$ (right
panel). The red solid (dashed) lines correspond to the $50$ab$^{-1}$
($5$ab$^{-1}$) Belle II future bounds, at 95\% C.L. . The blue solid
line is the prospected bound for the LHCb Upgrade II, on Br$(B_{s}\to\tau\tau)/$Br$(B_{s}\to\tau\tau)_{{\rm SM}}$.}}
\label{fig:BelleIIprosp}
\end{figure}

In Fig.~\ref{fig:BelleIIprosp} we show how the preferred parameter-space
regions for the models $S_{1}+S_{3}^{\,(LH)}$ (left) and $S_{1}+S_{3}^{\,(all)}$
(right) map in the plane of the branching fractions of the LFV decay
$B^{+}\to K^{+}\tau\mu$ and the decay $B^{+}\to K^{+}\tau\tau$ (normalised
to the SM value).\footnote{It should be noted that at tree-level in our model this ratio is the
same for all decays involving the $b\to s\tau\tau$ transition, e.g.
$B_{s}\to\tau\tau$ (see Ref. \cite{Gherardi:2020qhc}).} The red horizontal lines correspond to the Belle II future bounds
at $95\%$ C.L. on Br$(B^{+}\to K^{+}\tau^{+}\tau^{-})$ at $5$ab$^{-1}$
(dashed lines) and $50$ab$^{-1}$ (solid lines), while the vertical
ones represent the Belle II $50$ab$^{-1}$ prospect for Br$(B^{+}\to K^{+}\tau\mu)$.
One can see that, in both scenarios, the predictions for both the
non-LFV and LFV semileptonic $B$ decay into $\tau$ are in the ballpark
of the future Belle II sensitivity at $50$ab$^{-1}$, while the expected
bounds at $5$ab$^{-1}$ are still too weak to set significant constraints
on the models. Furthermore, one can notice that the future measurements
of $b\to s\tau\tau$ observables are constraining more strongly the
parameter space of the $S_{1}+S_{3}~^{{\rm (LH)}}$ model than the
one of the $S_{1}+S_{3}~^{{\rm (all)}}$ model. For the leptonic decay
$B_{s}\to\tau^{+}\tau^{-}$ at Belle II only the prospect at luminosity
of $5$ab$^{-1}$ is available; it is not shown in the plots since
it is weaker with respect to the semileptonic decays and correspond
to a horizontal line at $\sim1250$. On the other hand, for the Upgrade
II of LHCb, the prospected bound on Br$(B_{s}\to\tau^{+}\tau^{-})$
(blue horizontal lines) is stronger and leads to constraints similar
to the ones that we obtain from the $B^{+}\to K^{+}\tau^{+}\tau^{-}$
decay measured at $5$ab$^{-1}$ Belle II. In order to evaluate the
constraining power of future Br$(B_{s}\to\tau\mu)$ measurements,
in Fig.~\ref{fig:BelleIIprosp}, one could keep in mind that in our
model we have Br$(B_{s}\to\tau\mu)/$Br$(B^{+}\to K^{+}\tau\mu)\approx0.89$,
at tree-level \cite{Gherardi:2020qhc}.

\section{\label{sec:S1S3-Light-generations}Light generations}

For the purpose of addressing $B$-anomalies and $(g-2)_{\mu}$, $S_{1}$
and $S_{3}$ are only required to couple to second and third generation
quarks and leptons, whereas couplings to first generation could in
principle be very small. As already mentioned, the exact vanishing
of first generation couplings assumed in Ref. \cite{Gherardi:2020qhc}
was a mere working hypothesis, whereas in more realistic models one
would expect non-zero couplings, perhaps suppressed by some mechanism,
\emph{e.g.} flavour symmetry. 

In Ref. \cite{Marzocca:2021miv}, the authors consider the $S_{1}+S_{3}$
model with left-handed couplings only (cf. Subsec. \ref{sec:S1S3-models-S1S3LH})
and with an assumed $\text{U}(2)^{\text{5}}$ flavour symmetry, which
dictates the magnitude of first generation couplings. As shown in
the reference, the resulting constraints from Kaon observables prevent
a fit of $R(D^{(*)})$ beyond the $2\sigma$ level, in contrast with
our previous findings in Subsec. \ref{sec:S1S3-models-S1S3LH}, where
bounds from Kaon physics were neglected. Ref. \cite{Marzocca:2021miv}
also studies generic predictions for Kaon observables such as $\text{Br}(K_{L}\to\pi^{0}\nu\nu)$
and $\text{Br}(K_{S}\to\mu^{+}\mu^{-})$, which could be potentially
probed by KOTO stage-1 or LHCb are found to be allowed, and for electronic
observables such as $\mu\to e$ conversion in nuclei and $\mu\to3e$
decay, whose bounds are also expected to improve by several orders
of magnitude in the near future \cite{Moritsu:2021fns,Ankenbrandt:2006zu,mu2e:2013szc,Mu2e:2014fns,Mu3e:2020gyw}.

We see thus that in the $S_{1}+S_{3}$ model there exist important
correlations between Kaon observables, electron number violating observables,
and the $B$-anomalies. These findings allow us to anticipate the
\emph{leitmotiv} of the next Chapter: if the $B$-anomalies are experimentally
confirmed, a thorough understanding of how the anomalous $b\to c\ell\nu$
and $b\to s\ell\ell$ observables correlate with other observables
will be crucial, both for model validation and for prospects. 

\section{Conclusions}

\label{sec:conclusions}

\begin{figure}[t]
\centering \includegraphics[height=13cm]{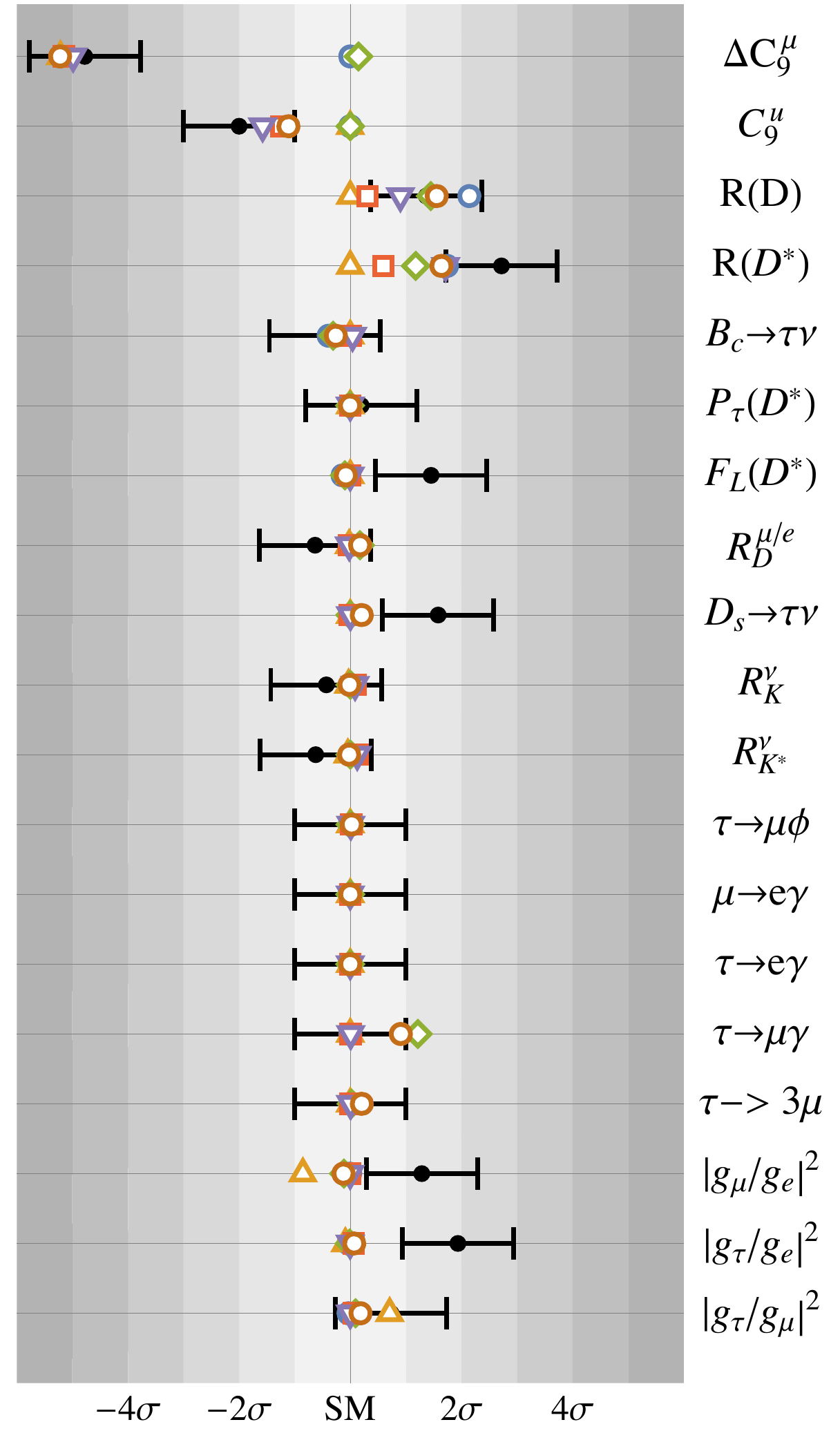}\quad{}\includegraphics[height=13cm]{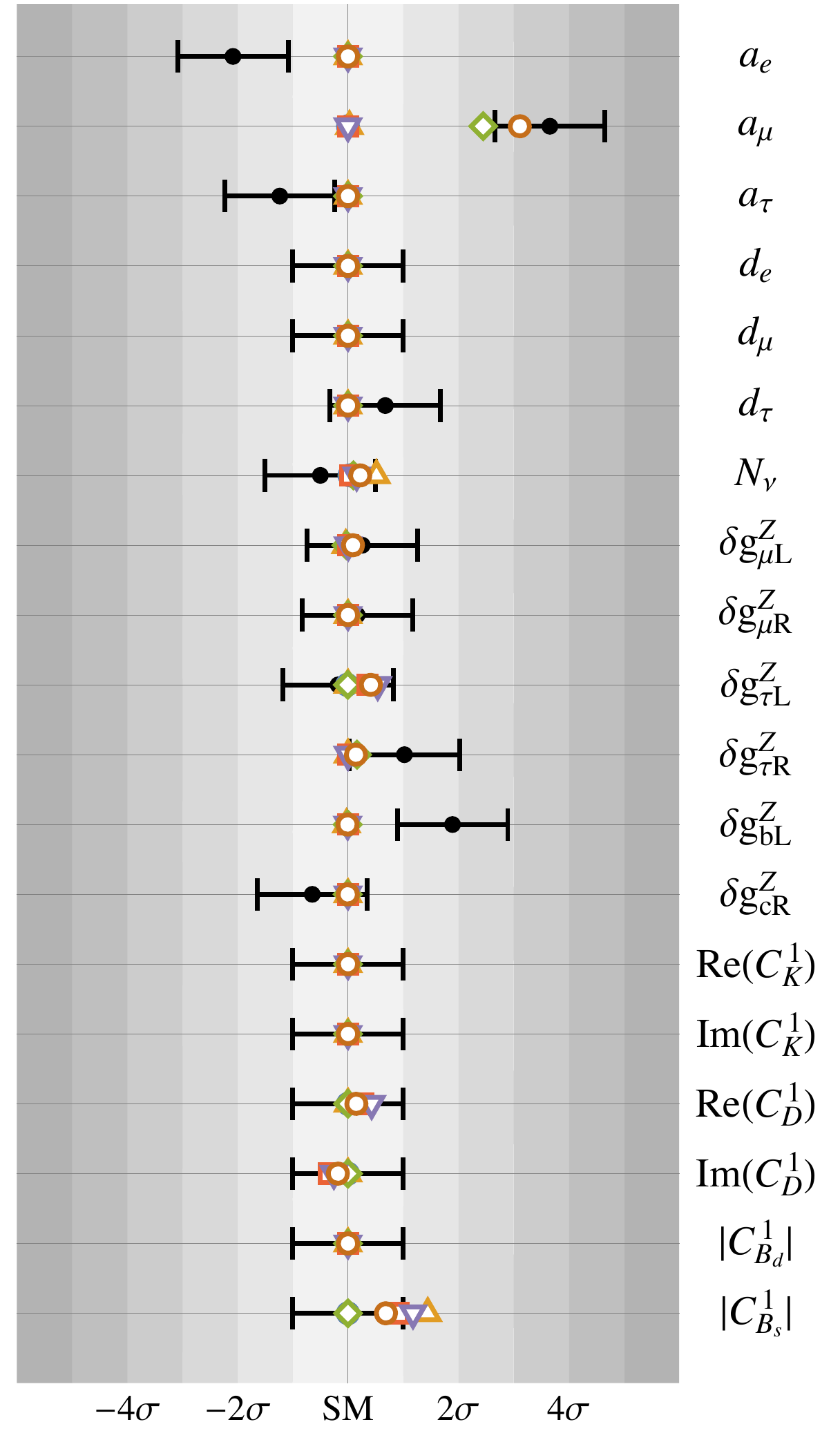}\\[10pt]
\includegraphics[width=8cm]{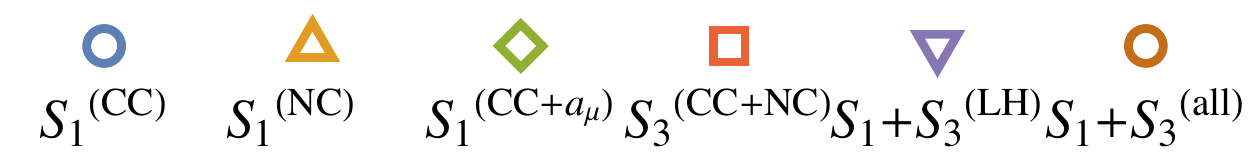} \caption{{\small{}For the best-fit points in each model studied in this Chapter,
we show the relative deviations from the Standard Model prediction
in all observables, in terms of number of sigmas given by the experimental
precision in that observable. The black intervals represent the experimental
measurements}}
\label{fig:summary}
\end{figure}

In this Chapter we examined in detail and at one-loop accuracy the
phenomenology of Standard Model extensions involving the two leptoquarks
$S_{1}$ and $S_{3}$, motivated by the experimental discrepancies
observed in $B$-meson decays and in the muon anomalous magnetic moment
$(g-2)_{\mu}$.

To this aim, we performed global fits for several benchmark models
to a comprehensive list of flavor and electroweak precision observables,
each computed at one-loop accuracy, leveraging on our previous work
\cite{Gherardi:2020det}. For each model, we identify best-fit regions
and major sources of tension, when present, and also provide prospects
for $B$-decays to $\tau\tau$ and $\tau\mu$, in the experimental
scope of Belle II and LHCb.

It is found that models involving only the $S_{1}$ leptoquark can
consistently address $R(D^{(*)})$ and $(g-2)_{\mu}$ anomalies, while
a fully-satisfactory solution for $b\to s\mu\mu$ anomalies is prevented
by the combination of constraints from $B_{s}$-mixing and LFU in
$\tau$ decays. Conversely, the $S_{3}$ leptoquark when taken alone
can only address neutral-current $B$-meson anomalies. A model with
both $S_{1}$ and $S_{3}$, and only left-handed couplings for $S_{1}$,
can address both $B$-anomalies but not the muon magnetic moment.
Finally, allowing for right handed $S_{1}$ couplings makes it possible
to fit also $(g-2)_{\mu}$. Concerning the prospects for both the
LF conserving branching fraction $\text{Br}(B\to K\tau\tau)$ and
the LFV one $\text{Br}(B\to K\tau\mu)$, they are found to be in the
ballpark of the future expected sensitivity of Belle-II and LHCb.

A quick glance summary of the various models is provided by Fig.~\ref{fig:summary}
where we show, for the best-fit point of each model, the deviations
from the SM prediction of each of the most relevant observables studied
in the global fit. The black dots and intervals represent the experimentally
preferred values and uncertainties, and for each observable we normalize
the $x$-axis to the corresponding uncertainty (i.e. we count the
number of standard deviations). Detailed informations for each model
can be found in Subsections~\ref{sec:S1S3-models-S1S3amu}-\ref{sec:S1S3-models-S1S3amu}.
A separate analysis is provided in Sec.~\ref{sec:S1S3Higgs} for
Higgs physics observables and electroweak oblique corrections, which
put constraints on leptoquark-Higgs couplings.

To conclude, we find that the combination of $S_{1}$ and $S_{3}$
provides a good combined explanation of several experimental anomalies:
charged and neutral-current $B$-meson anomalies as well as the muon
magnetic moment. Their mass is necessarily close to the $1\,\text{TeV}$
scale, particularly to address charged-current anomalies $R(D^{(*)})$,
and is thus in the region that could still show some signals at HL-LHC,
if they are light enough, but that will definitely be tested at future
hadron colliders.

In the next few years, several experiments are expected to provide
concluding answers as to the nature of all these puzzles. While at
this time it is still very possible that some, or all, of these will
turn out to be only statistical fluctuations and will be shown to
be compatible with SM predictions, the possibility that even only
one will instead be confirmed is real. Such an event would have profound
and revolutionary implications for our understanding of Nature at
the smallest scales. The scalar leptoquarks considered here are very
good candidates for combined explanations and could thus be the heralds
of a new physics sector lying at the TeV scale.

\chapter{\label{sec:ROFV}Rank One Flavor Violation}

The $B$-anomalies, if experimentally confirmed, could shed some light
on the NP flavour structure, which can be probed by studying the correlations
between the anomalies and other flavour observables. This type of
information is, generically speaking, impossible to extract from few
data such as the $R(D^{(*)})$ or $R(K^{(*)})$ discrepancies, and
without any specific model in mind. However, under some generic flavour
assumptions, one is sometimes able to draw model independent conclusions,
which can be useful either to exclude the initial guess itself, or
to make theoretically well-grounded predictions. This is the point
of view taken in Ref. \cite{Gherardi:2019zil}, a work in collaboration
with D. Marzocca, M. Nardecchia and A. Romanino, which I describe
in this Chapter.

In \cite{Gherardi:2019zil}, we assume that the NP sector giving rise
to the $R(K^{(*)})$ anomalies couples only to muons and to a single
direction in quark flavour space. In other words, the NP lagrangian
takes the form\footnote{Semileptonic operators including right handed quarks do not contribute
to $R(K^{(*)})$ \cite{DAmico:2017mtc}, and do not have any impact
on the bounds studied in our work, so that the right handed quark
fields $d$ and $u$ are not considered in Eq. (\ref{eq:L_ROFV})
and what follows.}\footnote{In order to avoid confusions between SM and low-energy quark and lepton
fields, in this Chapter we shall denote SM chiral fields by $q_{iL},\,u_{iR},\,d_{iR},\,\ell_{\alpha L}$
and $e_{\alpha R}$, whereas the notation without any explicit chirality
(such as $\mu$, $d,$ $u$, etc.) refers to the Dirac (non-chiral)
quark and lepton fields in the broken electroweak symmetry phase.}:
\begin{equation}
\mathcal{L}_{\text{NP}}=\mathcal{L}_{\text{NP}}\left[\hat{n}_{i}^{*}q^{i},\,\ell_{2L},\,\mu_{R}\right],\label{eq:L_ROFV}
\end{equation}
where $\hat{n}$ is a unit complex vector. In this case, the Wilson
coefficient matrices of the semileptonic operators contributing to
$R(K^{(*)})$, which at the low-energy (\emph{i.e.} GeV scale) level
are:
\begin{equation}
O_{ij}^{\chi}=(\overline{d_{i}}\gamma_{\rho}P_{L}d_{j})(\overline{\mu}\gamma^{\rho}P_{\chi}\mu)\qquad(\chi=L,\,R)\label{eq:LEFT Ops for R(K)}
\end{equation}
are all of rank one and proportional:
\begin{equation}
C_{ij}^{\chi}=C^{\chi}\hat{n}_{i}\hat{n}_{j}^{*}\quad(\chi=L,\,R),\label{eq:ROFV WCs}
\end{equation}
with $C^{\chi}\in\mathbb{R}$. An equation similar to (\ref{eq:ROFV WCs})
holds for the Wilson coefficients of SMEFT operators, as we will discuss
in more details below.

We dub the scenario described above as Rank One Flavor Violation (ROFV).
Although the ROFV assumption might appear quite ad-hoc, it is actually
automatically realized in a handful of models addressing $R(K^{(*)})$:
single leptoquark models, models in which the quark doublets mixes
with a single generation of vector-like fermions, one-loop models
of linear flavour violation \cite{Gripaios:2015gra}, to give some
examples. It is worth to notice that, as part of the ROFV hypothesis,
it is assumed that the $R(K^{(*)})$ deviations are entirely due to
the muonic channel, \emph{i.e.} the numerators in Eq. (\ref{eq:R(K)}),
which as we already commented is supported by data.

Under the ROFV assumption, one can establish correlations between
$R(K^{(*)})$ and other flavour observables, which mainly depend on
the NP direction $\hat{n}$. For instance, in the low energy setting
of Eqs. (\ref{eq:LEFT Ops for R(K)}), once the unit vector $\hat{n}$
and the ratios $C^{L}\colon C^{R}$ are fixed, one can fit the overall
energy scale of the Wilson coefficients in Eq. (\ref{eq:ROFV WCs})
to the $R(K^{(*)})$ data; this, in turn, allows to compute NP contributions
to all observables of the form $d_{i}\to d_{j}\mu\mu$. The latter
information can be used in a two-fold way: either to constrain the
possible NP directions $\hat{n}$ using the bounds coming from $d_{i}\to d_{j}\mu\mu$
processes (the main approach advocated in Ref. \cite{Gherardi:2019zil}),
or to obtain predictions for these processes for a given (motivated,
in some way) direction $\hat{n}$.

In our original work \cite{Gherardi:2019zil}, we address only the
neutral LFU ratio anomalies (\emph{i.e. }$R(K^{(*)})$), but it should
be noted that even if an assumption analogous to Eq. (\ref{eq:ROFV WCs})
held within the $b\to c\tau\nu_{\tau}$ sector, the corresponding
unit vectors for neutral and charged current Wilson coefficients would,
in general, be entirely uncorrelated.

Apart from identifying a set of relevant observables correlating with
$R(K^{(*)})$ under ROFV, the main conclusion of Ref. \cite{Gherardi:2019zil}
is that, in models satisfying the rank-one condition, NP couplings
must be closely aligned to the third family of quarks. Such an alignment
can be naturally explained in the framework of flavour symmetries,
in which case the quark $\text{U}(2)_{\mathcal{Q}}^{3}$ emerges as
a promising candidate \cite{Gherardi:2019zil}. In the absence of
flavour symmetries, the phenomenological constraints analysed in Ref.
\cite{Gherardi:2019zil} call either for an alternative mechanism
naturally providing the required alignment, or for a mere fine-tuning
along the heavy quark family.

\section{The ROFV framework}

We start by listing the relevant effective operators for our analysis,
both at the low-energy scale of $B$-anomalies, and at the high-energy
scale of SMEFT, \emph{i.e.} the GeV and TeV scales respectively. In
what follows we consider only tree-level contributions to Wilson coefficients,
so that the link between SMEFT and low-energy operators amounts to
a simple projection.

Model-independent analyses of neutral-current anomalies hint towards
NP coupling to quark and lepton vectorial currents \cite{Hiller:2014yaa,Descotes-Genon:2015uva,Altmannshofer:2017fio,Capdevila:2017bsm,DAmico:2017mtc,Altmannshofer:2017yso,Geng:2017svp,Ciuchini:2017mik,Hiller:2017bzc,Alok:2017sui,Hurth:2017hxg,Alguero:2019aa,Alok:2019ufo,Ciuchini:2019usw,Aebischer:2019mlg}.
As a matter of fact, the vast majority of NP explanations of the anomalies
boils down, at low energy, to one of the following muonic operators:
\begin{align}
\mathcal{O}_{L} & =(\overline{s}\gamma_{\rho}P_{L}b)(\overline{\mu}\gamma^{\rho}P_{L}\mu), & \mathcal{O}_{9} & =(\overline{s}\gamma_{\rho}P_{L}b)(\overline{\mu}\gamma^{\rho}\mu),\label{eq:O_L and O_9}
\end{align}
although it has been pointed out that allowing for NP in both muons
and electrons provides a slight improvement in the fits \cite{Alguero:2018nvb,Alguero:2019pjc,Datta:2019zca}. 

\begin{table}[t]
\centering %
\begin{tabular}{|c|c|c|}
\hline 
$R_{K}~[1.1,~6]\,\text{GeV}^{2}$ & $0.846\pm0.062$ & LHCb \cite{Aaij:2014ora,Aaij:2019aa}\tabularnewline
\hline 
\multirow{2}{*}{$R_{K^{*}}~[0.045,~1.1]\,\text{GeV}^{2}$} & $0.66\pm0.11$ & LHCb \cite{Aaij:2017vbb}\tabularnewline
 & $0.52_{-0.26}^{+0.36}$ & Belle \cite{2019:BelleRKst}\tabularnewline
\hline 
\multirow{2}{*}{$R_{K^{*}}~[1.1,~6]\,\text{GeV}^{2}$} & $0.69\pm0.12$ & LHCb \cite{Aaij:2017vbb}\tabularnewline
 & $0.96_{-0.29}^{+0.45}$ & Belle \cite{2019:BelleRKst}\tabularnewline
\hline 
$R_{K^{*}}~[15,~19]\,\text{GeV}^{2}$ & $1.18_{-0.32}^{+0.52}$ & Belle \cite{2019:BelleRKst}\tabularnewline
\hline 
\multirow{2}{*}{$\text{Br}(B_{s}^{0}\to\mu\mu)$} & $(3.0_{-0.63}^{+0.67})\times10^{-9}$ & LHCb \cite{Aaij:2017vad}\tabularnewline
 & $(2.8_{-0.7}^{+0.8})\times10^{-9}$ & ATLAS \cite{Aaboud:2018mst}\tabularnewline
\cline{2-3} \cline{3-3} 
\end{tabular}\caption{\label{tab:bsmumuObs} Clean observables sensitive to $bs\mu\mu$
contact interactions.}
\end{table}

The two low-energy operators in Eq.~(\ref{eq:O_L and O_9}) can be
thought to be part of an effective lagrangian involving all the three
quark families 
\begin{equation}
\mathcal{L}_{\text{NP}}^{\text{EFT}}=C_{L}^{ij}(\overline{d_{i}}\gamma_{\rho}P_{L}d_{j})(\overline{\mu}\gamma^{\rho}P_{L}\mu)+C_{R}^{ij}(\overline{d_{i}}\gamma_{\rho}P_{L}d_{j})(\overline{\mu}\gamma^{\rho}P_{R}\mu)\;,\label{eq:ROFV-NPEFT}
\end{equation}
where the coefficient of the $\mathcal{O}_{L}$ operator is identified
with $C_{L}^{sb}$, the coefficient of the $\mathcal{O}_{9}$ operator
with $C_{L}^{sb}+C_{R}^{sb}$, and we have focussed on muon processes
on the leptonic side. 

From the SMEFT point of view, the operators that can contribute to
(\ref{eq:ROFV-NPEFT}) at the tree-level are collected in the following
lagrangian: 
\begin{equation}
\mathcal{L}_{\text{NP}}^{\text{SMEFT}}=C_{S}^{ij}(\overline{q_{iL}}\gamma_{\rho}q_{jL})(\overline{\ell_{2L}}\gamma^{\rho}\ell_{2L})+C_{T}^{ij}(\overline{q_{iL}}\gamma_{\rho}\tau^{a}q_{jL})(\overline{\ell_{2L}}\gamma^{\rho}\tau^{a}\ell_{2L})+C_{R}^{ij}(\overline{q_{iL}}\gamma_{\rho}q_{jL})(\overline{\mu_{R}}\gamma^{\rho}\mu_{R})\label{eq:ROFV-SMEFTlagrangian}
\end{equation}
with $C_{L}^{ij}=C_{S}^{ij}+C_{T}^{ij}$. In the previous equation,
$\ell_{L}^{i}=\left(\nu_{L}^{i},e_{L}^{i}\right)^{t}$ and $q_{L}^{i}=\left(V_{ji}^{*}u_{L}^{j},d_{L}^{i}\right)^{t}$
are the lepton and quark doublets, in the charged-lepton and down
quarks mass basis respectively, and $V$ is the CKM matrix.

Our key assumption is that the NP sector responsible of the $R_{K^{(*)}}$
signal couples to a single direction in the quark flavour space (as
mentioned, we focus here on muon processes on the leptonic side),
which requires the Wilson coefficient matrices $C_{S,T,R}^{ij}$ in
Eq.~(\ref{eq:ROFV-SMEFTlagrangian}) (and consequently $C_{L,R}^{ij}$
in (\ref{eq:ROFV-NPEFT})) to be rank-one and proportional:
\begin{equation}
C_{S,T,R,L}^{ij}=C_{S,T,R,L}\hat{n}_{i}\hat{n}_{j}^{*}\label{eq:ROFV-WC parametrization}
\end{equation}
 where $C_{S,T,R,L}\in\mathbb{R}$, $C_{L}=C_{S}+C_{T}$, and $\hat{n}_{i}$
is a unitary vector in $\text{U}(3)_{q}$ flavour space. We dub this
scenario \emph{Rank-One Flavor Violation} (ROFV). Rather than being
an assumption on the flavour symmetry and its breaking terms (such
as Minimal Flavor Violation \cite{DAmbrosio:2002vsn}, for example),
this is an assumption on the dynamics underlying these semileptonic
operators. We refer to \cite{Glashow:2014iga,Bhattacharya:2014wla}
for similar approaches in different contexts.

It is perhaps worth emphasizing that our analysis does not rely upon
any particular assumption concerning NP effects in the $\tau$ sector,
as far as observables with muons are concerned. Such effects could
become relevant only when considering observables with neutrinos,
whose flavour is not observed, or loop-generated ones such as $\Delta F=2$
processes in leptoquark models.\footnote{Since experimental limits on semi-tauonic operators are much weaker
than those on semi-muonic ones, couplings of new physics to tau leptons
can be much larger than to muons, which is consistent with theoretical
expectations from NP coupled preferentially to the third family, and
for example allows combined explanations of both neutral and charged-current
$B$-meson anomalies, see e.g. the analysis in \cite{Buttazzo:2017ixm}.} On the other hand, we do assume negligible NP effects in the electron
sector. This is, by itself, a reasonable assumption since it is supported
by data and it is also well motivated in scenarios where NP couplings
to leptons follows the same hierarchy as SM Yukawas (such as SU(2)$^{5}$
flavour symmetries or partial compositeness).

The ROFV assumption is well motivated. For example, it is automatically
realised in all single leptoquark models generating the operators
in Eq.~(\ref{eq:O_L and O_9}) at low energy\footnote{To be precise, the correlations discussed in the present Chapter apply
to all single leptoquark models in which the coupling to electrons
is suppressed with respect to the one to muons.} (see e.g. Ref.~\cite{Angelescu:2018tyl} for a recent comprehensive
analysis). Furthermore, Eq.~(\ref{eq:ROFV-WC parametrization}) is
automatically satisfied in all cases where a single linear combination
of SM quark doublets couples to NP. This condition is actually stronger
than strictly required by ROFV, since not only semimuonic operators
have rank-one coefficients, but all operators involving quark doublets.
This scenario finds realization in several UV models, such as models
with single vector-like fermion mediators, and one-loop models with
linear flavour violation \cite{Gripaios:2015gra}. Contrary to the
MFV or the minimally broken $\text{U}(2)^{5}$ scenarios, which predict
the flavour structure of all NP contributions, the ROFV assumption
is specific to the set of semimuonic operators in Eq.~(\ref{eq:ROFV-SMEFTlagrangian}).
On the other hand, while those scenarios require strong assumptions
on the flavour symmetry and its symmetry-breaking terms, ROFV can
be accidentally realised from the underlying dynamics, see e.g. Refs.~\cite{Cline:2017aed,Cline:2017qqu}.

From a theoretical point of view it might be natural to expect the
direction of the unitary vector $\hat{n}$ to be close to the third
generation of SM quarks. This case is studied in more detail in Sec.~\ref{sec:ROFV-FlavourSymmetry}.
In the following, instead, we abandon any theory prejudice on $\hat{n}$
and study what are the experimental constraints on its possible directions.
We parametrize $\hat{n}$ as 
\begin{equation}
\hat{n}=\begin{pmatrix}\sin\theta\cos\phi e^{i\alpha_{bd}}\\
\sin\theta\sin\phi e^{i\alpha_{bs}}\\
\cos\theta
\end{pmatrix},\label{eq:VersorDef}
\end{equation}
where the angles and phases can be chosen to lie in the following
range: 
\begin{equation}
\theta\in\left[0,\frac{\pi}{2}\right],\qquad\phi\in\left[0,2\pi\right),\quad\alpha_{bs,bd}\in\left[-\frac{\pi}{2},+\frac{\pi}{2}\right]\label{eq:ROFV-Ranges}
\end{equation}
The values of the angles and phases associated to specific directions
in flavour space (up and down quarks) are collected in Table~\ref{tab:quarkDir}
and shown in the corresponding Figure.
\begin{table}[t]
\centering %
\begin{tabular}{c|c|cccc}
quark & $\hat{n}$ & $\phi$ & $\theta$ & $\alpha_{bd}$ & $\alpha_{bs}$\tabularnewline
\hline 
\textcolor{blue}{down} & $(1,0,0)$ & $0$ & $\pi/2$ & $0$ & $0$\tabularnewline
\textcolor{green}{strange} & $(0,1,0)$ & $\pi/2$ & $\pi/2$ & $0$ & $0$\tabularnewline
bottom & $(0,0,1)$ & $0$ & $0$ & $0$ & $0$\tabularnewline
\textcolor{orange}{up} & $e^{i\arg(V_{ub})}(V_{ud}^{*},V_{us}^{*},V_{ub}^{*})$ & $0.23$ & $1.57$ & $-1.17$ & $-1.17$\tabularnewline
\textcolor{purple}{charm} & $e^{i\arg(V_{cb})}(V_{cd}^{*},V_{cs}^{*},V_{cb}^{*})$ & $1.80$ & $1.53$ & $-6.2\times10^{-4}$ & $-3.3\times10^{-5}$\tabularnewline
\textcolor{teal}{top} & $e^{i\arg(V_{tb})}(V_{td}^{*},V_{ts}^{*},V_{tb}^{*})$ & $4.92$ & $0.042$ & $-0.018$ & $0.39$\tabularnewline
\end{tabular}\\
 \includegraphics{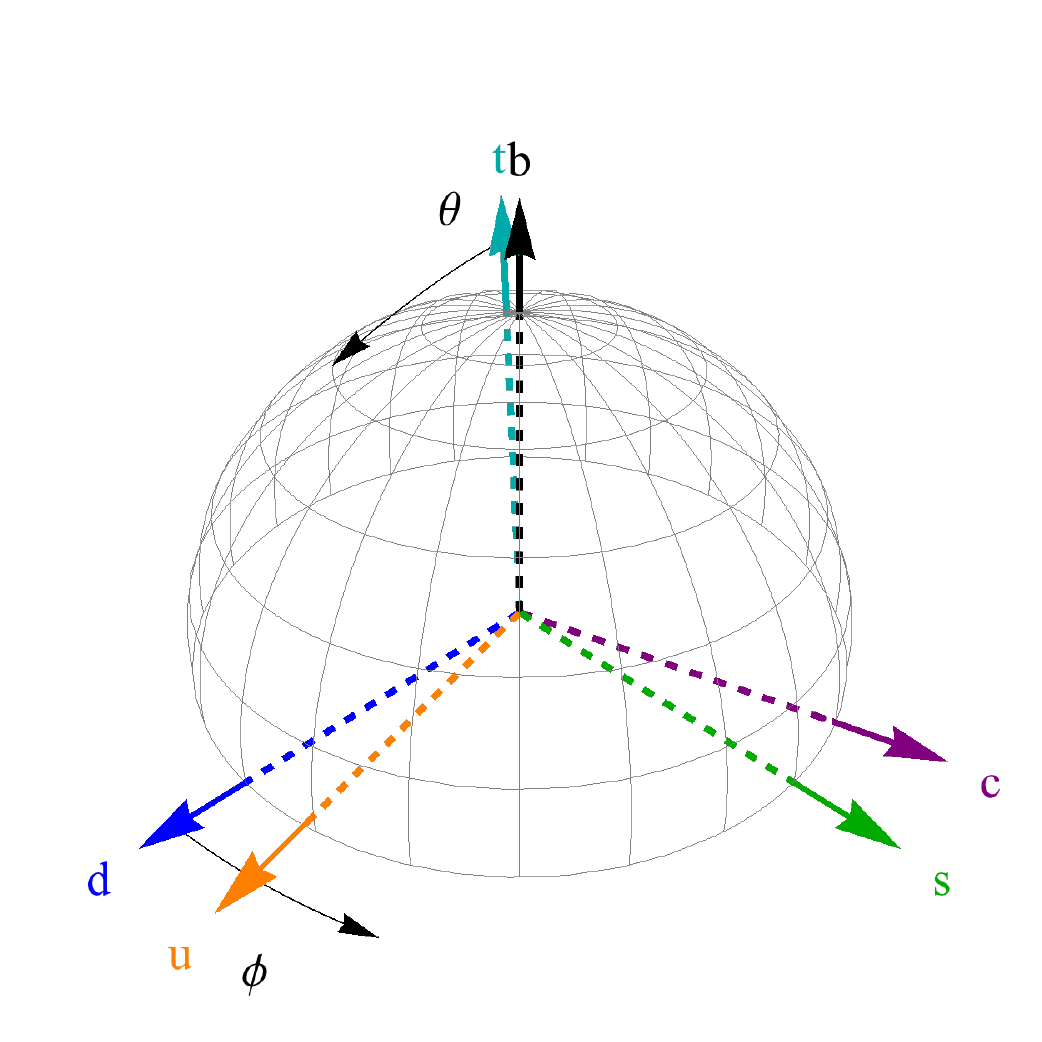} \caption{\label{tab:quarkDir} SM quark directions of the unitary vector $\hat{n}_{i}$.
The plot shows the corresponding directions in the semi-sphere described
by the two angles $(\theta,\phi)$.}
\end{table}

\begin{table}[t]
\begin{centering}
\begin{tabular}{cc}
\toprule 
Channel & Coefficient dependencies\tabularnewline
\midrule
\midrule 
$d_{i}\to d_{j}\mu^{+}\mu^{-}$ & $C_{S}+C_{T},\ C_{R}$\tabularnewline
\midrule 
$u_{i}\to u_{j}\overline{\nu_{\mu}}\nu_{\mu}$ & $C_{S}+C_{T}$\tabularnewline
\midrule 
$u_{i}\to u_{j}\mu^{+}\mu^{-}$ & $C_{S}-C_{T},\ C_{R}$\tabularnewline
\midrule 
$d_{i}\to d_{j}\overline{\nu_{\mu}}\nu_{\mu}$ & $C_{S}-C_{T}$\tabularnewline
\midrule 
$u_{i}\to d_{j}\mu^{+}\nu_{\mu}$ & $C_{T}$\tabularnewline
\bottomrule
\end{tabular}
\par\end{centering}
\caption{\label{tab:ROFV-WC dependencies}Dependencies of various semileptonic
processes on the three coefficients $C_{S,T,R}$ (cf. Eq. (\ref{eq:ROFV-WC parametrization})).
Here and in the text, a given quark level process represents all processes
obtained through a crossing symmetry from the shown one.}
\end{table}

The ROFV structure of the semileptonic operators, Eq.~(\ref{eq:ROFV-WC parametrization}),
implies the existence of correlations between the NP contributions
to $b\to s\mu\mu$ anomalous observables and to other observables.
In the SMEFT, additional correlations follow from the SU(2)$_{L}$
invariance of the lagrangian in (\ref{eq:ROFV-SMEFTlagrangian}).
We can then take advantage of the experimental constraints on those
additional observables to constrain the flavour directions $\hat{n}$
accounting for the anomalies. In order to do that, we proceed as follows:
for a given direction $\hat{n}$, we fix (some combination of) the
overall coefficients in Eq. (\ref{eq:ROFV-WC parametrization}) by
matching with the best-fit value of the $C_{L}^{sb}$ (or $C_{9}^{sb}$)
coefficient obtained from global fits. Once this is done, we can compute
NP contributions to other semileptonic processes as functions of $\hat{n}$,
and compare with the corresponding experimental values/bounds. By
this procedure, we are able to narrow down considerably the space
of allowed flavour directions $\hat{n}$.\footnote{We checked explicitly that the results obtained in this way, i.e by
fixing $C_{L,9}^{bs}$ to its best-fit point, or by performing a global
$\chi^{2}$ analysis to get the $95\%$CL excluded region agree very
well with each other.}

We analyse the constraints on the direction $\hat{n}$ under different
assumptions. We begin in Sec.~\ref{sec:GeneralCorrelations} by using
the effective description in (\ref{eq:ROFV-NPEFT}) and focussing
on the case $C_{R}=0$. This allows us to derive general correlations
with other $d_{i}d_{j}\mu\mu$ observables. In Sec.~\ref{sec:ROFV-Mediators}
we extend the analysis to $\text{SU}(2)_{L}\times\text{U}(1)_{Y}$
invariant operators, thus enabling us to consider also observables
with up-quarks and/or muon neutrinos. Tab.~\ref{tab:ROFV-WC dependencies}
shows the dependencies of the various types of process upon the three
coefficients $C_{S,T,R}$. In particular, we consider specific combinations
of $C_{S,T,R}$ obtained in some single-mediator simplified models:
$S_{3}$ and $U_{1}^{\mu}$ leptoquarks, as well as of a $Z^{\prime}$
coupled to the vector-like combination of muon chiralities. In Sec.~\ref{sec:ROFV-FlavourSymmetry}
we study the connection of our rank-one assumption with $\text{U}(3)^{5}$
and $\text{U}(2)^{5}$ flavour symmetries. A discussion on the impact
of future measurements is presented in Sec.~\ref{sec:ROFV-Prospects}.
The simplified fit of the $R_{K}$ and $R_{K^{*}}$ anomalies used
below, as well as details on the flavour observables considered, are
collected in the two Appendices of the original reference \cite{Gherardi:2019zil}.

\section{\label{sec:GeneralCorrelations}General correlations in $V-A$ solutions}

In this Section, we study the correlations that follow directly from
the rank-one condition, for all models in which NP couples only to
left-handed fermions. We begin by using the effective description
in (\ref{eq:ROFV-NPEFT}). For $C_{R}=0$, and for fixed $\theta$
and $\phi$ in the ranges specified by Eq. (\ref{eq:ROFV-Ranges}),
the coefficient $C_{L}=C_{S}+C_{T}$ and the phase $\alpha_{bs}$
are univocally determined by the $b\to s\mu^{+}\mu^{-}$ anomalies
fit: 
\begin{equation}
C_{L}\sin\theta\cos\theta\sin\phi e^{i\alpha_{bs}}=C_{L}^{bs}\equiv\frac{e^{i\alpha_{bs}}}{\Lambda_{bs}^{2}}.\label{eq:ROFV-CLmatchingfit}
\end{equation}
From a fit of the observables listed in Table~\ref{tab:bsmumuObs},
we find that the phase $\alpha_{bs}$ has an approximately flat direction
in the range $|\alpha_{bs}|\lesssim\pi/4$. Since a non-zero phase
necessarily implies a lower $\Lambda_{bs}$ scale in order to fit
the anomalies, to be conservative we fix $\alpha_{bs}=0$. In this
case the best-fit point for the NP scale is:
\begin{equation}
(\Lambda_{bs})_{\text{best-fit}}=38.5\,\text{TeV}\label{eq:ROFV-CLfitArgZero}
\end{equation}
\begin{table}[t]
\begin{centering}
\begin{tabular}{|c|c|c|c|}
\hline 
Observable & Experimental value/bound & $\text{SM}$ prediction & References\tabularnewline
\hline 
\hline 
$\text{Br}(B_{d}^{0}\to\mu^{+}\mu^{-})$ & $<2.1\times10^{-10}\ \text{(95\% CL)}$ & $(1.06\pm0.09)\times10^{-10}$ & \cite{Aaboud:2018mst,Bobeth:2013uxa}\tabularnewline
\hline 
$\text{Br}(B^{+}\to\pi^{+}\mu^{+}\mu^{-})_{\left[1,6\right]}$ & $(4.55_{-1.00}^{+1.05}\pm0.15)\times10^{-9}$ & $(6.55\pm1.25)\times10^{-9}$ & \cite{Aaij:2015nea,Du:2015tda,Khodjamirian:2017fxg}\tabularnewline
\hline 
$\text{Br}(K_{S}\to\mu^{+}\mu^{-})$ & $<1.0\times10^{-9}\ \text{(95\% CL)}$ & $(5.0\pm1.5)\times10^{-12}$ & \cite{Aaij:2017tia}\tabularnewline
\hline 
$\text{Br}(K_{L}\to\mu^{+}\mu^{-})_{\text{SD}}$ & $<2.5\times10^{-9}\ $ & $\approx0.9\times10^{-9}$ & \cite{Ambrose:2000gj,Isidori:2003ts}\tabularnewline
\hline 
$\text{Br}(K_{L}\to\pi^{0}\mu^{+}\mu^{-})$ & $<3.8\times10^{-10}\ \text{(90\% CL)}$ & $1.41_{-0.26}^{+0.28}(0.95_{-0.21}^{+0.22})\times10^{-11}$ & \cite{AlaviHarati:2000hs,DAmbrosio:1998gur,Buchalla:2003sj,Isidori:2004rb,Mescia:2006jd}\tabularnewline
\hline 
\end{tabular}
\par\end{centering}
\caption{\label{tab:ROFV-Correlated-observables}Observables with direct correlation
with $bs\mu\mu$.}
\end{table}

\begin{figure}[t]
\centering \includegraphics[scale=0.5]{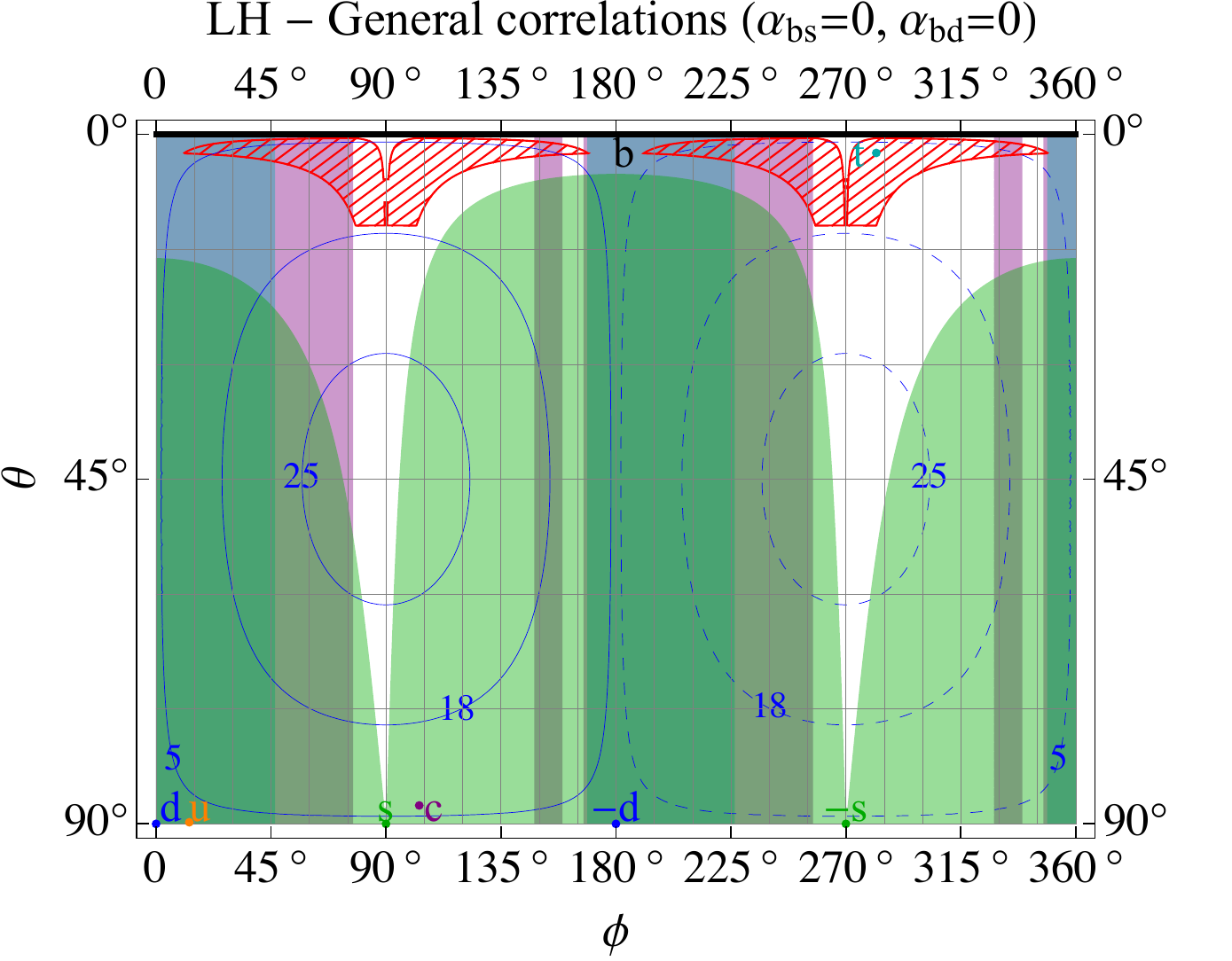} \includegraphics[scale=0.5]{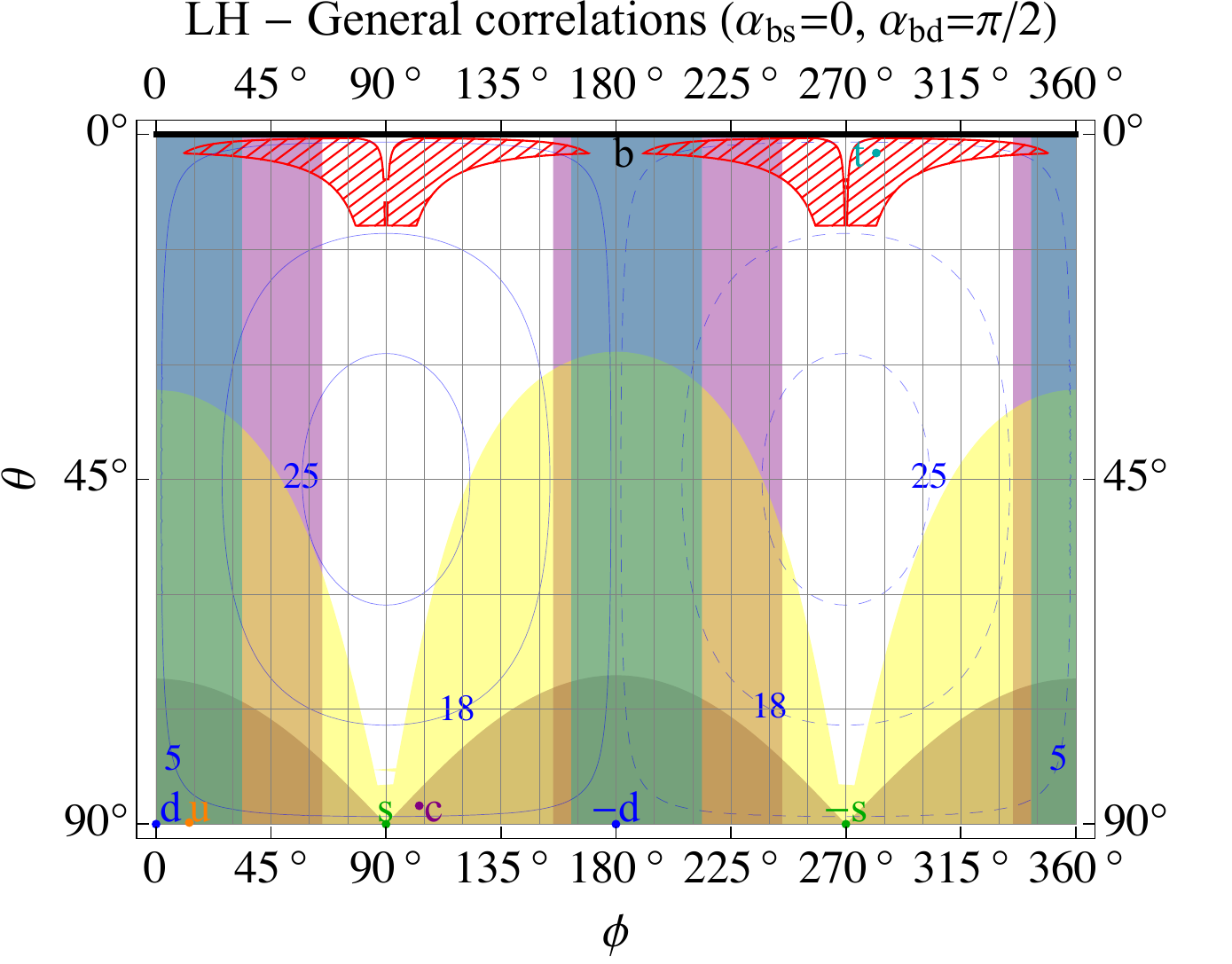}\\
 \includegraphics[scale=0.5]{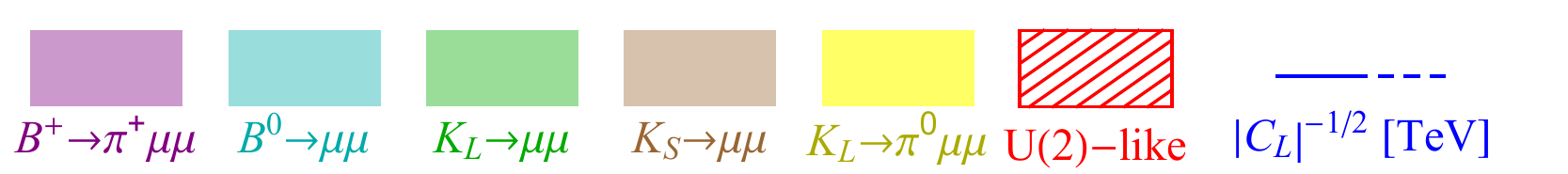} \caption{\label{fig:ROFV-Minimal}Limits in the plane $(\phi,\theta)$ for
two choices of the phases $\alpha_{bs}$ and $\alpha_{bd}$ from observables
with direct correlation with $R_{K^{(*)}}$. The blue contours correspond
to the value of $|C_{L}|^{-1/2}$ in TeV, where solid (dashed) lines
are for positive (negative) $C_{L}$. The meshed red region correspond
to the one suggested by partial compositeness or $SU(2)_{q}$-like
flavour symmetry, Eq.~(\ref{eq:TopLikeDirection}) with $|a_{bd,bs}|\in[0.2-5]$.}
\end{figure}

We now constrain $\hat{n}$ (or, more precisely, $\theta$ and $\phi$
for given $\alpha_{bs}$) using the other observables correlated with
$R_{K}$ by the relation $C_{L}^{ij}=C_{L}\hat{n}_{i}\hat{n}_{j}^{*}$.
Such observables are associated to the quark-level transitions 
\begin{equation}
d_{i}\to d_{j}\mu^{+}\mu^{-}\label{eq:minimalconstraints}
\end{equation}
and cross-symmetric counterparts. The most relevant among those observables
are listed in Tab.~\ref{tab:ROFV-Correlated-observables}, and the
corresponding allowed regions for $\theta$, $\phi$ are shown in
Fig.~\ref{fig:ROFV-Minimal} in the two cases $(\alpha_{bd},\alpha_{bs})=(0,0)$
and $(\alpha_{bd},\alpha_{bs})=(\pi/2,0)$. As can be seen from the
plots, the most severe bounds arise from $B^{+}\to\pi^{+}\mu^{+}\mu^{-}$
(LHCb \cite{Aaij:2015nea}) and $K_{L}\to\mu^{+}\mu^{-}$ (E871 \cite{Ambrose:2000gj,Isidori:2003ts}).
However, the latter observable does not yield any bound for $\alpha_{bd}-\alpha_{bs}=\pi/2$,
i.e. for $\text{Re}\,C_{L}^{ds}=0$. The imaginary part of that coefficient
can instead be tested by $K_{S}\to\mu^{+}\mu^{-}$ (LHCb \cite{Aaij:2012rt})
and $K_{L}\to\pi^{0}\mu^{+}\mu^{-}$ (KTeV \cite{AlaviHarati:2000hs}).
More details on the observables and their NP dependence can be found
in Ref. \cite{Gherardi:2019zil}.

As a final remark, let us stress that here and in the following we
are ignoring possible NP contributions to (pseudo)scalar, tensor,
or dipole operators. While these are known to be too constrained to
give significant contributions to $R_{K^{(*)}}$ (see e.g. \cite{Hiller:2014yaa}),
they may nonetheless produce important effects in other observables,
so that some of the bounds discussed here may be relaxed, if some
degree of fine-tuning is allowed.

\section{\label{sec:ROFV-Mediators}SMEFT and simplified mediators}

Let us now assume that the effective operators in (\ref{eq:ROFV-NPEFT})
originate from the SM-invariant ones in (\ref{eq:ROFV-SMEFTlagrangian}),
as expected. The SU(2)$_{L}$ invariance then relates $d_{i}\to d_{j}\mu^{+}\mu^{-}$
processes to processes involving up quarks and muon neutrinos listed
in Tab. \ref{tab:ROFV-WC dependencies}. Using the experimental constraints
on those, we can impose further constraints on $\hat{n}$. These,
though, are model dependent even in the $C_{R}=0$ case, as they depend
on the relative size of the two operators in (\ref{eq:ROFV-SMEFTlagrangian})
contributing to $C_{L}$, i.e. $C_{S}$ and $C_{T}$. The origin of
the model dependence can be clarified taking advantage of a phenomenological
observation. Our analysis (see below) shows that the the most relevant
constraints come from the processes 
\begin{equation}
d_{i}\to d_{j}\mu^{+}\mu^{-}\quad\text{and}\quad d_{i}\to d_{j}\overline{\nu_{\mu}}\nu_{\mu}\;.\label{eq:generalconstraints}
\end{equation}
As Table \ref{tab:ROFV-WC dependencies} shows, those two classes
of processes are associated respectively to the two operators $\mathcal{O}^{\pm}$
whose Wilson coefficients are $C_{\pm}=C_{S}\pm C_{T}$ (note that
$C_{+}\equiv C_{L}$), i.e. 
\begin{align}
\mathcal{O}_{ij}^{+} & =\frac{\mathcal{O}_{ij}^{S}+\mathcal{O}_{ij}^{T}}{2}=\left(\overline{q_{iL}}\gamma_{\mu}\ell_{2L}\right)\left(\bar{\ell}_{L}^{2}\gamma^{\mu}q_{L}^{j}\right)\\
\mathcal{O}_{ij}^{-} & =\frac{\mathcal{O}_{ij}^{S}-\mathcal{O}_{ij}^{T}}{2}=2\left(\overline{q_{iL}^{c}}\ell_{2L}\right)\left(\overline{\ell_{L}^{2}}q_{jL}^{c}\right)\;.\label{eq:Opm}
\end{align}
The model-independent constraints shown in Fig. \ref{fig:ROFV-Minimal}
only take into account the $d_{i}\to d_{j}\mu^{+}\mu^{-}$ processes
and as such only depend on $C_{+}$, which is thus the only combination
fixed by $R_{K^{(*)}}$. On the other hand, the model-dependent weight
of the $d_{i}\to d_{j}\overline{\nu_{\mu}}\nu_{\mu}$ constraints
depends on the relative size of $C_{-}$. In this context, the results
in \ref{fig:ROFV-Minimal} correspond to a SMEFT with $C_{-}=0$,
i.e.\ to the $U_{1}$ case in Tab. \ref{tab:Mediators}. 

Note that the experimental constraints on the processes involving
neutrinos do not distinguish among the three neutrino flavours. In
order to get constraints on the muon neutrino operators we consider,
one should then make an assumption on the relative size of the operators
with different neutrino flavours. Below, we will conservatively assume
that only the muon neutrino operators contribute to the neutrino processes. 

In order to reduce the number of free parameters, we focus in this
Section on single-mediator simplified models, which generate specific
combinations of the three operators when integrated out at the tree-level.
Some relevant benchmarks are shown in Table \ref{tab:Mediators},
where in the last column we list the ratios: 
\begin{equation}
c_{X}\equiv\dfrac{C_{X}}{C_{S}+C_{T}}=\frac{C_{X}}{C_{+}}\qquad(X=S,T,R)~.\label{eq:c_x ratios}
\end{equation}
Notice that the exclusions shown in Fig.~\ref{fig:ROFV-Minimal}
hold in all models in Table \ref{tab:Mediators}, except for the $Z_{V}'$
which has vector-like coupling to muons. We find that the most relevant
bounds, beyond those already analized, arise from the FCNC observables
$\text{Br}(K^{+}\to\pi^{+}\nu_{\mu}\overline{\nu_{\mu}})$ and $\text{Br}(K_{L}\to\pi^{0}\nu_{\mu}\overline{\nu_{\mu}})$,
reported in \ref{tab:Correlated-observables-simplified}. The connection
of these observables with the $B$-meson anomalies has also been emphasised
in Ref.~\cite{Bordone:2017lsy}. The effect of constraints from rare
kaon decays on LQ models addressing instead the $\epsilon^{\prime}/\epsilon$
anomaly have been studied in Ref.~\cite{Bobeth:2017ecx}. Some simplified
models also allow to compute neutral meson mixing amplitudes, which
we include in the analysis when appropriate.

\begin{table}[t]
\centering 
\[
\begin{array}{c|c|c|c}
\textrm{Simplified model} & \textrm{Spin} & \textrm{SM irrep} & (c_{S},c_{T},c_{R})\\
\hline S_{3} & 0 & (\overline{3},3,1/3) & (3/4,1/4,0)\\
U_{1} & 1 & (3,1,2/3) & (1/2,1/2,0)\\
U_{3} & 1 & (3,3,2/3) & (3/2,-1/2,0)\\
V' & 1 & (1,3,0) & (0,1,0)\\
Z' & 1 & (1,1,0) & (1,0,c_{R})
\end{array}
\]
\vspace{-0.5cm}
 \caption{\label{tab:Mediators} Wilson coefficients ratios (cf. Eq. (\ref{eq:c_x ratios}))
for some single-mediator simplified models.}
\end{table}

\begin{table}[t]
\centering %
\begin{tabular}{|c|c|c|c|}
\hline 
Observable & Experimental value/bound & $\text{SM}$ prediction & References\tabularnewline
\hline 
\hline 
$\text{Br}(K^{+}\to\pi^{+}\nu_{\mu}\overline{\nu_{\mu}})$ & $(17.3_{-10.5}^{+11.5})\times10^{-11}$ & $(8.4\pm1.0)\times10^{-11}$ & \cite{Artamonov:2008qb,Buras:2015qea}\tabularnewline
\hline 
$\text{Br}(K_{L}\to\pi^{0}\nu_{\mu}\overline{\nu_{\mu}})$ & $<3.0\times10^{-9}\ \text{(90\% CL)}$ & $(3.4\pm0.6)\times10^{-11}$ & \cite{Ahn:2018mvc,Buras:2015qea}\tabularnewline
\hline 
\end{tabular}\caption{\label{tab:Correlated-observables-simplified}$R_{K^{(*)}}$-correlated
observables for single-mediator models.}
\end{table}

Some comments are in order regarding the phenomenological relevance
of the various processes listed in \ref{tab:ROFV-WC dependencies}.
Flavor observables of the type $u_{i}\to u_{j}\overline{\nu_{\mu}}\nu_{\mu}$
or $u_{i}\to u_{j}\mu^{+}\mu^{-}$ are much less constrained than
their $d_{i}\to d_{j}$ counterparts, from the experimental point
of view. On the other hand, the charged current processes $u_{i}\to d_{j}\mu^{+}\nu_{\mu}$
(which could in principle yield correlations between $C_{+}$ and
$C_{-}$), being unsuppressed in the SM, receive only tiny corrections
in the present framework. It turns out that all these observables
lead to weaker bounds than those arising from other sectors, so that
we omit them altogether from our analysis (as possibly relevant observables,
we examined $\text{Br}(J/\psi\to\text{invisible})$, $\text{Br}(D^{0}\to\mu^{+}\mu^{-})$
and $\text{Br}(K^{+}(\pi^{+})\to\mu^{+}\nu_{\mu})$, for the three
kind of quark level processes mentioned above, respectively.). Instead,
for the purpose of comparison, we display in this Section the collider
bounds arising from the high-$p_{T}$ tails of muonic Drell-Yan process
measured at LHC \cite{Aaboud:2017buh}, for which we follow the analysis
of Ref. \cite{Greljo:2017vvb}. As it can be seen from the plots below,
the collider bounds are outmatched by FCNC bounds in a large part
of parameter space. The only region where LHC searches are the most
relevant constraint is close to the bottom quark direction, i.e. for
$\theta\ll1$, as it can be seen directly in the top-left panel of
Fig.~\ref{fig:ROFV-S3limits} in the case of the $S_{3}$ leptoquark.

In the rest of this Section we focus on the following models:
\begin{enumerate}
\item Scalar leptoquark $S_{3}$. This is the simplest \textit{renormalizable}
model explaining the $R_{K^{(*)}}$ anomalies with NP in muons \cite{Gripaios:2014tna,Hiller:2014yaa,Varzielas:2015iva,Hiller:2017bzc,Dorsner:2017ufx,Cline:2017aed,Cline:2017qqu}. 
\item Vector leptoquark $U_{1}$. Besides having some theoretical motivation
(from Pati-Salam SM extensions), this is the \textit{only} single-mediator
simplified model for which a combined explanation of $R_{K^{(*)}}$
and $R_{D^{(*)}}$ anomalies is possible \cite{Barbieri:2015yvd,Alonso:2015sja,Barbieri:2016las,Buttazzo:2017ixm,DiLuzio:2017vat,Barbieri:2017tuq,Bordone:2017bld,DiLuzio:2018zxy,Bordone:2018nbg,Crivellin:2018yvo,Baker:2019sli}. 
\item Vector singlet $Z'$ with vector-like coupling to muons and a single
vectorlike partner for quark doublets. Arguably the most compelling
$\mathcal{O}_{9}$-type solution, it is relevant to some interesting
proposals such as gauged $L_{\mu}-L_{\tau}$ or $B_{3}-3L_{\mu}$,
see for example Refs.~\cite{Altmannshofer:2014cfa,Crivellin:2015mga,Bonilla:2017lsq,Biswas:2019twf}.
\end{enumerate}

\subsection{Scalar leptoquark $S_{3}$}

\begin{figure}[t]
\centering \hspace{-0.5cm}\includegraphics[scale=0.5]{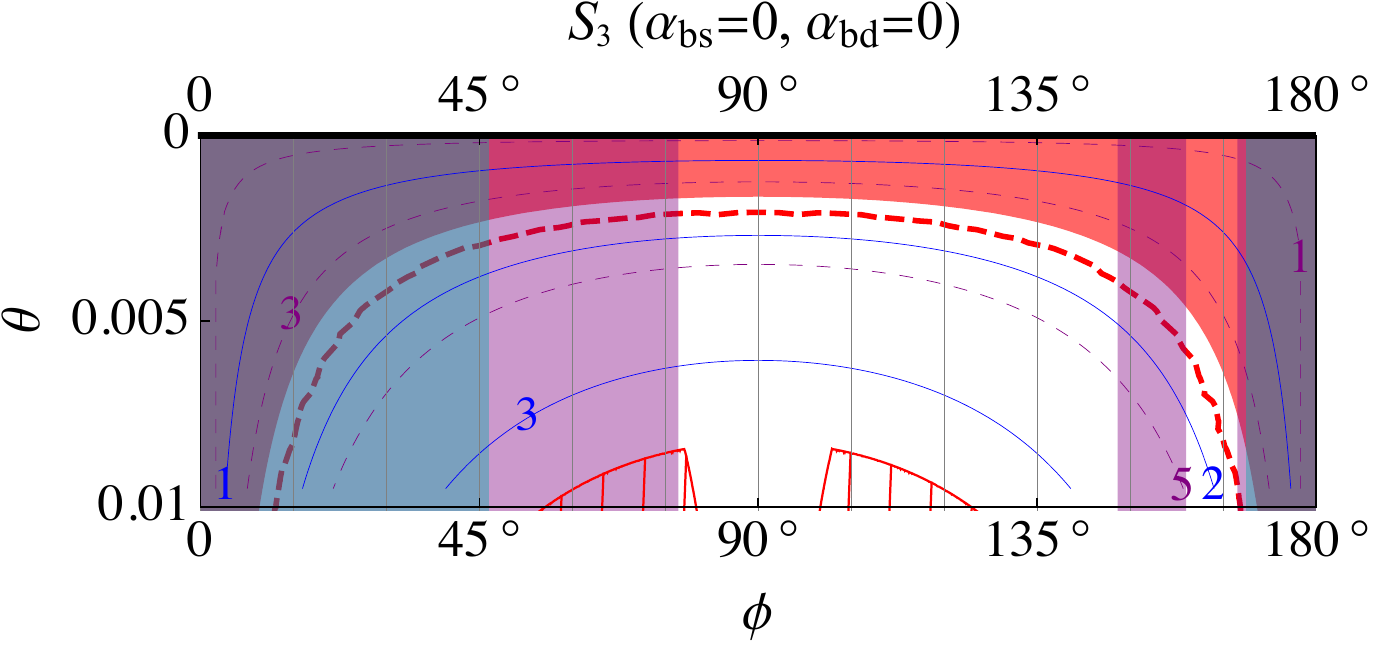}
\quad{}\includegraphics[scale=0.5]{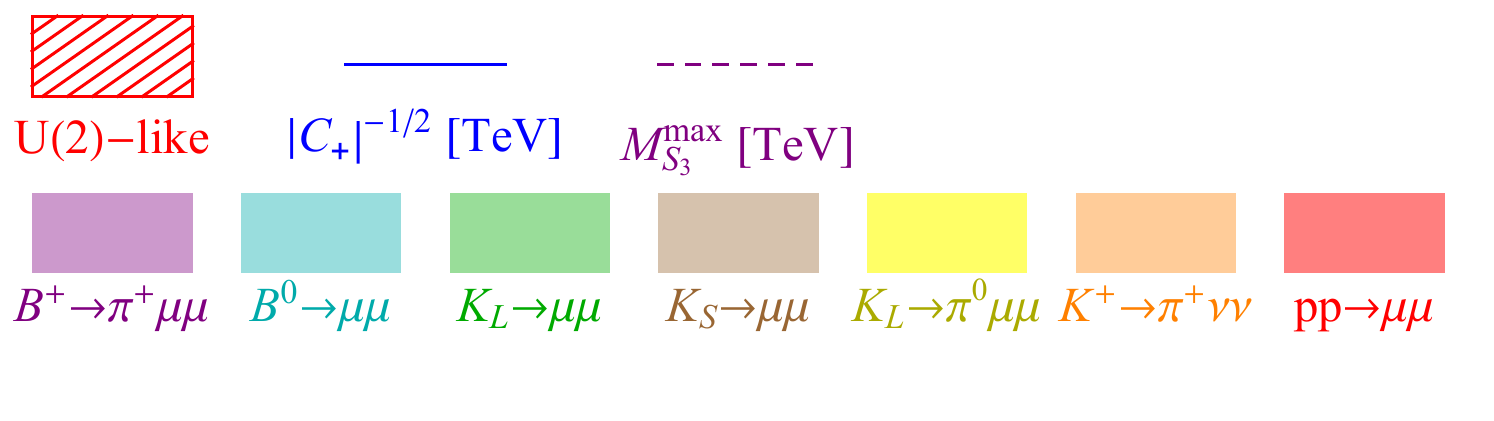}\\
 \includegraphics[scale=0.5]{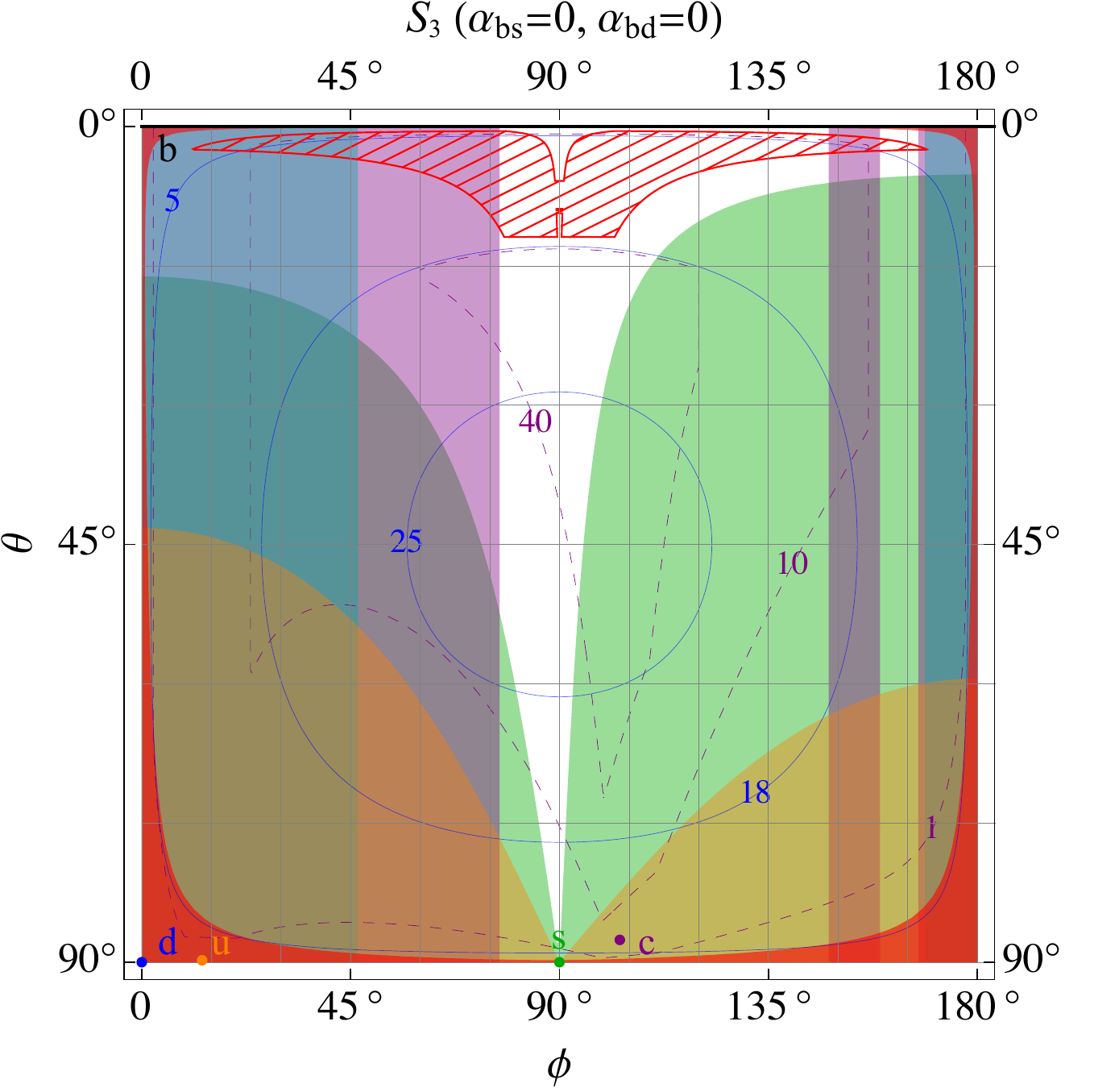} \includegraphics[scale=0.5]{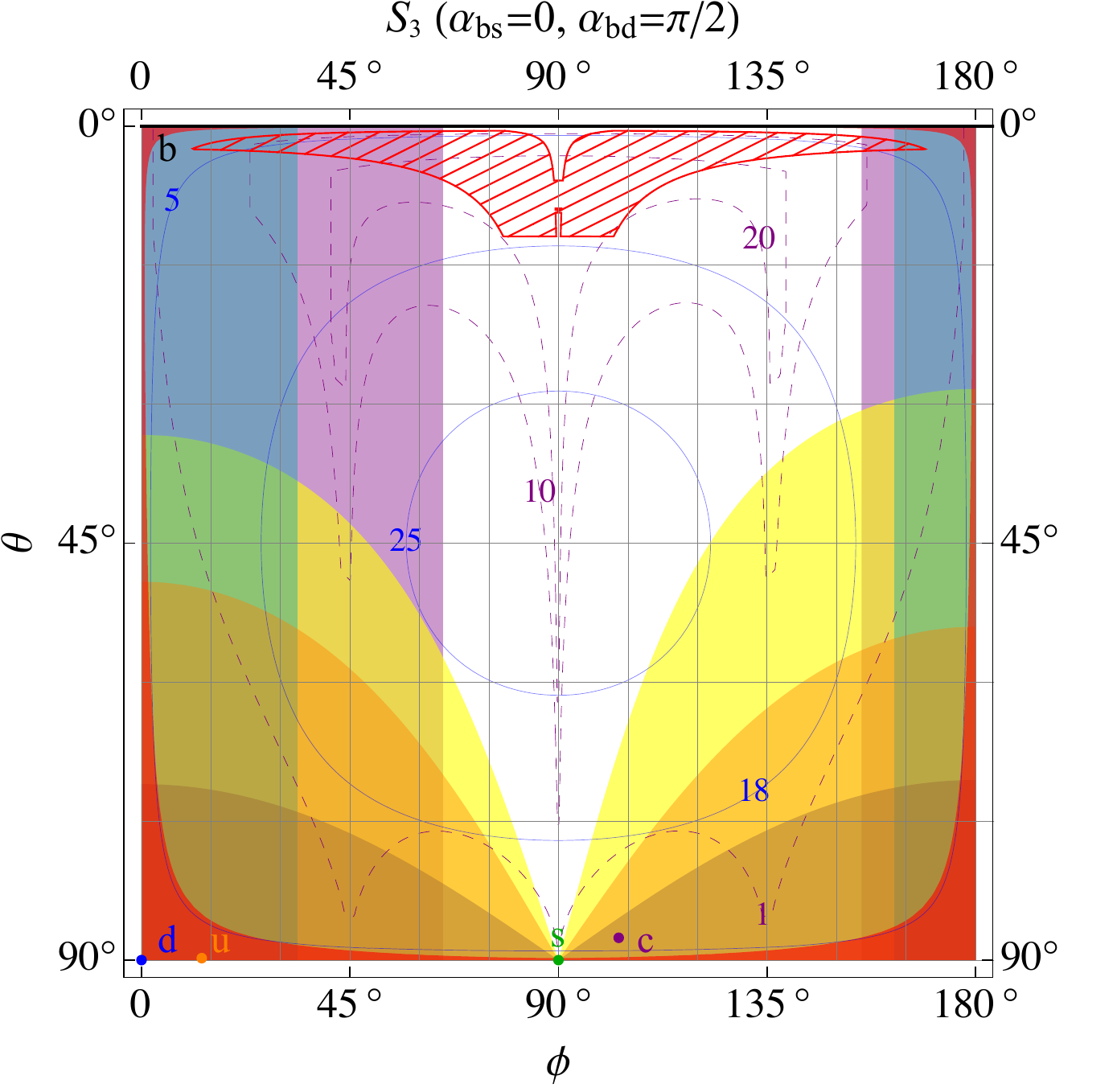}
\caption{\label{fig:ROFV-S3limits}Limits in the plane $(\phi,\theta)$ for
the scalar leptoquark $S_{3}$ and two choices of the phases $\alpha_{bs}$
and $\alpha_{bd}$. In addition to the limits in Fig.1, the orange
bound is from $K\to\pi\nu\nu$ while the red one is from the high-$p_{T}$
tail of $pp\to\mu^{+}\mu^{-}$ at the LHC \cite{Greljo:2017vvb}.
The top-left panel is a zoom of the region $\theta\ll10$ of the bottom-left
one, which shows in more detail the region excluded by LHC dimuon
searches. The dashed purple contour lines are the upper limits (in
TeV) on the leptoquark mass from $\Delta F=2$ processes.}
\end{figure}

The relevant interaction of the $S_{3}$ leptoquark with SM quarks
and leptons can be described by the Lagrangian:
\begin{equation}
\mathcal{L}_{\text{NP}}\supset\beta_{i\mu}(\overline{q_{iL}^{c}}\epsilon\sigma^{a}\ell_{2L})S_{3}^{a}+\text{h.c.}\label{eq:ROFV-S3Lagr}
\end{equation}
where we focussed on the interaction with muons. Integrating out $S_{3}$
at the tree-level, the effective operators in Eq.~(\ref{eq:ROFV-SMEFTlagrangian})
are generated, with 
\begin{equation}
C_{S}^{ij}=\frac{3}{4}\frac{\beta_{i\mu}^{*}\beta_{j\mu}}{M_{S_{3}}^{2}},\qquad C_{T}^{ij}=\frac{1}{4}\frac{\beta_{i\mu}^{*}\beta_{j\mu}}{M_{S_{3}}^{2}},\qquad C_{R}^{ij}=0\label{eq:ROFV-S3SMEFTmatch}
\end{equation}
We can match to our parametrization by writing the coupling in (\ref{eq:ROFV-S3Lagr})
as $\beta_{i\mu}^{*}\equiv\beta^{*}\,\hat{n}_{i}$, giving $C_{+}=|\beta|^{2}/M_{S_{3}}^{2}>0$,
$c_{S}=3/4$, $c_{T}=1/4$, and $c_{R}=0$. Since in this case $C_{+}=C_{L}>0$
and the r.h.s. of Eq.~(\ref{eq:ROFV-CLmatchingfit}) is also positive,
the angle $\phi$ is restricted to the range $\phi\in[0,\pi]$. The
constraints on $\phi$ and $\theta$ we obtain are shown in Fig. \ref{fig:ROFV-S3limits}.

The scalar LQ $S_{3}$ generates a contribution to $\Delta F=2$ processes
at the one-loop level. The relevant diagrams are finite and the contribution
from muonic loops is given by:
\begin{equation}
\Delta\mathcal{L}_{\Delta F=2}=-\frac{5}{128\pi^{2}}C_{+}^{2}M_{S_{3}}^{2}\left[(\hat{n}_{i}\hat{n}_{j}^{*}\overline{d_{iL}}\gamma^{\alpha}d_{jL})^{2}+(V_{ik}\hat{n}_{k}\hat{n}_{l}^{*}V_{jl}^{*}\;\overline{u_{iL}}\gamma^{\alpha}u_{jL})^{2}\right]~\label{eq:ROFV-DeltaF2_S3}
\end{equation}
Given a direction in quark space, i.e.\ a fixed $\hat{n}$, and fixing
$C_{+}$ to reproduce $R_{K^{(*)}}$, the experimental bounds on $K-\bar{K}$,
$B_{d,s}-\bar{B}_{d,s}$, and $D_{0}-\bar{D}_{0}$ mixing can be used
to set an \emph{upper limit} on the LQ mass, assuming the muonic contributions
shown in Eq.~(\ref{eq:ROFV-DeltaF2_S3}) to be dominant compared
to other possible NP terms. For the sake of clarity, it is worth remarking
that loops involving $\tau$ leptons could in general also give substantial
contributions to Eq.~(\ref{eq:ROFV-DeltaF2_S3}), possibly making
the bounds on $M_{S_{3}}$ qualitatively stronger or weaker than those
shown in Fig.~\ref{fig:ROFV-S3limits}, depending on the specific
flavour structure of leptoquark couplings. Another upper limit on
its mass, for a given value of $C_{+}$, can be set by requiring that
the coupling $\beta$ does not exceed the perturbative unitarity limit
$|\beta^{{\rm max}}|^{2}=(8\pi)/(3\sqrt{3})$ \cite{DiLuzio:2017chi}.
The contours of the strongest of these two upper limits on $M_{S_{3}}$
are shown as dashed purple lines (in TeV) in the plots of Fig.~\ref{fig:ROFV-S3limits}.
The perturbativity limit is never stronger than the one from $\Delta F=2$
processes in this scenario. Direct searches at the LHC of pair-produced
leptoquarks, on the other hand, set \emph{lower limits} on its mass,
which are now in the $\sim1\ \text{TeV}$ range.

\subsection{Vector leptoquark $U_{1}$}

\begin{figure}[t]
\centering \includegraphics[scale=0.5]{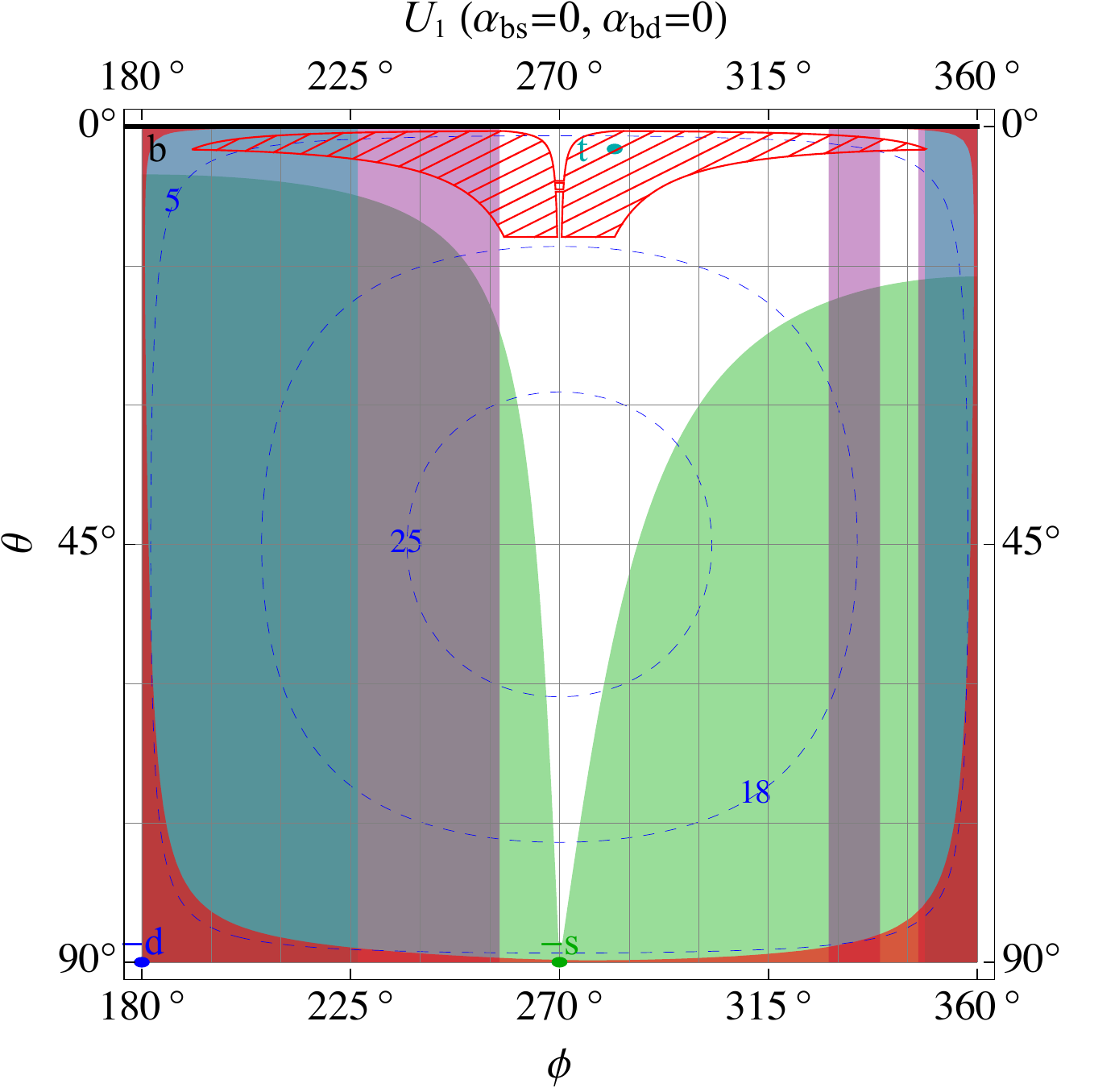} \includegraphics[scale=0.5]{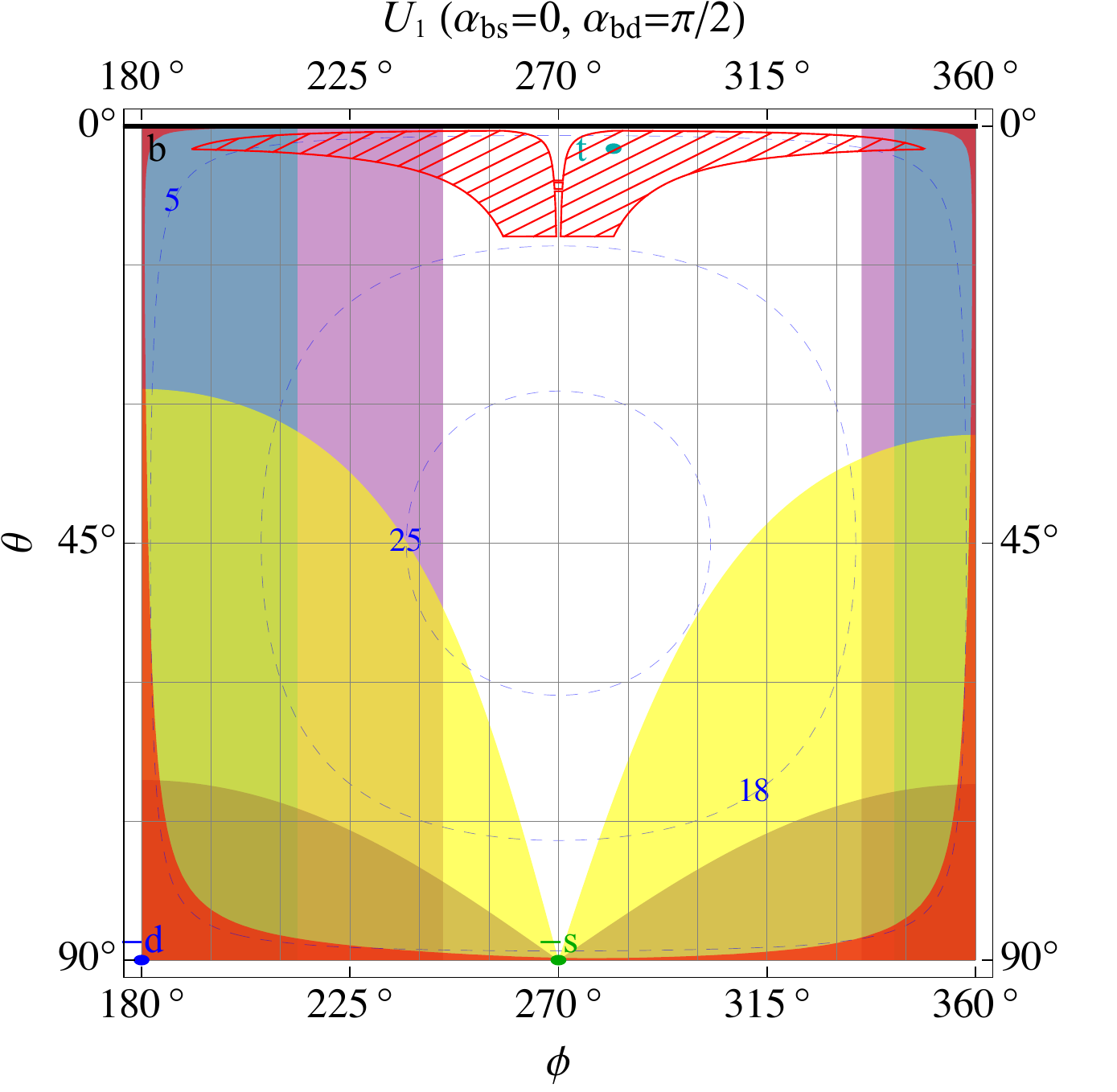}\\
 \includegraphics[scale=0.5]{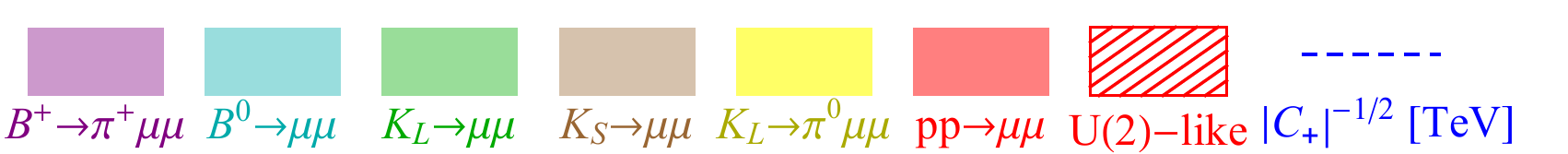} \caption{Limits in the plane $(\phi,\theta)$ for the vector leptoquark $U_{1}$
and two choices of the phases $\alpha_{bs}$ and $\alpha_{bd}$. The
red region is excluded by the high-$p_{T}$ tail of $pp\to\mu^{+}\mu^{-}$
at the LHC \cite{Greljo:2017vvb}.}
\label{fig:U1limits}
\end{figure}

The interaction lagrangian of the vector leptoquark $U_{1}$ is
\begin{equation}
\mathcal{L}_{\text{NP}}\supset\gamma_{i\mu}(\overline{q_{iL}}\gamma_{\alpha}\ell_{2L})U_{1}^{\alpha}+\text{h.c.}\label{eq:ROFV-U1Lagr}
\end{equation}
The matching to the SMEFT operators generated by integrating out $U_{1}$
at the tree-level is given by:
\begin{equation}
C_{S}^{ij}=C_{T}^{ij}=-\frac{1}{2}\frac{\gamma_{i\mu}\gamma_{j\mu}^{*}}{M_{U_{1}}^{2}},\qquad C_{R}=0.\label{eq:ROFV-U1SMEFTmatch}
\end{equation}
We can match to our parametrization by defining $\gamma_{i\mu}\equiv\gamma\,\hat{n}_{i}$,
corresponding to $C_{+}=-|\gamma|^{2}/M_{U_{1}}^{2}<0$, $c_{S}=1/2$,
$c_{T}=1/2$, and $c_{R}=0$. Contrary to the $S_{3}$ model, the
$U_{1}$ LQ implies $C_{+}=C_{L}<0$. Therefore, Eq.~(\ref{eq:ROFV-CLmatchingfit}),
whose r.h.s. is positive, restricts the angle $\phi$ to the range
$[\pi,2\pi)$. The constraints on $\phi$ and $\theta$ we obtain
are shown in \ref{fig:U1limits}. As anticipated, they coincide with
the constraints (in the $\pi<\phi<2\pi$ part) of \ref{fig:ROFV-Minimal}.

Like $S_{3}$, also the $U_{1}$ vector LQ contributes to meson anti-meson
mixing at one-loop. Such a contribution is however UV-divergent and,
in order to be calculable, requires a UV-completion of the simplified
model. In general such UV completions contains other contributions
to the same processes, which must also be taken into account \cite{Barbieri:2016las,DiLuzio:2017vat,Barbieri:2017tuq,Bordone:2017bld,DiLuzio:2018zxy,Bordone:2018nbg,Baker:2019sli}.

\subsection{Vector singlet $Z^{\prime}$ with vector-like couplings to muons}

\begin{figure}[t]
\centering \includegraphics[scale=0.5]{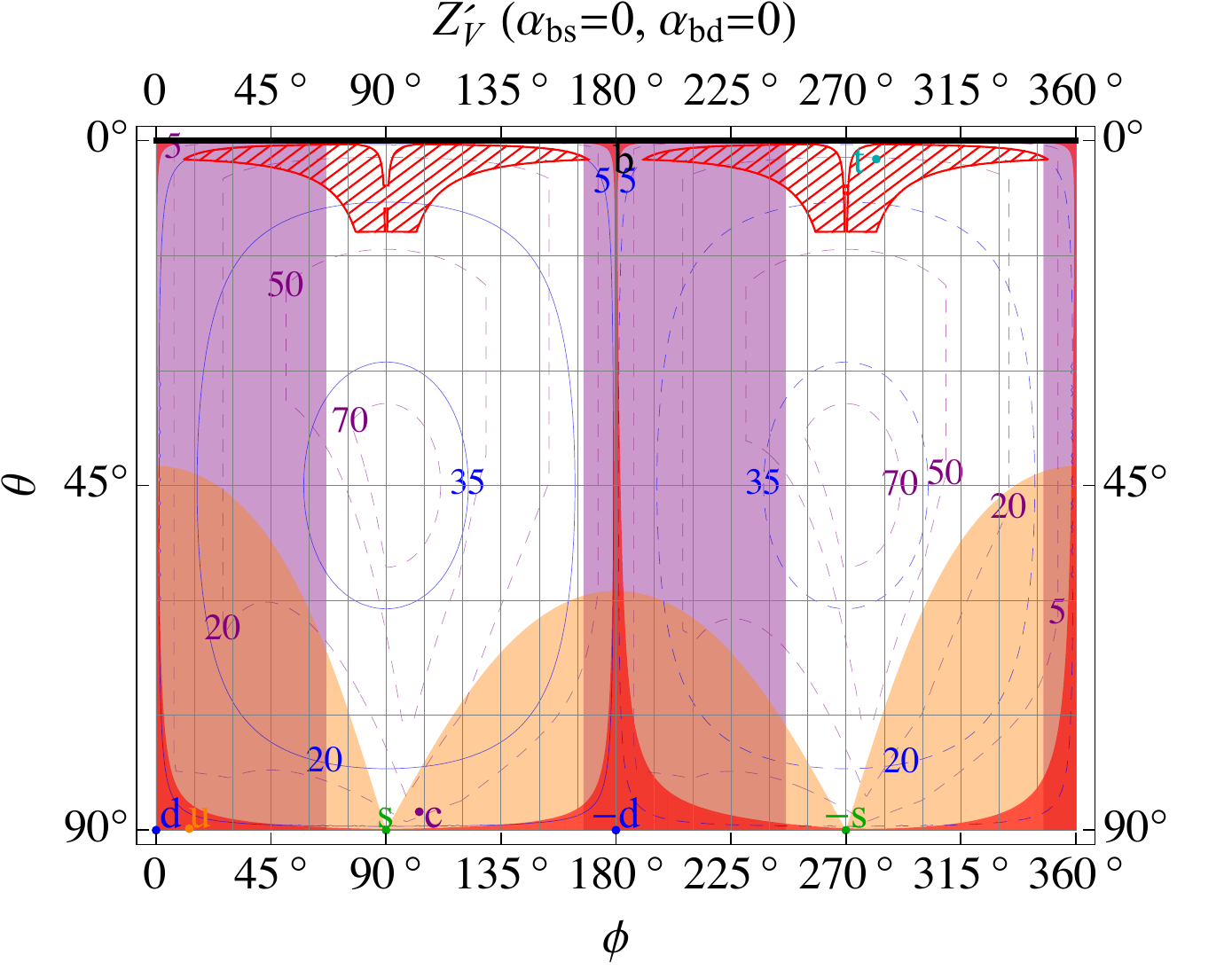}\!\!\!
\includegraphics[scale=0.5]{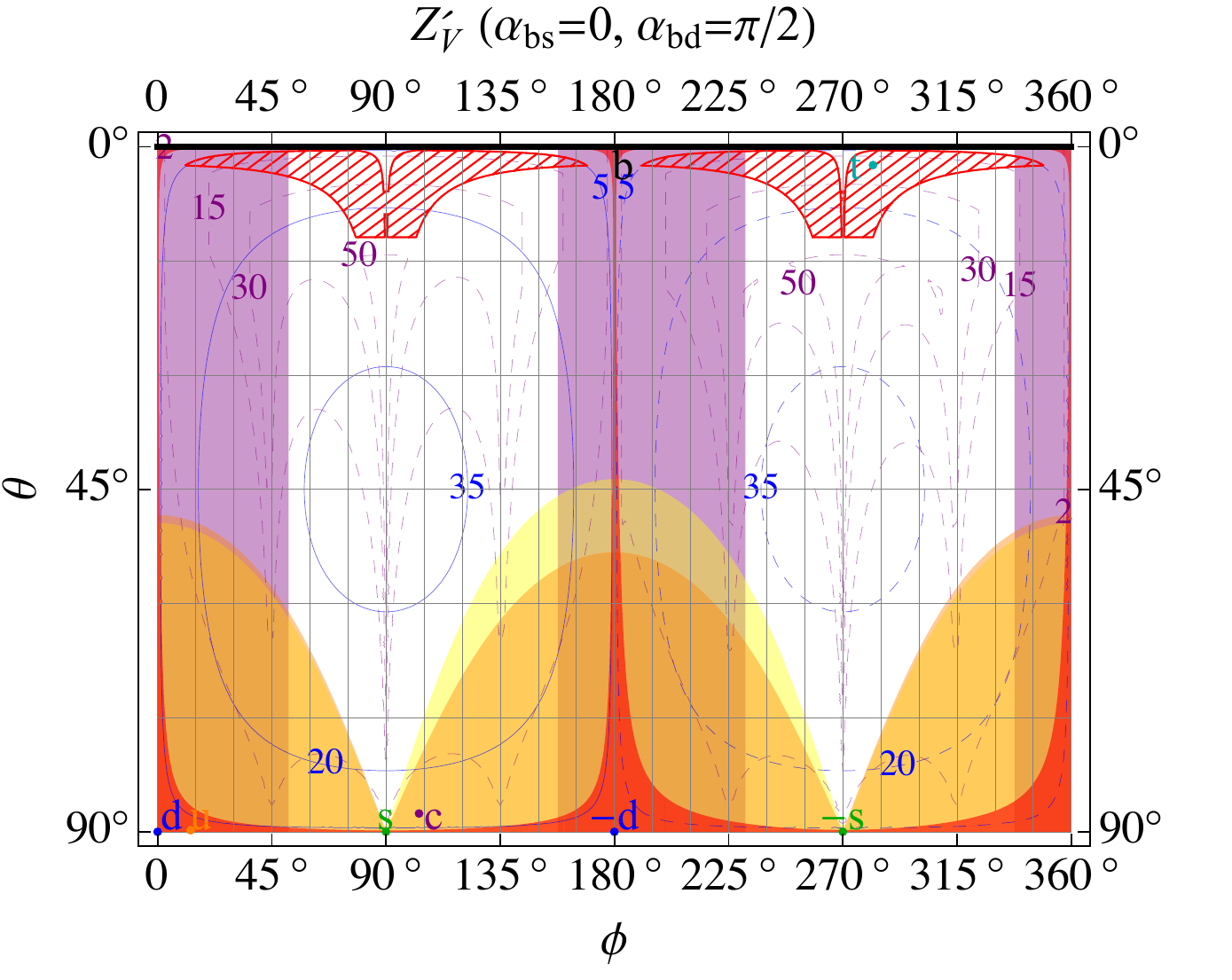}\\
 \includegraphics[scale=0.5]{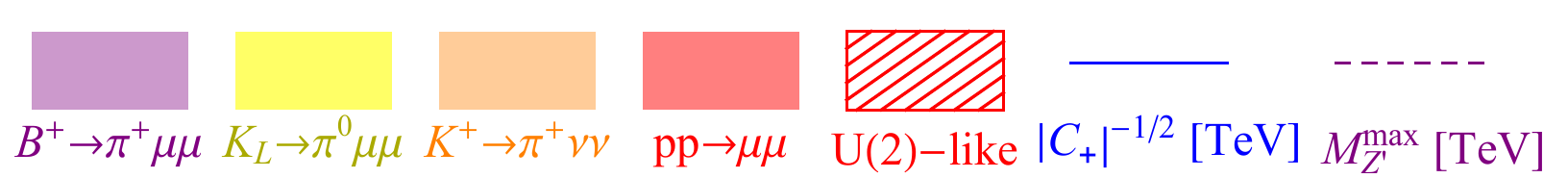} \caption{Limits in the plane $(\phi,\theta)$ for the vector singlet $Z^{\prime}$
with vector-like couplings to muons and two choices of the phases
$\alpha_{bs}$ and $\alpha_{bd}$. The dashed purple contour lines
are upper limits on the $Z^{\prime}$ mass {[}TeV{]} from $\Delta F=2$
processes using Eq.~(\ref{eq:MaxZpMass}).}
\label{fig:Zplimits}
\end{figure}

Let us consider a heavy singlet vector $Z^{\prime}$ with couplings:
\begin{equation}
\mathcal{L}_{\text{NP}}\supset\left[g_{q}\hat{n}_{i}\hat{n}_{j}^{*}(\overline{q_{iL}}\gamma^{\alpha}q_{jL})+g_{\mu}(\overline{\ell_{2L}}\gamma^{\alpha}\ell_{2L}+\overline{\mu_{R}}\gamma^{\alpha}\mu_{R})\right]Z_{\alpha}^{\prime}\label{eq:ROFV-ZpLagr}
\end{equation}
Such a flavour structure of the $Z^{\prime}$ couplings to quarks
could arise, for example, by assuming that they couple to $Z^{\prime}$
only via the mixing with a heavy vector-like quark doublet $Q$ in
the form 
\begin{equation}
M_{i}\overline{Q}\,q_{iL}+\text{h.c.}\;.\label{eq:mixing}
\end{equation}
In such a case, $\hat{n}_{i}\propto M_{i}^{*}$. The matching with
the SMEFT operators in this case is given by:
\begin{equation}
C_{S}^{ij}=C_{R}^{ij}=-\frac{g_{q}g_{\mu}}{M_{Z^{\prime}}^{2}}\hat{n}_{i}\hat{n}_{j}^{*},\quad C_{T}=0,\label{eq:ROFV-ZpSMEFTmatch}
\end{equation}
corresponding to $C_{+}=-g_{q}g_{\mu}/(M_{Z^{\prime}}^{2})$, $c_{S}=1$,
$c_{R}=1$, and $c_{T}=0$. The matching to the operators relevant
for the $b\to s\mu\mu$ anomalies is now 
\begin{equation}
C_{+}\sin\theta\cos\theta\sin\phi e^{i\alpha bs}=\frac{G_{F}\alpha}{\sqrt{2}\pi}V_{tb}V_{ts}^{*}\Delta C_{9}^{\mu}.\label{eq:ROFV-C9matchZp}
\end{equation}
Note that in this scenario the overall coefficient $C_{+}$ can take
any sign. It is worth noting that all purely leptonic meson decays
such as $K_{L,S}$ or $B^{0}$ to $\mu\mu$ vanish in this setup since
the leptonic current is vector-like \footnote{The $J=0$ constraint forces the final muon pair to be in a state
with $C=+1$, whereas the vectorial current $\overline{\mu}\gamma\mu$
has negative $C$-parity.}. The only relevant limits then arise from $B^{+}\to\pi^{+}\mu\mu$,
$K^{+}\to\pi^{+}\nu\nu$, and from LHC dimuon searches, as shown in
Fig.~\ref{fig:Zplimits}.

This model also generates at the tree-level four quark operators which
contribute to $\Delta F=2$ observables: 
\begin{equation}
\Delta\mathcal{L}_{\Delta F=2}=-\frac{g_{q}^{2}}{2M_{Z^{\prime}}^{2}}\left[(\hat{n}_{i}\hat{n}_{j}^{*}\overline{d_{iL}}\gamma^{\alpha}d_{jL})^{2}+(V_{ik}\hat{n}_{k}\hat{n}_{l}^{*}V_{jl}^{*}\;\overline{u_{iL}}\gamma^{\alpha}u_{jL})^{2}\right]
\end{equation}
For a fixed direction in quark space, $\hat{n}$, and a fixed value
of $R_{K^{(*)}}$, we can use $\Delta F=2$ constraints to put an
upper limit on the ratio $r_{q\mu}\equiv|g_{q}/g_{\mu}|$. We can
then assign a maximum value to $g_{\mu}$ and derive an upper limit
for the $Z^{\prime}$ mass: 
\begin{equation}
M_{Z^{\prime}}^{\text{lim}}=\min\left[\sqrt{\frac{r_{q\mu}^{\text{lim}}}{\left|C_{+}\right|}}\left|g_{\mu}^{\text{max}}\right|,\sqrt{\left|\frac{g_{\mu}^{\text{max}}g_{q}^{\text{max}}}{C_{+}}\right|}\right]\label{eq:MaxZpMass}
\end{equation}
where the first limit is from $\Delta F=2$ observables while the
second is from perturbativity. For the maximum values of the couplings
we use the limits from perturbative unitarity from Ref.~\cite{DiLuzio:2017chi},
$|g_{\mu}^{{\rm max}}|^{2}=2\pi$ and $|g_{q}^{{\rm max}}|^{2}=2\pi/3$.

\section{\label{sec:ROFV-FlavourSymmetry}ROFV and flavour symmetries}

In the previous Sections we have been agnostic about the structure
of the rank one coefficients of the NP interactions, and parameterised
it in terms of the unit vector $\hat{n}$. Here we would like to illustrate,
with a flavour symmetry example, the possible theoretical expectations
on the direction in flavour space at which $\hat{n}$ points.

In the SM, the gauge lagrangian flavour group 
\begin{equation}
\text{U}(3)^{5}\equiv\text{U}(3)_{q}\times\text{U}(3)_{\ell}\times\text{U}(3)_{u}\times\text{U}(3)_{d}\times\text{U}(3)_{e}\;\label{eq:U(3)5 group}
\end{equation}
is explicitly broken by the Yukawa couplings $Y_{u,d,e}$. Here, the
unit vector $\hat{n}$, and the UV couplings from which it originates,
represent an additional source of explicit breaking. In fact, we can
formally assign the UV couplings introduced in the previous Section
(and the SM Yukawa couplings) quantum numbers under $\text{U}(3)^{5}$:
\begin{equation}
\text{SM}:\,\begin{cases}
Y_{u}\sim3_{q}\otimes\bar{3}_{u},\\
Y_{d}\sim3_{q}\otimes\bar{3}_{d},\\
Y_{e}\sim3_{\ell}\otimes\bar{3}_{e}.
\end{cases}\qquad\qquad\hat{n}_{i}\propto\begin{cases}
\beta_{i\mu}\sim\overline{3}_{q}\otimes\overline{3}_{\ell} & S_{3},\\
\gamma_{i\mu}\sim\overline{3}_{q}\otimes3_{\ell} & U_{1}^{\mu},\\
M_{i}\sim\overline{3}_{q} & Z',\,V'
\end{cases}\label{eq:Spurion Reps}
\end{equation}
Therefore, different models can be characterised not only in terms
of the SM quantum numbers of the messengers, but also in terms of
the flavour quantum numbers of the couplings.

Correlations between the two sets of couplings in (\ref{eq:Spurion Reps})
can arise if they share a common origin. This may be the case, for
example, if we assume a subgroup $\mathcal{G}\subseteq\text{U}(3)^{5}$
to be an actual symmetry of the complete UV lagrangian, and the above
couplings to originate from its spontaneous breaking by means of a
common set of ``flavon'' fields.

Correlations cannot arise if $\mathcal{G}$ coincides with the full
$\text{U}(3)^{5}$. The quantum numbers of the relevant flavons coincide
in this case with the transformation properties in (\ref{eq:Spurion Reps}).
Therefore, the flavons entering the Yukawas and the NP couplings are
in this case entirely independent. In particular, the ROFV assumption
is not compatible with the Minimal Flavor Violation one~\cite{DAmbrosio:2002vsn}.
We therefore need to consider proper subgroups of $\text{U}(3)^{5}$.
Among the many possibilities, let us consider the $\mathcal{G}=\text{U}(2)^{5}$
subgroup of transformations on the first two fermion families. The
latter extends the quark $\text{U}(2)^{3}$~\cite{Barbieri:2011ci}
to the leptons, relevant for two of the three NP couplings in (\ref{eq:Spurion Reps}).
Some of the conclusions we will draw hold for a generic extension
to $\text{U}(2)^{3}\times\mathcal{G}_{l}$, where $\mathcal{G}_{l}$
only acts on leptons.

The fact that correlations can arise in the U(2) case is not a surprise.
In the unbroken limit, the versor $\hat{n}$ and the SM Yukawas must
leave the same $\text{U}(2)_{q}$ subgroup invariant, and are therefore
aligned (although no flavour violation would be generated in such
a limit). In order to investigate them, we write all the $\mathcal{G}$-violating
couplings as VEVs of flavons with irreducible $\mathcal{G}$ quantum
numbers, and assume the UV theory to contain at most one flavon of
each type. The predictions that follow then depend on the structure
of the flavon sector, as we now discuss.

Let us first consider the case, which we will refer to as ``minimally
broken'' $\text{U}(2)^{5}$, in which no flavon is charged under
both the quark $\text{U}(2)^{3}$ and the lepton $\text{U}(2)^{2}$,
or $\mathcal{G}_{l}$. In such a case one finds a precise correlation
between the first two components of the unit vector $\hat{n}$ and
of the third line of the CKM matrix: $\hat{n}_{1}/\hat{n}_{2}=V_{td}^{*}/V_{ts}^{*}$,
up to corrections of order $m_{s}/m_{b}$. We then have 

\begin{equation}
\hat{n}_{U2}\propto\left(c_{U2}e^{i\gamma}V_{td}^{*},\,c_{U2}e^{i\gamma}V_{ts}^{*},\,1\right)\label{eq:ROFV-U2direction}
\end{equation}
where $c_{U2}\sim\mathcal{O}(1)$ and the normalisation is fixed by
the condition $|\hat{n}|^{2}=1$. Comparing with the parametrization
in Eq.~(\ref{eq:VersorDef}), one gets: 
\begin{equation}
\tan\phi=\frac{|V_{ts}|}{|V_{td}|}\;,\quad\tan\theta\approx c_{U2}|V_{ts}|\;,\quad\alpha_{bd}=-\arg(V_{td})+\gamma\;,\quad\alpha_{bs}=-\arg(V_{ts})+\gamma\;.\label{eq:U2minimalpar}
\end{equation}
This prediction also applies in the case of $Z'$ or $V'$ messengers,
independent on whether $\text{U}(2)^{5}$ is minimally broken or not.
\begin{figure}[t]
\centering \includegraphics[scale=0.5]{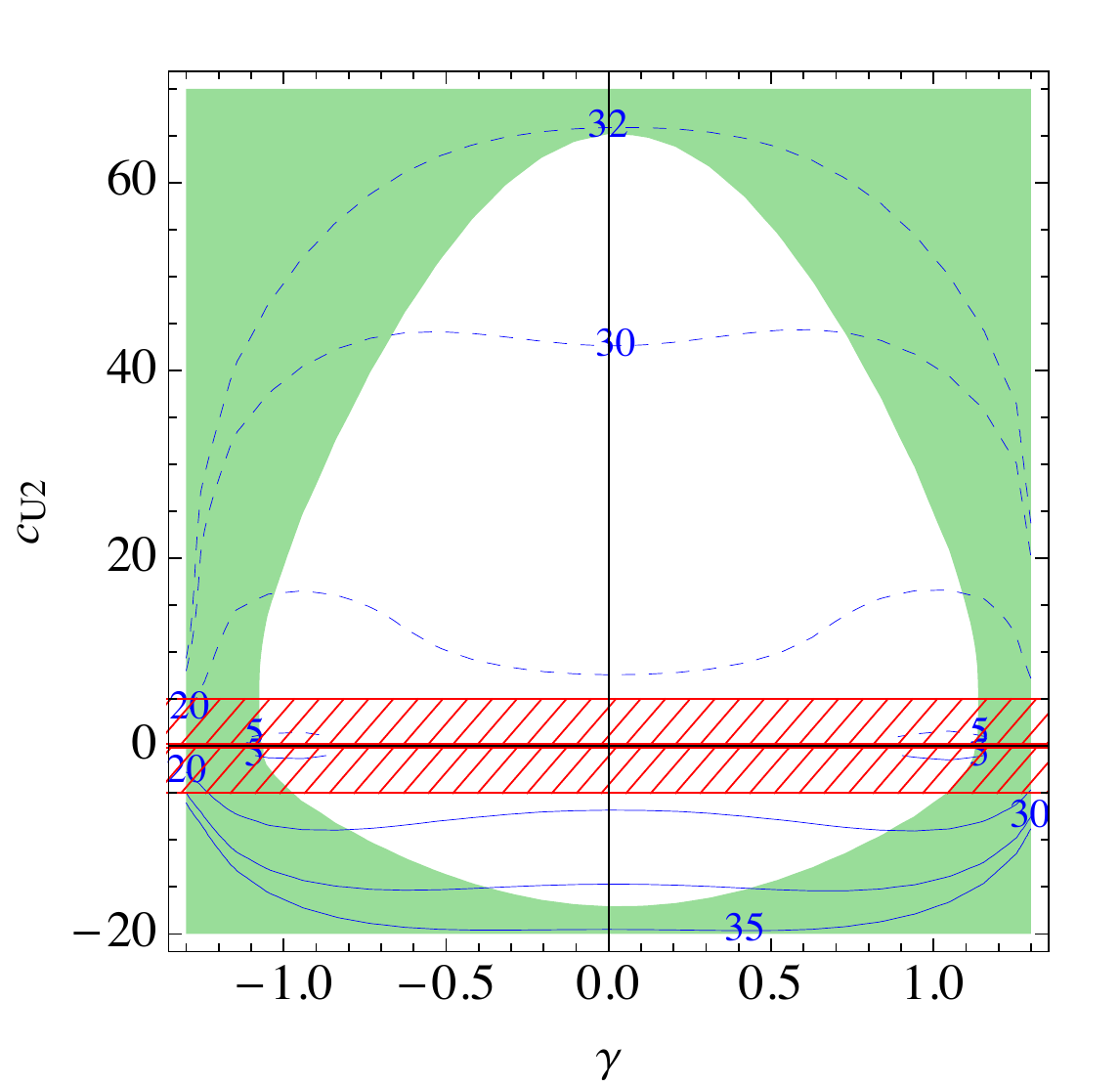} \caption{Region excluded at $95\%$CL in a global fit of $bs\mu\mu$ clean
observables and the other $d_{i}d_{j}\mu\mu$ ones listed in \ref{tab:ROFV-Correlated-observables},
assuming the structure of Eq.~(\ref{eq:ROFV-U2direction}), in the
plane $(\gamma,c_{U2})$. The relevant observable in the excluded
region is $K_{L}\to\mu^{+}\mu^{-}$. The range $|c_{U2}|\in[0.2-5]$
is highlighted as the meshed red region. Blue lines indicate the size
of the overall coefficient $|C_{L}|^{-1/2}$ (in TeV), as extracted
from the fit. Solid (dashed) lines are for positive (negative) values
of $C_{L}$.}
\label{fig:minimalU2}
\end{figure}

In the presence of flavons charged both under the quark and lepton
part of the flavour group, or in the presence of more flavons with
the same flavour quantum numbers, the above prediction does not need
to hold. On the other hand, with reasonable assumptions on the size
of the flavons, one obtains a generic correlation between the unit
vector $\hat{n}$ and the third line of the CKM matrix, which holds
up to $\mathcal{O}(1)$ factors: $\hat{n}=(\mathcal{O}(V_{td}),\mathcal{O}(V_{ts}),\mathcal{O}(1))$.
We can parametrize such a scenario in full generality as 
\begin{equation}
\hat{n}\propto\left(a_{bd}e^{i\alpha_{bd}}\left|V_{td}\right|,\,a_{bs}e^{i\alpha_{bs}}\left|V_{ts}\right|,\,1\right)\label{eq:TopLikeDirection}
\end{equation}
where $a_{bd}$ and $a_{bs}$ are $\mathcal{O}(1)$ real parameters.
The area in the $(\phi,\theta)$ plane corresponding to values $|a_{bs,bd}|\in[0.2-5]$
is shown as a meshed-red one in the plots of Figs.~\ref{fig:ROFV-Minimal},
\ref{fig:ROFV-S3limits}, \ref{fig:U1limits}.

The correlation in (\ref{eq:TopLikeDirection}) is also found with
different flavour groups, and in models with partial compositeness
(and no flavour group). In the limit in which the top quark exchange
dominates FCNC processes, the SM itself satisfies the ROFV condition,
with $\hat{n}=(V_{td}^{*},V_{ts}^{*},V_{tb}^{*})$ also in the form
above.

One comment is in order about the role of the lepton flavour sector.
The latter can play a twofold role. On the one hand, it can affect
the prediction for the direction of $\hat{n}$. This can be the case
for $S_{3}$ and $U_{1}^{\mu}$ messengers, for which $\hat{n}$ is
associated to the muon row of the $\beta$ and $\gamma$ matrices
in (\ref{eq:Spurion Reps}). On the other hand, the lepton flavour
breaking can affect the overall size of the effect. The anomalies
require in fact a breaking of $\mu$-$e$ lepton universality, whose
size is associated to the size of $\text{U}(2)_{l}$ breaking. A sizeable
breaking is necessary in order to account for a NP effect as large
as suggested by the $B$-meson anomalies. A detailed analysis of the
implications of the anomalies for lepton flavour breaking and for
processes involving other lepton families is outside the scope of
this work.\par We now focus on the case of minimally broken $\text{U}(2)^{5}$,
(\ref{eq:U2minimalpar}). The 95\%CL limit in the plane $(\gamma,c_{U2})$,
from our global fit of $bs\mu\mu$ clean observables and the other
$d_{i}d_{j}\mu\mu$ ones (Tab. \ref{tab:ROFV-Correlated-observables})
is shown in Fig.~\ref{fig:minimalU2}-Left. The relevant observable
in the excluded region is $K_{L}\to\mu^{+}\mu^{-}$. For positive
(negative) values of $C_{+}$ we obtain a limit $c_{U2}\gtrsim-20$
($\lesssim65$), which are well outside the natural region predicted
by the flavour symmetry.

Another interesting point is that, within the parametrization (\ref{eq:ROFV-U2direction}),
one has 
\begin{equation}
R_{K}\approx R_{\pi}\label{eq:ROFV-RKRpiSU2}
\end{equation}
up to $\mathcal{O}(m_{s}/m_{b})$ corrections, where: 
\begin{equation}
R_{H}\equiv\dfrac{\text{Br}(B\to H\mu^{+}\mu^{-})_{[1,6]}}{\text{Br}(B\to He^{+}e^{-})_{[1,6]}}\quad\quad(H=K,\pi)\label{eq:ROFV-LFUratios}
\end{equation}
so that a comparison between the two observables could in principle
rule out, in this context, the $V'$ and $Z'$ cases, as well as minimally
broken $\text{U}(2)^{5}$. Assuming no NP in the electron channels,
we also have\footnote{Using the LFU ratios (\ref{eq:ROFV-LFUratios}) is of course advisable
from the theoretical point of view. Unfortunately, there are no measurements
of $\text{Br}(B\to\pi e^{+}e^{-})$, at present.}: 
\begin{equation}
R_{\pi}=\dfrac{\text{Br}(B\to\pi\mu^{+}\mu^{-})_{\left[1,6\right]}}{\text{Br}(B\to\pi\mu^{+}\mu^{-})_{\left[1,6\right]}^{\text{SM}}}\label{eq:ROFV-RpiAssumption}
\end{equation}
The RHS of Eq. (\ref{eq:ROFV-RpiAssumption}) is, experimentally,
(cf. Table \ref{tab:ROFV-Correlated-observables}): 
\begin{equation}
\dfrac{\text{Br}(B\to\pi\mu^{+}\mu^{-})_{\left[1,6\right]}^{\text{exp}}}{\text{Br}(B\to\pi\mu^{+}\mu^{-})_{\left[1,6\right]}^{\text{SM}}}=0.70\pm0.30,\label{eq:RpiExpt}
\end{equation}
showing no tension neither with the SM prediction, nor with the $\text{U(2)}^{5}$
prediction (\ref{eq:ROFV-RKRpiSU2}). Another prediction of this setup
is for the branching ratio of $B_{d}^{0}\to\mu^{+}\mu^{-}$ with respect
to $B_{s}\to\mu^{+}\mu^{-}$: 
\begin{equation}
\dfrac{\text{Br}(B_{s}^{0}\to\mu^{+}\mu^{-})}{\text{Br}(B_{s}^{0}\to\mu^{+}\mu^{-})_{\text{SM}}}\approx\dfrac{\text{Br}(B_{d}^{0}\to\mu^{+}\mu^{-})}{\text{Br}(B_{d}^{0}\to\mu^{+}\mu^{-})_{\text{SM}}}\label{eq:B0BsmumuSU2}
\end{equation}
The two predictions (\ref{eq:ROFV-RKRpiSU2}) and (\ref{eq:B0BsmumuSU2})
are independent on the specific chiral structure of the muon current.
If the operators responsible for $R_{K^{(*)}}$ are left-handed only,
the two ratios in Eq.~(\ref{eq:B0BsmumuSU2}) are also predicted
to be of the same size as $R_{K}$ and $R_{\pi}$, up to $\mathcal{O}(2\%)$
corrections. Such corrections are however negligible when compared
to the expected precision in the measurements of these relations,
which is at best of $\approx4\%$, cf. \ref{tab:Prospects}. It is
perhaps worth pointing out that the predictions in Eqs.~(\ref{eq:ROFV-RKRpiSU2},\ref{eq:B0BsmumuSU2})
are a consequence of the minimally broken $\text{U}(2)^{5}$ flavour
symmetry alone, independently of the ROFV assumption. This can be
understood from the fact that the $b-s$ and $b-d$ transitions are
related by $\text{U}(2)^{5}$ symmetry as $C_{S,T}^{bd}/C_{S,T}^{bs}=(V_{q})^{1}/(V_{q})^{2}=V_{td}^{*}/V_{ts}^{*}=n_{U2}^{1}/n_{U2}^{2}$,
where $V_{q}$ is the spurion doublet under $\text{U}(2)_{q}$.

\section{\label{sec:ROFV-Prospects}Prospects}

\begin{table}[t]
\centering %
\begin{tabular}{|c|clc|}
\hline 
Observable & Expected sensitivity & Experiment & Reference\tabularnewline
\hline 
\hline 
\multirow{2}{*}{$R_{K}$} & $0.7\;(1.7)\%$ & LHCb 300 (50) fb$^{-1}$ & \cite{Bediaga:2018lhg}\tabularnewline
 & $3.6\;(11)\%$ & Belle II 50 (5) ab$^{-1}$ & \cite{Kou:2018nap}\tabularnewline
\hline 
\multirow{2}{*}{$R_{K^{*}}$} & $0.8\;(2.0)\%$ & LHCb 300 (50) fb$^{-1}$ & \cite{Bediaga:2018lhg}\tabularnewline
 & $3.2\;(10)\%$ & Belle II 50 (5) ab$^{-1}$ & \cite{Kou:2018nap}\tabularnewline
\hline 
$R_{\pi}$ & $4.7\;(11.7)\%$ & LHCb 300 (50) fb$^{-1}$ & \cite{Bediaga:2018lhg}\tabularnewline
\hline 
\multirow{2}{*}{$\text{Br}(B_{s}^{0}\to\mu^{+}\mu^{-})$} & $4.4\;(8.2)\%$ & LHCb 300 (23) fb$^{-1}$ & \cite{Bediaga:2018lhg,Cerri:2018ypt}\tabularnewline
 & $7\;(12)\%$ & CMS 3 (0.3) ab$^{-1}$ & \cite{Cerri:2018ypt}\tabularnewline
\hline 
\multirow{2}{*}{$\text{Br}(B_{d}^{0}\to\mu^{+}\mu^{-})$} & $9.4\;(33)\%$ & LHCb 300 (23) fb$^{-1}$ & \cite{Bediaga:2018lhg,Cerri:2018ypt}\tabularnewline
 & $16\;(46)\%$ & CMS 3 (0.3) ab$^{-1}$ & \cite{Cerri:2018ypt}\tabularnewline
\hline 
$\text{Br}(K_{S}\to\mu^{+}\mu^{-})$ & $\sim10^{-11}$ & LHCb 300fb$^{-1}$ & \cite{Bediaga:2018lhg,Cerri:2018ypt}\tabularnewline
\hline 
\multirow{3}{*}{$\text{Br}(K_{L}\to\pi^{0}\nu\nu)$} & $\sim1.8\times10^{-10}$ & KOTO phase-I $^{7}$ & \tabularnewline
 & $20\%$ & KOTO phase-II $^{7}$ & \tabularnewline
 & $20\%$ & KLEVER & \cite{Ambrosino:2019qvz}\tabularnewline
\hline 
$\text{Br}(K^{+}\to\pi^{+}\nu\nu)$ & $10\%$ & NA62 goal & \cite{Ruggiero:2017hjh}\tabularnewline
\hline 
\end{tabular}\caption{\label{tab:Prospects} Future prospects for the precision reach in
various flavour observables. The expected sensitivity in percent are
quoted with respect to the Standard Model prediction.}
\end{table}

\begin{figure}[t]
\centering \includegraphics[scale=0.5]{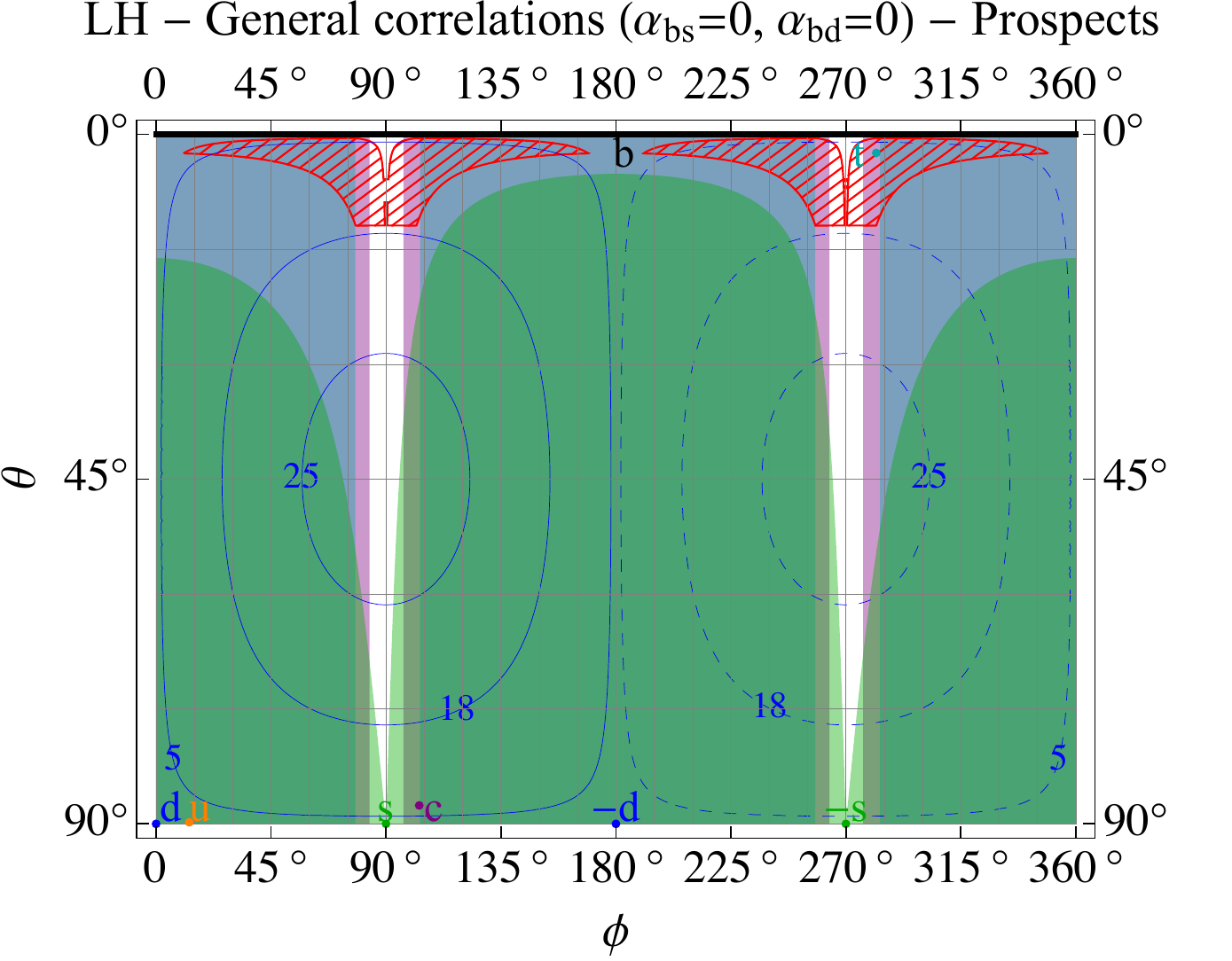}
\includegraphics[scale=0.5]{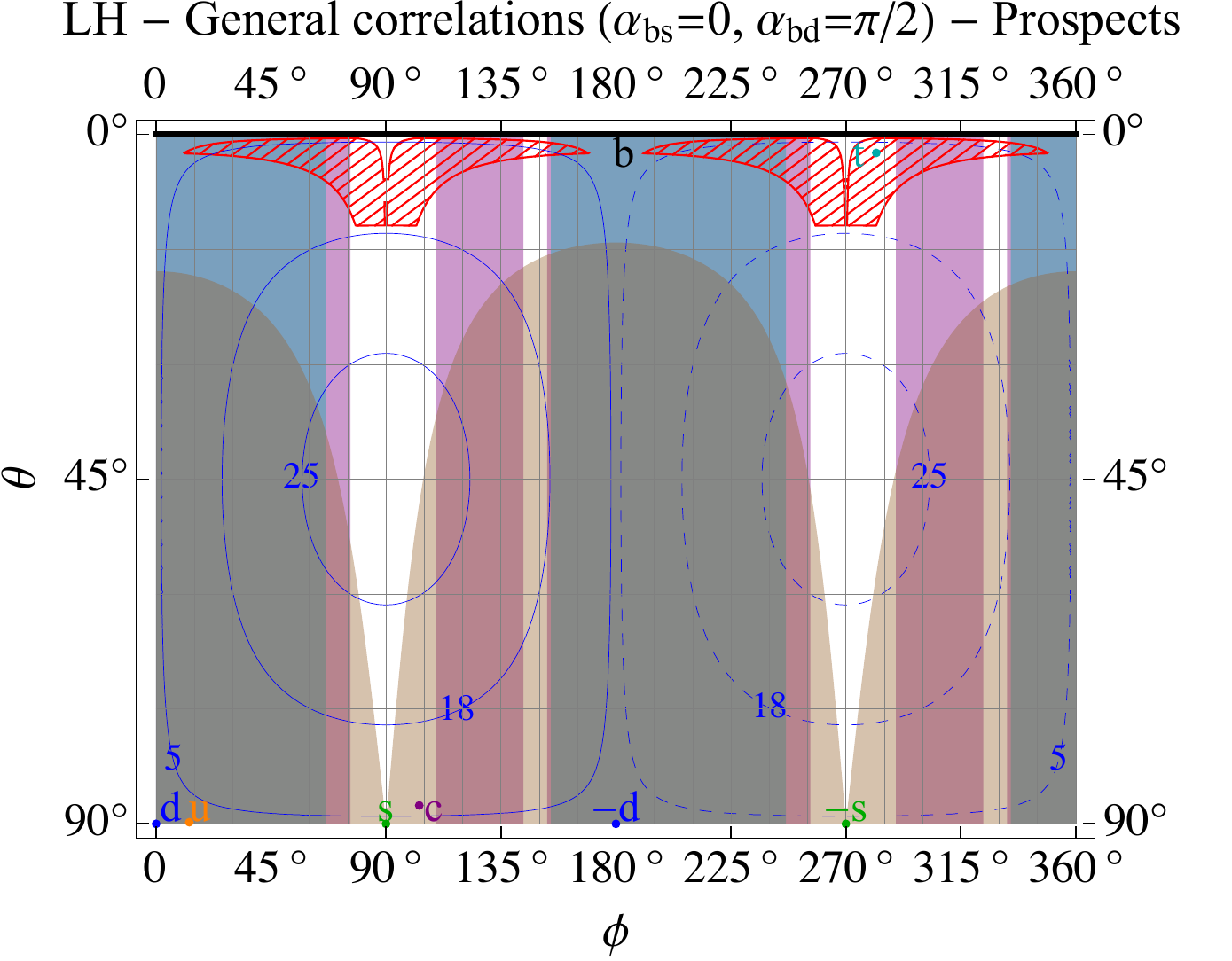}\\
 \includegraphics[scale=0.5]{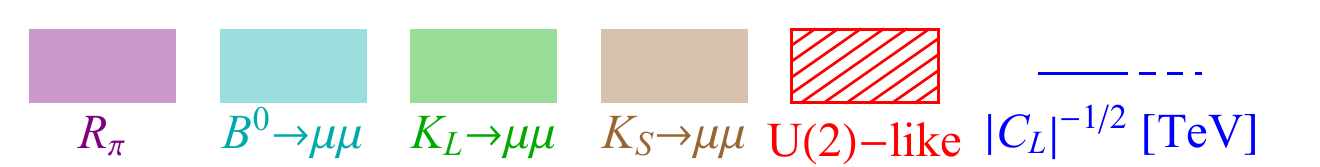} \caption{\label{fig:ROFV-Minimal_prospect}Future prospects for the exclusion
limits in the plane $(\phi,\theta)$ for two choices of the phases
$\alpha_{bs}$ and $\alpha_{bd}$ from observables with direct correlation
with $R_{K^{(*)}}$. For $K_{L}\to\mu\mu$ we use the present bound.}
\end{figure}

Future measurements by LHCb, Belle-II, and other experiments are expected
to improve substantially the precision of most of the observables
studied in the present Chapter. We collect in Table~\ref{tab:Prospects}
the relevant prospects.

First of all, the anomalous observables themselves, $R_{K}$ and $R_{K^{*}}$,
are expected to be tested with sub-percent accuracy by LHCb with 300fb$^{-1}$
of luminosity. Furthermore, a larger set of observables sensitive
to the same partonic transition $b\to s\mu^{+}\mu^{-}$ will be measured
(such as $R_{\phi}$, $R_{pK}$ and $Q_{5}$ for example \cite{Bediaga:2018lhg}).
This will allow to confirm or disprove the present anomalies and to
pinpoint the size of the New Physics contribution with high accuracy.

The leptonic decays $B_{(d,s)}^{0}\to\mu^{+}\mu^{-}$ will be crucial
for discriminating between the $\mathcal{O}_{9}$ and $\mathcal{O}_{L}$
scenarios. As to the $B\to\pi\ell^{+}\ell^{-}$ channels, we note
that the power of the muon-specific $\text{Br}(B^{+}\to\pi^{+}\mu^{+}\mu^{-})$
as a probe of NP is, already at present, limited by theoretical uncertainties
\cite{Khodjamirian:2017fxg}. The situation improves substantially
for the LFU ratio $R_{\pi}$ (cf. Eq. (\ref{eq:ROFV-LFUratios})),
for which, as already noted, $\text{U}(2)^{5}$ flavour symmetry predicts
$R_{\pi}=R_{K}$ and for which LHCb is expected to reach a $\sim4.7\%$
sensitivity with 300 fb$^{-1}$ of luminosity \cite{Bediaga:2018lhg}.
As can be seen in Fig.~\ref{fig:ROFV-Minimal_prospect}, these channels
will be able to cover almost the complete parameter space of the setup
studied here, particularly if the phase $\alpha_{bd}$ is small.
\begin{figure}[t]
\centering \includegraphics[scale=0.5]{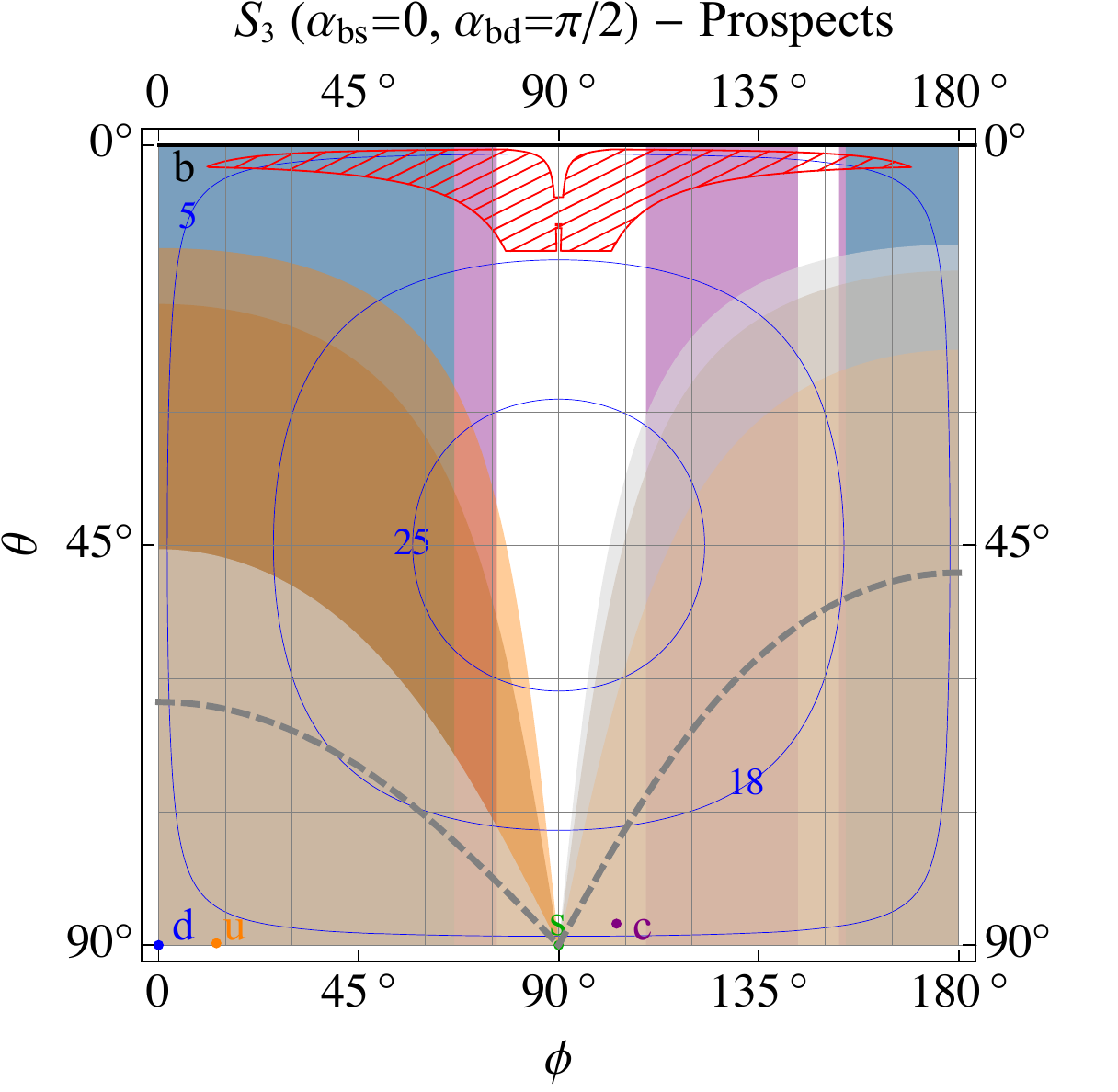}\!\!\!\!\!
\includegraphics[scale=0.5]{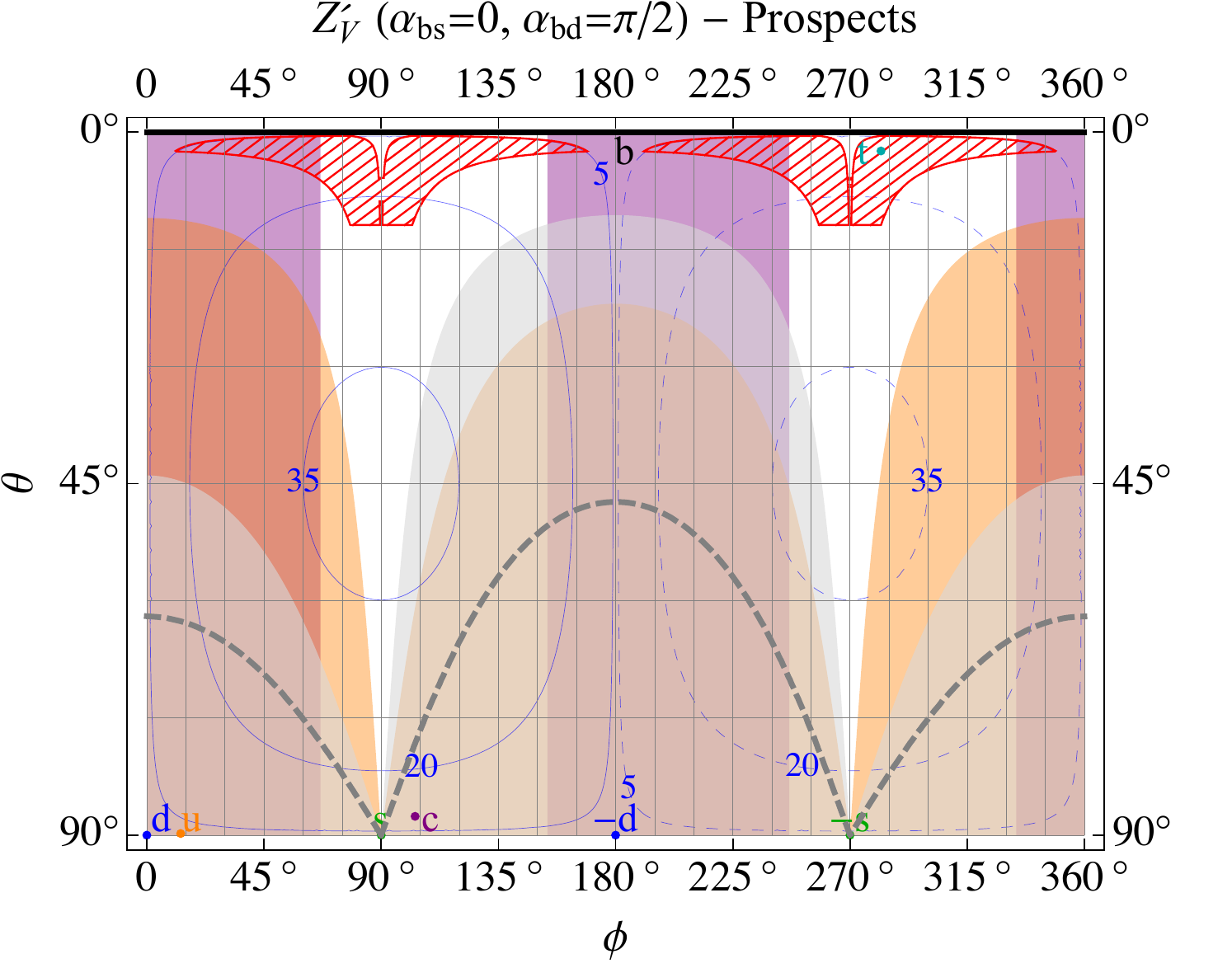}\\
 \includegraphics[scale=0.5]{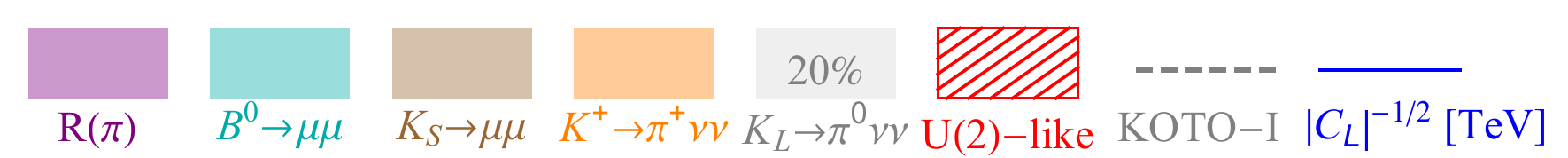} \caption{Future prospects for the exclusion limits in the $(\phi,\theta)$
plane for the scalar leptoquark $S_{3}$ (left) and the vector singlet
$Z^{\prime}$ with vector-like couplings to muons (right), for a choice
of phases. The orange and gray regions correspond to the future expected
limits from $K^{+}\to\pi^{+}\nu\nu$ (NA62 \cite{Ruggiero:2017hjh})
and $K_{L}\to\pi^{0}\nu\nu$ (KLEVER \cite{Ambrosino:2019qvz} or
KOTO phase-II), respectively. The dashed gray line corresponds to
the expected limit on $K_{L}\to\pi^{0}\nu\nu$ from KOTO phase-I.}
\label{fig:prospectsKpinunu}
\end{figure}

In all cases where $C_{S}\neq C_{T}$, such as in the $S_{3}$ and
$Z^{\prime}$ models, other relevant channels which will improve substantially
in sensitivity are $\text{Br}(K^{+}\to\pi^{+}\bar{\nu}\nu)$ and $\text{Br}(K_{L}\to\pi^{0}\bar{\nu}\nu)$.
The former is expected to be measured with a 10\% accuracy by NA62
\cite{Ruggiero:2017hjh} in the next few years, while, for the latter,
the KOTO experiment at JPARC \cite{Ahn:2018mvc} should reach a single-event
sensitivity at the level of the SM branching ratio, with a signal
to background ratio $\sim1$, which translates to a projected $95\%$CL
limit of $\sim5.4$ times the SM value, i.e. $\sim1.8\times10^{-10}$.
A possible future upgrade of the whole KOTO experiment (stage-II)\footnote{From a CERN EP seminar given in February 2019 (slides at \url{https://indico.cern.ch/event/799787/attachments/1797668/2939627/EPSeminar_YuChen.pdf})
and of a talk presented at the Rencontres de Moriond 2019 (slides
at \url{http://moriond.in2p3.fr/2019/EW/slides/1_Sunday/1_morning/5_Nanjo.pdf}).}, or the proposed KLEVER experiment at CERN SPS \cite{Ambrosino:2019qvz},
could both reach a $\sim20\%$ sensitivity of the $\text{Br}(K_{L}\to\pi^{0}\bar{\nu}\nu)$
SM value. An example for the prospects due to these observables for
a particular choice of phases in the two simplified models are shown
in Fig.~\ref{fig:prospectsKpinunu}.

\section{Conclusions}

If the flavor anomalies in $b\to s\mu\mu$ transitions are experimentally
confirmed, they will provide important information about the flavor
structure of the underlying New Physics. The latter can be tested
by studying possible correlations with other measurements in flavor
physics. In this Chapter we assumed that the putative NP, responsible
for the anomalous effects, couples to SM left-handed down quarks in
such a way to generate a rank-one structure in the novel flavor-violating
sector. We dub such a scenario \emph{Rank-One Flavor Violation} (ROFV).
Such a structure can result from a number of well motivated UV completions
for the explanation of the flavor anomalies, in which a single linear
combination of SM quark doublets couples to the relevant NP sector.
This automatically includes all single-leptoquark models, and models
where LH quarks mix with a single vector-like fermion partner. As
these examples reveal, the ROFV condition might not originate from
symmetry but rather as a feature of the UV dynamics.\par Varying
the direction associated to the NP ($\hat{n}$) in $\text{U}(3)_{q}$
flavor space, we identified the most important observables that can
be correlated to the flavor anomalies. The more model-independent
correlations are with $d_{i}\to d_{j}\mu\mu$ transitions (and their
crossed symmetric processes). A large part of the parameter space
is probed by the measurement of the branching ratio of $B^{+}\to\pi^{+}\mu\mu$.
While the sensitivity to NP effects in this channel is limited by
the large hadronic uncertainty of the SM prediction, future measurements
of the theoretically clean ratio $R_{\pi}$ are going to provide further
information on $b\to d$ flavor violations. Among the transitions
involving the first two generations of quarks ($s\to d$), the $K_{L}\to\mu\mu$
decay rate has a major impact and it is particularly sensitive to
the phases of our parametrization. Unfortunately, future prospects
in this channel are limited by theory uncertainties in the SM prediction
of the long-distance contribution to the decay. A sizeable improvement
by LHCb is instead expected in the limit on the $K_{S}\to\mu\mu$
decay rate.\par While the former conclusions rely only on our rank-one
hypothesis, more model dependent correlations can be established once
the relevant effective operators are embedded into the SMEFT or in
the presence of specific mediators. An example of such correlations
is given by $d_{i}\to d_{j}\nu\nu$ processes, and we have in fact
shown that present data from $K^{+}\to\pi^{+}\nu\nu$ are particularly
relevant to the leptoquark $S_{3}$ or vector $Z'$ simplified models.\par From
a more theoretical point of view, we investigated whether the flavor
violation associated to the NP can be connected to the one present
in the SM Yukawa sector. A generic expectation is that the leading
source of $\text{U}(3)_{q}$ breaking in the NP couplings is provided
by a direction in flavor space close to the one identified by the
top quark. Indeed, we showed in a concrete example based on a flavor
symmetry that the vector $\hat{n}$ turns out to be correlated to
the third line of the CKM matrix, as in Eq.~(\ref{eq:TopLikeDirection}).
Remarkably, a large portion of the theoretically favoured region (red
meshed lines region in our plots) survive the bounds from current
flavor physics data. Our order of magnitude predictions can be narrowed
down under further theoretical assumptions. For example a minimally
broken $\text{U}(2)^{5}$ flavor symmetry predicts $R_{K}=R_{\pi}$
and the ratio $\text{Br}(B_{s}\to\mu\mu)/\text{Br}(B_{d}\to\mu\mu)$
to be SM-like (up to small corrections of a few percent).\par In
our last section we explored future prospects for the exclusion limits
in the ROFV framework. In the near future a series of experiments
will be able to cover almost all of the parameter space identified
by our ansatz. Indeed, in the next few years, significant information
will be provided by the NA62 and KOTO experiments, thanks to precise
measurements of the $K^{+}\to\pi^{+}\nu\nu$ and $K_{L}\to\pi^{0}\nu\nu$
decays, while on a longer time scale results from LHCb and Belle II
will almost completely cover our parameter space (and test the minimally
broken $\text{U}(2)^{5}$ model).\par A confirmation of the $B$-meson
anomalies would open a new era in high energy physics. In this enticing
scenario, studying correlations of the anomalies to other observables
would provide a powerful mean to investigate the flavor structure
of New Physics.

\part{Explaining the SM flavour structure}

\chapter{\label{chap:Explaining-the-SM}The SM flavour puzzle}

Providing a theoretical explanation for the peculiarities observed
in the SM fermion mass spectra and mixing patterns is arguably one
of the greatest challenges of modern Particle Physics. What determines
the mass hierarchies between the three families of charged fermions
(quarks and charged leptons)? What is the nature of neutrino masses,
and why are they so small? How do the quark and lepton mixing patterns,
embodied by the CKM and PMNS matrices respectively, originate? These
and many other questions constitute the so-called SM flavour puzzle.
This introductory Chapter provides an overview of the various subproblems
of which the SM flavour puzzle is composed.

\section{\label{sec:The-SM-flavour-puzzle}Pieces of the puzzle}

The SM as it stands has a fairly high number of tunable inputs, both
at the discrete and continuous levels. Although many of these are
univocally fixed by experiment\footnote{A well recognized exception is the QCD $\theta$ angle, for which
only an upper bound exists \cite{Kim:2008hd}.}, from a BSM perspective one has the hope that, at a sufficiently
high scale, the relevant SM UV completion will provide theoretical
explanations for at least some of the patterns observed at low energies.
This kind of theoretical prejudice, when applied to the SM fermionic
matter content, motivates what is known as the SM flavour puzzle \cite{Feruglio:2015jfa}.

At the discrete level, SM fermions are characterized by their representation
under the SM gauge group, which is a direct sum of fifteen irreducible
representations. Out of these, there are five inequivalent representations,
$q$, $u$, $d$, $\ell$ and $e$ (cf. Table \ref{tab:SM fields}),
each of which is repeated exactly three times. For this reason, the
SM fermionic matter content is usually depicted as being constituted
by a certain number\emph{ }(which happens to be three) of repetitions
of a single reducible representation, which is referred to as a ``family'',
or ``generation''.

At the continuous level, flavour gets its shape from the Yukawa part
of the SM lagrangian:
\begin{equation}
\mathcal{L}_{\text{SM}}^{\text{yuk}}=-(y_{U})_{ij}\overline{q}_{i}\widetilde{H}u_{j}-(y_{D})_{ij}\overline{q}_{i}Hd_{j}-(y_{E})_{\alpha\beta}\overline{\ell}_{\alpha}He_{\beta}+\text{h.c.},\label{eq:SMFLAV-SM Yukawa}
\end{equation}
which gives rise to quark masses and mixings, and lepton masses (such
a description is clearly insufficient for any phenomenological discussion
of lepton flavour, as discussed below). The following two Subsections
separetely describe the flavour structures of the quark and lepton
sectors.

\subsection{\label{subsec:SMFLAV-Quark Sector}Quark sector}

Following the conventions of Sec. \ref{sec:The-SM-flavor structure},
without loss of generality, we can use the so-called ``down-quark
mass basis'' of SM quark fields, in which (cf. Eq. (\ref{eq:y_UD in down basis})):

\begin{equation}
y_{D}=\frac{1}{v}\text{diag}(m_{d},\,m_{s},\,m_{b}),\qquad y_{U}=\frac{1}{v}V^{\dagger}\text{diag}(m_{u},\,m_{c},\,m_{t}).\label{eq:SMFLAV-y_UD in down basis}
\end{equation}
Here $m_{\bullet}$ are the quark running masses at the electroweak
scale $v$, and $V$ is the CKM matrix in the standard phase convention.

All quark masses but the top's one, $m_{t}\sim v$, are well below
the electroweak scale. In terms of quark-Higgs couplings, we have
$y_{t}\sim\mathcal{O}(1)$, whereas the remaining couplings are all
suppressed by several orders of magnitudes. Some approximate values
for $m_{\bullet}/m_{t}$ (or equivalently $y_{\bullet}/y_{t}$) are
\cite{Zyla:2020zbs}:
\begin{align}
\frac{1}{m_{t}}\left(m_{d},\,m_{s},\,m_{b}\right) & \approx\left(2.9\times10^{-5},\,5.8\times10^{-4},\,2.6\times10^{-2}\right),\label{eq:Y_D / yt}\\
\frac{1}{m_{t}}\left(m_{u},\,m_{c},\,m_{t}\right) & \approx\left(1.3\times10^{-5},\,7.8\times10^{-3},\,1\right).\label{eq:Y_U / yt}
\end{align}
Both down and up quark mass spectra are hierarchical, with a further
overall suppression of third and second generation down quark masses
relative to up quark ones.

Intriguing patterns are also observed in the quark mixing matrix $V$.
These are most transparently illustrated in the Wolfenstein parametrization
of the CKM matrix \cite{Zyla:2020zbs}, \emph{viz.}:
\begin{equation}
V=\begin{pmatrix}1-\frac{\lambda^{2}}{2} & \lambda & A\lambda^{3}(\rho-i\eta)\\
-\lambda & 1-\frac{\lambda^{2}}{2} & A\lambda^{2}\\
A\lambda^{3}(1-\rho-i\eta) & -A\lambda^{2} & 1
\end{pmatrix}+O(\lambda^{4}),\label{eq:SMFLAV-Wolfenstein}
\end{equation}
where the Wolfenstein parameters are defined by:
\begin{align}
\lambda & \equiv\frac{\left|V_{us}\right|}{\sqrt{\left|V_{ud}\right|^{2}+\left|V_{us}\right|^{2}}}\approx0.22,\label{eq:SMFLAV-Wolfenstein lambda}\\
A & \equiv\frac{\left|V_{cb}\right|}{\lambda\left|V_{us}\right|}\approx0.83,\label{eq:SMFLAV-Wolfenstein A}\\
\rho+i\eta & \equiv\frac{V_{ub}^{*}}{A\lambda^{3}}\approx0.15+0.36i.\label{eq:SMFLAV-Wolfenstein rho+ieta}
\end{align}
It is seen that non-diagonal entries, which govern the strength of
$u_{i}\to W^{+}d_{j}$ and $d_{j}\to W^{-}u_{i}$ transitions between
different quark generations, are suppressed by powers of $\lambda$
as follows:
\begin{equation}
1\leftrightarrow2\sim\lambda,\qquad2\leftrightarrow3\sim\lambda^{2},\qquad3\leftrightarrow1\sim\lambda^{3}.\label{eq:SMFLAV-Wolfenstein hierarchies}
\end{equation}
Notice that $\lambda$ can be identified with the Cabibbo sine, for
which the well-known numerical relationship \cite{Gatto:1968ss}:
\begin{equation}
\lambda\approx\sqrt{\frac{m_{d}}{m_{s}}},\label{eq:Cabibbo Sine and numerical relationship}
\end{equation}
holds at the percent level.

To conclude, we remark that the quark flavour structure is (at least
phenomenologically) fully accounted by the SM lagrangian (cf. Eq.
(\ref{eq:SMFLAV-SM Yukawa})), and that quark running masses and CKM
matrix elements can be either computed or experimentally determined
with relatively high accuracy, so that the corresponding uncertainties
often play little role in BSM discussions addressing the origins of
quark flavour. We now turn to lepton flavour, which presents important
qualitative differences with respect to the quark case discussed so
far.

\subsection{\label{subsec:Lepton-sector}Lepton sector}

In contrast to quark flavour, lepton flavour is in large part a BSM
issue. On the one hand, the SM can account, through its Yukawa couplings
(\ref{eq:SMFLAV-SM Yukawa}), for massive charged lepton and massless,
left-handed neutrinos, whose mass and flavour eigenstates can be made
to coincide. On the other one, it is an experimental fact that neutrinos
have non-zero masses, and that their mass eigenstates are linear combinations
of flavour eigenstates. Therefore, in order to discuss the experimental
data on neutrino flavour, we need to move a step beyond the pure SM
description and take a phenomenological approach.

Neutrino oscillation experiments measure neutrino squared mass (absolute)
differences and mixing angles. Most of the experimental results can
be coherently interpreted in the context of three light active neutrinos
and CPT invariance \cite{Feruglio:2019ktm}, which we assume in what
follows. Two standard independent mass differences are:
\begin{align}
\delta m^{2} & =m_{2}^{2}-m_{1}^{2},\label{eq:SMFLAV-dm=0000B2}\\
\Delta m^{2} & =m_{3}^{2}-\frac{m_{1}^{2}+m_{2}^{2}}{2},\label{eq:SMFLAV-Dm=0000B2}
\end{align}
where $m_{1-3}$ are neutrino masses in the usual labeling conventions
\cite{Zyla:2020zbs}\footnote{In the standard convention, $m_{1}$ and $m_{2}$ are defined to be
the mass eigenstates with smallest mass difference $\left|m_{2}-m_{1}\right|$,
with $m_{1}<m_{2}$. Labeling the remaining eigenstate as $m_{3}$,
there are two possibilities: either $m_{1}<m_{2}<m_{3}$ or $m_{3}<m_{1}<m_{2}$,
which correspond to a normal and inverted ordered mass spectrum, respectively.}; notice that $\Delta m^{2}$ is positive (negative) for a mass spectrum
with normal (inverted) ordering. The measured mixing angles can be
explained in terms of a $3\times3$ mixing matrix intervening in $l_{i}^{-}W^{+}\to\nu_{j}$,
the so called PMNS matrix $U$, defined by:
\begin{equation}
\mathcal{L}_{\text{CC}}=-\frac{g}{\sqrt{2}}W_{\rho}^{+}\left(\overline{e_{L}},\,\overline{\mu_{L}},\,\overline{\tau_{L}}\right)\gamma^{\rho}U\left(\nu_{1},\,\nu_{2},\,\nu_{3}\right)^{T}.\label{eq:SMFLAV-defining PMNS}
\end{equation}
The standard parametrization of $U$, \emph{viz.}:
\begin{equation}
U=\begin{pmatrix}1 & 0 & 0\\
0 & c_{23} & s_{23}\\
0 & -s_{23} & c_{23}
\end{pmatrix}\begin{pmatrix}c_{13} & 0 & s_{13}e^{-i\delta}\\
0 & 1 & 0\\
-s_{13}e^{i\delta} & -s_{23} & c_{13}
\end{pmatrix}\begin{pmatrix}c_{12} & s_{12} & 0\\
-s_{12} & c_{12} & 0\\
0 & 0 & 1
\end{pmatrix}\begin{pmatrix}e^{i\eta_{1}} & 0 & 0\\
0 & e^{i\eta_{2}} & 0\\
0 & 0 & 1
\end{pmatrix},\label{eq:SMFLAV-parametrization PMNS}
\end{equation}
is defined in terms of three mixing angles $\theta_{12}$, $\theta_{13}$
and $\theta_{23}$, a ``Dirac'' phase $\delta$ and two ``Majorana''
phases $\eta_{1,2}$. Majorana phases are well-defined only if neutrinos
are Majorana particles \cite{Zyla:2020zbs}, since otherwise these
phases can always be reabsorbed in the definition of the mass eigenstate
fields; moreover, neutrino oscillation experiments are insensitive
to $\eta_{1,2}$ \cite{Giunti:2010ec}.

\begin{table}[t]
\centering{}%
\begin{tabular}{|c|cc|}
\hline 
\texttt{Observable} & \multicolumn{2}{c|}{\texttt{Best-fit value with $1\sigma$ error}}\tabularnewline
\hline 
\hline 
$m_{e}/m_{\mu}$ & \multicolumn{2}{c|}{$0.0048_{-0.0002}^{+0.0002}$}\tabularnewline
$m_{\mu}/m_{\tau}$ & \multicolumn{2}{c|}{$0.0565_{-0.0045}^{+0.0045}$}\tabularnewline
\hline 
 & NO & IO\tabularnewline
$\delta m^{2}\left[10^{-5}~\text{eV}^{2}\right]$ & \multicolumn{2}{c|}{$7.42_{-0.20}^{+0.21}$}\tabularnewline
$\Delta m^{2}\left[10^{-3}~\text{eV}^{2}\right]$ & $2.480_{-0.028}^{+0.026}$ & $-2.461_{-0.028}^{+0.028}$\tabularnewline
$r\equiv\delta m^{2}/|\Delta m^{2}|$ & $0.0299_{-0.0009}^{+0.0009}$ & $0.0301_{-0.0009}^{+0.0009}$\tabularnewline
$\sin^{2}\theta_{12}$ & $0.304_{-0.012}^{+0.012}$ & $0.304_{-0.012}^{+0.013}$\tabularnewline
$\sin^{2}\theta_{13}$ & $0.02219_{-0.00063}^{+0.00062}$ & $0.02238_{-0.00062}^{+0.00063}$\tabularnewline
$\sin^{2}\theta_{23}$ & $0.573_{-0.020}^{+0.016}$ & $0.575_{-0.019}^{+0.016}$\tabularnewline
$\delta/\pi$ & $1.09_{-0.13}^{+0.15}$ & $1.57_{-0.17}^{+0.14}$\tabularnewline
\hline 
\end{tabular}\caption{\label{tab:FLAVOR-NUFit}Best-fit values of the charged lepton mass
ratios and the neutrino oscillation parameters with the corresponding
1$\sigma$ errors. For the charged lepton mass ratios we have used
the values given in ref.~\cite{Ross:2007az}, averaged over $\tan\beta$
as described in the text, whereas for the neutrino parameters we have
used the results obtained in refs.~\cite{Esteban:2020cvm,NuFITv50}
(with Super-Kamiokande atmospheric data).}
\end{table}

Currently, the mass differences $\delta m^{2}$ and $\Delta m^{2}$,
and the three mixing angles $\theta_{12}$, $\theta_{13}$ and $\theta_{23}$
have been measured with percent-level accuracy, whereas there is still
large ($\mathcal{O}(10\%-15\%)$) uncertainty on the Dirac phase $\delta$.
We collect these observables in the lower part of Tab. \ref{tab:FLAVOR-NUFit}
for the two ordering cases. The upper part of the Table reports the
ratios of running lepton masses at the GUT scale\footnote{We remark that such ratios have very weak scale dependencies, so that
the precise scale is largely irrelevant for the purpose of our discussion.
We follow Ref. \cite{Feruglio:2017spp} and employ the updated GUT
scale determination \cite{Ross:2007az} for these quantities.}. A number of lepton flavour observables is still unknown: 
\begin{itemize}
\item The neutrino mass ordering (normal vs. inverted). A preference (currently
at the level of about $3\sigma$) for a normal mass ordering is starting
to emerge from data \cite{Feruglio:2019ktm}, but is still far from
conclusive evidence.
\item The absolute scale of neutrino masses. Since neutrino oscillation
experiments only measure squared mass differences, they give no direct
information on the absolute values of neutrino masses. At present,
the thightest upper bound on the sum of neutrino masses is the one
from cosmology \cite{Zyla:2020zbs}: 
\begin{equation}
\sum_{i}m_{i}<0.12\div0.68\,\text{eV}\label{eq:SMFLAV-Cosmology bound neutrino}
\end{equation}
\item The Majorana phases, if neutrinos are Majorana particles.
\end{itemize}
The picture which emerges from Tab. \ref{tab:FLAVOR-NUFit} is quite
distinct from the one found in the quark sector. While charged lepton
masses are hierarchical as the quark ones (and indeed comparable in
size with those of down quarks), it is not clear whether the same
holds for the neutrino spectrum, and, most importantly, neutrino masses
are several order of magnitudes smaller than the ones of all other
SM fermions. Also, the PMNS mixing matrix is radically different from
its quark counterpart: while the CKM matrix is close to diagonal,
with mixing angles $\theta_{12}^{\text{CKM}},\,\theta_{23}^{\text{CKM}}$
and $\theta_{13}^{\text{CKM}}$ suppressed by $\lambda$, $\lambda^{2}$
and $\lambda^{3}$ respectively (cf. Eq. (\ref{eq:SMFLAV-Wolfenstein})),
two of the three PMNS mixing angles are seen to be $\mathcal{O}(1)$.
All these features strongly suggest that the mechanisms behind the
generation of lepton and, in particular, neutrino masses and mixings
are quite different from the corresponding ones underlying the quark
sector.

To conclude (compare with the conclusions of Subsec. \ref{subsec:SMFLAV-Quark Sector}),
we remark that our low-energy description of lepton flavour is necessarily
of phenomenological kind, and the simple framework of three light
active neutrinos with PMNS mixing provides a perfectly valid explanation
of experimental data. Such a description has some important unknowns,
most prominently the neutrino mass ordering, the absolute scale of
neutrino masses and the two Majorana phases if neutrinos are Majorana
particles.

\section{\label{sec:SMFLAV-Flavor-from-symmetries}Theoretical ideas}

It is a widespread position that the special features of the SM fermionic
spectrum discussed in the previous Section strongly suggests the existence
of some non-trivial high-energy mechanism, or mechanisms, from which
they would result in a natural way. This is, indeed, the essence of
the SM flavour puzzle, which, according to our previous description,
can be decomposed into a series of subproblems:
\begin{enumerate}
\item \emph{Explaining the SM fermionic representation}. This includes the
basic representation (\emph{i.e. $q\oplus u\oplus d\oplus\ell\oplus e$)}
of a single SM family, and the actual number (three) of replicæ of
this representation.
\item \emph{Explaining quark masses and mixings. }What especially calls
for a theoretical explanation are the hierarchies in down and up quark
mass spectra, the overall down \emph{vs.} up mass suppression, and
the suppression and hierarchies in CKM matrix off-diagonal terms.
Moreover, a natural question is whether (some of) these features are
interconnected (see \emph{e.g. }Eq. (\ref{eq:Cabibbo Sine and numerical relationship})).
\item \emph{Explaining lepton masses and mixings. }Here, the most salient
features are the hierarchies in the charged lepton mass spectrum,
the huge suppression of neutrino masses and the structure of the PMNS
matrix, with two large and one small mixing angles. There is also
a possible hint to a hierarchy in the solar-to-atmospheric ratio $r\equiv\delta m^{2}/\Delta m^{2}\approx0.03$
\footnote{One should however realize that such a ratio is, in some sense, a
small quantity by construction. For instance, performing the simple
experiment of generating three neutrino masses uniformly at random
in a fixed interval (say $\left[0,\,1\right]$), I find that $\approx25\%$
of the times $r<0.1$; with a positive gaussian and an exponential
distribution, instead, the first quantiles of the $r$ distribution
are found for $r\lesssim0.07$ and $r\lesssim0.05$, respectively.
Clearly, these numbers indicate that the parameter space region producing
a solar-to-atmospheric ratio as small as its actual value might not
be small at all, depending on the particular measure adopted (cf.
also the remarks at the end of the next Section).}.
\end{enumerate}
All these questions have inspired for a long time many theoretical
investigations. Without any attempt at completeness (for a more comprehensive
review, see \emph{e.g. }\cite{Feruglio:2019ktm} and refereinces therein),
let us cite some of the most prominent ideas which have been put forward
on the various fronts.

\medskip{}

\textbf{Grand Unification.} \cite{Langacker:1980js} One of the main
appeals of Grand Unified Theories (GUTs) is their potential to explain
the SM basic representation in terms of a few (ideally a single one)
irreducible representations of the unification group. To cite a classical
example, the SM basic representation can be embedded into a single
spinorial representation of $\text{SO}(10)$: the latter decomposes
under the electroweak group into the five irreducible representations
of quarks and leptons, with an additional SM singlet, which can be
identified with a singlet ``sterile'' neutrino. As the latter can
in principle take part in the see-saw mechanism for generating neutrino
masses (see below), this example also shows how the solutions of different
parts of the flavour puzzle may well be interconnected.

\textbf{Flavour symmetries.} The theoretical reach of flavour symmetries
spans from the explanation of the horizontal structure of the SM fermion
content (\emph{i.e. }the existence of three identical generations),
to the prediction of mass and mixing hierarchies. For example, in
the Froggatt-Nielsen mechanism \cite{Froggatt:1978nt}, the hierarchies
in quark masses and mixings originate from the soft spontaneous breaking
of a $\text{U}(1)_{\text{FN}}$ global symmetry, under which quark
fields are charged. Also discrete flavour symmetries have been used
extensively to address the SM flavour structures, in particular in
the lepton sector, since the PMNS mixing structure is more straight-forwardly
reproduced (or approximated) within this context \cite{Feruglio:2015jfa}.

\textbf{Accidental symmetries.} These class of symmetries is, conceptually,
very different from the postulated flavour symmetries discussed in
the previous point. Accidental symmetries are usually enforced by
structural requirements such as gauge invariance and renormalizability.
A particularly relevant example is provided by the $\text{U}(1)_{\ell}$
global symmetry of the renormalizable SM, whose conserved charge is
total lepton number. This symmetry implies, in particular, that SM
neutrinos are massless at the renormalizable level, which is clearly
a good leading order approximation; from a SMEFT point of view, neutrino
mass generation is post-poned to dimension five, through the Weinberg
operator (\ref{eq:Weinberg operator}), which can be originated at
tree-level in models realizing one of the three types of See-Saw mechanisms\cite{Mohapatra:2004zh}\footnote{A simple minded SM extension with a (type I) See-Saw mechanism consists
of the SM itself with the addition of three singlet right-handed fermions
$N_{1,2,3}$ with a large Majorana mass term $M_{\alpha\beta}$. The
relevant terms in the model's lagrangian are:
\[
\mathcal{L}\supset-(y_{N})_{\alpha\beta}\overline{\ell_{\alpha}}\widetilde{H}N_{\beta}-\frac{M_{\alpha\beta}}{2}\overline{N_{\alpha}^{c}}N_{\beta},
\]
which, after integrating out the heavy singlets $N$, give rise to
the Weinberg operator with Wilson coefficient $C\sim\mathcal{O}(y_{N}^{2}v^{2}/M)$.
If $y_{N}$ has $\mathcal{O}(1)$ singular values, the resulting neutrino
masses turn out to be in the $\text{meV}\div\text{eV}$ range if the
$M$ eigenvalues lie around the GUT scale. Unfortunately, the singlets'
Majorana mass term also has the side effect of destabilizing the electroweak
scale, making the model incur in the (in)famous hierarchy problem
\cite{Farina:2013mla}. }. The general approach to fermion masses and mixings using accidental
symmetries\emph{ }was originally proposed in Ref. \cite{Ferretti:2006df}.

\textbf{Special field theory frameworks.} Some of the features of
the SM flavour structure might be enforced by a structural property
of a class of SM extensions. For example, in (type II) Two Higgs Doublet
Models (and, in particular, in the Supersymmetric SM), the large $m_{t}/m_{b}$
ratio can be ``explained'' by a large $v_{u}/v_{d}$ ratio (\emph{i.e.
}a large $\tan\beta$), where $v_{u,d}$ are the electroweak breaking
expectation values of the up and down Higgses $H_{u,d}$, respectively.
Even though at a first glance this might appear to be a vacuous explanation
(after all, we exchange a large ratio, $m_{t}/m_{b}$, for a similarly
large one, $v_{u}/v_{d}$), notice that we have managed to transform
the $m_{t}/m_{b}$ problem into a dynamical one, since the Higgs expectation
values are now controlled by the model's scalar potential. Another
special framework which can shed light on the $m_{t}/m_{b}$ ratio
is that of partial compositeness \cite{Panico:2015jxa}, if the $t$-quark
has a significantly higher degree of compositeness than the other
quarks.

\textbf{Anarchy. }Completely at odds with the approaches mentioned
so far, one could argue that there is, in fact, no organizing principle
underlying the SM flavour structure. This idea has been formalized
in the context of neutrino mixing using a probabilistic/measure theoretic
approach \cite{Hall:1999sn,Haba:2000be,deGouvea:2003xe,Espinosa:2003qz,deGouvea:2012ac}.
Even though such a position can be under many respects less appealing
than the more theoretically driven approaches previously mentioned,
one must concede that the lack of strong hierarchies in the PMNS mixing
matrix (except perhaps for the small $\theta_{13}$ angle) and, possibly,
in the neutrino mass spectrum, provides a logical motivation for it
(on the other hand, I find the same position hard to mantain when
applied as a whole to the SM flavour puzzle).

\medskip{}

The common denominator to all the frameworks mentioned in the previous
Section (including, to a certain extent, also Anarchy) is \emph{Naturalness,
}in the sense that for the models in question there exists a reasonably
large parameter space region for which the special (discrete or continuous)
phenomenological features of SM flavour are realized. Furthermore,
the examples also illustrate how the SM flavour puzzle is entangled
with other theoretical enigmas surrounding the SM, gauge coupling
unification and the Higgs mass hierarchy problem being just a few,
to a degree which depends on the particular theoretical framework.
A solution to the former might also shed light on some of these further
fundamental questions.

\chapter{\label{Chap:Are-neutrino-masses}Are lepton masses modular forms?}

Modular invariance has recently emerged in the context of neutrino
physics \cite{Feruglio:2017spp}\footnote{The title of this Chapter is a generalization to the question ``Are
neutrino masses modular forms?'', posed by the seminal paper \cite{Feruglio:2017spp}.}, as a powerful generalization of discrete linear symmetry. In supersymmetric
modular invariant models, the action is invariant under a non-linear
realization of the $\text{SL}(2,\mathbb{Z})$ group, which is inevitably
broken by the vacuum expectation value of a superfield, the \emph{modulus}
$\tau$. The latter is acted on by the modular group by linear fractional
transformations:
\begin{equation}
\tau\mapsto\frac{a\tau+b}{c\tau+d}\qquad(a,\,b,\,c,\,d\in\mathbb{Z},\qquad ad-bc=1),\label{eq:MODULAR-Linear fractional transformation}
\end{equation}
and, by modular invariance, couplings of chiral fields (which transform
non-linearly under $\text{SL}(2,\mathbb{Z})$) must be functions of
the modulus $\tau$ with well-defined transformation properties under
the modular group. 

Specifically, supersymmetry forces the (holomorphic) superpotential
couplings to be modular forms of $\text{SL}(2,\mathbb{Z})$, which,
at a fixed level $N$ of the principal congruence subgroup employed
in the model's construction, form a linear space of finite dimension.
This fact, combined with superpotential non-renormalization theorems
\cite{Weinberg:1998uv}, highly constrains the superpotential part
of the supersymmetric action, which is bound to depend upon a reduced
set of couplings, in addition to the complex expectation value of
$\tau$. Such a strict determination significantly enhances the predictive
power of this class of models, which goes without saying it is of
the greatest importance for the task of explaining the SM flavour
structure.

This Chapter presents results from my work \cite{Feruglio:2021dte}
with F. Feruglio, A. Romanino and A. Titov, in which we attempt to
address charged lepton mass hierarchies through modular invariance.
We propose that the hierarchies might originate from a residual $\mathbb{Z}_{4}$
symmetry subgroup of the full modular group, which is exact at the
symmetric point $\tau=i$, and whose breaking is governed by the (assumed
small) departure $\tau-i$. In fact, earlier model building attempts
\cite{Feruglio:2017spp,Kobayashi:2018scp,Criado:2018thu,Novichkov:2018ovf,Novichkov:2019sqv,Okada:2019uoy,Criado:2019tzk,Kobayashi:2019gtp,Okada:2020rjb,Novichkov:2020eep,Wang:2020dbp,Okada:2020brs}
have shown that, for several modular invariant models of lepton flavour,
a good agreement with experimental data is obtained in the neighbourood
of $\tau\approx i$, with deviation$\left|\tau-i\right|\sim\mathcal{O}(10^{-2}\div10^{-1})$,
and somewhat independently of the actual levels $N$ and specific
lepton field representations. This phenomenological observation led
us to investigate in greater detail the structure of modular invariant
theories in the symmetric limit, in which one of the two generators
of the modular group, as well as the $CP$ (charge-parity conjugation)
generator, are not broken, and to point out some of its consequences
for model building.

A second motivation for our work was to generalize the previous investigations
by allowing for non-minimal terms in the model's Kähler potential:
as a matter of fact, in most of the literature a certain ``minimal''
form of Kähler potential is assumed without special mention; such
a form arises, in fact, in certain string theory compactifications
in the large volume limit, that is $\text{Im}\tau\to\infty$ (see,
e.g.\ \cite{Chen:2019ewa,Asaka:2020tmo} and references therein),
which however contradicts the findings of explicit model building
attempts discussed so far. For $\tau\approx i$, the minimal Kähler
potential can in principle receive large non-perturbative corrections,
so that the original assumption appears to be somewhat unjustified.
This is actually very relevant also from a low-energy point of view,
since, in contrast to the superpotential, the Kähler potential does
not have any holomorphicity or no-renormalization constraint, and
it is in fact easy to show that modular invariance alone allows for
an infinite number of functionally independent contributions to the
Kähler metric of any chiral field.

If the Kähler potential is regarded to as completely arbitrary, much
of the predictive power mentioned previously gets apparently lost.
On the other hand, as we illustrate in Ref. \cite{Feruglio:2021dte}
through explicit models, some \emph{qualitative} predictions of the
minimal models continue to be valid also in the presence of Kähler
corrections, as long as the latters are at most $\mathcal{O}(1)$
modifications of the minimal potential. In our special case, we are
particularly concerned with charged lepton mass hierarchies, whose
size turns out to be essentially governed by the soft breaking of
$\mathbb{Z}_{4}$ symmetry in the neighbourood of $\tau\approx i$.

\section{Modular invariant models}

In the simplest case, modular invariance arises from the compactification
of a higher dimensional theory on a torus or an orbifold. Size and
shape of the compact space are parametrised by a modulus $\tau$ living
in the upper-half complex plane, up to modular transformations. These
can be interpreted as discrete gauge transformations, related to the
redundancy of the description. The low-energy effective theory, relevant
to the known particle species, has to obey modular invariance and
Yukawa couplings become functions of $\tau$. In this section we shortly
review the formalism of supersymmetric modular invariant theories~\cite{Ferrara:1989bc,Ferrara:1989qb}
applied to flavour physics~\cite{Feruglio:2017spp}.

The theory depends on a set of chiral supermultiplets $\varphi$ comprising
the dimensionless modulus $\tau\equiv\varphi_{0}/\Lambda$ (${\tt Im}~\tau>0$)
and other superfields $\varphi_{i}$~$(i\geq1)$. Here $\Lambda$
represents the cut-off of our effective theory, and can be interpreted
as the relevant mass scale of an underlying fundamental theory. In
the case of rigid supersymmetry, the Lagrangian $\mathscr{L}$~\footnote{Up to terms with at most two derivatives in the bosonic fields.}
is fully specified by the Kähler potential $K(\varphi,\bar{\varphi})$,
a real gauge-invariant function of the chiral multiplets and their
conjugates, by the superpotential $W(\varphi)$, a holomorphic gauge-invariant
function of the chiral multiplets, and by the gauge kinetic function
$f(\varphi)$, a dimensionless holomorphic gauge-invariant function
of the chiral superfields. Neglecting gauge interactions, we have:
\[
\mathcal{L}=\int d^{2}\theta d^{2}\bar{\theta}~K(\varphi,\bar{\varphi})+\int d^{2}\theta~W(\varphi)+\int d^{2}\bar{\theta}~\overline{W}(\bar{\varphi}).
\]
The Lagrangian is invariant under transformations $\gamma$ of the
homogeneous modular group $\Gamma=\text{SL}(2,Z)$:
\begin{equation}
\tau\to\frac{a\tau+b}{c\tau+d},\qquad\varphi_{i}\to(c\tau+d)^{-k_{i}}\rho(\tilde{\gamma})_{ij}\varphi_{j}\quad(i,j\geq1)\label{eq:MODULAR-modular transformation}
\end{equation}
where $a$, $b$, $c$, $d$ are integers obeying $ad-bc=1$. Such
transformations are generated by the two elements of $\Gamma$: 
\begin{equation}
S=\begin{pmatrix}0 & 1\\
-1 & 0
\end{pmatrix}\quad\quad\text{and}\quad\quad T=\begin{pmatrix}1 & 1\\
0 & 1
\end{pmatrix}.\label{eq:MODULAR-SandT}
\end{equation}
The matrix $\rho(\tilde{\gamma})$ is a unitary representation of
the group $\Gamma_{N}=\Gamma/\Gamma(N)$, obtained as a quotient between
the group $\Gamma$ and a principal congruence subgroup $\Gamma(N)$,
the positive integer $N$ being the level of the representation. The
level $N$ is kept fixed in the construction, and $\tilde{\gamma}$
represents the equivalence class of $\gamma$ in $\Gamma_{N}$. In
general $\rho(\tilde{\gamma})$ is a reducible representation and
all superfields belonging to the same irreducible component should
have the same weight $k_{i}$, here assumed to be integer\footnote{We restrict to integer modular weights. Fractional weights are in
general allowed, but require a suitable multiplier system \cite{Liu:2020msy,Nilles:2020nnc}.}. In the following, we denote by $(\varphi_{i},\psi_{i})$ the spin-$(0,1/2)$
components of the chiral superfields $\varphi_{i}$ $(i\geq1)$~\footnote{The distinction between superfields and their scalar components should
be clear from the context.}. The terms bilinear in the fermion fields read~\cite{Brignole:1996fn}:
\begin{equation}
\mathscr{L}_{F}=\mathscr{L}_{F,K}+\mathscr{L}_{F,2},
\end{equation}
with~\footnote{The covariant derivative is $D_{\mu}\psi^{i}=\partial_{\mu}\psi^{i}+\left(K^{-1}\right)_{m}^{i}K_{kl}^{m}\partial_{\mu}\varphi^{k}\psi^{l}$.}:
\begin{equation}
\mathscr{L}_{F,K}=i\,K_{i}^{j}\,\overline{\psi}_{j}\bar{\sigma}^{\mu}D_{\mu}\psi^{i},~~~~~\mathscr{L}_{F,2}=-\frac{1}{2}\left[W_{ij}-W_{l}(K^{-1})_{m}^{l}K_{ij}^{m}\right]\psi^{i}\psi^{j}+\text{h.c.},\label{eq:Lag_modu}
\end{equation}
where lower (upper) indices in $K$ and $W$ stand for derivatives
with respect to holomorphic (anti-holomorphic) fields. When the scalar
fields in eq.~(\ref{eq:Lag_modu}) take their VEVs, we can move to
the basis where matter fields are canonically normalised, through
a transformation:  
\begin{equation}
\psi_{i}\to(z^{-1/2})_{ij}\psi_{j},\label{eq:MODULAR-KahlerRenorm}
\end{equation}
where the matrix $(z^{1/2})_{ij}$ satisfies: $K_{i}^{j}=[(z^{1/2})^{\dagger}]^{jl}(z^{1/2})_{li}$~\footnote{Notice that this transformation mixes holomorphic and anti-holomorphic
indices, and there is no more fundamental distinction between upper
and lower components of the matrix $(z^{1/2})$.}. We can identify the fermion mass matrix as: 
\begin{equation}
m_{kn}=\left[W_{ij}-W_{l}(K^{-1})_{m}^{l}K_{ij}^{m}\right](z^{-1/2})_{ik}(z^{-1/2})_{jn}\label{eq:MODULAR-FermionMassMatFull}
\end{equation}
where VEVs are understood. In the previous equation, the second term
in the square bracket vanishes when supersymmetry is unbroken and
the VEV of $W_{l}$ is zero. When we turn on supersymmetry breaking
effects, the first term is expected to dominate over the second one,
provided there is a sufficient gap between the sfermion masses $m_{\mathrm{SUSY}}$
and the messenger/cutoff scale $M$. This holds both for vector-like
and for chiral fermions. Indeed, up to loop factors or other accidental
factors, the VEVs of $W_{l}$, $W_{ij}$ and $K_{ij}^{m}$ are of
the order of $m_{\mathrm{SUSY}}M$, $M$ and $1/M$, respectively,
when fermions are vector-like. When chiral fermions are considered,
$W_{ij}$ and $K_{ij}^{m}$ are both depleted by $v/M$ with respect
to the vector-like case, $v$ denoting the gauge symmetry breaking
scale. Thus we have a relative suppression between the two contributions
of order $m_{\mathrm{SUSY}}/M$, which can be made tiny (cf. ref.~\cite{Criado:2018thu}).
If we work under this assumption, the mass matrix is well approximated
by:
\begin{equation}
m_{kn}=W_{ij}~(z^{-1/2})_{ik}(z^{-1/2})_{jn}.\label{eq:MODULAR-FermionMassMat}
\end{equation}
The supersymmetry breaking terms neglected here can be useful to give
masses to light fermions, which otherwise would remain massless in
the exact supersymmetry limit. We will come back to this point when
discussing concrete models, in Section~\ref{sec:MODULAR-Models}.
Due to the conservation of the electric charge, the equality of eq.~(\ref{eq:MODULAR-FermionMassMat})
holds separately in any charge sector. By focusing on the lepton sector
$(E^{c},L)$ and by assuming that the neutrino masses arise from the
Weinberg operator, we have: 
\begin{equation}
W=-E_{i}^{c}\,\mathcal{Y}_{ij}^{e}(\tau)L_{j}H_{d}-\frac{1}{2\Lambda_{L}}L_{i}\,\mathcal{C}_{ij}^{\nu}(\tau)L_{j}H_{u}H_{u},\label{eq:MODULAR-superpotential}
\end{equation}
where $H_{u,d}$ are the Higgs chiral multiplets and $\Lambda_{L}$
is the scale where lepton number is broken. The general relation~(\ref{eq:MODULAR-FermionMassMat})
specialises into: 
\begin{equation}
m_{e}=(z_{E^{c}}^{-1/2})^{T}\mathcal{Y}^{e}(\tau)(z_{L}^{-1/2})v_{d},~~~~~~m_{\nu}=(z_{L}^{-1/2})^{T}\mathcal{C}^{\nu}(\tau)(z_{L}^{-1/2})v_{u}^{2}/\Lambda_{L},\label{eq:MODULAR-LeptonMassMatrices}
\end{equation}
where we have absorbed the renormalisation factors for $H_{u,d}$
in the definition of their VEVs. In Section~\ref{sec:MODULAR-Models},
we will also comment on the special limit where $z_{E^{c},L}^{-1/2}$
are universal, i.e.\ proportional to the unit matrix. The mass matrices
obtained in this case will be referred to as ``bare'' matrices and
denoted by $m_{e,\nu}^{(0)}$. An important consequence of modular
invariance is the special functional dependence of $\mathcal{Y}^{e}(\tau)$
and $\mathcal{C}^{\nu}(\tau)$ on the modulus $\tau$. Under a transformation
of $\Gamma$, the chiral multiplets $(E_{i}^{c},L_{i},H_{u,d})$ transform
as in eq.~(\ref{eq:MODULAR-modular transformation}), with weights
$(k_{E_{i}^{c}},\,k_{L_{i}},\,k_{H_{u,d}})$ and representations $(\rho_{E^{c}}(\tilde{\gamma}),\,\rho_{L}(\tilde{\gamma}),\,\boldsymbol{1})$.
For the superpotential $W$ to be modular invariant, $\mathcal{Y}^{e}(\tau)$
and $\mathcal{C}^{\nu}(\tau)$ should obey:  
\[
\mathcal{Y}^{e}(\gamma\tau)=(c\tau+d)^{k_{e}}~\rho_{E^{c}}^{*}(\tilde{\gamma})~\mathcal{Y}^{e}(\tau)\rho_{L}^{\dagger}(\tilde{\gamma})~,~~~~~\mathcal{C}^{\nu}(\gamma\tau)=(c\tau+d)^{k_{\nu}}~\rho_{L}^{*}(\tilde{\gamma})\mathcal{C}^{\nu}(\tau)\rho_{L}^{\dagger}(\tilde{\gamma}),~
\]
where the weights $k_{e,\nu}$ are matrices satisfying: $(k_{e})_{ij}=k_{E_{i}^{c}}+k_{L_{j}}+k_{H_{d}}$
and $(k_{\nu})_{ij}=k_{L_{i}}+k_{L_{j}}+2k_{H_{u}}$. Thus $\mathcal{Y}^{e}(\tau)$
and $\mathcal{C}^{\nu}(\tau)$ are modular forms of given level and
weight. Since the linear space of such modular forms is finite dimensional,
the choices for $\mathcal{Y}^{e}(\tau)$ and $\mathcal{C}^{\nu}(\tau)$
are limited. If neutrino masses originate from a type I seesaw mechanism,
Eqs.~(\ref{eq:MODULAR-superpotential}) and (\ref{eq:MODULAR-FermionMassMat})
hold with the identification:
\begin{equation}
\frac{\mathcal{C}^{\nu}(\tau)}{\Lambda_{L}}=-(\mathcal{Y}^{\nu}(\tau))^{T}~\mathcal{M}(\tau)^{-1}\mathcal{Y}^{\nu}(\tau),\label{eq:MODULAR-SeesawFormula}
\end{equation}
where $\mathcal{Y}^{\nu}(\tau)$ and $\mathcal{M}(\tau)$ denote the
matrix of neutrino Yukawa couplings and the mass matrix of the heavy
electroweak singlets $N^{c}$, respectively. Notice that there is
no dependence on the renormalisation factor $(z_{N^{c}}^{-1/2})$
of the heavy modes. In some cases $\mathcal{Y}^{e}(\tau)$ and/or
$\mathcal{C}^{\nu}(\tau)$ are completely determined as a function
of $\tau$ up to an overall constant, thus providing a strong potential
constraint on the mass spectrum, Eq.~(\ref{eq:MODULAR-FermionMassMat}).

Unfortunately, such property does not extend to the Kähler potential
$K$ and to the renormalisation factors $(z_{E^{c},L}^{-1/2})$. Minimal
choices of $K$, appropriate for a perturbative regime, can receive
large non-perturbative corrections in the region of the moduli space
we will consider. Without a control over the non-perturbative dynamics,
in a generic point of the moduli space the factors $(z_{E^{c},L}^{-1/2})$
remain unknown. If we allowed for completely arbitrary $(z_{E^{c},L}^{-1/2})$,
under mild conditions any mass matrix could be predicted. From eq.~(\ref{eq:MODULAR-FermionMassMat})
we see that, given $\mathcal{Y}^{e}(\tau)$ and $(z_{L}^{-1/2})$,
we could reproduce any desired matrix $m_{e}$, by selecting a particular
$(z_{E^{c}}^{-1/2})^{T}$: 
\begin{equation}
(z_{E^{c}}^{-1/2})^{T}=m_{e}~(\mathcal{Y}^{e}(\tau)v_{d}(z^{-1/2}))^{-1}\label{eq:MODULAR-ArbitraryMe}
\end{equation}
An arbitrary matrix $m_{e}$ would result in a completely unconstrained
lepton mixing matrix. Similar considerations would apply to the neutrino
mass matrix $m_{\nu}$.

The loss of predictability associated to the Kähler corrections may
however be less severe than eq.~(\ref{eq:MODULAR-ArbitraryMe}) might
suggest, for two reasons. First, note that the above solution requires
a non-singular $\mathcal{Y}^{e}(\tau)$. A singular $\mathcal{Y}^{e}(\tau)$
can only give rise to a singular $m_{e}$. Correspondingly, a hierarchical
$\mathcal{Y}^{e}(\tau)$ can only correspond to a hierarchical $m_{e}$,
unless the eigenvalues of the matrix $(z_{E^{c}}^{-1/2})^{T}$ in
eq.~(\ref{eq:MODULAR-ArbitraryMe}) come in very large ratios. Although
we cannot exclude the latter possibility, here we focus on the class
of models where the corrections associated to the Kähler potential
do not alter the ``bare'' limit by more than about one order of
magnitude. Hence a singular or nearly singular $\mathcal{Y}^{e}(\tau)$
will tame the loss of predictability associated with the Kähler potential.
Needless to say, a hierarchical $\mathcal{Y}^{e}(\tau)$ is needed
to reproduce the mass spectrum in the charged lepton sector. Different
considerations apply to the neutrino sector, where a singular $\mathcal{C}^{\nu}(\tau)$
might not be a good first order approximation of the data.

A second constraint on the effect of the Kähler corrections arises
in the vicinity of the fixed points of $\Gamma$, $\tau=i$, $\tau=-1/2+i\sqrt{3}/2$
and $\tau=i\infty$, invariant under the action of the elements $S$,
$ST$ and $T$, respectively. In the following, we will assume the
modulus to be in the vicinity of the point $\tau=i$, as suggested
by several models to correctly reproduce the data. The invariance
under $S$ provides a constraint on the possible form of the Kähler
potential at $\tau=i$ and in its vicinity.

\section{Residual symmetry near $\tau=i$}

The residual symmetry of the theory at $\tau=i$ is the cyclic group
$Z_{4}$ generated by the element $S$, whose action on the chiral
multiplets $\varphi_{i}$ in $\tau=i$ can be read from eq.~(\ref{eq:MODULAR-modular transformation}):
\begin{equation}
\varphi_{i}\to\sigma_{ij}\varphi_{j},~~~~~~~~~~~~~~\sigma_{ij}=i^{k_{i}}\rho(\tilde{S})_{ij}~~~~(i,j>0),\label{eq:MODULAR-Sigma}
\end{equation}
where $\sigma$ is unitary, $\sigma^{2}$ is a parity operator and
$\sigma^{4}=1$. To analyse the neighbourhood of $\tau=i$, we expand
both the Kähler potential and the superpotential in powers of the
matter fields $\varphi_{i}$ $(i>0)$~\footnote{Electrically neutral multiplets whose scalar component acquires a
VEV, like $H_{u,d}$, might mix in the kinetic term with the modulus
$\tau$. The mixing is parametrically suppressed by $v/\Lambda$ and
will be ignored in the following.}: 
\begin{equation}
\begin{aligned}W & =\sum_{i_{1},...,i_{n}}Y_{i_{1},\dots,i_{n}}(\tau)~\varphi_{i_{1}}\dots\varphi_{i_{n}}+\dots,\\
K & =
\overline{\varphi}_{i}\,z_{~j}^{i}(\tau,\bar{\tau})\,\varphi_{j}+\dots.
\end{aligned}
\label{expan}
\end{equation}
In the vicinity of $\tau=i$, it is possible to cast the theory as
an ordinary $Z_{4}$ invariant theory, where the symmetry acts linearly
on the fields, slightly broken by the spurion $(\tau-i)$. When we
depart from $\tau=i$, the $S$ elements acts on the fields as: 
\begin{equation}
\tau\to-\frac{1}{\tau}~~~,~~~~~~\varphi_{i}\to(-\tau)^{-k_{i}}\rho(\tilde{S})_{ij}\varphi_{j}~~~~(i,j>0).\label{eq:MODULAR-S action on phi}
\end{equation}
We perform the field redefinition: 
\begin{equation}
\begin{cases}
\tau=i\frac{i+\frac{\epsilon}{2}}{i-\frac{\epsilon}{2}}\\
\tilde{\varphi}_{j}=(1-i\frac{\epsilon}{2})^{-k_{j}}\varphi_{j}
\end{cases}\label{eq:MODULAR-Z4FieldRedefinition}
\end{equation}
mapping the upper-half complex plane into the disk $|\epsilon|<2$.
In the linear approximation: 
\begin{equation}
\epsilon=(\tau-i)+O\left((\tau-i)^{2}\right).
\end{equation}
Under the $S$ transformation in~(\ref{eq:MODULAR-S action on phi}),
the new fields transform as: 
\begin{equation}
\left\{ \begin{aligned} & \epsilon\to-\epsilon\\[1mm]
 & \tilde{\varphi}_{i}\to\sigma_{ij}~\tilde{\varphi}_{j}
\end{aligned}
\right.\;.\label{eq:MODULAR-linearizedZ4}
\end{equation}
We see that the action of the $Z_{4}$ symmetry is linear in the new
field basis, even when $\tau\ne i$. In particular $\epsilon$ behaves
as a spurion with $Z_{4}$ charge $+2$. In the new field basis, the
coefficients of the field expansion (\ref{expan}) read: 
\begin{equation}
\begin{aligned}\tilde{Y}_{i_{1},\dots,i_{n}} & =\left(1-i\frac{\epsilon}{2}\right)^{k_{i_{1}}+\ldots+k_{i_{n}}}Y_{i_{1},\dots,i_{n}}\\
\tilde{z}_{~j}^{i} & =\left(1+i\frac{\bar{\epsilon}}{2}\right)^{k_{i}}z_{~j}^{i}\left(1-i\frac{\epsilon}{2}\right)^{k_{j}}
\end{aligned}
~~~.
\end{equation}
The invariance of the theory under $Z_{4}$ requires $\tilde{Y}_{i_{1},...,i_{n}}(\epsilon)$
and $\tilde{z}_{~j}^{i}(\epsilon,\bar{\epsilon})$ to satisfy: 
\begin{equation}
\begin{aligned}\tilde{Y}_{i_{1},...,i_{n}}(\epsilon) & =\sigma_{j_{1}i_{1}}~...~\sigma_{j_{n}i_{n}}\tilde{Y}_{j_{1},...,j_{n}}(-\epsilon)\\[0.2cm]
\tilde{z}_{~j}^{i}(\epsilon,\bar{\epsilon}) & =\sigma^{\dagger ik}~\tilde{z}_{~l}^{k}(-\epsilon,-\bar{\epsilon})~\sigma_{lj}
\end{aligned}
~~~.\label{eq:MODULAR-Z4constraints}
\end{equation}
In particular, setting $\epsilon=0$, the above equations express
the necessary conditions for the invariance of the theory at the symmetric
point $\tau=i$. By expanding $\tilde{z}_{~j}^{i}(\epsilon,\bar{\epsilon})$
in powers of $\epsilon$ we see that the terms of first order vanish,
up to possible non-diagonal terms relating fields with opposite value
of $\sigma$. We conclude that in a neighbourhood of the fixed point
$\tau=i$, and in the absence of any information about the Kähler
potential, the theory reduces to a linearly realised $Z_{4}$ flavour
symmetric theory, in the presence of a (small) spurion with charge
$+2$. 

\section{Models\label{sec:MODULAR-Models}}

In this section, we present two models making use of the results of
the previous section to account for the observed hierarchies in the
lepton spectrum, namely the smallness of the charged lepton mass ratios,
$m_{e}/m_{\tau}$ and $m_{\mu}/m_{\tau}$ and of the neutrino mass
ratio $r\equiv\delta m^{2}/|\Delta m^{2}|$, where $\delta m^{2}\equiv m_{2}^{2}-m_{1}^{2}$
and $\Delta m^{2}\equiv m_{3}^{2}-(m_{1}^{2}+m_{2}^{2})/2$ (with
the standard neutrino labeling). The hierarchies will be naturally
accounted for by the small breaking of $Z_{4}$, $|\epsilon|\ll1$,
i.e.\ by the closeness of the modulus $\tau$ to the $Z_{4}$ symmetric
point $\tau=i$, while the parameters in the superpotential will be
$\mathcal{O}(1)$, and the corrections to the minimal Kähler will
not be larger than $\mathcal{O}(1)$. In Table~\ref{tab:MODULAR-Best-fit-values},
we collect the best-fit values of the leptonic parameters with the
corresponding $1\sigma$ uncertainties. 
\begin{table}[t]
\centering{}%
\begin{tabular}{|c|cc|}
\hline 
\texttt{Observable} & \multicolumn{2}{c|}{\texttt{Best-fit value with $1\sigma$ error}}\tabularnewline
\hline 
\hline 
$m_{e}/m_{\mu}$ & \multicolumn{2}{c|}{$0.0048_{-0.0002}^{+0.0002}$}\tabularnewline
$m_{\mu}/m_{\tau}$ & \multicolumn{2}{c|}{$0.0565_{-0.0045}^{+0.0045}$}\tabularnewline
\hline 
 & NO & IO\tabularnewline
$\delta m^{2}\left[10^{-5}~\text{eV}^{2}\right]$ & \multicolumn{2}{c|}{$7.42_{-0.20}^{+0.21}$}\tabularnewline
$\Delta m^{2}\left[10^{-3}~\text{eV}^{2}\right]$ & $2.480_{-0.028}^{+0.026}$ & $-2.461_{-0.028}^{+0.028}$\tabularnewline
$r\equiv\delta m^{2}/|\Delta m^{2}|$ & $0.0299_{-0.0009}^{+0.0009}$ & $0.0301_{-0.0009}^{+0.0009}$\tabularnewline
$\sin^{2}\theta_{12}$ & $0.304_{-0.012}^{+0.012}$ & $0.304_{-0.012}^{+0.013}$\tabularnewline
$\sin^{2}\theta_{13}$ & $0.02219_{-0.00063}^{+0.00062}$ & $0.02238_{-0.00062}^{+0.00063}$\tabularnewline
$\sin^{2}\theta_{23}$ & $0.573_{-0.020}^{+0.016}$ & $0.575_{-0.019}^{+0.016}$\tabularnewline
$\delta/\pi$ & $1.09_{-0.13}^{+0.15}$ & $1.57_{-0.17}^{+0.14}$\tabularnewline
\hline 
\end{tabular}\caption{\label{tab:MODULAR-Best-fit-values}Best-fit values of the charged
lepton mass ratios and the neutrino oscillation parameters with the
corresponding 1$\sigma$ errors. For the charged lepton mass ratios
we have used the values given in ref.~\cite{Ross:2007az}, averaged
over $\tan\beta$ as described in the text, whereas for the neutrino
parameters we have used the results obtained in refs.~\cite{Esteban:2020cvm,NuFITv50}
(with Super-Kamiokande atmospheric data).}
\end{table}

For the charged lepton mass ratios we use the results of Ref.~\cite{Ross:2007az},
where for $m_{\mu}/m_{\tau}$ we take an average between the values
obtained for $\tan\beta=10$ and $\tan\beta=38$. For the neutrino
oscillation parameters we employ the results of the global analysis
performed in refs.~\cite{Esteban:2020cvm,NuFITv50}. In what follows,
when fitting models to the data, we use five dimensionless observables
that have been measured with a good precision, i.e.\ two mass ratios~\footnote{In the models presented below, $m_{e}=0$ by construction, so we do
not include the ratio $m_{e}/m_{\mu}$ here. See subsection~\ref{subsec:generating_me}
for possible ways of generating non-zero $m_{e}$.}, $m_{\mu}/m_{\tau}$ and $r$, and three leptonic mixing angles,
$\sin^{2}\theta_{12}$, $\sin^{2}\theta_{13}$, $\sin^{2}\theta_{23}$.
Regarding the Dirac CPV phase, $\delta$, values between $\pi$ and
$2\pi$ (approximately) are currently allowed at $3\sigma$ for both
neutrino mass spectrum with normal ordering~(NO) and that with inverted
ordering (IO). Moreover, in ref.~\cite{Novichkov:2018ovf}, it has
been shown that under the transformation $\tau\to-\tau^{\ast}$ and
complex conjugation of couplings present in the superpotential, CPV
phases change their signs, whereas masses and mixing angles remain
the same. In fact, this reflects CP properties of modular invariant
models~\cite{Novichkov:2019sqv} (see also \cite{Kobayashi:2019uyt}).
As a consequence, the Dirac phase $\delta$ is not particularly constraining
for our fits, and we do not include it in the list of input observables,
regarding the obtained values as predictions.

\subsection{\label{sec:MODULAR-model_1}Model 1: Weinberg operator and inverted
ordering}

We work at level 3, and the relevant finite modular group is $\Gamma_{3}$.
In this subsection, we assume that neutrino masses are generated by
the Weinberg operator. The field content of the model along with the
assignment of $\Gamma_{3}$ representations and modular weights $k$
is shown in Table~\ref{tab:MODULAR-model_1}. The corresponding charges
under $Z_{4}$, obtained using $\sigma$, are shown in Table~\ref{tab:MODULAR-model_1_Z4}.
We work in a real basis for the elements of $\Gamma_{3}$ where $\rho(\tilde{S})=\text{diag}(+1,-1,-1)$
for the irreducible three-dimensional representation.

\begin{table}[t]
\centering{}%
\begin{tabular}{|c|c|c|c|c|c|c|}
\hline 
 & $L$ & $E_{1}^{c}$ & $E_{2}^{c}$ & $E_{3}^{c}$ & $H_{u}$ & $H_{d}$\tabularnewline
\hline 
\hline 
$SU(2)_{L}\times U(1)_{Y}$ & $(\boldsymbol{2},-1/2)$ & $(\boldsymbol{1},+1)$ & $(\boldsymbol{1},+1)$ & $(\boldsymbol{1},+1)$ & $(\boldsymbol{2},+1/2)$ & $(\boldsymbol{2},-1/2)$\tabularnewline
\hline 
$\Gamma_{3}$ & $\boldsymbol{3}$ & $\boldsymbol{1}$ & $\boldsymbol{1}$ & $\boldsymbol{1}^{\prime}$ & $\boldsymbol{1}$ & $\boldsymbol{1}$\tabularnewline
\hline 
$k$ & $1$ & $3$ & $3$ & $3$ & $0$ & $0$\tabularnewline
\hline 
\end{tabular}\caption{\label{tab:MODULAR-model_1}Assignment of representations and modular
weights in Model 1.}
\end{table}

\begin{table}[t]
\centering{}%
\begin{tabular}{|c|c|c|c|c|c|c|c|c|c|}
\hline 
 & $\tilde{L}_{1}$ & $\tilde{L}_{2}$ & $\tilde{L}_{3}$ & $\tilde{E}_{1}^{c}$ & $\tilde{E}_{2}^{c}$ & $\tilde{E}_{3}^{c}$ & $\tilde{H}_{u}$ & $\tilde{H}_{d}$ & $\epsilon$\tabularnewline
\hline 
\hline 
$Z_{4}$ & $1$ & $-1$ & $-1$ & $-1$ & $-1$ & $-1$ & $0$ & $0$ & 2\tabularnewline
\hline 
\end{tabular}\caption{\label{tab:MODULAR-model_1_Z4}$Z_{4}$ charges (\texttt{mod} 4) in
Model 1.}
\end{table}

The quantum number assignments have immediate consequences for the
charged lepton mass spectrum: 
\begin{enumerate}
\item At $\tau=i$, the charged lepton mass matrix $m_{e}$ has rank one.
This follows from the $Z_{4}$ charges in Table~\ref{tab:MODULAR-model_1_Z4},
forcing: 
\begin{equation}
m_{e}=\begin{pmatrix}\alpha & 0 & 0\\
\beta & 0 & 0\\
\gamma & 0 & 0
\end{pmatrix}.\label{eq:MODULAR-me1}
\end{equation}
\item For a generic $\tau\neq i$, $m_{e}$ has rank two. While $Z_{4}$
alone would allow $m_{e}$ to have rank three, the underlying modular
invariance forces the coefficients of the first and second rows of
$m_{e}$ to be proportional, thus reducing the rank. In fact, modular
invariance requires the coupling of $E_{1}^{c}$ and $E_{2}^{c}$
to $L$ to be proportional to the same modular form multiplet, namely,
the triplet of weight four. The Kähler corrections cannot modify the
rank condition. Thus, in the considered model, the electron has zero
mass. 
\end{enumerate}
For $\tau\approx i+\epsilon$, with $\left|\epsilon\right|\ll1$,
we obtain the prediction
\begin{equation}
m_{e}\,\colon\,m_{\mu}\,\colon\,m_{\tau}=0\,\colon\,\mathcal{O}(\epsilon)\,\colon\,1\,.\label{eq:Ratios mE - model 1}
\end{equation}
Concerning the neutrino mass spectrum, from the charges of the lepton
doublets $L_{i}$ in Table~\ref{tab:MODULAR-model_1_Z4}, we deduce
that $m_{\nu}$ in $\tau=i$ takes the following general form: 
\begin{equation}
m_{\nu}=\begin{pmatrix}0 & a & b\\
a & 0 & 0\\
b & 0 & 0
\end{pmatrix}.\label{eq:General neutrino Weinberg}
\end{equation}
This matrix has rank two and two degenerate non-zero eigenvalues.
Notice that, while a generic $Z_{4}$ model would not account for
the values of the parameters $a$ and $b$, here the underlying modular
invariance fixes the relative values, before Kähler corrections. With
the $Z_{4}$ assignment of Table \ref{tab:MODULAR-model_1_Z4}, we
are implicitly using the basis where $S$ is diagonal for the irreducible
triplet of $\Gamma_{3}$ and we find $a/b=Y_{3}(i)/Y_{2}(i)$, where
$Y^{(2)}(\tau)\equiv(Y_{1}(\tau),Y_{2}(\tau),Y_{3}(\tau))^{T}$ denotes
the weight-two triplet of modular forms. On the other hand, generic
Kähler corrections could mix $L_{2}$ and $L_{3}$, as they have the
same $Z_{4}$ charge (see (\ref{eq:MODULAR-Z4constraints})), leading
to arbitrary $a/b$, as in generic $Z_{4}$ models. For $\tau\approx i+\epsilon$,
the rank of $m_{\nu}$ becomes three, and we obtain the neutrino mass
spectrum with inverted ordering of the form
\begin{equation}
m_{1}\,\colon\,m_{2}\,\colon\,m_{3}=1\,\colon\left(1+\mathcal{O}(\epsilon)\right)\colon\,\mathcal{O}(\epsilon)\,,\label{eq:ratios mN - model 1}
\end{equation}
and, in particular:
\begin{equation}
r=\mathcal{O}(\epsilon)\,.\label{eq:ratio r - model 1}
\end{equation}
Clearly, both qualitative relations (\ref{eq:Ratios mE - model 1})
and (\ref{eq:ratio r - model 1}) are phenomenologically intriguing.
They are consequences of modular invariance alone, and are thus independent
from the parameters in the superpotential or the Kähler potential
(provided these are non-hierarchical by themselves). In subsection~\ref{subsec:generating_me},
we discuss two possible mechanisms to generate a naturally small electron
mass.

While this model successfully accounts for the observed mass hierarchies
(with a non-vanishing electron mass still to be generated), it is
not satisfactory when it comes to the mixing angles. The point is
that in order for (\ref{eq:General neutrino Weinberg}) to lead to
a reasonable leading order approximation, the tau lepton should correspond
to a linear combination of $L_{2}$ and $L_{3}$, while eq.~(\ref{eq:MODULAR-me1})
forces the tau lepton to be mainly $L_{1}$. Indeed, the prediction
for the mixing angles at $\tau=i$ is 
\begin{equation}
\sin^{2}\theta_{13}=\cos^{2}\theta_{12}^{e}\,,\qquad\sin^{2}\theta_{12}=\frac{1}{2}\,,\qquad\sin^{2}\theta_{23}=1\,,\label{eq:anglesLO}
\end{equation}
where $\theta_{12}^{e}$ is an arbitrary angle related to the presence
of two vanishing eigenvalues in $m_{e}$, to be fixed by the $Z_{4}$
breaking. These predictions imply that in order to generate the correct
mixing angles $\sin^{2}\theta_{23}\approx0.6$ and $\sin^{2}\theta_{12}\approx0.3$,
large hierarchical deviations from the minimal Kähler metrics are
required~\footnote{The need for non-minimal Kähler metrics stems not only from the leading
order predictions for the mixing angles, but also from the mass spectrum
of the model. In the vicinity of $\tau=i$ we found, both numerically
and through an approximate analytical study, that $m_{\mu}/m_{\tau}$
is smaller than $r$, while data require the opposite.}, as $|\epsilon|\ll1$ cannot give rise to such large corrections.
This is clearly an unpleasant feature, since it introduces a source
of hierarchy in the Lagrangian parameters. We carried out a full numerical
study of the model, after adding a non-minimal Kähler potential depending
on four new real parameters. The outcome confirms the above qualitative
considerations. More precisely, we gauge the degree of hierarchy related
to a non-canonical Kähler potential $K$ by means of the condition
number 
\begin{equation}
\kappa(K)=\lambda_{\text{max}}(z)/\lambda_{\text{min}}(z),\label{Eq:MODULAR-condnum}
\end{equation}
the ratio between the maximum and minimum eigenvalues of $z_{~j}^{i}$
at the best-fit point. We find that all Kähler metrics providing a
good fit near $\tau=i$ turn out to have $\kappa(K_{L,\,E^{c}})$
very large, typically in the range $10^{3}\div10^{4}$. We discuss
in the next subsection a seesaw variant of the present model which
allows to mitigate the need of hierarchical Kähler metrics.

\subsection{Model 2: seesaw mechanism and normal ordering}

The main phenomenological obstructions in the model discussed above
are the leading order predictions for the mixing angles. In this subsection,
we show how to evade them by introducing electroweak singlet neutrinos
$N^{c}$ and generating the Weinberg operator through the type I seesaw
mechanism. This widens the class of possible neutrino mass matrices
that can be obtained, if the singlet neutrino mass matrix becomes
singular in the limit $\tau\to i$. In this case, for the standard
analysis of the seesaw mechanism to be valid, singlet neutrino masses
are required to be large compared to the electroweak scale. In the
example discussed below, this is easily achieved outside of a neighbourhood
of $\tau=i$, provided the overall singlet neutrino mass scale is
large enough.

To be concrete, we augment the field content of Table~\ref{tab:MODULAR-model_1}
with electroweak singlets $N^{c}\sim\boldsymbol{3}$ under $\Gamma_{3}$,
with weight $k_{N^{c}}=1$. As before, we denote by $Y^{(2)}$ the
weight $2$ modular form triplet, and by $Y^{(4)}\equiv(Y^{(2)}Y^{(2)})_{\boldsymbol{3}_{S}}$
the weight 4 triplet of modular forms. We denote by $\boldsymbol{3}_{S}$
and $\boldsymbol{3}_{A}$ the symmetric and antisymmetric triplet
contractions of two $\Gamma_{3}$ triplets, respectively. The superpotential
$W=W_{e}+W_{\nu}$ of the lepton sector reads: 
\begin{align}
W_{e} & =-\left[\alpha E_{1}^{c}\left(LY^{(4)}\right)_{\boldsymbol{1}}+\beta E_{2}^{c}\left(LY^{(4)}\right)_{\boldsymbol{1}}+\gamma E_{3}^{c}\left(LY^{(4)}\right)_{\boldsymbol{1}^{\prime\prime}}\right]H_{d}\,,\label{eq:Electron superpotential}\\[0.2cm]
W_{\nu} & =-\kappa\,\left(\left[\left(N^{c}L\right)_{\boldsymbol{3}_{S}}+g\left(N^{c}L\right)_{\boldsymbol{3}_{A}}\right]Y^{(2)}\right)_{\boldsymbol{1}}H_{u}-\Lambda\left(N^{c}N^{c}Y^{(2)}\right)_{\boldsymbol{1}}\,.\label{eq:See saw superpotential}
\end{align}
The parameters $\kappa$ and $\Lambda$ can be made real without loss
of generality, whereas $g$ is complex in general. In the real basis
for $\boldsymbol{3}$ of $\Gamma_{3}$, this superpotential leads
to the following matrices $\mathcal{Y}^{e}(\tau)$, $\mathcal{Y}^{\nu}(\tau)$
and $\mathcal{M}(\tau)$: 
\begin{align}
\mathcal{Y}^{e}(\tau) & =2\begin{pmatrix}\alpha Y_{2}Y_{3} & \alpha Y_{1}Y_{3} & \alpha Y_{1}Y_{2}\\
\beta Y_{2}Y_{3} & \beta Y_{1}Y_{3} & \beta Y_{1}Y_{2}\\
\gamma Y_{2}Y_{3} & \gamma\omega Y_{1}Y_{3} & \gamma\omega^{2}Y_{1}Y_{2}
\end{pmatrix},\label{eq:bare_me}\\[2mm]
\mathcal{Y}^{\nu}(\tau) & =\kappa\left[\begin{pmatrix}0 & Y_{3} & Y_{2}\\
Y_{3} & 0 & Y_{1}\\
Y_{2} & Y_{1} & 0
\end{pmatrix}+g\begin{pmatrix}0 & Y_{3} & -Y_{2}\\
-Y_{3} & 0 & Y_{1}\\
Y_{2} & -Y_{1} & 0
\end{pmatrix}\right],\label{eq:Ynu0}\\[2mm]
\mathcal{M}(\tau) & =2\Lambda\begin{pmatrix}0 & Y_{3} & Y_{2}\\
Y_{3} & 0 & Y_{1}\\
Y_{2} & Y_{1} & 0
\end{pmatrix}.\label{eq:M0}
\end{align}
The matrix $\mathcal{C}^{\nu}(\tau)$ of eq. (\ref{eq:MODULAR-superpotential})
is now given by the seesaw formula of eq.~(\ref{eq:MODULAR-SeesawFormula}).

Some analytical considerations easily follow from the previous equations
for the ``bare'' quantities, i.e.\ those corresponding to the minimal
Kähler potential. We make use of the following $\epsilon$-expansion
of $Y^{(2)}~$\footnote{One can prove, in general, that $\left.\frac{\text{d}}{\text{d}\tau}\right|_{i}Y_{2,3}=i\,Y_{2,3}(i)$.
Moreover, we can rephase $Y^{(2)}$ in such a way that $(Y_{3}(i))^{*}=Y_{2}(i)\equiv y$.
In this basis, we find that $\left.\frac{\text{d}}{\text{d}\tau}\right|_{i}Y_{1}\in-i\,\mathbb{R}^{+}$.}: 
\begin{equation}
Y_{1}=-ix\epsilon\,,\qquad Y_{2}=y\left(1+i\epsilon\right)\,,\qquad Y_{3}=y^{\ast}\left(1+i\epsilon\right),\label{eq:eps-expansion}
\end{equation}
where, up to an overall constant, $x\approx1.49087$ and $y=\sqrt{3}/2+i(3/2-\sqrt{3})$.
To first order in $\epsilon$ we obtain: 
\begin{equation}
\frac{\mathcal{Y}_{\nu}(\tau)}{\kappa}=\begin{pmatrix}0 & \left(1+g\right)y^{\ast} & \left(1-g\right)y\\
\left(1-g\right)y^{\ast} & 0 & 0\\
\left(1+g\right)y & 0 & 0
\end{pmatrix}+i\,\epsilon\begin{pmatrix}0 & \left(1+g\right)y^{\ast} & \left(1-g\right)y\\
\left(1-g\right)y^{\ast} & 0 & -\left(1+g\right)x\\
\left(1+g\right)y & -\left(1-g\right)x & 0
\end{pmatrix},\label{eq:eps-expansion_Ynu0}
\end{equation}
\begin{equation}
\frac{\mathcal{M}(\tau)}{2\Lambda}=\begin{pmatrix}0 & y^{\ast} & y\\
y^{\ast} & 0 & 0\\
y & 0 & 0
\end{pmatrix}+i\,\epsilon\begin{pmatrix}0 & y^{\ast} & y\\
y^{\ast} & 0 & -x\\
y & -x & 0
\end{pmatrix}.\label{eq:eps-expansion_M0}
\end{equation}
Notice that the bare Majorana mass matrix has one eigenvalue proportional
to $\epsilon$, thus vanishing in the limit $\tau\to i$. This corresponds
to the case of single right-handed neutrino dominance, in which one
of the electroweak singlet neutrinos is massless in the symmetric
limit~\footnote{As already observed, in order for the standard seesaw analysis to
be valid, we must require the product $\left|\epsilon\right|\Lambda$
(that is the order of magnitude of the lightest right-handed neutrino
mass) to be large with respect to the electroweak scale. For the present
model this does not pose any practical restriction, since the best-fit
region (see below) is achieved for values of $\vert\epsilon\vert\sim10^{-2}\div10^{-1}$.}. Inverting the matrix in eq.~(\ref{eq:eps-expansion_M0}) and using
the seesaw relation for the bare light neutrino mass matrix $m_{\nu}^{(0)}$
we find to $\mathcal{O}(\epsilon)$: 
\begin{equation}
m_{\nu}^{(0)}=\begin{pmatrix}\frac{2ig^{2}|y|^{2}}{x}\frac{1}{\epsilon}-\frac{4g^{2}|y|^{2}}{x} & -\left(1+g^{2}\right)y^{\ast} & -\left(1+g^{2}\right)y\\
-\left(1+g^{2}\right)y^{\ast} & 0 & 0\\
-\left(1+g^{2}\right)y & 0 & 0
\end{pmatrix}+\mathcal{O}(\epsilon)\,.\label{eq:seesaw at LO}
\end{equation}
The leading order form of the charged lepton mass matrix is as in
eq.~(\ref{eq:MODULAR-me1}).

The leading order predictions for neutrino masses and mixing angles
strongly depend upon the parameter $g$.
\begin{itemize}
\item A neutrino mass spectrum with IO can be realised when $|g|^{2}\ll|\epsilon|$.
Then, $m_{\nu}^{(0)}$ has approximately the same form as in the model
with the Weinberg operator considered in subsection~\ref{sec:MODULAR-model_1}.
We get the neutrino mass spectrum with IO:
\begin{equation}
m_{1}\,\colon\,m_{2}\,\colon\,m_{3}=1\,\colon\approx1\colon\,\mathcal{O}(\epsilon)\label{eq:neutrino_masses_g2lleps}
\end{equation}
and the predictions for the mixing angles reported in eq.~(\ref{eq:anglesLO}),
in particular, $\sin^{2}\theta_{23}=1$. 
\item A neutrino mass spectrum with NO can be realised when $|\epsilon|\ll|g|^{2}$.
In this case,
\begin{equation}
m_{\nu}^{(0)}=\begin{pmatrix}c(g^{2}/\epsilon) & a & b\\
a & 0 & 0\\
b & 0 & 0
\end{pmatrix}+\mathcal{O}(\epsilon)\,,
\end{equation}
with $|a|$, $|b|$, $|c|$ being $\mathcal{O}(1)$ numbers. Therefore,
we have the neutrino mass spectrum with NO: 
\begin{equation}
m_{1}\,\colon\,m_{2}\,\colon\,m_{3}=\mathcal{O}\left(\epsilon^{2}/g^{2}\right)\,\colon\,\mathcal{O}\left(\epsilon^{2}/g^{4}\right)\,\colon\,\mathcal{O}(1)
\end{equation}
implying $r=O(\epsilon^{4}/g^{4})$. The mixing angles are:
\begin{equation}
\sin^{2}\theta_{12}=\mathcal{O}(1)\,,\qquad\sin^{2}\theta_{23}\approx\sin^{2}\theta_{13}\approx r^{1/2}\,.\label{eq:anglesLO1}
\end{equation}
Again, the leading order prediction for $\sin^{2}\theta_{23}$ is
far away from its measured value and requires significant corrections
from the Kähler potential. 
\end{itemize}
To verify the viability of the model we have performed a full numerical
study, also allowing for a non-minimal form of the Kähler potential
for the matter fields. In general, modular invariance allows many
terms in the Kähler potential~\cite{Feruglio:2017spp,Chen:2019ewa}.
In the considered bottom-up approach, there seems to be no way of
reducing the number of these terms. However, this may change if modular
symmetry is augmented by a traditional finite flavour symmetry~\cite{Nilles:2020kgo}
or, perhaps, if some other top-down principle is in action. In what
follows, to be concrete, we adopt three simplifying assumptions:
\begin{itemize}
\item The new terms in $K$ are quadratic in $Y^{(2)}$. This is sufficient
to illustrate our results. 
\item The minimal form (up to overall normalisation) is restored at $\text{Im}\tau\to\infty$.
This assumption is inspired by the minimal form of the Kähler potential
arising in certain string theory compactifications in the large volume
limit, corresponding to $\text{Im}\tau\to\infty$ (see, e.g.\ \cite{Chen:2019ewa,Asaka:2020tmo}
and references therein). 
\item The diagonal entries, already controlled by the minimal Kähler potential,
are not affected by the new terms. 
\end{itemize}
Under these assumptions and with the assignment of representations
and weights given in Table~\ref{tab:MODULAR-model_1}, we find~\footnote{We present the full expressions for $K_{L}$ and $K_{E^{c}}$ quadratic
in $Y^{(2)}$ in Appendix~\ref{app:Kahler}.}:
\begin{equation}
K=L^{\dagger}K_{L}L+E^{c\dagger}K_{E^{c}}E^{c}\,,\label{eq:Kahler potential}
\end{equation}
where 
\begin{equation}
K_{L}=\frac{1}{2\text{Im}\tau}\begin{pmatrix}1 & 0 & 0\\
0 & 1 & 0\\
0 & 0 & 1
\end{pmatrix}+2\text{Im}\tau\begin{pmatrix}0 & \left(\alpha_{5}+i\alpha_{6}\right)X_{12} & \left(\alpha_{5}-i\alpha_{6}\right)X_{13}\\
\left(\alpha_{5}-i\alpha_{6}\right)X_{12}^{\ast} & 0 & \left(\alpha_{5}+i\alpha_{6}\right)X_{23}\\
\left(\alpha_{5}+i\alpha_{6}\right)X_{13}^{\ast} & \left(\alpha_{5}-i\alpha_{6}\right)X_{23}^{\ast} & 0
\end{pmatrix}.\label{eq:KL_two_param}
\end{equation}
Here $\alpha_{5}$ and $\alpha_{6}$ are real coefficients and
\begin{equation}
X_{12}=Y_{1}^{\ast}Y_{2}-Y_{1}Y_{2}^{\ast}\,,\qquad X_{13}=Y_{1}^{\ast}Y_{3}-Y_{1}Y_{3}^{\ast}\,,\qquad X_{23}=Y_{2}^{\ast}Y_{3}-Y_{2}Y_{3}^{\ast}\,.\label{eq:Xij}
\end{equation}
In the $E^{c}$ sector, we obtain:
\begin{equation}
K_{E^{c}}=\frac{1}{8\left(\text{Im}\tau\right)^{3}}\begin{pmatrix}1 & 0 & 0\\
0 & 1 & 0\\
0 & 0 & 1
\end{pmatrix}+\frac{1}{2\text{Im}\tau}\begin{pmatrix}0 & 0 & c_{13}X\\
0 & 0 & c_{23}X\\
c_{13}^{\ast}X^{\ast} & c_{23}^{\ast}X^{\ast} & 0
\end{pmatrix},\label{eq:KEc_two_param}
\end{equation}
where $c_{13}$ and $c_{23}$ are complex coefficients and
\begin{equation}
X=Y_{1}^{\ast}Y_{1}+\omega Y_{2}^{\ast}Y_{2}+\omega^{2}Y_{3}^{\ast}Y_{3}\sim\boldsymbol{1}^{\prime\prime}\,.\label{eq:X}
\end{equation}
Noteworthy, the seesaw formula~(\ref{eq:MODULAR-SeesawFormula})
does not depend on the renormalisation factor $z_{N^{c}}^{-1/2}$
of the heavy fields $N^{c}$, so that we will not need to specify
the Kähler metric of $N^{c}$ in what follows.
\begin{table}[t]
\centering{}{\footnotesize{}}%
\begin{tabular}[t]{|c|c|}
\hline 
\multicolumn{2}{|c|}{\texttt{\footnotesize{}Input parameters}}\tabularnewline
\hline 
\hline 
{\footnotesize{}$\text{Re}\tau$} & {\footnotesize{}$\pm0.0235$}\tabularnewline
{\footnotesize{}$\text{Im}\tau$} & {\footnotesize{}$1.080$}\tabularnewline
{\footnotesize{}$\beta/\alpha$} & {\footnotesize{}$0.1459$}\tabularnewline
{\footnotesize{}$\gamma/\alpha$} & {\footnotesize{}$5.955$}\tabularnewline
{\footnotesize{}$\text{Re}g$} & {\footnotesize{}$-0.1494$}\tabularnewline
{\footnotesize{}$\text{Im}g$} & {\footnotesize{}$\mp0.3169$}\tabularnewline
{\footnotesize{}$\alpha_{5}$} & {\footnotesize{}$-0.2071$}\tabularnewline
{\footnotesize{}$\alpha_{6}$} & {\footnotesize{}$-0.1437$}\tabularnewline
{\footnotesize{}$c_{13}$} & {\footnotesize{}$-0.2656$}\tabularnewline
{\footnotesize{}$c_{23}$} & {\footnotesize{}$0.0145$}\tabularnewline
{\footnotesize{}$v_{u}^{2}\kappa^{2}/\Lambda$~{[}eV{]}} & {\footnotesize{}$0.0189$}\tabularnewline
{\footnotesize{}$|\epsilon|\approx|\tau-i|$} & {\footnotesize{}$0.0830$}\tabularnewline
\hline 
\end{tabular}{\footnotesize{}\hspace{0.2cm} }%
\begin{tabular}[t]{|c|c|}
\hline 
\multicolumn{2}{|c|}{\texttt{\footnotesize{}Observables}}\tabularnewline
\hline 
\hline 
{\footnotesize{}$m_{e}/m_{\mu}$} & {\footnotesize{}$0$}\tabularnewline
{\footnotesize{}$m_{\mu}/m_{\tau}$} & {\footnotesize{}$0.0565$}\tabularnewline
{\footnotesize{}$r$} & {\footnotesize{}$0.0299$}\tabularnewline
{\footnotesize{}$\sin^{2}\theta_{12}$} & {\footnotesize{}$0.304$}\tabularnewline
{\footnotesize{}$\sin^{2}\theta_{13}$} & {\footnotesize{}$0.02219$}\tabularnewline
{\footnotesize{}$\sin^{2}\theta_{23}$} & {\footnotesize{}$0.573$}\tabularnewline
{\footnotesize{}$\delta m^{2}~[10^{-5}~\text{eV}^{2}]$} & {\footnotesize{}$7.42$}\tabularnewline
{\footnotesize{}$\Delta m^{2}~[10^{-3}~\text{eV}^{2}]$} & {\footnotesize{}$2.480$}\tabularnewline
\hline 
\end{tabular}{\footnotesize{}\hspace{0.2cm} }%
\begin{tabular}[t]{|c|c|}
\hline 
\multicolumn{2}{|c|}{\texttt{\footnotesize{}Predictions}}\tabularnewline
\hline 
\hline 
{\footnotesize{}$m_{1}$~{[}eV{]}} & {\footnotesize{}$0.0062$}\tabularnewline
{\footnotesize{}$m_{2}$~{[}eV{]}} & {\footnotesize{}$0.0106$}\tabularnewline
{\footnotesize{}$m_{3}$~{[}eV{]}} & {\footnotesize{}$0.0506$}\tabularnewline
{\footnotesize{}$\delta/\pi$} & {\footnotesize{}$\pm0.92$}\tabularnewline
{\footnotesize{}$\alpha_{21}/\pi$} & {\footnotesize{}$\pm0.97$}\tabularnewline
{\footnotesize{}$\alpha_{31}/\pi$} & {\footnotesize{}$\pm0.93$}\tabularnewline
{\footnotesize{}$|m_{ee}|$~{[}eV{]}} & {\footnotesize{}0}\tabularnewline
{\footnotesize{}$\sum_{i}m_{i}$~{[}eV{]}} & {\footnotesize{}0.0673}\tabularnewline
{\footnotesize{}Ordering} & {\footnotesize{}NO}\tabularnewline
{\footnotesize{}$M_{1}/\Lambda$} & {\footnotesize{}0.225}\tabularnewline
{\footnotesize{}$M_{2}/\Lambda$} & {\footnotesize{}2.298}\tabularnewline
{\footnotesize{}$M_{3}/\Lambda$} & {\footnotesize{}2.524}\tabularnewline
\hline 
\end{tabular}{\footnotesize{}\caption{{\footnotesize{}\label{tab:MODULAR-bfp_model_2}}First pair of best-fit
points in a vicinity of $\tau=i$ found considering the Kähler potential
in eqs.~(\ref{eq:KL_two_param})--(\ref{eq:X}).}
}
\end{table}

\begin{table}[t]
\centering{}{\footnotesize{}}%
\begin{tabular}[t]{|c|c|}
\hline 
\multicolumn{2}{|c|}{\texttt{\footnotesize{}Input parameters}}\tabularnewline
\hline 
\hline 
{\footnotesize{}$\text{Re}\tau$} & {\footnotesize{}$\pm0.0328$}\tabularnewline
{\footnotesize{}$\text{Im}\tau$} & {\footnotesize{}$1.137$}\tabularnewline
{\footnotesize{}$\beta/\alpha$} & {\footnotesize{}$0.2388$}\tabularnewline
{\footnotesize{}$\gamma/\alpha$} & {\footnotesize{}$7.854$}\tabularnewline
{\footnotesize{}$\text{Re}g$} & {\footnotesize{}$-0.2234$}\tabularnewline
{\footnotesize{}$\text{Im}g$} & {\footnotesize{}$\pm0.4469$}\tabularnewline
{\footnotesize{}$\alpha_{5}$} & {\footnotesize{}$-0.1865$}\tabularnewline
{\footnotesize{}$\alpha_{6}$} & {\footnotesize{}$-0.1116$}\tabularnewline
{\footnotesize{}$c_{13}$} & {\footnotesize{}$-0.2405$}\tabularnewline
{\footnotesize{}$c_{23}$} & {\footnotesize{}$-0.0959$}\tabularnewline
{\footnotesize{}$v_{u}^{2}\kappa^{2}/\Lambda$~{[}eV{]}} & {\footnotesize{}$0.0191$}\tabularnewline
{\footnotesize{}$|\epsilon|\approx|\tau-i|$} & {\footnotesize{}$0.1408$}\tabularnewline
\hline 
\end{tabular}{\footnotesize{}\hspace{0.2cm} }%
\begin{tabular}[t]{|c|c|}
\hline 
\multicolumn{2}{|c|}{\texttt{\footnotesize{}Observables}}\tabularnewline
\hline 
\hline 
{\footnotesize{}$m_{e}/m_{\mu}$} & {\footnotesize{}$0$}\tabularnewline
{\footnotesize{}$m_{\mu}/m_{\tau}$} & {\footnotesize{}$0.0565$}\tabularnewline
{\footnotesize{}$r$} & {\footnotesize{}$0.0299$}\tabularnewline
{\footnotesize{}$\sin^{2}\theta_{12}$} & {\footnotesize{}$0.304$}\tabularnewline
{\footnotesize{}$\sin^{2}\theta_{13}$} & {\footnotesize{}$0.02219$}\tabularnewline
{\footnotesize{}$\sin^{2}\theta_{23}$} & {\footnotesize{}$0.573$}\tabularnewline
{\footnotesize{}$\delta m^{2}~[10^{-5}~\text{eV}^{2}]$} & {\footnotesize{}$7.42$}\tabularnewline
{\footnotesize{}$\Delta m^{2}~[10^{-3}~\text{eV}^{2}]$} & {\footnotesize{}$2.480$}\tabularnewline
\hline 
\end{tabular}{\footnotesize{}\hspace{0.2cm} }%
\begin{tabular}[t]{|c|c|}
\hline 
\multicolumn{2}{|c|}{\texttt{\footnotesize{}Predictions}}\tabularnewline
\hline 
\hline 
{\footnotesize{}$m_{1}$~{[}eV{]}} & {\footnotesize{}$0.0063$}\tabularnewline
{\footnotesize{}$m_{2}$~{[}eV{]}} & {\footnotesize{}$0.0107$}\tabularnewline
{\footnotesize{}$m_{3}$~{[}eV{]}} & {\footnotesize{}$0.0506$}\tabularnewline
{\footnotesize{}$\delta/\pi$} & {\footnotesize{}$\pm0.91$}\tabularnewline
{\footnotesize{}$\alpha_{21}/\pi$} & {\footnotesize{}$\pm0.98$}\tabularnewline
{\footnotesize{}$\alpha_{31}/\pi$} & {\footnotesize{}$\pm0.88$}\tabularnewline
{\footnotesize{}$|m_{ee}|$~{[}eV{]}} & {\footnotesize{}0}\tabularnewline
{\footnotesize{}$\sum_{i}m_{i}$~{[}eV{]}} & {\footnotesize{}0.0675}\tabularnewline
{\footnotesize{}Ordering} & {\footnotesize{}NO}\tabularnewline
{\footnotesize{}$M_{1}/\Lambda$} & {\footnotesize{}0.353}\tabularnewline
{\footnotesize{}$M_{2}/\Lambda$} & {\footnotesize{}2.130}\tabularnewline
{\footnotesize{}$M_{3}/\Lambda$} & {\footnotesize{}2.483}\tabularnewline
\hline 
\end{tabular}{\footnotesize{}\caption{{\footnotesize{}\label{tab:MODULAR-bfp2_model_2}}Second pair of best-fit
points in a vicinity of $\tau=i$ found considering the Kähler potential
in eqs.~(\ref{eq:KL_two_param})--(\ref{eq:X}).}
}
\end{table}

The inclusion of a non-minimal Kähler potential, even within the above
restrictive assumptions, brings in several additional free parameters:
$\alpha_{5,6}$, ${\tt Re}(c_{13,23})$ and ${\tt Im}(c_{13,23})$~\footnote{In our numerical analysis, we have set ${\tt Im}(c_{13,23})=0$.}.
Adding them to $\beta/\alpha$, $\gamma/\alpha$, ${\tt Re}(g)$,
${\tt Im}(g)$, ${\tt Re}(\tau)$ and ${\tt Im}(\tau)$, we have a
total of 12 dimensionless input parameters, more than the number of
observables. Thus the focus of our analysis cannot be on predictability.
Rather, we are interested in accounting for the mass hierarchies in
terms of the $Z_{4}$ parameter $\epsilon$, in the context of a model
reproducing all lepton masses and mixings. While the mass hierarchies
alone can be easily accommodated without the need of hierarchical
Lagrangian parameters, some degree of hierarchy turns out to be required
by the need to fix the mixing parameters. Useful parameters to estimate
such hierarchies in the Kähler potential are the condition numbers
of eq.~(\ref{Eq:MODULAR-condnum}). To establish the possibility
to reproduce all the relevant observables, and the role of $Z_{4}$
breaking in setting the mass hierarchies, we have selected several
benchmark points with slightly different features. We show the results
of two (pairs) of the benchmark points in Tables~\ref{tab:MODULAR-bfp_model_2}
and \ref{tab:MODULAR-bfp2_model_2}. In all such benchmark points,
all five dimensionless observables take exactly their experimental
best-fit values (for the time being we set $m_{e}=0$). In addition,
the model predicts a normal ordered neutrino mass spectrum and the
values of the CPV phases. Interestingly, for both pairs of the benchmark
points, the predicted value of $\delta$ (the one with minus sign)
matches its experimental best-fit value. Notice also the interesting
result $\vert m_{ee}\vert=0$ which at the leading order can be seen
as a simple consequence of the matrix patterns~(\ref{eq:seesaw at LO})
and (\ref{eq:MODULAR-me1})~\footnote{Given the column ordering in eq.~(\ref{eq:MODULAR-me1}), $\vert m_{ee}\vert$
is given at the leading order by $\big(m_{\nu}^{(0)}\big)_{33}=0+O(\epsilon)$.}. Finally, we report in the last column the masses $M_{i}$, $i=1,2,3$,
of the heavy neutrinos in the units of $\Lambda$. Although the value
of $\Lambda$ cannot be uniquely fixed, it can be estimated as (see
the first column of the tables) $\Lambda\approx v_{u}^{2}\kappa^{2}/(0.02~\text{eV})\approx10^{15}\sin^{2}\beta$~GeV,
where we have used $v_{u}=v\sin\beta$, with $v=174$~GeV, and $\kappa\sim\mathcal{O}(1)$.
This implies that for $\tan\beta\gtrsim1$, the scale $\Lambda\gtrsim5\times10^{14}$~GeV.
Let us stress once again that in the considered model, $M_{1}\sim\left|\epsilon\right|\Lambda$,
and thus, it is generated by a small departure of $\tau$ from $i$.

Our analysis shows that the mass hierarchies are indeed governed by
$Z_{4}$ breaking, whereas, in general, Kähler corrections reflect
on the lepton mass spectrum through $\mathcal{O}(1)$ changes. For
example, in the first pair of benchmark points (see Table~\ref{tab:MODULAR-bfp_model_2}),
we verified numerically that Kähler corrections only affect the mass
ratios by about a factor of 2~\footnote{For the minimal Kähler potential, \textit{i.e.} setting $\alpha_{5}=\alpha_{6}=c_{13}=c_{23}=0$,
we find $m_{\mu}/m_{\tau}=0.0520$ and $r=0.0637$, whereas the angles
$\sin^{2}\theta_{12}=0.228$, $\sin^{2}\theta_{13}=0.03751$ and $\sin^{2}\theta_{23}=0.256$
are far away from their experimental values.}; on the other hand, at these points, the Kähler metrics are by themselves
somewhat hierarchical, as shown by the condition numbers $\kappa(K_{L})\approx12$
and $\kappa(K_{E^{c}})\approx16$. In the second pair of benchmark
points shown in Table~\ref{tab:MODULAR-bfp2_model_2}, the hierarchies
in the Kähler metrics are both reduced (the condition numbers are
$\kappa(K_{L})\approx6$ and $\kappa(K_{E^{c}})\approx12$), and points
with even milder hierarchies may potentially be found. These observations
lead us to conclude that the deviations from the canonical Kähler
metric present in the best-fit points, have little to do with the
mass spectrum hierarchies; rather, they are necessary in order to
reproduce the correct PMNS mixing angles.
\begin{figure}[t]
\includegraphics[width=0.48\textwidth]{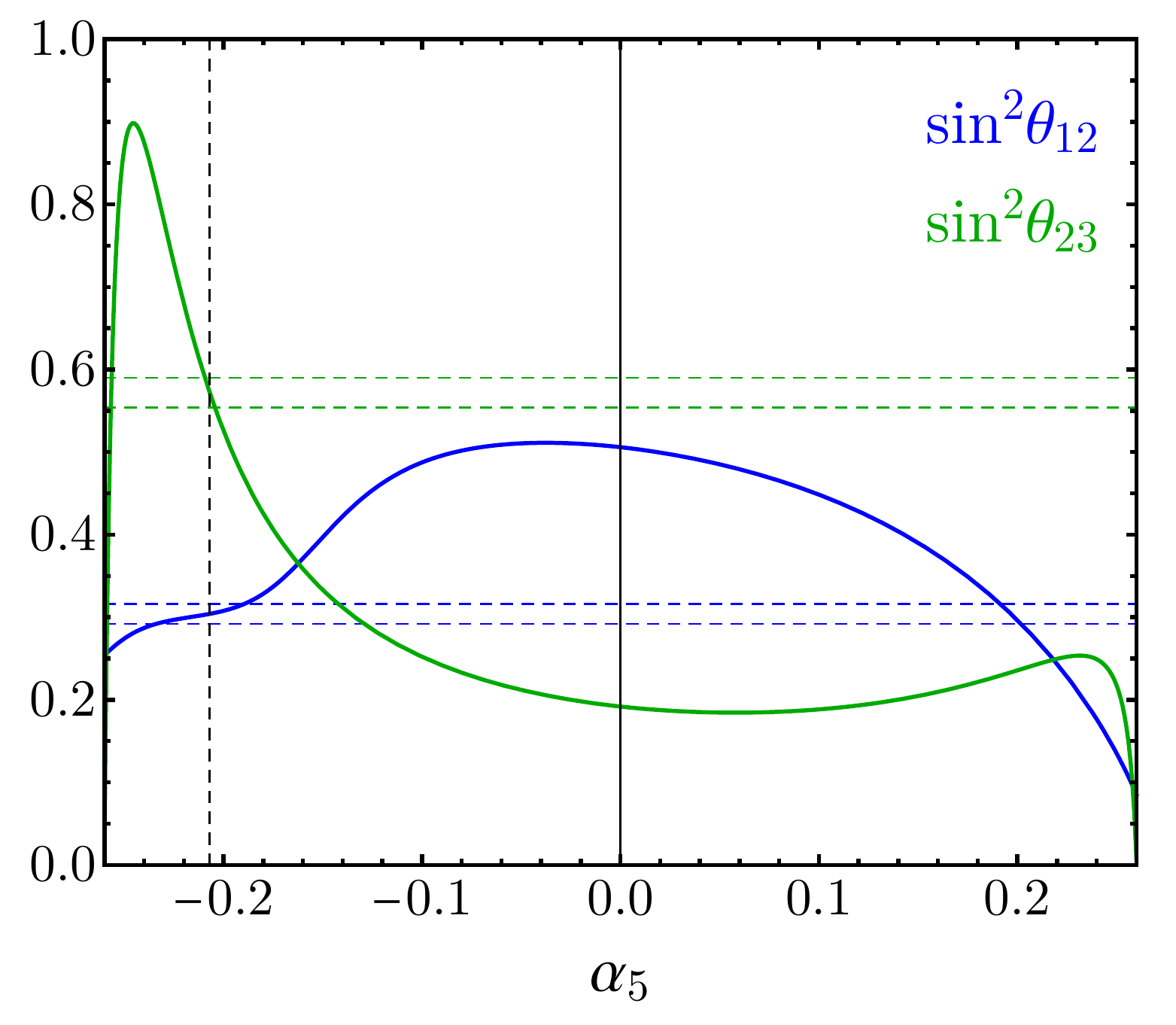} \hspace{0.02\textwidth}
\includegraphics[width=0.48\textwidth]{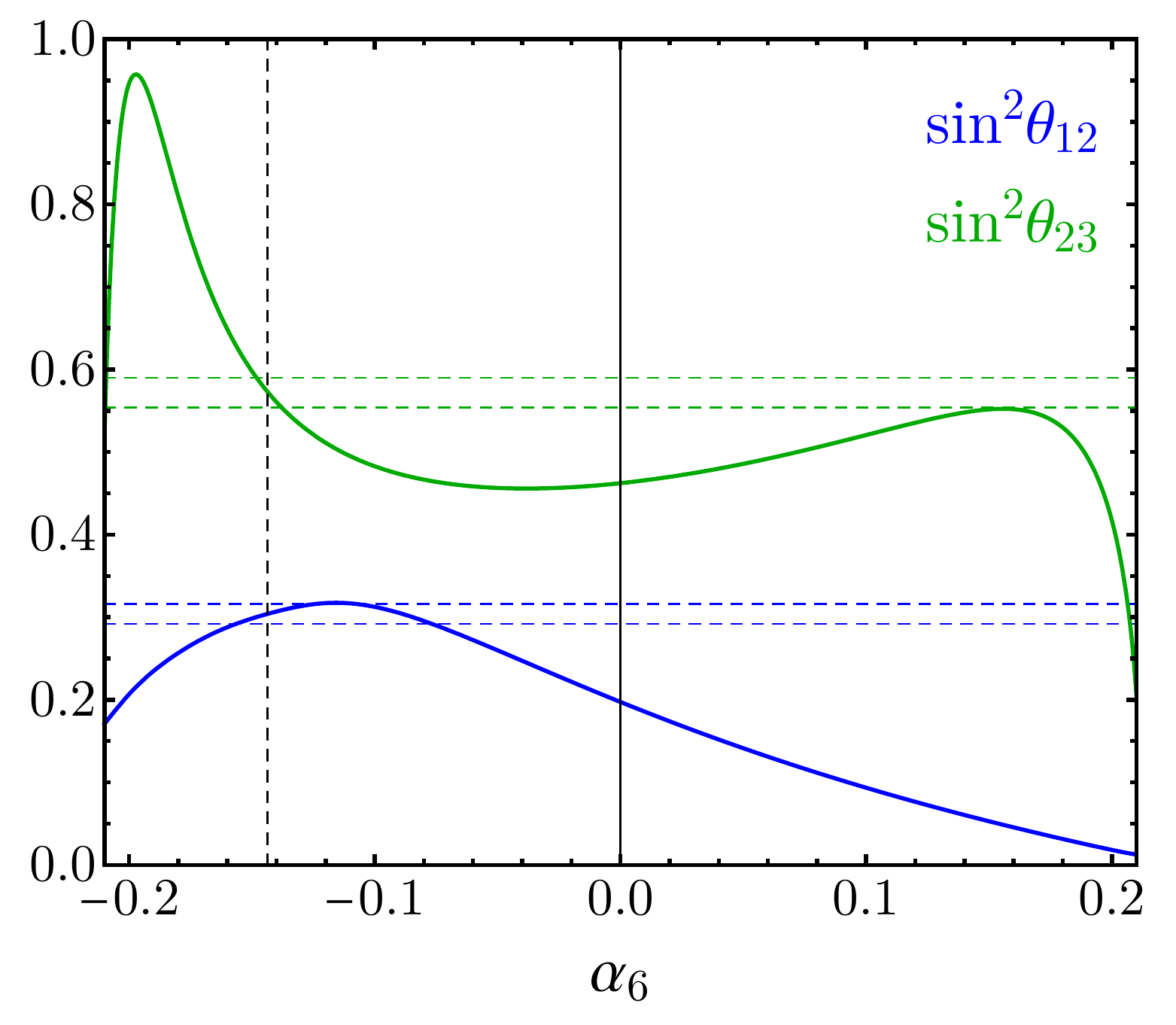} \includegraphics[width=0.48\textwidth]{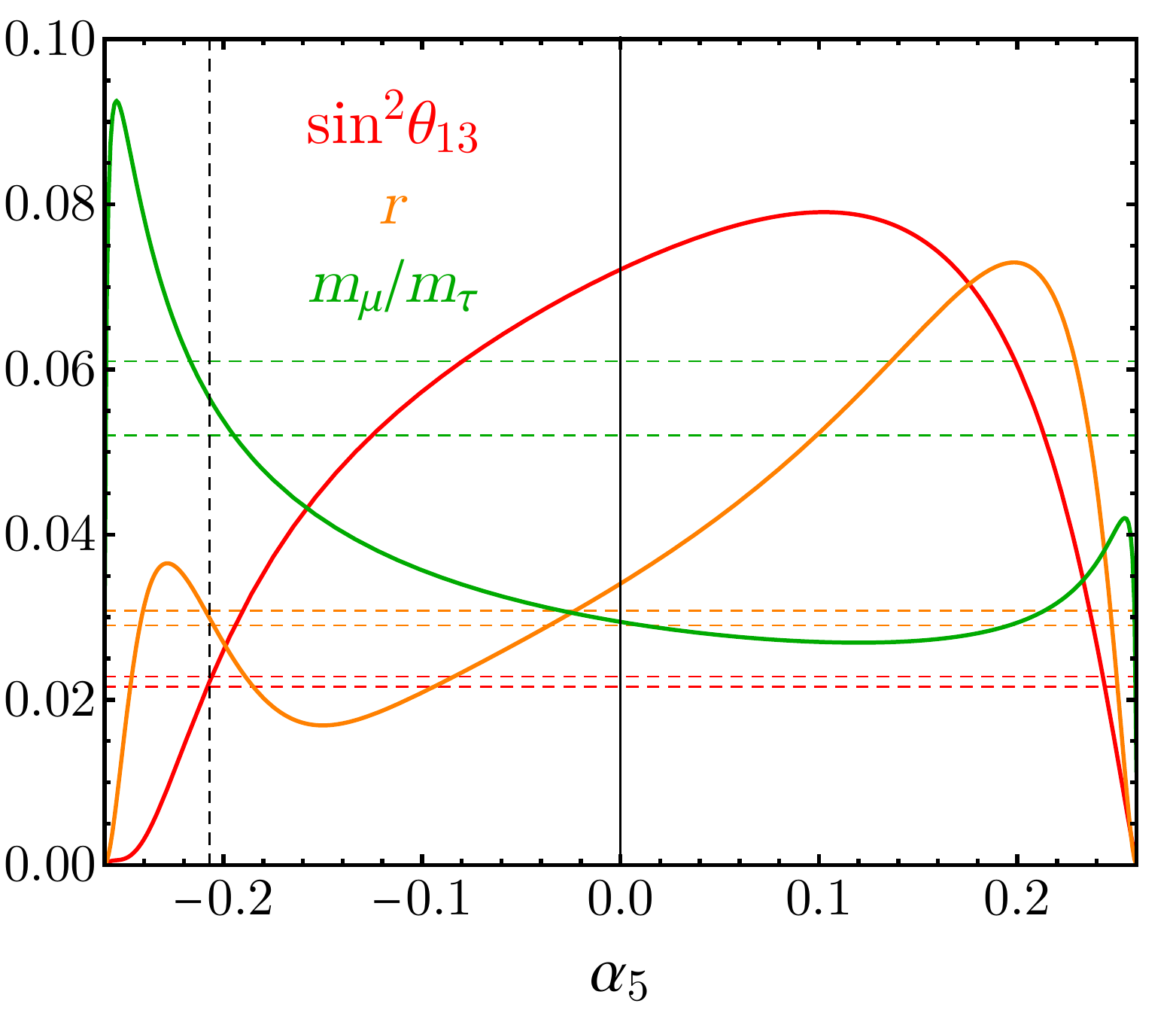}
\hspace{0.02\textwidth} \includegraphics[width=0.48\textwidth]{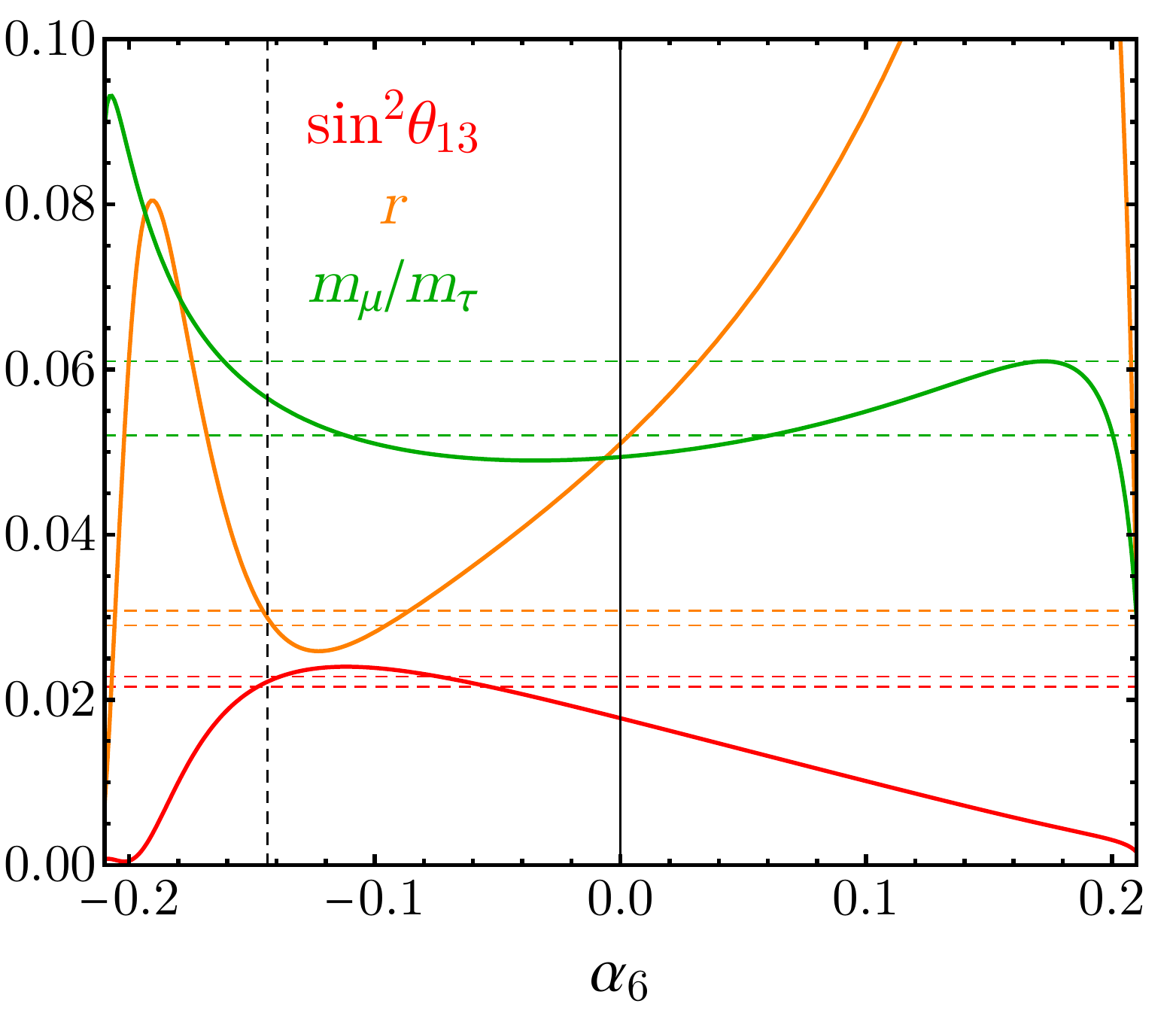}
\caption{Dependence of the mixing angles and two mass ratios on $\alpha_{5}$
(left) and $\alpha_{6}$ (right), fixing all other input parameters
to their best-fit values from Table~\ref{tab:MODULAR-bfp_model_2}.
The \textit{horizontal dashed lines} indicate the boundaries of the
respective $1\sigma$ ranges from Table~\ref{tab:MODULAR-Best-fit-values}.
The \textit{vertical dashed line} in the left (right) panels stands
for the best-fit value of $\alpha_{5}$ ($\alpha_{6}$) from Table~\ref{tab:MODULAR-bfp_model_2}.}
\label{fig:KL}
\end{figure}

\begin{figure}[t]
\includegraphics[width=0.48\textwidth]{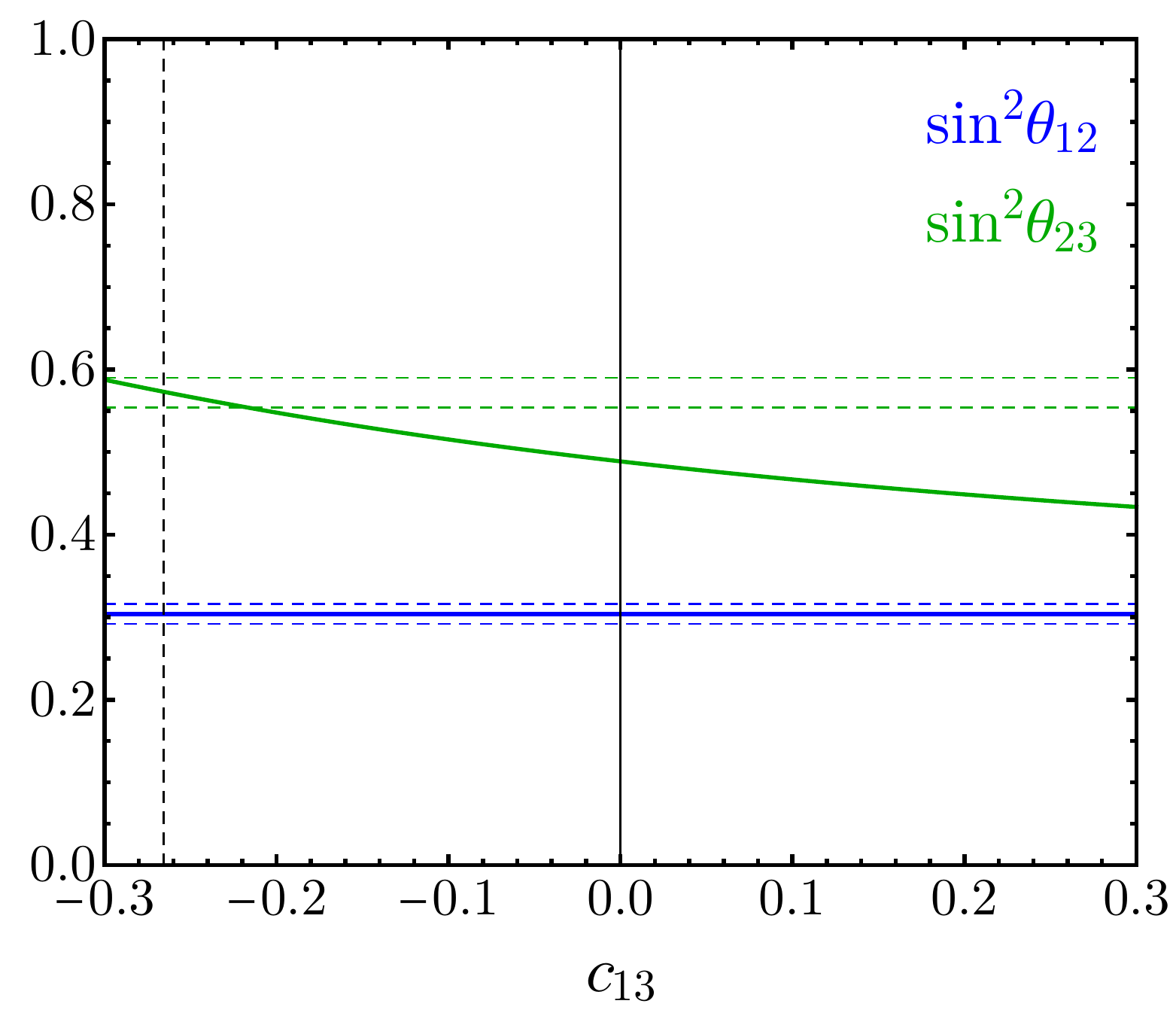} \hspace{0.02\textwidth}
\includegraphics[width=0.48\textwidth]{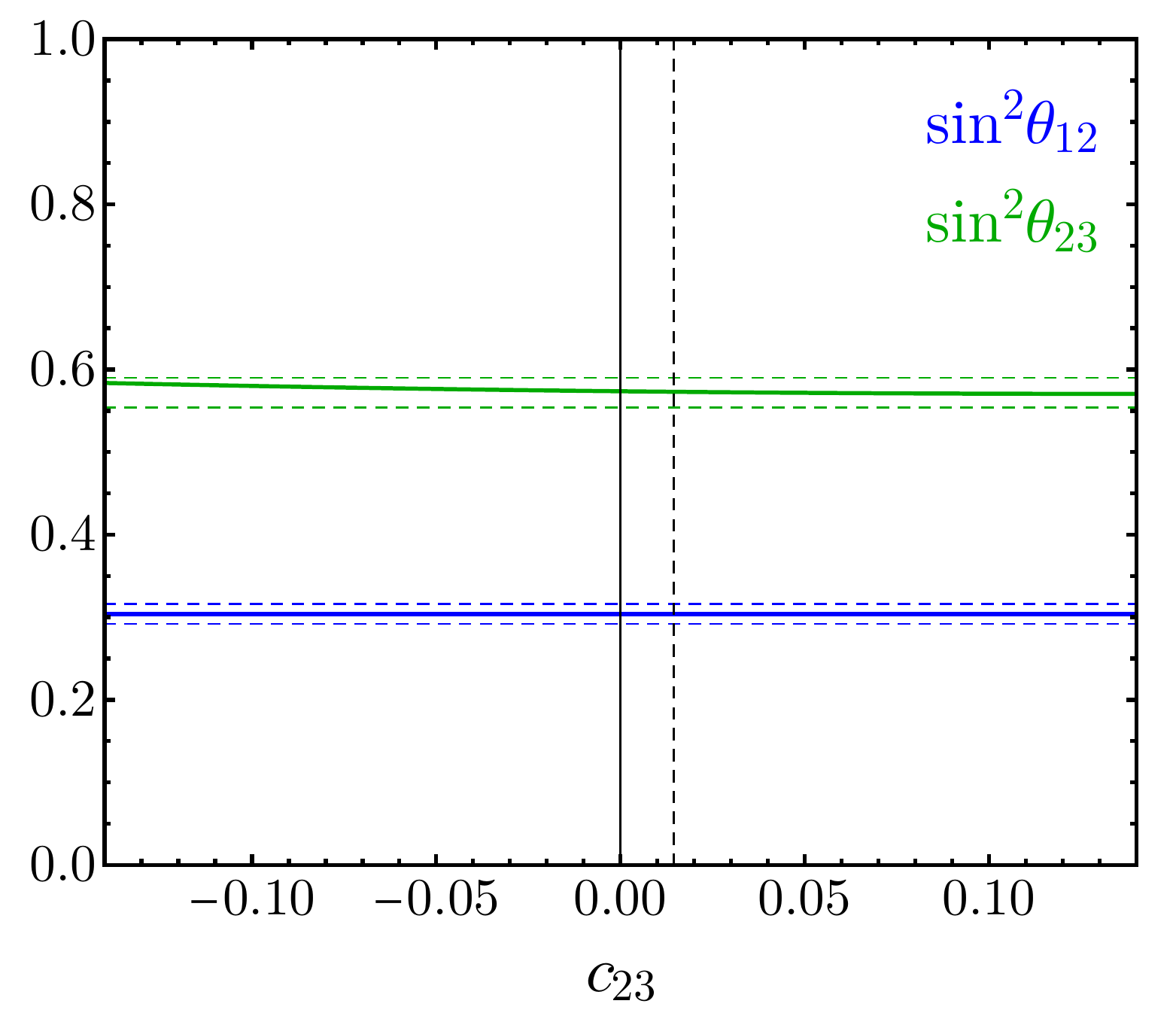} \includegraphics[width=0.48\textwidth]{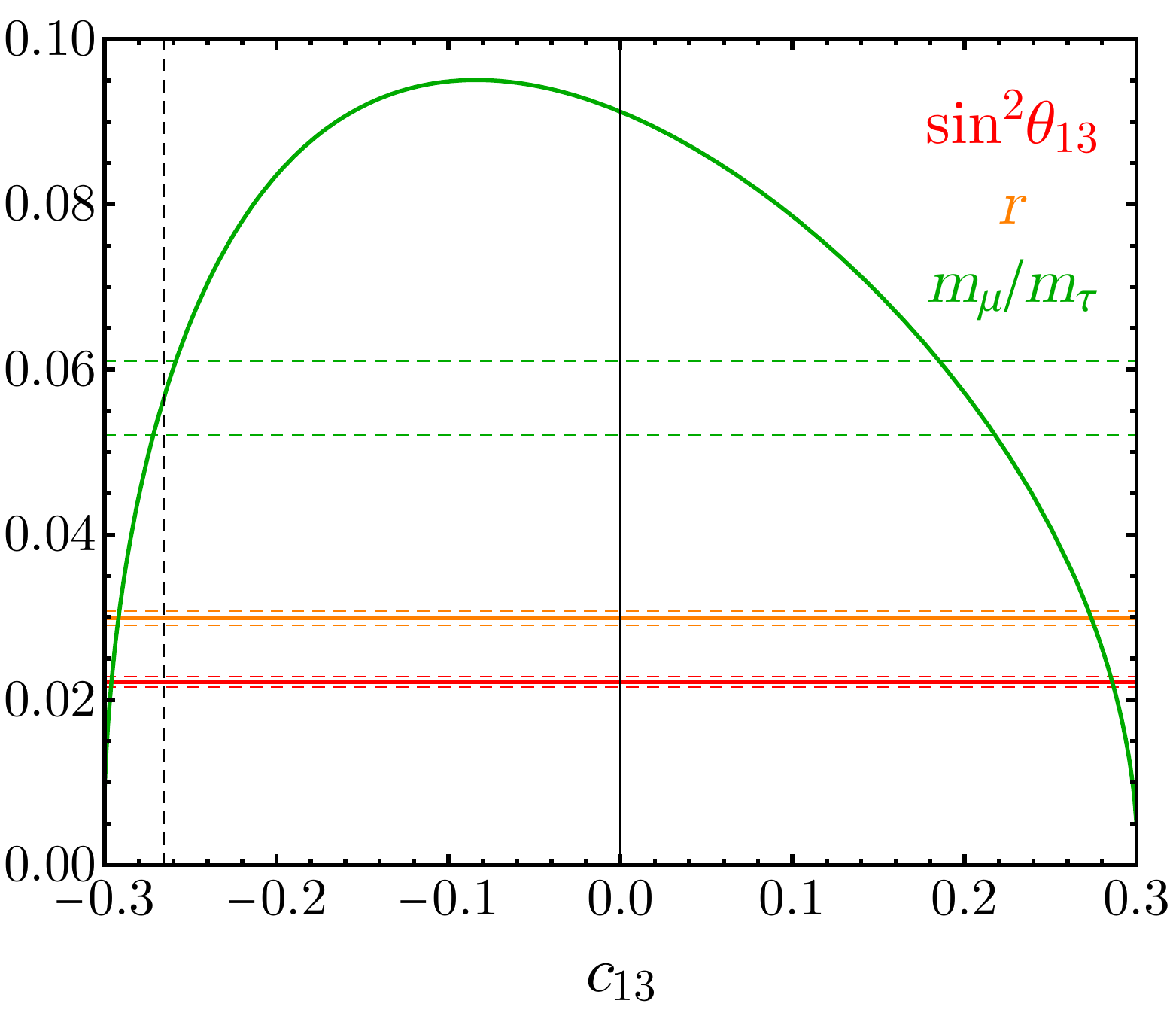}
\hspace{0.02\textwidth} \includegraphics[width=0.48\textwidth]{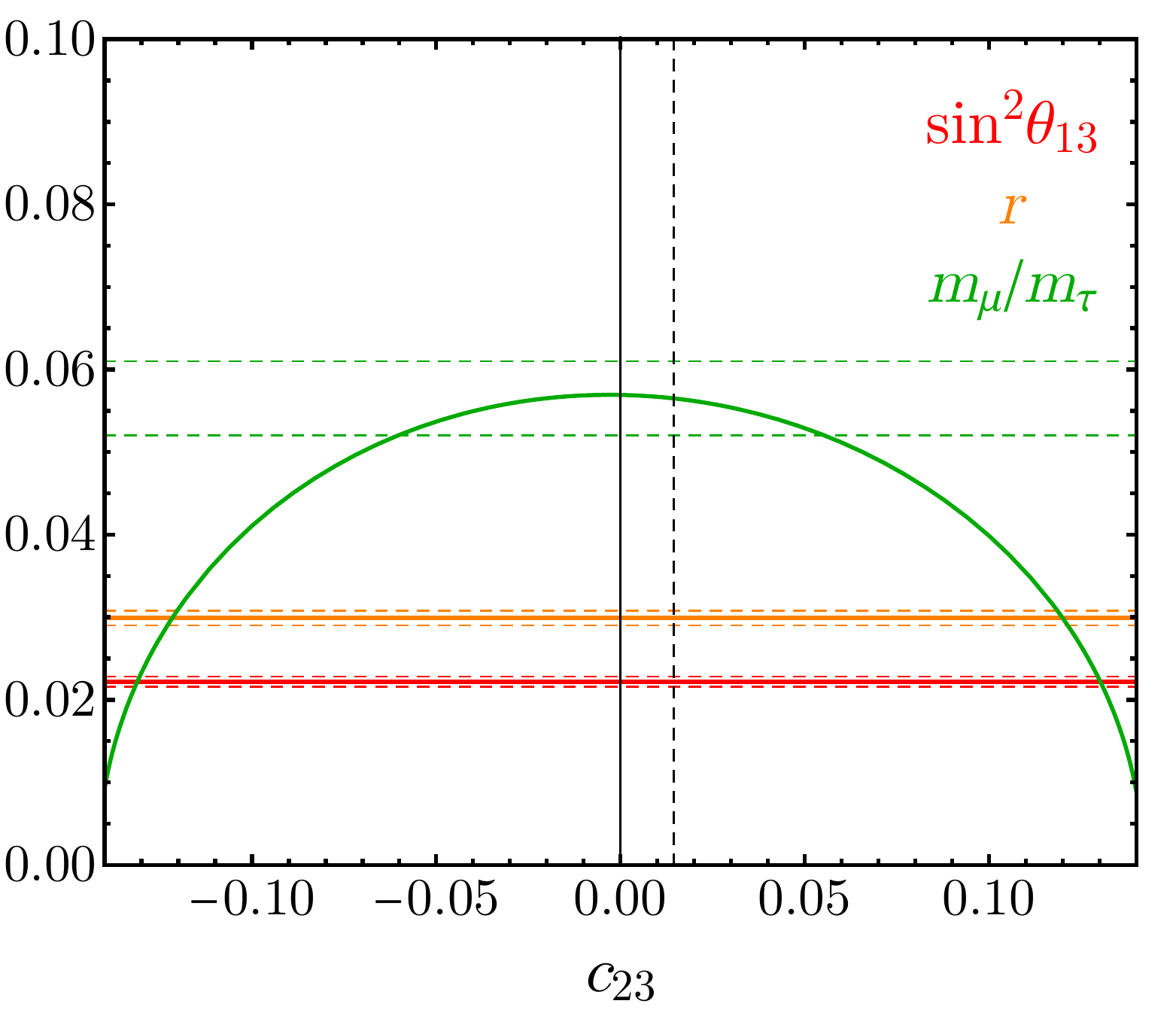}
\caption{Dependence of the mixing angles and two mass ratios on $c_{13}$ (left)
and $c_{23}$ (right), fixing all other input parameters to their
best-fit values from Table~\ref{tab:MODULAR-bfp_model_2}. The \textit{horizontal
dashed lines} indicate the boundaries of the respective $1\sigma$
ranges from Table~\ref{tab:MODULAR-Best-fit-values}. The \textit{vertical
dashed line} in the left (right) panels stands for the best-fit value
of $c_{13}$ ($c_{23}$) from Table~\ref{tab:MODULAR-bfp_model_2}.}
\label{fig:KEc}
\end{figure}

We have analysed more in detail the dependence of the fitted observables
on the parameters of the Kähler potential. In Figs.~\ref{fig:KL}
and \ref{fig:KEc}, we plot the values of the five dimensionless observables
versus $\alpha_{5,6}$ and $c_{13,23}$, respectively~\footnote{For the values of $\alpha_{5,6}$ ($c_{13,23}$) beyond the range
displayed in the $x$-axis, the matrix $K_{L}$ ($K_{E^{c}}$) is
not positive definite, and thus, the corresponding Kähler metric is
well-defined only for the displayed range of $\alpha_{5,6}$ ($c_{13,23}$).}. All other input parameters are fixed to their best-fit values as
in Table~\ref{tab:MODULAR-bfp_model_2}. We see that the parameters
$\alpha_{5}$ and $\alpha_{6}$ strongly impact the predictions for
the mixing angles and the two mass ratios, $r$ and $m_{\mu}/m_{\tau}$,
whereas $c_{13}$ and $c_{23}$ in $K_{E^{c}}$ mainly affect the
predictions for $\sin^{2}\theta_{23}$ and $m_{\mu}/m_{\tau}$.

In conclusion, we see that a mass matrix of reduced rank at the self-dual
point $\tau=i$ can explain the observed mass hierarchies in terms
of $\mathcal{O}(1)$ Lagrangian parameters. At the same time, at least
in the model considered here, moderately hierarchical Kähler and superpotential
parameters are needed to fix the mixing angle predictions. Whether
or not this is a general feature of this class of models is a question
which definitely requires further investigation, but is beyond the
scope of the present Chapter. On the other hand, in order to get a
fully realistic description of lepton masses we should still generate
a non-vanishing electron mass, without perturbing too much the results
achieved so far. We discuss this point in the next subsection.

\subsection{Generating $m_{e}\protect\neq0$}

\label{subsec:generating_me}Both the models discussed above yield,
by construction, $m_{e}=0$. One can easily concoct mechanisms to
generate the small electron mass without spoiling the other predictions.
We give below two examples, where $m_{e}$ is generated by supersymmetry
breaking and by dimension six operators, respectively.

If supersymmetry is broken by some $F$-term, fermion masses get corrected
by the second term of eq.~(\ref{eq:MODULAR-FermionMassMatFull}),
which, as discussed below the same equation, scales as $m_{\mathrm{SUSY}}\,v/M$
for SM fermion masses (where $M$ is the SUSY breaking messenger scale).
For instance, a Kähler interaction of the form: 
\[
K\supset\frac{1}{\Lambda^{2}}\chi^{\dagger}E_{i}^{c}\left[a_{i}(\tau)H_{d}+b_{i}(\tau)\widetilde{H}_{u}\right]L+\text{h.c.},
\]
where the superfield $\chi$ gets a supersymmetry breaking expectation
value $\langle\chi\rangle=F\theta^{2}$, gives a contribution to the
charged lepton Yukawa matrix proportional to $F/\Lambda^{2}$, which
in turn generically induces an electron Yukawa coupling of the same
order.

As a second possibility, one may generate $m_{e}\neq0$ through the
dimension six operator: 
\begin{equation}
(E_{i}^{c}LH_{d})(H_{u}H_{d}),\label{eq:MODULAR-dim6 superpotential}
\end{equation}
whose Wilson coefficient should be a modular form of the appropriate
weight. In order for this mechanism to work, we need to generalise
the weight assignments in Table~\ref{tab:MODULAR-model_1}. We make
the following requirements: 
\begin{align}
k_{L} & =1-k_{u}\,,\label{eq:generalised_weights}\\
k_{E^{c}} & =3+k_{u}-k_{d}\,,\\
k_{u}+k_{d} & \neq0\,.
\end{align}
The first two conditions ensure that the superpotentials discussed
in the previous subsections have weight zero; the last condition implies
that the operator (\ref{eq:MODULAR-dim6 superpotential}) has different
weight from the corresponding renormalisable Yukawa term~\footnote{Curiously, the same condition can be exploited to make the Higgs $\mu$-term
vanish at $\tau=i$. The Higgs $\mu$-term, being a $\Gamma_{3}$
singlet modular form of weight $k_{u}+k_{d}$, vanishes by ~eq.~(\ref{eq:MODULAR-Z4constraints})
at $\tau=i$ if $k_{u}+k_{d}\neq0~(\texttt{mod}~4)$, since all $\Gamma_{3}$
singlets have $\rho(\widetilde{S})=1$.}, so that it couples to a functionally independent modular form multiplet
(making the resulting charged lepton mass matrix of rank three). Such
a mechanism thus generates $m_{e}\sim v_{u}v_{d}^{2}/\Lambda^{2}$,
where $\Lambda$ is the scale at which the operator in eq.~(\ref{eq:MODULAR-dim6 superpotential})
is generated.

While in some flavour models, the ratios $m_{e}/m_{\tau}$ and $m_{\mu}/m_{\tau}$
are associated to different powers of the same expansion parameter,
we note that here the two ratios are associated to independent parameters.

\section{Conclusions}

\label{S5} Supersymmetric modular invariant theories offer an attractive
framework to address the flavour puzzle. The role of flavour symmetry
is played by modular invariance, regarded as a discrete gauge symmetry,
thus circumventing the obstruction concerning fundamental global symmetries.
The arbitrary symmetry breaking sector of the conventional models
based on flavour symmetries is replaced by the moduli space. Yukawa
couplings become modular forms, severely restricted by the matter
transformation properties. So far this framework has delivered interesting
preliminary results especially in the lepton sector, where neutrino
masses and lepton mixing parameters can be efficiently described in
terms of a limited number of input parameters.

Weak points in most of the existing constructions are the need of
independent hierarchical parameters to describe charged lepton masses,
the reduced predictability caused by a general form of the Kähler
potential, and the absence of a reliable dynamical mechanism to determine
the value of $\tau$ in the vacuum. As a matter of fact, in several
models reproducing lepton masses and mixing parameters, the required
value of $\tau$ is close to $i$, the self-dual point where the generator
$S$ of the modular group and CP (if the Lagrangian is CP invariant)
are unbroken. A small departure of $\tau$ from $i$ suffices to generate
sizeable CP-violating effects in the lepton sector.

For these reasons, we were led to analyse more in detail the vicinity
of $\tau=i$. Our goal was to show that a small deviation from the
self-dual point can be responsible for the observed mass hierarchy
$m_{e}\ll m_{\mu}\ll m_{\tau}$ and $\delta m^{2}\ll\Delta m^{2}$.
At $\tau=i$, the theory has an exact $Z_{4}$ symmetry, generated
by the element $S$ of the modular group. In the neighbourhood of
$\tau=i$, the breaking of $Z_{4}$ can be fully described by the
(small) spurion $\epsilon\approx\tau-i$, that flips its sign under
$Z_{4}$. We explained how to exploit this residual $Z_{4}$ symmetry
in order to obtain lepton mass matrices having reduced rank at $\tau=i$.
This can be easily done with a suitable assignment of modular weights
and representations for matter fields. There is a twofold advantage
in this strategy. First, mass ratios that are forced to vanish at
$\tau=i$ by the $Z_{4}$ symmetry are expected to acquire small values
$\propto|\epsilon|^{n}$ $(n>0)$ near the self-dual point. Second,
the reduced rank of the mass matrices can tame the contribution from
a non-minimal Kähler potential, provided the metrics of the matter
fields do not display large hierarchies.

To see whether this strategy can be successfully realised or not,
we built a concrete model at level 3, where neutrinos get masses through
the type I seesaw mechanism. The model predicts a normal mass ordering.
The number of parameters exceeds the number of fitted observables
and we cannot claim predictability. However, with $\tau$ being near
$i$, mass ratios and mixing angles are reproduced with input parameters
nearly of the same order of magnitude and matter kinetic terms display
only a moderate hierarchy. We saw that the main contribution to the
mass hierarchy can be induced by the singular mass matrix at the $Z_{4}$
symmetric point. In the model we considered, the Kähler potential
and the other Lagrangian parameters are crucial in order to correctly
reproduce the values of the mixing angles. While the $Z_{4}$ symmetry
plays a fundamental role in all our discussion, we notice that our
model could not have been realised in the context of a $Z_{4}$ flavour
symmetry alone. In particular, the electron mass vanishes in the models
we have considered due to the correlations among generic $Z_{4}$-invariant
operators provided by the underlying modular invariance. Also the
leading order values of the mixing angles are dictated by $Z_{4}$.
We have discussed possible sources of a non-vanishing electron mass.
While the models we formulated have clearly room for improvement,
we consider them as a good starting point to naturally accommodate
the observed fermion mass hierarchies within a modular invariant framework.

\section{Appendix: Kähler potential quadratic in $Y^{(2)}$}

\label{app:Kahler} In the real basis for the $\Gamma_{3}$ generators
$S$ and $T$ in the 3-dimensional representation we employ in this
Chapter, $L^{\ast}$ and $Y^{(2)\ast}$ transform as triplets, i.e.\ $L^{\ast}\to\rho_{\boldsymbol{3}}(\tilde{\gamma})L^{\ast}$
and $Y^{(2)\ast}\to\rho_{\boldsymbol{3}}(\tilde{\gamma})Y^{(2)\ast}$.
Thus, we can contract first $L^{\ast}$ with $L$ and $Y^{(2)\ast}$
with $Y^{(2)}$, and after that perform contractions of the obtained
multiplets. Proceeding in this way, we obtain:
\begin{align}
\left(L^{\ast}L\right)_{\boldsymbol{1}} & =L^{\dagger}L\,,\qquad\left(L^{\ast}L\right)_{\boldsymbol{1}^{\prime}}=L^{\dagger}M_{\boldsymbol{1}^{\prime}}L\,,\qquad\left(L^{\ast}L\right)_{\boldsymbol{1}^{\prime\prime}}=L^{\dagger}M_{\boldsymbol{1}^{\prime\prime}}L\,,\\
\left(L^{\ast}L\right)_{\boldsymbol{3}_{S}} & =\begin{pmatrix}L^{\dagger}M_{\boldsymbol{3}_{S}}^{(1)}L\\
L^{\dagger}M_{\boldsymbol{3}_{S}}^{(2)}L\\
L^{\dagger}M_{\boldsymbol{3}_{S}}^{(3)}L
\end{pmatrix},\qquad\left(L^{\ast}L\right)_{\boldsymbol{3}_{A}}=\begin{pmatrix}L^{\dagger}M_{\boldsymbol{3}_{A}}^{(1)}L\\
L^{\dagger}M_{\boldsymbol{3}_{A}}^{(2)}L\\
L^{\dagger}M_{\boldsymbol{3}_{A}}^{(3)}L
\end{pmatrix},
\end{align}
with the matrices $M_{\bullet}$ being:
\begin{align}
M_{\boldsymbol{1}^{\prime}} & =\begin{pmatrix}1 & 0 & 0\\
0 & \omega^{2} & 0\\
0 & 0 & \omega
\end{pmatrix},\qquad M_{\boldsymbol{1}^{\prime\prime}}=\begin{pmatrix}1 & 0 & 0\\
0 & \omega & 0\\
0 & 0 & \omega^{2}
\end{pmatrix},\label{eq:M1p_M1pp}\\
M_{\boldsymbol{3}_{S}}^{(1)} & =\begin{pmatrix}0 & 0 & 0\\
0 & 0 & 1\\
0 & 1 & 0
\end{pmatrix},\qquad M_{\boldsymbol{3}_{S}}^{(2)}=\begin{pmatrix}0 & 0 & 1\\
0 & 0 & 0\\
1 & 0 & 0
\end{pmatrix},\qquad M_{\boldsymbol{3}_{S}}^{(3)}=\begin{pmatrix}0 & 1 & 0\\
1 & 0 & 0\\
0 & 0 & 0
\end{pmatrix},\\
M_{\boldsymbol{3}_{A}}^{(1)} & =\begin{pmatrix}0 & 0 & 0\\
0 & 0 & 1\\
0 & -1 & 0
\end{pmatrix},\qquad M_{\boldsymbol{3}_{A}}^{(2)}=\begin{pmatrix}0 & 0 & -1\\
0 & 0 & 0\\
1 & 0 & 0
\end{pmatrix},\qquad M_{\boldsymbol{3}_{A}}^{(3)}=\begin{pmatrix}0 & 1 & 0\\
-1 & 0 & 0\\
0 & 0 & 0
\end{pmatrix},
\end{align}
and $\omega=e^{2\pi i/3}$. The same equations hold for $\left(Y^{(2)\ast}Y^{(2)}\right)_{\bullet}$.
Taking further invariant contractions of the obtained multiplets,
we find 
\begin{align}
K_{L}=(2\text{Im}\tau)^{-k_{L}} & +\nonumber \\
+(2\text{Im}\tau)^{k_{Y}-k_{L}}\bigg\lbrace & \alpha_{1}Y^{(2)\dagger}Y^{(2)}\mathbb{1}+\alpha_{2}\left[\left(Y^{(2)\dagger}M_{\boldsymbol{1}^{\prime\prime}}Y^{(2)}\right)M_{\boldsymbol{1}^{\prime}}+(Y^{(2)\dagger}M_{\boldsymbol{1}^{\prime}}Y^{(2)})M_{\boldsymbol{1}^{\prime\prime}}\right]\nonumber \\
 & +\alpha_{3}\,i\left[\left(Y^{(2)\dagger}M_{\boldsymbol{1}^{\prime\prime}}Y^{(2)}\right)M_{\boldsymbol{1}^{\prime}}-\left(Y^{(2)\dagger}M_{\boldsymbol{1}^{\prime}}Y^{(2)}\right)M_{\boldsymbol{1}^{\prime\prime}}\right]\nonumber \\
 & +\alpha_{4}\sum_{n=1}^{3}\left(Y^{(2)\dagger}M_{\boldsymbol{3}_{S}}^{(n)}Y^{(2)}\right)M_{\boldsymbol{3}_{S}}^{(n)}+\alpha_{5}\sum_{n=1}^{3}\left(Y^{(2)\dagger}M_{\boldsymbol{3}_{A}}^{(n)}Y^{(2)}\right)M_{\boldsymbol{3}_{A}}^{(n)}\nonumber \\
 & +\alpha_{6}\sum_{n=1}^{3}i\left(Y^{(2)\dagger}M_{\boldsymbol{3}_{A}}^{(n)}Y^{(2)}\right)M_{\boldsymbol{3}_{S}}^{(n)}+\alpha_{7}\sum_{n=1}^{3}i\left(Y^{(2)\dagger}M_{\boldsymbol{3}_{S}}^{(n)}Y^{(2)}\right)M_{\boldsymbol{3}_{A}}^{(n)}\bigg\rbrace\,,\label{eq:KL}
\end{align}
where $\alpha_{j}$, $j=1\,,\dots\,,7$, are real coefficients, which
accompany hermitian matrices. (We have used the fact that $M_{\boldsymbol{1}^{\prime}}^{\dagger}=M_{\boldsymbol{1}^{\prime\prime}}$,
$M_{\boldsymbol{3}_{S}}^{(n)\dagger}=M_{\boldsymbol{3}_{S}}^{(n)}$,
and $M_{\boldsymbol{3}_{A}}^{(n)\dagger}=-M_{\boldsymbol{3}_{A}}^{(n)}$.)

One of our assumptions is that the canonical form of $K_{L}$ is restored
at $\text{Im}\tau\to\infty$. The $q$-expansions of $Y_{i}$ in the
real basis read: 
\begin{align}
Y_{1}(\tau) & =\frac{1}{\sqrt{3}}\left(1-6q^{1/3}-18q^{2/3}+12q+\dots\right),\\
Y_{2}(\tau) & =\frac{1}{\sqrt{3}}\left(1-6\omega q^{1/3}-18\omega^{2}q^{2/3}+12q+\dots\right),\\
Y_{3}(\tau) & =\frac{1}{\sqrt{3}}\left(1-6\omega^{2}q^{1/3}-18\omega q^{2/3}+12q+\dots\right),
\end{align}
where $q=e^{2\pi i\tau}$. Thus, at $\text{Im}\tau\to\infty$ 
\begin{equation}
Y_{1}=Y_{2}=Y_{3}\sim\frac{1}{\sqrt{3}}\,,
\end{equation}
and $K_{L}$ has the following form: 
\begin{equation}
K_{L}\sim\frac{1}{2\text{Im}\tau}+\frac{2}{3}\text{Im}\tau\begin{pmatrix}\frac{3}{2}\alpha_{1} & \alpha_{4}+i\alpha_{7} & \alpha_{4}-i\alpha_{7}\\
\alpha_{4}-i\alpha_{7} & \frac{3}{2}\alpha_{1} & \alpha_{4}+i\alpha_{7}\\
\alpha_{4}+i\alpha_{7} & \alpha_{4}-i\alpha_{7} & \frac{3}{2}\alpha_{1}
\end{pmatrix},\label{eq:KLinf}
\end{equation}
where we have used $k_{L}=1$ and $k_{Y}=2$. To satisfy our assumption
of the asymptotic behaviour of $K_{L}$, the coefficients $\alpha_{1}=\alpha_{4}=\alpha_{7}=0$.
Thus, the number of free parameters in $K_{L}$ is reduced from seven
to four. Then, the elements of $K_{L}$ from eq.~(\ref{eq:KL}) read:
\begin{align}
\left(K_{L}\right)_{11} & =\frac{1}{2\text{Im}\tau}+2\text{Im}\tau\left[2\alpha_{2}\left|Y_{1}\right|^{2}-\left(\alpha_{2}+\sqrt{3}\alpha_{3}\right)\left|Y_{2}\right|^{2}-\left(\alpha_{2}-\sqrt{3}\alpha_{3}\right)\left|Y_{3}\right|^{2}\right],\label{eq:KL11}\\
\left(K_{L}\right)_{22} & =\frac{1}{2\text{Im}\tau}+2\text{Im}\tau\left[-\left(\alpha_{2}-\sqrt{3}\alpha_{3}\right)\left|Y_{1}\right|^{2}+2\alpha_{2}\left|Y_{2}\right|^{2}-\left(\alpha_{2}+\sqrt{3}\alpha_{3}\right)\left|Y_{3}\right|^{2}\right],\\
\left(K_{L}\right)_{33} & =\frac{1}{2\text{Im}\tau}+2\text{Im}\tau\left[-\left(\alpha_{2}+\sqrt{3}\alpha_{3}\right)\left|Y_{1}\right|^{2}-\left(\alpha_{2}-\sqrt{3}\alpha_{3}\right)\left|Y_{2}\right|^{2}+2\alpha_{2}\left|Y_{3}\right|^{2}\right],\\
\left(K_{L}\right)_{12} & =2\text{Im}\tau\left(\alpha_{5}+i\alpha_{6}\right)\left[Y_{1}^{\ast}Y_{2}-Y_{1}Y_{2}^{\ast}\right],\\
\left(K_{L}\right)_{13} & =2\text{Im}\tau\left(\alpha_{5}-i\alpha_{6}\right)\left[Y_{1}^{\ast}Y_{3}-Y_{1}Y_{3}^{\ast}\right],\\
\left(K_{L}\right)_{23} & =2\text{Im}\tau\left(\alpha_{5}+i\alpha_{6}\right)\left[Y_{2}^{\ast}Y_{3}-Y_{2}Y_{3}^{\ast}\right].\label{eq:KL23}
\end{align}
For the sake of simplicity, we set further $\alpha_{2}=\alpha_{3}=0$.
In this case, the diagonal entries of $K_{L}$ are not affected by
the contributions containing modular forms, on the contrary to the
off-diagonal elements. Thereby, we arrive at the form of $K_{L}$
in eqs.~(\ref{eq:KL_two_param}) and (\ref{eq:Xij}).

What concerns $K_{E^{c}}$, with the assignment of representations
and weights given in Table~\ref{tab:MODULAR-model_1}, the most general
Kähler potential quadratic in $Y^{(2)}$ reads
\begin{align}
K_{E^{c}} & =\frac{1}{8\left(\text{Im}\tau\right)^{3}}\begin{pmatrix}c_{11}^{0} & c_{12}^{0} & 0\\
c_{12}^{0\ast} & c_{22}^{0} & 0\\
0 & 0 & c_{33}^{0}
\end{pmatrix}\nonumber \\
 & +\frac{1}{2\text{Im}\tau}\begin{pmatrix}c_{11}Y^{(2)\dagger}Y^{(2)} & c_{12}Y^{(2)\dagger}Y^{(2)} & c_{13}Y^{(2)\dagger}M_{\boldsymbol{1}^{\prime\prime}}Y^{(2)}\\
c_{12}^{\ast}Y^{(2)\dagger}Y^{(2)} & c_{22}Y^{(2)\dagger}Y^{(2)} & c_{23}Y^{(2)\dagger}M_{\boldsymbol{1}^{\prime\prime}}Y^{(2)}\\
c_{13}^{\ast}Y^{(2)\dagger}M_{\boldsymbol{1}^{\prime}}Y^{(2)} & c_{23}^{\ast}Y^{(2)\dagger}M_{\boldsymbol{1}^{\prime}}Y^{(2)} & c_{33}Y^{(2)\dagger}Y^{(2)}
\end{pmatrix}\,,
\end{align}
with $c_{ii}^{(0)}$ being real and $c_{ij}^{(0)}$, $i\neq j$, complex
coefficients.

Taking into account that at $\text{Im}\tau\to\infty$, the invariant
combination $Y^{(2)\dagger}Y^{(2)}\sim1$, whereas $X\equiv Y^{(2)\dagger}M_{\boldsymbol{1}^{\prime\prime}}Y^{(2)}$
and $X^{\ast}=Y^{(2)\dagger}M_{\boldsymbol{1}^{\prime}}Y^{(2)}$ decay
exponentially, we find 
\begin{equation}
K_{E^{c}}\sim\frac{1}{8\left(\text{Im}\tau\right)^{3}}\begin{pmatrix}c_{11}^{0} & c_{12}^{0} & 0\\
c_{12}^{0\ast} & c_{22}^{0} & 0\\
0 & 0 & c_{33}^{0}
\end{pmatrix}+\frac{1}{2\text{Im}\tau}\begin{pmatrix}c_{11} & c_{12} & 0\\
c_{12}^{\ast} & c_{22} & 0\\
0 & 0 & c_{33}
\end{pmatrix}.
\end{equation}
In order to restore the canonical form of $K_{E^{c}}$ in the considered
limit, $c_{12}^{0}=c_{12}=0$. Furthermore, we set $c_{ii}=0$ for
simplicity. Finally, we can always make $c_{ii}^{0}=1$ by independent
rescalings of $E_{i}^{c}$, $i=1,2,3$. Thus, we recover $K_{E^{c}}$
given by eqs.~(\ref{eq:KEc_two_param}) and (\ref{eq:X}).

\newpage{}

\bibliographystyle{Bib/JHEP}
\bibliography{Bib/Biblio1}

\end{document}